\documentclass[12pt, a4paper]{article}
\usepackage{amsfonts}
\usepackage{amsmath}
\usepackage{times}
\usepackage{graphics,xcolor}
\usepackage{amsfonts,slashed}
\usepackage{url}
\usepackage{latexsym}
\usepackage{epsfig}
\usepackage{latexsym,amssymb}
\usepackage{amsmath,amssymb,amsthm}
\usepackage{slashbox}
\usepackage{multirow}
\usepackage{colortbl}
\usepackage{color}
\usepackage{array}
\usepackage{float}
\usepackage{subfig}
\usepackage{ifpdf}
\ifx\pdfoutput\undefined
   \pdffalse
   \usepackage{cite}
 \else
   \pdfoutput=1
   \pdftrue
  \usepackage[pdftex]{hyperref}
\fi
\DeclareFontFamily{OMS}{rsfs}{\skewchar\font'60}
\DeclareFontShape{OMS}{rsfs}{m}{n}{<-5>rsfs5 <5-7>rsfs7 <7->rsfs10 }{}
\DeclareSymbolFont{rsfs}{OMS}{rsfs}{m}{n}

\theoremstyle{plain} 
\theoremstyle{plain} 
\theoremstyle{plain} 
\theoremstyle{plain} 

\theoremstyle{definition}

\DeclareSymbolFontAlphabet{\Scr}{rsfs}

\numberwithin{equation}{section}
\def\be{\begin{equation}}
\def\ee{\end{equation}}
\def\ba{\begin{array}}
\def\ea{\end{array}}

\newcommand{\bea}{\begin{eqnarray}}
\newcommand{\eea}{\end{eqnarray}}

\newcommand{\bbox}{\lower.2ex\hbox{$\Box$}}

\DeclareSymbolFontAlphabet{\Scr}{rsfs}

\newcommand{\Z}{\mathbb{Z}}

\begin{document}
\begin{center}
{\bf\LARGE Stationary $D=4$ Black Holes in Supergravity: The Issue of Real Nilpotent Orbits} \\
\vskip 2 cm
{\bf \large Daniele Ruggeri$^{1}$  and Mario Trigiante$^{2}$}
\vskip 8mm
 \end{center}

\noindent {\small $^{1}$ Universit\`a di Torino, Dipartimento di Fisica
and I.N.F.N. - sezione di Torino, Via P. Giuria 1, I-10125 Torino, Italy\\
    $^{2}$ \it DISAT, Politecnico di Torino, Corso Duca
    degli Abruzzi 24, I-10129 Turin\\}

\vskip 2 cm
\begin{center}
{\small {\bf Abstract}}
\end{center}
The complete classification of the nilpotent orbits of ${\rm SO}(2,2)^2$ in the representation ${\bf (2,2,2,2)}$, achieved in \cite{Dietrich:2016ojx}, is applied to the study of multi-center, asymptotically flat, extremal black hole solutions to the STU model. These real orbits provide an intrinsic characterization of regular single-center solutions, which is invariant with respect to the action of the global symmetry group ${\rm SO}(4,4)$, underlying the stationary solutions of the model, and provide stringent regularity constraints on multi-centered solutions. The known \emph{almost-BPS} and \emph{composite non-BPS} solutions are revisited in this setting. We systematically provide, for the relevant ${\rm SO}(2,2)^2$-nilpotent orbits of the global Noether charge matrix, regular representatives thereof. This analysis unveils a composition law of the orbits according to which those containing regular multi-centered solutions can be obtained as combinations of specific single-center orbits defining the constituent black holes. Some of the
${\rm SO}(2,2)^2$-orbits of the total Noether charge matrix are characterized as ``intrinsically singular'' in that they cannot contain any regular solution.
\vskip 1 cm
\vfill
\noindent {\small{\it
    E-mail:  \\
{\tt daniele.rug@gmail.com}; \\
{\tt mario.trigiante@polito.it}}}
   \eject
   \section{Introduction}
One of the most interesting aspects of (ungauged) extended supergravities is the  global symmetry of their field equations and Bianchi identities which was conjectured to encode all the known string/M-theory dualities. In four dimensions the on-shell global symmetry group $G_4$ (which is a non-compact Lie group at the classical level) acts on the scalar fields as the isometry group of the scalar manifold, and on the vector fields strengths $F^\Lambda_{\mu\nu}$ and their magnetic duals $G_{\Lambda\,\mu\nu}$ as symplectic electric-magnetic duality transformations \cite{Gaillard:1981rj}. In \cite{Breitenlohner:1987dg} it was found that a subset of all solutions to the four-dimensional theory, the stationary, (locally) asymptotically flat ones \cite{reviews}, actually feature a larger symmetry group $G$ which is not manifest in $D=4$, but rather in an effective Euclidean three-dimensional description which is formally obtained by compactifying the four-dimensional model along the time direction and dualizing the vector fields into scalars. Stationary four-dimensional, asymptotically-flat black hole solutions can be conveniently arranged in orbits with respect to this larger symmetry group  $G$ (to be dubbed \emph{duality group} in the following) whose action has proven to be a valuable tool for their classification \cite{Cvetic:1995kv,pioline,Gaiotto:2007ag,Bergshoeff:2008be,Chemissany:2009hq,Bossard:2009at,Bossard:2009we,Kim:2010bf,Chemissany:2010zp,bossard2,Fre:2012im,Fre:2011uy,Chemissany:2012nb} and for the definition of a  \emph{solution-generating technique}  \cite{Cvetic:1995kv} to construct new solutions from known ones \cite{Cvetic:2013cja,Andrianopoli:2013kya,Andrianopoli:2013jra,Chow:2013tia,Chow:2014cca}. More recently,
 it has found an application in the context of \emph{subtracted geometry} \cite{Cvetic:2013cja,Bertini:2011ga,Cvetic:2011dn,Virmani:2012kw}.\par
In the effective $D=3$ description, stationary, (locally) asymptotically flat  four-dimensional black holes are solutions to an Euclidean non-linear sigma-model coupled to gravity, the target space being a pseudo-Riemannian manifold ${\Scr M}$ of which $G$ is the isometry group. Such solutions are described by a set of scalar fields $\phi^I(x^i)$ parametrizing ${\Scr M}$, functions of the three spatial coordinates $x^i$, $i=1,2,3$ (in the axisymmetric solutions the dependence is restricted to the polar coordinates $r,\theta$ only). The asymptotic data defining the solution comprise the value $\phi_0\equiv (\phi^I_0)$ of the scalar fields at radial infinity and the Noether  charge matrix $Q$ associated with the global symmetry group the sigma-model which has value in the Lie algebra $\mathfrak{g}$ of $G$. If ${\Scr M}$ is homogeneous, we can always fix $G$ by mapping the point at infinity $\phi_0$ into the \emph{origin} $O$, where the invariance under the isotropy group $H^*$ is manifest. We shall restrict ourselves only to models in which ${\Scr M}$ is homogeneous symmetric of the form ${\Scr M}=G/H^*$. The solutions are therefore characterized, though in general not completely, by the properties of the
 Noether charge matrix $Q$, seen as an element of the tangent space to the manifold in $O$, with respect to the action of $H^*$. In other words they can be grouped in $H^*$-orbits of $Q$. This intrinsic geometric feature completely determines the physical properties of single-center solutions, while this is not the case for multi-centered solutions \cite{bossard2,BPSmc,Goldstein:2008fq,Bena:2009ev,Dall'Agata:2010dy,Galli:2010mg}, whose features crucially depend on their internal structure. Nevertheless we shall find that there exist $H^*$-orbits of $Q$ which do not contain single or multi-centered solution. Relating the properties of multi-centered solutions to the $H^*$-orbits of $Q$ and of the Noether charges of its constituents is the main object of the present paper. As far as axisymmetric single and multi-centered solutions are concerned, their rotation is encoded in another $\mathfrak{g}$-valued matrix $Q_\psi$, first introduced in \cite{Andrianopoli:2013kya,Andrianopoli:2012ee}, which contains the angular momentum of the solution as a characteristic component and vanishes in the static limit. Once we fix $\phi_0\equiv O$, both $Q$ and $Q_\psi$ transform in $H^*$-representations, that is the action of $G$ on the whole solution amounts to the adjoint action of $H^*$ on $Q$ and $Q_\psi$.\par
 Non-extremal (or extremal over-rotating) single-center solutions are characterized by matrices  $Q$ and $Q_\psi$  belonging to the same regular $H^*$-orbit which contains the Kerr (or the extremal-Kerr) solution.\par In the so-called (ungauged) STU model, which is an $\mathcal{N}=2$ supergravity coupled to three vector multiplets, the most general representative of the Kerr-orbit was derived in \cite{Chow:2013tia,Chow:2014cca}, and features all the duality-invariant properties of the most general solution to the maximal (ungauged) supergravity of which the STU model is a consistent truncation. On the other hand extremal static and \emph{under-rotating} solutions \cite{Rasheed,Larsen,Astefanesei} feature nilpotent $Q$ and $Q_\psi$  belonging to different orbits of $H^*$ \cite{Andrianopoli:2013kya,Andrianopoli:2012ee}.\par
The problem of classifying these solutions is therefore intimately related to the general (still open) mathematical problem of classifying the nilpotent orbits in a  given representation $\rho$ of a real non-compact, semisimple Lie group. In our case the representation $\rho$ is defined by the adjoint action of $H^*$ on the coset space ${\mathfrak{K}^*}$ (isomorphic to the tangent space to the manifold) which $Q$ and $Q_\psi$ belong to once we fix $\phi_0\equiv O$.\par Stationary extremal solutions have been studied in \cite{Bossard:2009we,bossard2} in terms of the nilpotent orbits of the complexification $H^{*\,\mathbb{C}}$ of $H^*$, which are known from the mathematical literature.\footnote{See \cite{deBoer:2014iba,Deger:2015tra} for recent applications of this classification to the study of supersymmetric string solutions.} These orbits are however large enough as to contain regular as well as singular solutions. Even fixing the $G_4$-orbit of the symplectic vector of quantized electric-magnetic charges $\Gamma=(p,q)$ and the $H^{*\,\mathbb{C}}$-orbit of $Q$, as we shall show, is not enough to single out a certain class regular single-center solutions. $H^*$-orbits of $Q$, on the other hand, provides an intrinsic characterization of all regular single-center solutions and thus provide stringent necessary conditions for the regularity of multi-center ones.\par
 A classification of real nilpotent orbits has been performed in specific $\mathcal{N}=2$ ungauged models \cite{Fre:2011uy,Fre:2011ns,Chemissany:2012nb}, in connection to the study of their extremal  four-dimensional solutions. There it is shown that, at least for single-center black holes, there is a one-to-one correspondence between the regularity of the solutions\footnote{\label{fot01} Here, somewhat improperly, we use the term regular also for \emph{small black holes}, namely solutions with vanishing horizon area. These are limiting cases of regular solutions with finite horizon-area (\emph{large black holes}).} (as well as their supersymmetry) and certain real nilpotent orbits. This allows to check the regularity of the single-center solution by simply inspecting the corresponding $H^*$-orbit. The classification procedure adopted in \cite{Fre:2011uy,Fre:2011ns,Chemissany:2012nb} is a  direct one
which combines the method of standard triples \cite{collingwood}
with new techniques based on the Weyl group: After a general group
theoretical analysis of the model this approach allows a
systematic construction of the various nilpotent orbits by solving
suitable matrix equations in nilpotent generators $e$. Solutions to
these equations belong to a same orbit of $H^\mathbb{*\,C}$ but to different orbits of $H^*$
and the final part of the analysis is to group them under the action of $H^*$. Solutions which are
 not  connected by the action of $H^*$ are then
found to be distinguished by certain $H^*$-invariants, which
comprise the signatures of suitable $H^*$-covariant symmetric tensors (\emph{tensor
classifiers}). In \cite{Dietrich:2016ojx} a more formal, general classification technique, based on the notion of \emph{carrier algebras},\footnote{The general classification methods presented in \cite{Dietrich:2016ojx} are alternative to the one developed in \cite{hvl}, whose practical implementation is more problematic.} was developed and applied, as an example, to the STU model which, in spite of its intrinsic simplicity, has played a special role in the black hole literature as a common universal truncation of a broad class of four-dimensional supergravities. These include all the extended (i.e. $\mathcal{N}\ge 2$) four-dimensional models whose scalar manifold is symmetric of the form ${\Scr M}_4=G_4/H_4$, and the isometry group $G_4\subset G$, which defines the global symmetry (or $D=4$ dimension duality) of the four-dimensional theory, is a \emph{non-degenerate group of type-${\rm E}_7$} \cite{brown}.\footnote{In the $\mathcal{N}=2$ case, the above condition in referred to the special K\"ahler manifold spanned by the scalar fields in the vector multiplets, since those in the hypermultiplets are not relevant to the black hole solutions under consideration. Moreover by specializing to the \emph{non-degenerate} case (see the second of references \cite{brown}), we are excluding those models with $G_4={\rm U}(p,q)$ and vector  field-strengths together with their magnetic duals  transforming in the ${\bf p+q}+\overline{\bf p+q}$, like the \emph{minimal coupling}  $\mathcal{N}=2$ models  with $G_4={\rm U}(1,q)$ or the $\mathcal{N}=3$ supergravity  with $G_4={\rm U}(3,q)$. } Those models typically have a $D=5$ uplift and include the maximal and half-maximal  supergravities ($\mathcal{N}=8,\, 4$), the so-called ``magical'' $\mathcal{N}=2$ supergravities and the infinite series of models with special K\"ahler manifold $\frac{{\rm SL}(2,\mathbb{R})}{{\rm SO}(2)}\times \frac{{\rm SO}(2,n)}{{\rm SO}(2)\times{\rm SO}(n)}$. At least as far as the single-center solutions are concerned, the $G$-orbits of regular black holes in all these models have a representative in the STU truncation.\par
The STU model, as mentioned earlier, describes $\mathcal{N}=2$ supergravity  coupled to three vector multiplets, whose three complex scalar fields span a manifold of the form ${\Scr M}^{(4)}_{scal}=(\frac{{\rm SL}(2,\mathbb{R})}{{\rm SO}(2)})^3$. Upon time-like dimensional reduction and dualization of vector fields into scalars, stationary solutions to the STU model are effectively described as solutions to a sigma-model with target manifold:
\begin{equation}
{\Scr M}_{scal}=\frac{G}{H^*}=\frac{{\rm SO}(4,4)}{{\rm SO}(2,2)^2}\,,
\end{equation}
coupled to gravity. The tangent space at the origin is isomorphic to the coset-space ${\mathfrak{K}^*}$ in the isometry algebra $\mathfrak{g}=\mathfrak{so}(4,4)$ which in turn supports a representation $\rho={\bf (2,2,2,2)}$ of the isotropy group $H^*={\rm SO}(2,2)^2=({\rm SL}(2,\mathbb{R})\times_{\mathbb{Z}_2} {\rm SL}(2,\mathbb{R}))^2$ with respect to its adjoint action. Extremal solutions to the STU model naturally fall within nilpotent orbits
of $\rho$ with respect to $H^*$, whose complete classification was achieved in \cite{Dietrich:2016ojx}.\footnote{Here we use a notation which is different from that of \cite{Dietrich:2016ojx}: $H^*$ corresponds to $G_0$ in \cite{Dietrich:2016ojx}, $H^{*\,\mathbb{C}}$ to $G_0^c$ in the same reference, ${\mathfrak{H}^*}$ to $\mathfrak{g}_0$, $\mathfrak{H}^{*\,\mathbb{C}}$ to $\mathfrak{g}_0^c$. the coset space here is denoted by ${\mathfrak{K}^*}$ while it is denoted by $\mathfrak{g}_1$ in \cite{Dietrich:2016ojx}. Similarly its complexification $\mathfrak{K}^{*\,\mathbb{C}}$ is denoted by $\mathfrak{g}_1^\mathbb{C}$ in the same reference. The maximal compact subalgebra of $\mathfrak{g}$ and its complement space of non-compact generators are denoted here by $\mathfrak{H}$ and $\mathfrak{K}$ and in \cite{Dietrich:2016ojx} by $\mathfrak{t}$ and $\mathfrak{p}$, respectively. } Viewing the ${\bf (2,2,2,2)}$ as a representation of the \emph{complexification} $H^{*\,\mathbb{C}}$ of $H^*$, the elements of its space ${\mathfrak{K}^{*\,\mathbb{C}}}$ are in one-to-one correspondence with states of a 4-qubit system. In fact  different orbits of regular extremal black holes were put in correspondence with states with different degree of entanglement \cite{Bergshoeff:2008be,Levay:2010ua} (see \cite{Borsten:2012fx} for an updated review on the subject). For the complete classification of the nilpotent $H^\mathbb{*\,C}$-orbits in the ${\bf (2,2,2,2)}$ see \cite{Bossard:2009we,Borsten:2010db} and \cite{Djo}.\par
With respect to the group $H^*$, we have found in \cite{Dietrich:2016ojx} a total of 101 nilpotent real orbits in the ${\bf (2,2,2,2)}$.
\footnote{The number of orbits with respect to ${\rm SL}(2,\mathbb{R})^4$, locally isomorphic to $H^*$, are 145.}Almost  all of them can be obtained by acting on representatives of the complex orbits by means of \emph{outer-automorphisms} of the isotropy algebra ${\mathfrak{H}^*}$. Indeed real forms of semisimple Lie algebras feature outer-automorphisms which correspond to inner- automorphisms of their complexifications. In the simple example of the Lie algebra $\mathfrak{sl}(2,\mathbb{R})$ we easily observe that conjugation by the matrix $S=i\,{\rm diag}(1,-1)\in {\rm SL}(2,\mathbb{C})$ is an automorphism in that it maps the algebra into itself, and it is outer since its effect cannot be offset by any inner automorphism.\footnote{ In extended supergravity theories outer-automorphisms of the four-dimensional duality group $G_4$ are related to \emph{parity transformations} \cite{parity,Trigiante:2016mnt}.}\par
For the orbits describing single-center solutions we define a frame in which the representative is ``simplest'', namely depends on the least number of independent parameters. This frame is defined by the so-called \emph{generating solution} which turns out to provide a convenient
 description of the orbits and of the effect of the outer-automorphisms of ${\mathfrak{H}^*}$ on the corresponding black hole solutions. In particular it makes apparent that this action in general spoils the regularity of the solution.\par
 In the present work we apply the orbit classification of \cite{Dietrich:2016ojx} to a systematic study of the stationary, asymptotically-flat, single and multi-center black hole solutions of the STU model, completing the analysis of \cite{Bossard:2009we,bossard2}. In particular we give an intrinsic, algebraic characterization in terms of $H^*$-orbits, of the regular single-center solutions. This provides a necessary, stringent condition for the regularity of the
multi-centered solutions: \emph{ Each center of a regular multi-centered solution must be itself a regular black hole, and thus its Noether charge should fall in the corresponding subset of real $H^*$-orbits.}\footnote{ We shall make this statement more precise by defining an \emph{intrinsic} $H^*$-orbit for each center, since, strictly speaking, the Noether charges of each constituent black hole do not belong to $H^*$-orbits. This is done by associating with each center an \emph{intrinsic Noether charge matrix} referred to the non-interacting configuration where the distances between the centers are sent to infinity.} We also give, for a representative selection of $H^*$-orbits of the total Noether charge $Q$, one or more examples of solutions (restricting to one or two-centers). These satisfy a system of solvable field equations associated with each $H^{*\,\mathbb{C}}$-orbit and derived, following \cite{bossard2}, using a corresponding characteristic nilpotent algebra.\par
Let us summarize the main points of our analysis:
\begin{itemize}
\item{{\bf General classification of the $H^*$-orbits.} The $H^*$-orbits of the solutions are conveniently classified by arranging them within larger orbits in a filtration structure starting from the largest. These are the nilpotent orbits in the complex algebra $\mathfrak{g}^{\mathbb{C}}\equiv \mathfrak{g}+i\,\mathfrak{g}$, with respect to the adjoint action of the complexification $G^\mathbb{C}$ (i.e. the Lie group generated by $\mathfrak{g}^{\mathbb{C}}$) of $G$, and are characterized by the $G^\mathbb{C}$-invariant $\alpha$-labels. In the STU model there are eleven $\alpha$-labels ($\alpha^{(\ell)}$, $\ell=1,\dots, 11$). The orbits of the single center solutions lie within the $G^\mathbb{C}$-orbits defined by $\alpha^{(1)},\dots, \alpha^{(6)}$, the last one, in particular contains the orbits of the black hole solutions with finite horizon area (large black holes). These split into BPS, non-BPS with $I_4(\Gamma)>0$ and non-BPS with $I_4(\Gamma)<0$, where $I_4(\Gamma)$ is the quartic invariant of the duality group $G_4$ written in terms of the quantized charges $\Gamma=(p^\Lambda,\,q_\Lambda)$. The small black holes \cite{Borsten:2011ai}
 are contained in the  $\alpha^{(1)},\dots, \alpha^{(5)}$ orbits. The only regular solutions described by the orbits $\alpha^{(7)},\dots, \alpha^{(11)}$ are multi-center non-BPS: $\alpha^{(11)}$ describes the (multi-center) \emph{almost-BPS} solutions of \cite{Goldstein:2008fq,Bena:2009ev}, while $\alpha^{(10)}$ the (multi-center) \emph{composite non-BPS} solutions first studied by \cite{bossard2}. Each $G^\mathbb{C}$-orbit further split into real orbits in $\mathfrak{g}$ with respect to the action of $G$. By the Kostant-Sechiguchi bijection, these orbits are completely classified in terms of the so-called $\beta$-labels and are in one to one correspondence with the nilpotent orbits of $H^{*\,\mathbb{C}}$ in the complexification $\mathfrak{K}^{*\,\mathbb{C}}\equiv {\mathfrak{K}^*}+i\,{\mathfrak{K}^*}$ of ${\mathfrak{K}^*}$, in turn described by the $\gamma$-labels. The sets of all possible $\gamma$ and $\beta$-labels coincide. For each $\alpha$-label, a nilpotent generator $e$ in the coset space ${\mathfrak{K}^*}$ can be simultaneously characterized as being in a certain $G$-orbit within $\mathfrak{g}$ ($\beta$-label) and in a certain $H^{*\,\mathbb{C}}$-orbit in $\mathfrak{K}^{*\,\mathbb{C}}$ ($\gamma$-label). This however does not completely characterize the $H^*$-orbit: Orbits in ${\mathfrak{K}^*}$ with given $\gamma$ and $\beta$-labels may further split into sub-orbits with respect to the action of $H^*$. When this happens, we describe this fine-structure using further labels $\delta^{(1)},\delta^{(2)},\dots$; }
 \item{{\bf Regular black holes and  $H^*$-orbits.} We pinpoint within these large complex orbits the $H^*$-ones containing regular solutions and write representatives of these as single center solutions or combinations thereof. As far as single-center solutions are concerned, since they are completely defined by the point on the scalar manifold at infinity (which we fix to coincide with the origin) and the Noether charge, the $H^*$-orbit of the latter only contains solutions connected by the global symmetry group $G$ and thus its representatives are either all regular or all singular. With an abuse of terminology, we shall dub the former as  as ``regular'' orbits, and the latter as ``singular'' ones. Regular single-center black holes belong to the orbits with $\alpha$-label $\alpha^{(\ell)}$ between $\alpha^{(1)}$ and $ \alpha^{(6)}$ and coinciding $\beta$ and $\gamma$ labels: $\beta^{(\ell;k)}=\gamma^{(\ell;k)}$. This is enough to completely fix the real orbit except for the $\beta^{(6;5)}=\gamma^{(6;5)}$ one, for which a further label ($\delta^{(1)}$) should be specified. This is the orbit of regular, static, single-center black holes with $I_4<0$. As a consequence of this, \emph{for the given $\gamma$-label $\gamma^{(6;5)}$ and charge vector $\Gamma$ in the $G_4$-orbit characterized by $I_4(\Gamma)<0$, which fixes the $\beta$-label to $\beta^{(6;5)}$, there are different inequivalent $H^*$-orbits, distinguished by $\delta^{(1)},\dots,\,\delta^{(4)}$,\footnote{In fact, modulo the triality symmetry of the STU model, there are only two distinct orbits.} only one of which (labeled by $\delta^{(1)}$) describes regular solutions}. An analogous situation occurs in the $\mathcal{N}=2$ model with $G={\rm F}_{4(4)}$ considered in \cite{Chemissany:2012nb}.
      In \cite{Bossard:2013nwa} a detailed analysis is made of the composite non-BPS solutions and a characterization of the regularity of each center (in the $G_4$-orbit $I_4<0$) is given as the requirement that a given charge-dependent, Jordan-algebra valued matrix be positive definite. This condition on the solution precisely singles out the  real orbit $\gamma^{(6;5)},\,\beta^{(6;5)},\,\delta^{(1)}$. We emphasize however that the formulation in terms of $H^*$-orbits represents an alternative, $G$-invariant characterization of the regularity of the single-center solutions.\par
     As for the orbits with $\alpha$-label $\alpha^{(7)},\dots, \alpha^{(11)}$, some of them will be characterized as intrinsically ``singular'' since they do not contain any regular composite solution.\footnote{To prove this we shall show that these orbits cannot be reached by combining any two orbits describing regular single-center solutions.} Aside from the ``singular'' ones, these $H^*$-orbits may describe regular as well as singular solutions. This is the case since regular solutions in these orbits can only be multi-center, which are no longer completely described by the overall Noether charge matrix, but also by the Noether charge matrices of each center. For a representative selection of these orbits we give one or more examples of axisymmetric 2-center systems. Being axisymmetric, each center describes a rotating solution whose angular momentum is parallel (or anti-parallel) to the axis connecting the two;}
\item{{\bf Regular representatives in $H^*$-orbits.} Our intrinsic algebraic characterization of the regular single-center solutions, allows to define a composition-law  mapping couples of regular-single-center orbits into double-center ones. The main results are summarized in Appendix \ref{sumrules}. In particular we find that regular 2-center composite non-BPS solutions can be obtained as combinations of two regular non-BPS black holes with $I_4<0$, consistently with the analysis of \cite{bossard2}, while regular 2-center almost-BPS solutions can be obtained combining one non-BPS center with $I_4<0$ and a BPS or non-BPS center with $I_4>0$.
 Combining a non-BPS black hole with $I_4<0$ with regular but small black holes we end up in the orbits orbits $\alpha^{(7)},\alpha^{(8)}, \alpha^{(9)}$, which are related by the STU triality symmetry. These can be obtained as limits of almost-BPS solutions in $\alpha^{(11)}$ and composite non-BPS solutions in $\alpha^{(10)}$ by setting to zero some of the charges associated with one of the centers which thus becomes small, belonging to one of the real orbits with $\alpha$-label $\alpha^{(1)},\,\dots, \alpha^{(5)}$. We give, for the first time, the sets of equations governing the extremal solutions in these three orbits and solve them. In general, for a representative sample of the ``non-singular'' $H^*$-orbits, we provide regular double-center solutions, proving their regularity property. There are orbits with $\alpha$-label between $\alpha^{(7)}$ and $\alpha^{(11)}$ which are never obtained combining representatives of the regular-single-center orbits. These are the intrinsically ``singular'' orbits mentioned above;}
   \item{{\bf Regularity conditions for multi-centered solutions from $H^*$-orbits.} For the sake of completeness, we also review the set of equations governing the composite non-BPS, discussed in \cite{bossard2} (associated with the orbit $\alpha^{(10)}$) and the almost -BPS, discussed in \cite{Goldstein:2008fq,Bena:2009ev,Dall'Agata:2010dy,Galli:2010mg} (associated with the orbit $\alpha^{(11)}$). The condition that these solutions be combinations of single-center ones in the $H^*$-orbits of the regular black holes provides a regularity constraint which is more stringent than the simple requirement that the solution be asymptotically well behaved, i.e. exhibit regular behavior near the centers and at spatial infinity.  In particular we show that, solutions corresponding to the ``singular'' $I_4(\Gamma)<0$ single-center orbits, in spite of exhibiting regularity of the metric near the centers and at infinity, feature singularities at finite distance;
     }
\item{{\bf Minimum value of the distance between the centers.} For each representative double-center solution, provided each center is separately regular, we find that
the distance $R$ between them should have a minimum value depending on the global charges. Below this value interaction terms, which manifest themselves in the solutions as powers of $1/R$, are large enough as to spoil the regularity of the total background and produce a singularity at finite $r$. This is consistent with the known fact that in the limit $R\rightarrow 0$, in which the two centers merge in a single one, there are no regular solutions in the orbits $\alpha^{(7)},\dots, \alpha^{(11)}$. This condition on $R$ adds to the \emph{bubble conditions} \cite{bossard2,BPSmc,Bena:2009ev} relating $R$ to the asymptotic data and which follow from the requirement that each center have vanishing NUT charge (absence of Misner strings). }
\item{In the following table we present the structure of all real orbits found by our analysis, in terms of $\alpha$-$\beta$-$\gamma$-labels. For each $\alpha$-label, the substructure encoded in the $\beta$-$\gamma$-labels is shown. Regular single-center solutions are described by  $\alpha$-labels up to $\alpha^{(6)}$; Beyond the vertical line, the regular solutions only have a multi-center description and are the main topic of this paper.}
\end{itemize}
\begin{figure}[H]
\centering
\includegraphics[width=17cm]{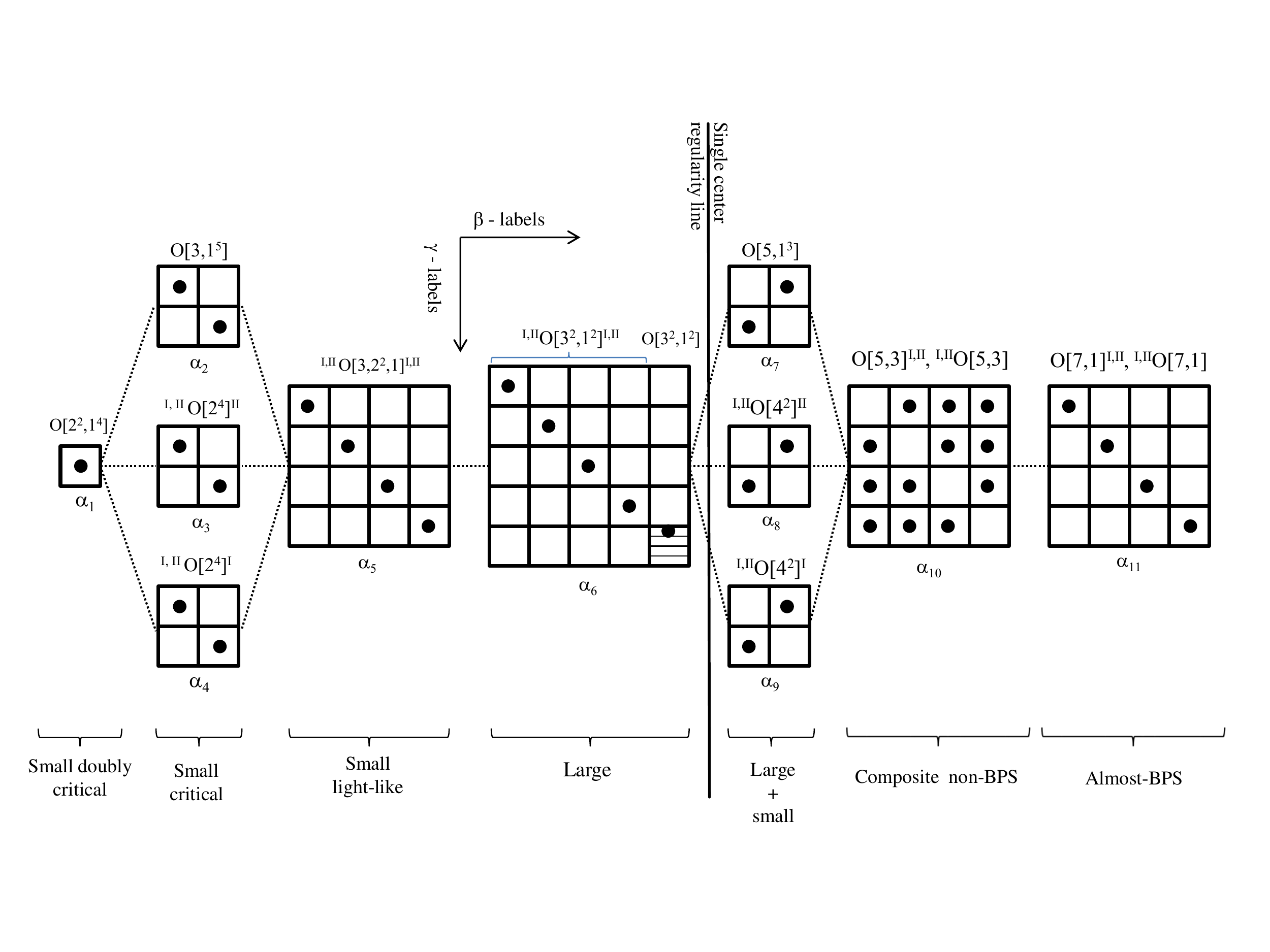}
 \caption{\small Schematic representations of the nilpotent orbits of $H^*={\rm SO}(2,2)^2$ on the coset space $\mathfrak{K}^*$ of ${\rm SO}(4,4)/{\rm SO}(2,2)^2$. Each square block represents an ${\rm SO}(4,4)^\mathbb{C}$-nilpotent orbit in its Lie algebra, while each column is in one-to -one correspondence with ${\rm SO}_0(4,4)$-nilpotent orbits in $\mathfrak{so}(4,4)$. We use for their description the notations of \cite{Bossard:2009we,collingwood,Borsten:2010db} (the trivial orbit $[1^8]$ is omitted). Thick vertical and horizontal lines separate orbits with distinct $\beta$ and $\gamma$-labels, respectively. The empty slots do not contain regular solutions and will be referred to as \emph{intrinsically singular}.  The orbit structure with respect to the ${\rm SL}(2,\mathbb{R})^4$ subgroup of ${\rm SO}(4,4)$ is the same except for a doubling of each cell in the $\alpha^{(7)},\dots, \alpha^{(11)}$ blocks, which yields a total of 145 orbits \cite{Dietrich:2016ojx}. For the $\alpha^{(7)},\dots, \alpha^{(9)}$ blocks, we mention below the fact that they contain, as instances of regular solutions,  combinations of a large and a small black hole. }\label{Figfigs}
\end{figure}

The paper is organized as follows. In Sec. \ref{sec1} we review the main facts about the effective $D=3$ description of four-dimensional stationary solutions to a four-dimensional supergravity model. General necessary conditions for the regularity of a multi-center system are stated in terms of $H^*$-orbits. In Sec. \ref{sec2} we focus on the STU model outlining its geometry and defining the corresponding Euclidean three-dimensional model describing its stationary solutions. In Sect. \ref{sec3} we address the issue of nilpotent $H^*$-orbits in the coset space of the $D=3$ scalar manifold and review one of the two classification methods pursued in \cite{Dietrich:2016ojx}. In Subsection \ref{generaloverv} we give a general overview of the real orbits for the STU model reviewing the notion of generating solution for single-center black holes. We define the orbits characterizing regular single-center solutions and formulate a necessary regularity condition for multi-center systems. In Subsection \ref{compositionlaw} we address the mathematical problem of defining combinations of representatives of two regular-single-center orbits yielding nilpotent elements of the higher-order non-BPS orbits with $\alpha$-label from $\alpha^{(7)}$ to $\alpha^{(11)}$. The problem is solved using a computer code and its solution, illustrated in  Appendix \ref{sumrules}, on the one hand defines a composition law of nilpotent orbits and on the other allows to characterize certain orbits as intrinsically singular since no multi-center solution with Noether charge in these orbits can be expressed as composite of regular black holes. Finally Sect. \ref{rnoca} is devoted to a case-by-case study of solutions with Noether charge in a representative set of real orbits with $\alpha$-label from $\alpha^{(7)}$ to $\alpha^{(11)}$. We start with writing the characteristic nilpotent algebra of the orbits $\alpha^{(7)}-\alpha^{(9)}$ yielding a set of graded field equations for the scalar fields of the $D=3$ model which are solved in general. Instances of solutions are then analyzed in some detail for a number of representative $H^*$-orbits, discussing their regularity. The same analysis is done for the $\alpha^{(10)},\,\alpha^{(11)}$-orbits of the composite non-BPS and almost-BPS solutions, partly reviewing the work in \cite{bossard2},\cite{Bena:2009ev}, to which a detailed case-by-case analysis of the solutions for a representative set of $H^*$-orbits is added.
For each orbit of the total Noether charge matrix we give a combination of two single-center orbits. This is consistent with the sum rules given in Appendix \ref{sumrules}. For the intrinsically singular orbits we also give a combination of single-center orbits, one of which  is necessarily associated with singular solutions, and discuss, in some representative cases, the corresponding two-center system. Generalizing the results of \cite{bossard2}, for each instance of axisymmetric system of two black holes the angular momentum is given and shown to be always expressed as the sum of the angular momenta associated with each center plus the contribution from the electromagnetic fields, proportional to the symplectic product of the electric-magnetic charge vectors of the two black holes. In the almost-BPS solutions realized as a system of a black hole with $I_4<0$ and one with $I_4>0$, a new phenomenon is observed: The interaction between the two centers induces an angular momentum  on the $I_4>0$ center which vanishes in the limit in which the two components are sent at infinite distance. We end with concluding remarks.

\section{Effective Description Stationary Solutions}\label{sec1}
In this section we review the basic facts about the effective three-dimensional description of four-dimensional stationary solutions, eventually restricting our analysis to axisymmetric field configurations only. We start with a $D=4$ extended (i.e. $\mathcal{N}>1$), ungauged supergravity,  whose bosonic sector consists in $n_s$
 scalar fields $\phi^r(x)$, $n_v$ vector fields $A^\Lambda_\mu(x)$, $\Lambda=1,\dots, n_v$, and the graviton $g_{\mu\nu}(x)$, which are described by the following Lagrangian
\footnote{Here we adopt the notations and conventions of \cite{Andrianopoli:2013kya,Andrianopoli:2012ee} (in particular we use the ``mostly plus'' convention and $8\pi G={\bf \mathtt{c}}=\hbar=1$).}:
\begin{equation}
\mathcal{L}_4={\tt e}\,\left(\frac{R}{2}-\frac{1}{2}\,G_{rs}(\phi^t)\,\partial_\mu \phi^r\,\partial^\mu \phi^s+\frac{1}{4}\,I_{\Lambda\Sigma}(\phi^r)\,F^\Lambda_{\mu\nu} \,F^{\Sigma\,\mu\nu}+ \frac{1}{8\,{\tt e}}\,R_{\Lambda\Sigma}(\phi^r)\,\epsilon^{\mu\nu\rho\sigma}\,F^\Lambda_{\mu\nu} \,F^{\Sigma}_{\rho\sigma}\right)\,,
\end{equation}
where ${\tt e}:=\sqrt{|{\rm det}(g_{\mu\nu})|}$. In \emph{symmetric supergravities}, as the STU model we shall restrict to,
 the scalar fields $\phi^s$ span a homogeneous, symmetric, Riemannian scalar manifold:
\begin{equation}
{\Scr M}^{(4)}_{scal}=\frac{G_4}{H_4}\,,\label{M4}
\end{equation}
where the isometry group  $G_4$ is the symmetry group of the whole theory provided its non-linear action on the
scalar fields is combined with a symplectic action, defining a representation ${\bf R}$ of $G$, on the vector
 field-strengths $F^\Lambda=dA^\Lambda$ and their magnetic duals $G_\Lambda$.\par
The space-time metric of a stationary,  asymptotically flat solution, in a suitable system of coordinates, has the general form:
 \begin{equation}
ds^2=-e^{2U}\,(dt+\omega)^2+e^{-2U}\,g_{ij}\,dx^i\,dx^j\,,\label{ds2}
\end{equation}
where $i,j=1,2,3$ label the spatial coordinates $x^i=(r,\theta,\varphi)$, $\omega=\omega_i\,d x^i$ and $U,\, \omega_i,\, g_{ij}$ are all
functions of $x^i$.\par
As mentioned in the introduction, these solutions can be given an effective description in an Euclidean $D=3$ model describing gravity coupled to $n=2+ n_s+2n_v$ scalar fields $\phi^I(x^i)$ comprising, besides the $D=4$ scalars $\phi^s$, the warp function $U$ and $2n_v+1$ scalars  $\mathcal{Z}^M=\{\mathcal{Z}^\Lambda,\,\mathcal{Z}_\Lambda\}$ and $a$ originating from the time-like dimensional reduction of the $D=4$ vectors and the dualization of the Kaluza-Klein vector $\omega_i$ into a scalar.
The precise relation between the scalars $a,\, \mathcal{Z}^M$ and the four-dimensional fields is \cite{Andrianopoli:2013kya}:
\begin{eqnarray}
A^M&=& \mathcal{Z}^M (dt + \omega)+ \tilde{A}^M \,,\quad \tilde{A}^M\equiv \tilde{A}^M_i dx^i\,,\label{AF1}\\
\mathbb{F}^M&=&\left(
                       \begin{array}{c}
                         F^\Lambda_{\mu\nu} \\
                         G_{\Lambda\,\mu\nu} \\
                       \end{array}
                     \right)\,\frac{dx^\mu\wedge dx^\nu}{2}
                     =d \mathcal{Z}^M \wedge (dt + \omega)+\tilde{F}^M=\nonumber\\&=&d \mathcal{Z}^M \wedge (dt + \omega) + e^{-2 U}\mathbb{C}^{MN}\mathcal{M}_{(4) NP} {}^{*_3} d\mathcal{Z}^P\,,\label{AF2}\\
da&=& - e^{4 U} \,{}^{*_3}d\omega - \mathcal{Z}^T \mathbb{C} d\mathcal{Z}\,,\label{AF3}
\end{eqnarray}
where $*_3$ is the Hodge duality operation in the $D=3$ Euclidean space, $\mathcal{M}_{(4)}$ the symmetric, symplectic matrix characterizing the symplectic structure over ${\Scr M}^{(4)}_{scal}$ (see Appendix \ref{stumodel} for an explicit construction).
In the above formulae we have used for the vector fields a symplectic-covariant notation in which $\mathcal{Z}^M$ are the time-components of the electric-magnetic vector potentials and $\tilde{A}^M_i$ the resulting $D=3$ vector fields. The field strengths of the latter are defined as follows:
\begin{equation}
\tilde{F}^M\equiv d\tilde{A}^M+\mathcal{Z}^M\,d\omega\,.
\end{equation}
In order to evaluate the $D=4$ vector fields from the $D=3$ solution, one first computes $\omega$ integrating (\ref{AF3}) and then derives $\tilde{A}^M$ integrating the following equation
\begin{equation}
*_3d\tilde{A}^M=-\mathcal{Z}^M\,*_3 d\omega+e^{-2 U}\mathbb{C}^{MN}\mathcal{M}_{(4) NP} \, d\mathcal{Z}^P\,,\label{AF4}
\end{equation}
which directly follows from (\ref{AF2}).\par
The effective $D=3$ Lagrangian describes a sigma-model coupled to gravity and reads:
\begin{align} \label{geodaction}
\frac{1}{{\tt e}^{(3)}}\,\mathcal{L}_{3} &= \frac{1}{2}\,\mathcal{R} - \frac{1}{2}
G_{IJ}(\phi)\partial_i{\phi}^I\partial^i{\phi}^J=\nonumber\\
&=\frac{1}{2}\, \mathcal{R} -[ \partial_i U \partial^i
U+\frac{1}{2}\,G_{rs}\,\partial_i{\phi}^r\,\partial^i{\phi}^s +
\frac{1}{2} e^{-2\,U}\,\partial_i{\mathcal{Z}}^T\,\mathcal{M}_{(4)}\,\partial^i{\mathcal{Z}} +\nonumber\\ &+
\frac{1}{4} e^{-4\,U}\,(\partial_i{a}+\mathcal{Z}^T\mathbb{C}\partial_i{\mathcal{Z}})(\partial^i{a}+\mathcal{Z}^T
\mathbb{C}\partial^i{\mathcal{Z}})]\,,
\end{align}
where ${\tt e}^{(3)}\equiv \sqrt{{\rm det}(g_{ij})}$ and $\mathbb{C}$ is the symplectic-invariant, antisymmetric matrix.
The scalar fields span a homogeneous, symmetric, pseudo-Riemannian manifold of the form
\begin{equation}
{\Scr M}_{scal}=\frac{G}{H^*}\,,
\end{equation}
containing ${\Scr M}^{(4)}_{scal}$ as a submanifold. The isometry group $G$ is a semisimple, non-compact Lie group
which defines the global symmetry of the model, while $H^*$ is a non-compact real form of the maximal compact subgroup $H$ of $G$. In particular $G$ contains, as a subgroup, ${\rm SL}(2,\mathbb{R})_E\times G_4$, ${\rm SL}(2,\mathbb{R})_E$ being the Ehlers group, with respect to which its adjoint representation branches as follows:
\begin{equation}
{\bf Adj(G)}\,\,\longrightarrow\,\,{\bf (3,1)}\oplus {\bf (1,Adj(G_4))}\oplus {\bf (2,R)}\,.\label{branchgg4}
\end{equation}
The coset geometry is defined by  the
 involutive automorphism $\sigma$ on the algebra
$\mathfrak{g}$ of $G$ which leaves the algebra ${\mathfrak{H}^*}$
generating $H^*$ invariant. All the formulas related to the group $G$ and its generators are referred to a matrix representation of $G$ (we shall in particular use the \emph{fundamental one}). The involution $\sigma$ in the chosen representation has the general action: $\sigma(M)=-\eta M^\dagger\eta$, $\eta$
being an $H^*$-invariant metric ($\eta=\eta^\dagger,\,\,\eta^2={\bf 1}$), and induces the pseudo-Cartan
decomposition\footnote{ This should be contrasted with the Cartan decomposition of the semisimple  Lie algebra  $\mathfrak{g}$ into compact and non-compact generators:
 \begin{equation}
 \mathfrak{g}=\mathfrak{H}\oplus \mathfrak{K}\,.\label{cdec}
 \end{equation}
 The action of the  corresponding involution (called Cartan involution) $\tau$ on a matrix $X$ can be defined in a given matrix representation as:
$\tau(X)=-X^\dagger$. In (\ref{cdec}) $\mathfrak{H}$ is the maximal compact subalgebra of $\mathfrak{g}$ while $\mathfrak{K}$ denotes the space of non-compact generators: $\tau(\mathfrak{H})=\mathfrak{H},\,\tau(\mathfrak{K})=-\mathfrak{K}$.
 The algebra $\mathfrak{H}$ is the compact real form of ${\mathfrak{H}^*}$ and generates the maximal compact subgroup $H$ of $G$.} of $\mathfrak{g}$ of the form:
\begin{equation}
\mathfrak{g}={\mathfrak{H}^*}\oplus {\mathfrak{K}^*}\,,\label{pseudoC}
\end{equation}
where $\sigma({\mathfrak{K}^*})=-{\mathfrak{K}^*}$, and the following
relations hold
\begin{equation}[{\mathfrak{H}^*},{\mathfrak{H}^*}]\subset{\mathfrak{H}^*},
\quad [{\mathfrak{H}^*},{\mathfrak{K}^*}]\subset {\mathfrak{K}^*},\quad
[{\mathfrak{K}^*},{\mathfrak{K}^*}] \subset
{\mathfrak{H}^*}.\label{HKrels}\end{equation}
The above relations imply that the coset space ${\mathfrak{K}^*}$ supports a representation $\rho$ of $H^*$, the action of the elements of $H^*$ on the generators in ${\mathfrak{K}^*}$ being the adjoint one.\par
With respect to the involution $\sigma$, the vielbein matrix $P$ and connection
$\mathcal{W}$ 1-forms on the manifold are computed, in terms of the coset-representative $\mathbb{L}(\phi^I)$  of $G/H^*$,  as the odd and even
components, respectively, of  the left-invariant one-form with
respect to $\sigma$:
\begin{equation}
\mathbb{L}^{-1}d \mathbb{L}=P+\mathcal{W}\,,
\end{equation}
where $P=\eta P^\dagger \eta=-\sigma(P)$, $\mathcal{W}=-\eta \mathcal{W}^\dagger
\eta=\sigma(\mathcal{W})$.\par
 The Maurer-Cartan equations imply
\begin{equation}
\mathcal{D} P\equiv d P+P\wedge \mathcal{W}+\mathcal{W}\wedge P=0\,\,;\,\,\,\,\,{\Scr R}[\mathcal{W}]\equiv d\mathcal{W}+\mathcal{W}\wedge \mathcal{W}=-P\wedge P\,,\label{MCs}
\end{equation}
where $\mathcal{D} $ is the $H^*$-covariant derivative and ${\Scr R}[\mathcal{W}]$ is the curvature 2-form of the scalar manifold with value in $\mathfrak{H}^*$. We can expand $P$ and $\mathcal{W}$ in bases $\{K_{\mathcal{A}}\}$ and $\{J_{\mathcal{I}}\}$ of $\mathfrak{K}^*$ and $\mathfrak{H}^*$, respectively:
$P=P^\mathcal{A}\,K_\mathcal{A},\,\mathcal{W}=\mathcal{W}^\mathcal{I}\,J_\mathcal{I}$. The metric on the scalar manifold $g_{\mathcal{AB}}$ at the origin $O$ is defined as:
\begin{equation}
g_{\mathcal{AB}}\equiv k\,{\rm Tr}(K_\mathcal{A}K_\mathcal{B})\,,
\end{equation}
where $k$ is a representation-dependent constant. The structure constants of the $\mathfrak{g}$-algebra only consist, according to (\ref{HKrels}), of the following non-vanishing components:
 $C_{\mathcal{IJ}}{}^{\mathcal{K}},\,C_{\mathcal{IA}}{}^{\mathcal{B}},\,C_{\mathcal{AB}}{}^{\mathcal{I}}$.
 In terms of $P$ the metric on the manifold reads:
\begin{equation}
dS^2_{(3)}=G_{IJ}(\phi)\,d{\phi}^I\,d{\phi}^J=P^\mathcal{A} P^\mathcal{B}\,g_{\mathcal{AB}}=k\,{\rm Tr}(P^2)\,.\label{geo3}
\end{equation}
 The scalar field Lagrangian density has therefore the form:
\begin{equation}
\mathcal{L}_3^{(s)}=\frac{{\tt e}^{(3)}}{2}\,{\Scr P}_i^\mathcal{A} {\Scr P}^{i\,\mathcal{B}}\,g_{\mathcal{AB}}={\tt e}^{(3)}\,\frac{k}{2}\,{\rm Tr}({\Scr P}_i{\Scr P}^i)\,,\label{Lscal}
\end{equation}
where ${\Scr P}_i\equiv \partial_i\phi^I\,P_I$ is the pull-back of the vielbein matrix on the Euclidean base-space through $\phi^I(x^i)$.
From the above Lagrangian we derive the scalar field equations:
\begin{equation}
\mathcal{D}*_3 {\Scr P}^\mathcal{A}=0\,\,\Leftrightarrow\,\,\,\,d*_3 {\Scr P}^\mathcal{A}+C_{\mathcal{IB}}{}^{\mathcal{A}}\,{\Scr W}^\mathcal{I}\wedge *_3{\Scr P}^\mathcal{B}=0\,,\label{feqscal}
\end{equation}
where we have defined ${\Scr W}_i\equiv \partial_i\phi^I\,\mathcal{W}_I$ as the pull-back of the connection matrix $\mathcal{W}$ on the same space through $\phi^I(x^i)$. The above equations can also be written in the more compact matrix form:
$$d*_3 {\Scr P}+[{\Scr W},\,*_3{\Scr P}]=0\,.$$

We choose the scalar fields $\phi^I$ to be defined by a  local \emph{solvable parametrization} of the coset, and the coset representative  is chosen to be
\begin{equation}
\mathbb{L}(\phi^I)=\exp(-a T_\bullet)\,\exp(\sqrt{2}
\mathcal{Z}^M\,T_M)\,\exp(\phi^r\,T_r)\,\exp(2\,U
H_0)\,,\label{cosetr3}
\end{equation}
where $T_{\mathcal{A}}=\{H_0,\,T_\bullet,\,T_s,\,T_M\}$ generate the solvable Lie group  defined by the Iwasawa decomposition of G with respect to its maximal compact subgroup $H$.
The structure of this solvable algebra is the following:
\begin{align}
[H_0,\,T_M]&=\frac{1}{2}\,T_M\,\,;\,\,\,[H_0,\,T_\bullet]=T_\bullet\,\,;\,\,\,[T_M\,T_N]=\mathbb{C}_{MN}\,T_\bullet\,,\nonumber\\
[H_0,T_r]&=[T_\bullet,T_r]=0\,\,;\,\,\,[T_r,T_M]=T_r{}^N{}_M\,T_N\,\,;\,\,\,[T_r,T_s]=-
T_{rs}{}^{s'} T_{s'}\,,\label{relc1}
\end{align}
where $T_r\in \mathfrak{g}_4$ ($\mathfrak{g}_4$ being the Lie algebra generating $G_4$) are the generators of the solvable Lie algebra defining the parametrization of ${\Scr M}^{(4)}_{scal}$, so that the coset representative of ${\Scr M}^{(4)}_{scal}$ is: $L_4(\phi^s)\equiv \exp(\phi^r\,T_r)$.  The quantities  $T_r{}^N{}_M$ in (\ref{relc1}) define the  matrix form of $T_r$ in the symplectic representation ${\bf R}$ on
contravariant vectors $d\mathcal{Z}^M$. The generators of the Ehlers group ${\rm SL}(2,\mathbb{R})_E$ are $H_0,T_\bullet,\,T_\bullet^\dagger$, while $\{T_M,\,T_M^\dagger\}$  define the ${\bf (2,R)}$ representation in (\ref{branchgg4}).
The representation-dependent constant in (\ref{geo3}) is given by: $k=1/(2{\rm Tr}(H_0H_0))$. In terms of $H_0$ we can also express the metric $\eta$ as follows: $\eta\equiv (-1)^{2\,H_0}$. \par
From this characterization of $\eta$ it immediately follows that the coset space ${\mathfrak{K}^*}$ contains compact generators belonging to the dimension-$2n_v$ subspace $\mathfrak{K}^{*\,(R)}={\mathfrak{K}^*}\cap \mathfrak{H}$ (see footnote 6 for the definition of $\mathfrak{K}, \,\mathfrak{H}$) generated by the following  matrices $K_M$:
\begin{equation}
\mathfrak{K}^{*\,(R)}={\rm Span}(K_M)\,\,;\,\,\,\,\,K_M=\frac{1}{2}(T_M+\eta T_M^\dagger\eta)=\frac{1}{2}(T_M- T_M^\dagger)\,.\label{KMdef}
\end{equation}
Similarly the non-compact generators of the algebra ${\mathfrak{H}^*}$ belong to the subspace $\mathfrak{H}^{*\,(R)}={\mathfrak{H}^*}\cap \mathfrak{K}$ generated by the following matrices $J_M$:
\begin{equation}
\mathfrak{H}^{*\,(R)}={\rm Span}(J_M)\,\,;\,\,\,\,\,J_M=\frac{1}{2}(T_M-\eta T_M^\dagger\eta)=\frac{1}{2}(T_M+ T_M^\dagger)\,.\label{JMdef}
\end{equation}
If we denote by  $H_c$ the maximal compact subgroup of $H^*$, generated by $\mathfrak{H}_{c}={\mathfrak{H}^*}\cap \mathfrak{H}$, $\mathfrak{H}^{*\,(R)}$ is the coset space of the symmetric Riemannian manifold $H^*/H_c$. It generates the so-called \emph{Harrison transformations}  \cite{Breitenlohner:1987dg}, namely $H^*$-transformations
 which play a special role in the solution generating techniques: They are not present
 among the global symmetries of the  $D=4$ theory and have the distinctive property of
 switching on electric or magnetic charges when acting on neutral solutions (like the
  Kerr or Schwarzshild ones). Their generators $J_M$ are indeed in one-to-one correspondence with the electric and magnetic charges $(\Gamma^M)=(p^\Lambda,\,q_\Lambda)$. It was shown in  \cite{Breitenlohner:1987dg} that the most general Kerr-Newman solution can be obtained by acting on the Kerr one by means of Harrison transformations.\par
The group $H_c$ has the general  form:  $H_c={\rm SO}(2)_E\times H_4={\rm SO}(2)_E\times {\rm SO}(2)^3$, where ${\rm SO}(2)_E$ is the maximal compact subgroup of the Ehlers group ${\rm SL}(2,\mathbb{R})_E$. Both spaces $\mathfrak{K}^{*\,(R)}$ in ${\mathfrak{K}^*}$ and $\mathfrak{H}^{*\,(R)}$ in ${\mathfrak{H}^*}$ support a same linear representation ${\bf R}'$ of this compact subgroup. With respect to $H_4={\rm SO}(2)^3$ alone, ${\bf R}'$ is nothing but the symplectic representation ${\bf R}$ of $G_4$, defining its electric-magnetic duality action, seen as a representation of the $H_4$-subgroup.\footnote{Recall that the nilpotent generators $T_M$ transform under the adjoint action of $G_4$ in the representation ${\bf R}$. Therefore their compact and non-compact components, in $\mathfrak{K}^{*\,(R)}$ and $\mathfrak{H}^{*\,(R)}$, respectively, only transform linearly under the maximal compact subgroup $H_4$ of $G_4$, the corresponding representation being denoted by ${\bf R}'$.}
\par
The Einstein equation for the Euclidean metric $g_{ij}$ is readily derived from (\ref{geodaction}) to be:
\begin{equation}
\mathcal{R}_{ij}=k\,{\rm Tr}({\Scr P}_i{\Scr P}_j)\,.\label{eeq}
\end{equation}
 Stationary axisymmetric solutions, taking  $Z$ to be the symmetry axis, feature the two Killing vectors  $\xi=\frac{\partial}{\partial t}$ and $\psi=\frac{\partial}{\partial \varphi}$ and all fields only depend on $r,\,\theta$, while  $\omega=\omega_\varphi d\varphi$. The corresponding solutions of the sigma model
are described by $n$ functions $\phi^I(r,\theta)$ and characterized by a unique ``initial point'' $\phi_0\equiv (\phi_0^I)$ at radial infinity
\begin{equation}
\phi_0^I=\lim_{r\rightarrow \infty} \phi^I(r,\theta)\,,
\end{equation}
and an ``initial velocity'' $Q$, at radial infinity, in the tangent space $T_{\phi_0}[{\Scr M}_{scal}]$, which is the Noether charge matrix of the solution.  Since the action of $G/H^*$ on $\phi_0$ is transitive, we can always fix $\phi_0$ to coincide with the origin $O$ (defined as the point in which the scalar fields vanish and where invariance under $H^*$ is manifest) and then classify the orbits of the solutions under the action of $G$ (i.e. in maximal sets of solutions connected through the action of $G$) in terms of the orbits of the velocity vector $Q\in T_{O}({\Scr M}_{scal})$ under the action of $H^*$.
The total Noether charge matrix $Q$ is computed, for a generic stationary solution, as:
\begin{equation}
Q=\frac{1}{4\pi}\,\int_{\Sigma}{}^{*_3}J\,,\label{nc}
\end{equation}
$J=J_i\,dx^i$ being the 1-form associated with the Noether current $J^i=g^{ij}\,J_j$ and $\Sigma$ is a 2-cycle encompassing all the centers of the solution. The explicit form of $J_i$ is given by the standard theory of sigma models on coset manifolds:\footnote{We use the short-hand notation $M^{-\dagger}\equiv (M^\dagger)^{-1}$, $M^{-T}\equiv (M^T)^{-1}$.}\begin{equation}
J_i\equiv \frac{1}{2}\partial_i \phi^I\,\mathcal{M}^{-1}\partial_I
\mathcal{M}=\mathbb{L}^{-\dagger}{\Scr  P}_i^\dagger\mathbb{L}^{\dagger}=\partial_i \phi^I\mathbb{L}^{-\dagger}P_I^\dagger\mathbb{L}^{\dagger}\,,\label{curr}
\end{equation}
where $\mathcal{M}(\phi^I)=\mathbb{L}(\phi^I)\eta \mathbb{L}(\phi^I)^\dagger$ is an $H^*$-invariant symmetric matrix built out of  the representative  $\mathbb{L}(\phi^I)$ at the point $\phi^I$ and $\eta$ is the $H^*$-invariant matrix defined earlier.
The sigma-model field equations (\ref{feqscal}) can also be cast in the form:
\begin{equation}
d\left({}^{*_3}J\right)=0\,\,\Leftrightarrow\,\,\,\,\,\partial_i\left(e^{(3)}J^i\right)=0\,.\label{sigmodfeq}
\end{equation}

Since the generators $T_M$ transform under the adjoint action of $G_4\subset G$ in the symplectic duality representation ${\bf R}$ of the electric-magnetic charges, we shall use for them the following notation: $(T_M)=(T_{q_\Lambda},\,T_{p^\Lambda})$.\par
As far as axisymmetric solutions are concerned, the Noether matrix $Q$ encodes all the conserved, global physical quantities, except the total angular momentum $M_\varphi$. In other words it contains no information about the rotation. In \cite{Andrianopoli:2012ee}  a new matrix $Q_\psi$ was defined which describes the global rotation of the axisymmetric solution:
\begin{equation}
Q_\psi =-\frac{3}{4\pi}\,\int_{S_2^\infty}\psi_{[i}\,J_{
j]}\,dx^i\wedge dx^j=\frac{3}{8\pi}\,\int_{S_2^\infty} g_{\varphi\varphi}\,J_\theta\,d\theta d\varphi\,,\label{qupsi}
\end{equation}
where $S_2^\infty$ is the 2-sphere at infinity.\par
The physical quantities globally characterizing the solution are then obtained as components of $Q$ and $Q_\psi$ \cite{Chemissany:2010zp,Andrianopoli:2013kya,Andrianopoli:2012ee}:\footnote{Eq.s (\ref{allthat}) hold also for generic values of the scalar fields at radial infinity, i.e. for $Q,\,Q_\psi\,\in \, T_{\phi_0}[{\Scr M}_{scal}]$.}
\begin{align}
m &=k\,{\rm
Tr}(H_0^\dagger\,Q)\,,\,\,\,n_{NUT}=-k\,{\rm
Tr}(T_\bullet^\dagger\,Q)\,,\,\,\,\Gamma^M=\sqrt{2}\,k\,\mathbb{C}^{MN}\,{\rm
Tr}(T_N^\dagger\,Q)\,,\,\,\,\Sigma_s=k\,{\rm
Tr}(T_s^\dagger\,Q)\nonumber\\
M_\varphi&=k\,{\rm Tr}(T_\bullet^\dagger\,Q_\psi)\,,\label{allthat}
\end{align}
where the angular momentum $M_\varphi$ along Z is normalized so that the leading term of $\omega_\varphi$ at spatial infinity reads:
\begin{equation}
\omega_\varphi=\dots+2\,M_\varphi\,\frac{\sin^2(\theta)}{r}\,.\label{omegaMphi}
\end{equation}
 The constant $m$ coincides with the  ADM-mass $M_{ADM}$ when $\phi_0\equiv O$, while $n_{NUT}$ is the NUT-charge, $\Sigma_s$ the scalar charges and  $\Gamma^M=(p^\Lambda,\,q_\Lambda)$ the  electric and magnetic charges:
 \begin{equation}
 \Gamma^M=\frac{1}{4\pi}\int_{\Sigma} \mathbb{F}^M\,\label{elmagch}.
 \end{equation}
Both $Q$ and $Q_\psi$ are matrices in the Lie algebra $\mathfrak{g}$ of $G$. More specifically they both belong to $T_{\phi_0}({\Scr M}_{scal})$. When $\phi_0=O$ the two matrices belong to $T_{O}({\Scr M}_{scal})$ which is isomorphic to the coset space ${\mathfrak{K}^*}$.
\par
Being $G$ the global symmetry group of the effective model, a generic element $g$ of it maps a solution $\phi^I(r,\theta)$ into another solution $\phi^{\prime\,I}(r,\theta)$ according to the matrix equation:
\begin{equation}
\mathcal{M}(\phi^{\prime I}(x^i))=g\,\mathcal{M}(\phi^{I}(x^i))\,g^\dagger\,.\label{sgt}
\end{equation}
From their definitions (\ref{nc}), (\ref{qupsi}), and from (\ref{sgt}), it follows that $Q$ and $Q_\psi$ transform under the
 adjoint action of $G$ as:
\begin{equation}
\forall g\in G\,\,:\,\,\,\,Q\rightarrow Q'=(g^{-1})^\dagger\,Q\,g^{\dagger}\,\,\,;\,\,\,\,\,Q_\psi\rightarrow
 Q'_\psi=(g^{-1})^\dagger\,Q_\psi\,g^{\dagger}\,.\label{QQpsitra}
\end{equation}
Eq.s (\ref{allthat}), and the last one
 in particular, allow to compute the angular momentum of the transformed solution without having to explicitly derive the
 latter from (\ref{sgt}) and to compute the corresponding Komar integral on it. Static solutions are characterized by the G-invariant condition $Q_\psi=0$. \par
 A generic stationary solution with asymptotic values $\phi_0=(\phi_0^I)$ of the scalar fields and Noether charge $Q$ can be mapped by means of $\mathbb{L}(\phi_0)^{-1}\in G/H^*$ into a solution with boundary values of the scalars corresponding to  the origin $O$ and Noether charge
 \begin{equation}
 Q_O=\mathbb{L}(\phi_0)^{\dagger}\,Q\,\mathbb{L}(\phi_0)^{-\dagger}=\frac{1}{4\pi}\int_{S_2^\infty}{}^{*_3}{\Scr P}\in T_{O}[{\Scr M}_{scal}]\sim {\mathfrak{K}^*}\,,
 \end{equation}
 where ${\Scr P}\equiv\partial_i\phi^I\,P_I\,dx^i$ is the pull-back of the vielbein 1-form $P$ on the solution.
 The electric and magnetic charges $\Gamma_O^M$ of this solution can be expressed in terms of the central and matter charges ${\bf  Z}(\phi_0,\,\Gamma)^M$ of the original one, computed at radial infinity. Indeed from (\ref{allthat}) and the structure of the solvable algebra it follows that:
 \begin{equation}
 \Gamma_O^M=(L_{4}(\phi^s_0))_N{}^M\,(\Gamma^N-n_{NUT}\,\mathcal{Z}_0^N)={\bf  Z}(\phi^s_0,\,\Gamma)^M-n_{NUT}\,(L_{4}(\phi_0))_N{}^M\mathcal{Z}_0^N\,.
 \end{equation}
 If $n_{NUT}=0$, $ \Gamma_O^M$ coincide with ${\bf  Z}(\phi^s_0,\,\Gamma)^M$, which thus represent the components of $Q_O$ along the generators $T_M$ or, equivalently, along the compact generators $K_M$ in ${\mathfrak{K}^*}$ defined in (\ref{KMdef}). The charges $\Gamma_O^M$ therefore naturally transform in the representation ${\bf R}'$ of $H_c$.

 We can then characterize an axisymmetric, single-center solution $\phi^I(r,\,\theta)$ by the set of data $[\phi_0,\,Q,\,Q_\psi]$ consisting in the  corresponding values $\phi_0^I$ of the scalar fields at radial infinity and the matrices $Q,\,Q_\psi\, \in  T_{\phi_0}[{\Scr M}_{scal}]$.\par \emph{We say that two single-center solutions $[\phi_0^{(1)},\,Q^{(1)},\,Q_\psi^{(1)}]$, $[\phi_0^{(2)},\,Q^{(2)},\,Q_\psi^{(2)}]$ belong to the same $G$-orbit if the matrices $Q^{(i)}_O,\,Q_{\psi\,O}^{(i)}$, $i=1,2$, describing the corresponding solutions with asymptotic values of the scalars at the origin $O$:
 \begin{align}
 Q^{(i)}_O &\equiv \mathbb{L}(\phi_0^{(i)})^\dagger\,Q^{(i)}\,\mathbb{L}(\phi_0^{(i)})^{-\dagger}\in T_{O}[{\Scr M}_{scal}]\sim \mathfrak{K}\,,\label{orbQ}\\
 Q_{\psi\,O}^{(i)}&\equiv \mathbb{L}(\phi_0^{(i)})^\dagger\,Q^{(i)}_\psi\,\mathbb{L}(\phi_0^{(i)})^{-\dagger}\in T_{O}[{\Scr M}_{scal}]\sim \mathfrak{K}\,,
 \end{align}
  belong to the same $H^*$-orbit:
  \begin{equation}
  Q^{(1)}_O\in (H^*)^{-1}\, Q^{(2)}_O\,H^*\,\,;\,\,\,\,Q_{\psi\,O}^{(1)}\in (H^*)^{-1}\, Q_{\psi\,O}^{(2)}\,H^*\,.
  \end{equation} }
  Stationary axisymmetric black holes can thus be grouped, with respect to the action of $G$, in orbits which are in one to one correspondence with the  $H^*$-orbits of the total Noether charge matrices $Q_O,\,Q_{\psi\,O}$, referred to the origin, according to Eq. (\ref{orbQ}).
This provides a complete classification of the single-center solutions and allows an intrinsic algebraic characterization of their physical properties, like \emph{regularity}, supersymmetry etc..\par
Multicenter solutions are also characterized by the Noether charge matrices $Q_k$ associated with each center. If $\Sigma_k$ denotes the  2-cycle surrounding the $k^{th}$- center, using the sigma-model field equations (\ref{sigmodfeq}) it follows that \cite{Bossard:2009we,bossard2}:
\begin{equation}
Q=\frac{1}{4\pi}\,\int_{\Sigma}{}^{*_3}J=\sum_k \frac{1}{4\pi}\,\int_{\Sigma_k}{}^{*_3}J=\sum_k Q_k\,.
\end{equation}
In terms of $Q_k$ the relevant quantities associated with each constituent of the system are computed using (\ref{allthat}).
Note that, as opposed to the total Noether charge matrix $Q$ which can always be mapped into an element of ${\mathfrak{K}^*}$ by means of
the coset representative computed at spatial infinity $\mathbb{L}(\phi_0)$, $Q_k$ are evaluated by an integration over a 2-cycle $\Sigma_k$ along which the scalar fields are not constant and thus it cannot in general be $G$-rotated into ${\mathfrak{K}^*}$. This would not be the case if the other centers were infinitely far away from the $k^{th}$ one, so that $\Sigma_k$ can be chosen to be a sphere at spatial infinity on which the scalar fields are uniform and $Q_k$ can be consistently mapped into an element in  ${\mathfrak{K}^*}$ belonging to some characteristic $H^*$-orbit. This amounts to associating with each center an ``intrinsic'' matrix $Q_k^{(0)}$, and thus an ``intrinsic'' $H^*$-orbit, which encodes its properties
 when the center is isolated from the others, namely in the limit of vanishing interactions. If $R_{ij}=|{\bf x}_i-{\bf x}_j|$ denotes the distance between the $i^{th}$ and the $j^{th}$ center in the solution, located at ${\bf x}_i$ and ${\bf x}_j$, respectively, we can thus define:
 \begin{equation}
 Q_k^{(0)}\equiv \lim_{R_{ki}\rightarrow \infty} Q_k\,\,\,\Leftrightarrow \,\,\,\,Q_k=Q_k^{(0)}+\sum O\left(\frac{1}{R_{ik}}\right)\,,\label{intrinsicQ}
 \end{equation}
 where the limit amounts to sending all the centers, different from the $k^{th}$-one under consideration, to spatial infinity. The $O\left(\frac{1}{R_{ik}}\right)$ terms represent the effect of the interactions. If the point $\phi_0$ on the moduli space at infinity is chosen to coincide with the origin, then $Q_k^{(0)}$, for any $k$, belongs to  ${\mathfrak{K}^*}$. This matrix only serves the purpose of characterizing the ``intrinsic'' regularity of each center: \emph{We shall say that the $k^{th}$ center of a solution is regular iff the corresponding $Q_k^{(0)}$ is associated with the $H^*$-orbit of a regular single center solution.} If a center is ``intrinsically'' singular, namely if  $Q_k^{(0)}$ belongs to an $H^*$-orbit corresponding to singular solutions, the corresponding single-center solution features singularities at finite $r$ which are unlikely to be offset by the interaction terms in the full multi-centered solution. Therefore we take as necessary condition for regularity that each center of a solution be ``intrinsically'' regular \cite{bossard2}. This is not a sufficient condition since, for instance, \emph{if the distance between the centers is small enough, interaction terms may produce singularities. We shall illustrate this in specific examples.}

\paragraph{Extremal solutions.} With the exception of the extremal Kerr solution and generalizations thereof, extremal solutions are characterized by a \emph{nilpotent} $Q$ \cite{pioline,Gaiotto:2007ag} and $Q_\psi$ \cite{Andrianopoli:2013kya,Andrianopoli:2012ee}. In fact the one-forms ${\Scr P}$ and ${\Scr W}$ take value in a nilpotent subalgebra of $\mathfrak{g}$ \cite{Bossard:2009at,Bossard:2009we,bossard2}. This implies ${\rm Tr}({\Scr P}_i{\Scr P}_j)=0$ and thus, from (\ref{eeq}), that $\mathcal{R}_{ij}=0$, that is $g_{ij}$ is the flat metric. Single-center, extremal solutions feature a characteristic     \emph{attractor behavior} at the horizon \cite{attractor}.
Known examples of multi-center extremal solutions are the \emph{BPS} ones \cite{BPSmc}, \emph{almost-BPS} \cite{Goldstein:2008fq,Bena:2009ev,Dall'Agata:2010dy,Galli:2010mg} and the \emph{composite non-BPS} ones \cite{bossard2}. Being $Q$ nilpotent, these extremal solutions fall in \emph{nilpotent orbits} of $Q_O$ with respect to the action of $H^*$. An intriguing feature of these (composite) black hole solutions is that they are  determined by systems of graded, first-order differential equations which are exactly solvable in a iterative way \cite{bossard2}.
In \cite{bossard2} a classification of the solutions  in terms of nilpotent orbits with respect to the complexification $H^{*\,\mathbb{C}}$ of $H^*$ was pursued. Such orbits are uniquely associated with nilpotent subalgebras $\mathfrak{n}$ of $\mathfrak{g}$, of which $Q$ is an element and which in turn determine the relevant graded system of first-order equations.
\paragraph{Regularity.} In order for a solution to be regular the following conditions should be satisfied:
\begin{itemize}
\item[i)]{\emph{Absence of Dirac-Misner (DM) string singularities}.\footnote{ A Dirac-Misner string is a gravitational Dirac string associated with the vector component $\omega$ of the four-dimensional metric.}
The presence of these objects extending to spatial infinity has to be excluded if we require, as we do here, the solution to be globally asymptotically flat. This does not exclude strings connecting the centers. However it is known that the region close to a DM string generically features unwanted closed time-like curves (CTC). We shall require the absence of any DM string in the four-dimensional background.\footnote{See \cite{Clement:2015cxa} for arguments in favor of relaxing this condition.} In an $N$-center solution this condition amounts to requiring the vanishing of the NUT-charge for each center or, equivalently, to the condition:
    \begin{align}
    n_{NUT}&\equiv \sum_{k=1}^Nn_{NUT,\,k}=0\,,\label{nonut1}\\
      n_{NUT,\,k}&=0\,\,,\,\,\,k=1,\dots, N-1\,,\label{nonut2}\\
    \end{align}
    where $ n_{NUT,\,k}$ are obtained applying the formulas (\ref{allthat}) to the Noether charges $Q_k$ at each center.
   The first implies the absence of a DM string stretching to infinity and is required by the condition of global asymptotic flatness. The last $N-1$ conditions exclude DM strings connecting the centers. They encode the so-called \emph{bubble equations} \cite{bossard2,BPSmc,Bena:2009ev} and constrain the distances between the constituent black holes relating them to the asymptotic data at spatial infinity;
     }
  \item[ii)]{\emph{No curvature singularities.} This in particular constrains the warp function $e^{-4U}$ to be everywhere positive. As pointed out earlier, we do not want to rule out  \emph{small} black holes, i.e. extremal solutions with vanishing horizon area, or composites thereof. These solutions feature a curvature singularity at the centers where the small black holes are located. We therefore require the warp factor $e^{-4U}$ to diverge near the center as $1/\Sigma_i^a$, $\Sigma$ being the distance from the $i^{th}$ center, with $a\le 4$, $a=4$ corresponding to a large black hole solution. }
\end{itemize}
Conditions $i)$ will be imposed directly on $Q$ and $Q_k$, $k=1,\dots, N-1$, see \cite{bossard2} while as we shall also illustrate in the explicit examples, a strong necessary condition for $ii)$ requires choosing the ``intrinsic'' matrices $Q_k^{(0)}$, characterizing each center, in the $H^*$-orbits associated with regular single-center solutions, \emph{provided} the distances between the centers be not too small. Therefore this is the point of our analysis where the study of the $H^*$-orbits enters the game. Alternatively condition $ii)$ can be imposed directly on the specific solution, whose expression may however be rather involved. As pointed out in the introduction, requiring regularity in the asymptotic regions near the centers and at spatial infinity is not enough. We shall indeed illustrate in specific examples that asymptotically well behaved solutions exist which feature singularities at finite $r$ and CTCs. Such solutions are obtained by choosing the matrix $Q_k^{(0)}$ of some of the centers in a ``wrong'' $H^*$-orbit, though being in the $H^{*\,\mathbb{C}}$ orbit which contains regular solutions with the same electric-magnetic charges.

As mentioned in the Introduction, the general problem of classifying $H^*$-orbits of nilpotent matrices in $\mathfrak{K}^*$ was pursued in a number of $\mathcal{N}=2$ models in \cite{Fre:2011uy,Fre:2011ns,Chemissany:2012nb} using a somewhat direct computational method. This approach was put on formal mathematical grounds in \cite{Dietrich:2016ojx} and applied to the STU model. It will be reviewed in Sect. \ref{sec3}. Let us now focus on the specific model under consideration.

\section{The STU Model and the $D=3$ Effective Description}\label{sec2}
\label{stumodel}
The STU model is an $\mathcal{N}=2$ supergravity coupled to three vector multiplets ($n_s=6,\,n_v=4$) and with:
\begin{equation}
{\Scr M}^{(4)}_{scal}=\frac{G_4}{H_4}=\left(\frac{{\rm SL}(2,\mathbb{R})}{{\rm SO}(2)}\right)^3\,.
\end{equation}
This manifold is a complex spacial K\"ahler space spanned by three complex scalar fields $\{z_i\}=\{\epsilon_i-i\,e^{\varphi_i}\}=\{s,t,u\}$, $i=1,2,3$.
The $D=4$ scalar metric for the STU model reads
\begin{equation}
dS^2_4=g_{rs}\,d\phi^sd\phi^r=2\,g_{i\bar{\jmath}}dz^i d\bar{z}^{\bar{\jmath}}=-2\,\sum_{i=1}^3\frac{dz_i
d\bar{z}_{\bar{\imath}}}{(z_i-\bar{z}_{\bar{\imath}})^2}=\sum_{{\bf I}=1}^3 e_i{}^{\bf I}\bar{e}_{\bar{\imath}}{}^{\bf I}\,dz^i\,d\bar{z}^{\bar{\imath}}\,.
\end{equation}
We also consider the real parametrization $\{\phi^s\}=\{\epsilon_i,\,\varphi_i\}$, related to the complex one by:
$z_i=\epsilon_i-i\,e^{\varphi_i}$. The K\"ahler potential has the simple form: $e^{-K}=8\,e^{\varphi_1+\varphi_2+\varphi_3}$. In the chosen symplectic frame (i.e. the special coordinate frame originating from Kaluza Klein reduction from $D=5$), the special geometry of ${\Scr M}^{(4)}_{scal}$ is characterized by a holomorphic prepotential $\mathcal{F}(z)=z_1 z_2 z_3$. The holomorphic $\Omega^M(z)$ section of the symplectic bundle reads:
\begin{equation}
\Omega^M(z)=\{1,z_1,z_2,z_3,-z_1 z_2 z_3,z_2 z_3,z_1 z_3,z_1
   z_2\}\,,\label{omega}
\end{equation}
while the covariantly holomorphic section is given by $V^M(z,\bar{z})=e^{\frac{K}{2}}\,\Omega^M(z)$.

Upon timelike reduction to $D=3$ the scalar manifold has the form $G/H^*$ with $G={\rm SO}(4,4)$ and $H^*={\rm SO}(2,2)^2$. We describe the generators of $\mathfrak{g}=\mathfrak{so}(4,4)$ in terms of Cartan $H_\alpha$ and shift generators $E_{\pm\alpha}$
in the fundamental representation, with the usual normalization convention:
\begin{equation}
[H_\alpha,\,E_{\pm \alpha}]=\pm 2\,E_{\pm \alpha}\,\,\,;\,\,\,\,[E_\alpha,\,E_{-\alpha}]=H_\alpha\,.
\end{equation}
In our notation $E_{-\alpha}=E_\alpha^\dagger=E_\alpha^T$ (being all matrices real). The positive roots of $\mathfrak{g}$ split into: the root $\alpha_0$ of the Ehlers subalgebra $\mathfrak{sl}(2,\mathbb{R})_E$ commuting with the algebra $\mathfrak{g}_4$ of $G_4$ inside $\mathfrak{g}$; the roots $\boldsymbol{\alpha}_i,\,(i=1,2,3)$, which coincide with the simple roots of $\mathfrak{g}_4$ when restricted to its Cartan subalgebra, and eight roots $\gamma_M$, $M=1,\dots,8$.\footnote{We denote the roots of $\mathfrak{g}_4$ by boldface Greek letters, to distinguish them form those of $\mathfrak{g}$. } The special coordinate parametrization of ${\Scr M}^{(4)}_{scal}$ corresponds to a solvable parametrization of the manifold in which the real coordinates $(\phi^s)=(\epsilon_i,\,\varphi_i)$ are parameters of a solvable Lie algebra generated by $(T_s)=(E_{\boldsymbol{\alpha}_i},\,\frac{1}{2}\,H_{\boldsymbol{\alpha}_i})$, $i=1,2,3$. The coset representative $L_4$ is an element of the corresponding solvable group defined by the following exponentialization prescription:
\begin{equation}
L_4(\phi^s)=\exp(\phi^s\,T_s)=\prod_{i=1}^3 e^{\epsilon_i E_{\boldsymbol{\alpha}_i}}e^{\varphi_i \frac{H_{\boldsymbol{\alpha}_i}}{2}}\,.\label{L4}
\end{equation}
The solvable (or Borel) subalgebra $Solv={\rm Span}(T_\mathcal{A})$, $\{T_\mathcal{A}\}=\{H_0,\,T_\bullet,\,T_s,\,T_M\}$ of $\mathfrak{g}$ used to define the parametrization of ${\Scr M}_{scal}$ in terms of the $D=3$ scalars $\phi^I$ through the coset representative (\ref{cosetr3}), is defined by the identification:
\begin{equation}
H_0=\frac{H_{\alpha_0}}{2}\,\,;\,\,\,\,T_\bullet=E_{\alpha_0}\,\,\,;\,\,\,\,T_M=E_{\gamma_M}\,.
\end{equation}
The symplectic representation of $T_s$ in the duality representation ${\bf R}={\bf (2,2,2)}$ of $G_4$ is defined through their adjoint action on $T_M$: $[T_s,\,T_M]=-T_{s\,M}{}^N\,T_N$. In order to reproduce the form of the $T_{s\,M}{}^N$ in the chosen special coordinate frame (\ref{omega}), the generators $T_M$ corresponding to the roots $\gamma_M$, have to be ordered according to (\ref{gammaord}). In this basis, the symplectic representation of  $L_4=(L_{4\,M}{}^N)$ defined in (\ref{L4}) allows to define the matrix $\mathcal{M}_{(4)}$:
\begin{equation}
\mathcal{M}_{(4)\,MN}=-\sum_{P=1}^8(L_{4\,M}{}^P)(L_{4\,N}{}^P)\,.\label{M4MN}
\end{equation}
The mathematical details of the model, including the explicit matrix form of $\mathcal{M}_{(4)\,MN}$, are given in Appendix \ref{mdstu}.\par
The simple roots of the $D_4$ algebra $\mathfrak{g}^\mathbb{C}$ generating $G^\mathbb{C}$, complexification  of $G={\rm SO}(4,4)$, are denoted by $\alpha_1,\dots, \alpha_4$, and their numbering corresponds to the following labeling of the Dynkin diagram:
\begin{center}
\scalebox{1.4}{\begin{picture}(70,17)
  \put(18,0){\circle{6}}
\put(15,4){{\tiny 1}}
  \put(33,0){\circle{6}}
  \put(48,0){\circle{6}}
  \put(33,15){\circle{6}}
\put(45,4){{\tiny 4}}
\put(29,4){{\tiny 2}}
\put(36,15){{\tiny 3}}
  \put(21,0){\line(1,0){9}}
  \put(36,0){\line(1,0){9}}
  \put(33,3){\line(0,1){9}}
\end{picture}}\label{d4}
\end{center}
The simple roots $\boldsymbol{\alpha}_i,\,(i=1,2,3)$ of $\mathfrak{g}_4$ coincide with the $\mathfrak{so}(4,4)$-roots $\alpha_1,\,\alpha_3,\,\alpha_4$. The \emph{STU triality}, which amounts to interchanging the role of the three complex scalars, that is the three factors in ${\Scr M}^{(4)}_{scal}$, is defined by the outer-automorphisms permuting the legs of the $D_4$ diagram, i.e. the roots $\alpha_1,\,\alpha_3,\,\alpha_4$.

The complexification $\mathfrak{H}^{*\,\mathbb{C}}$ of the Lie algebra  ${\mathfrak{H}^*}=\mathfrak{sl}(2,\mathbb{R})^4$ is defined by simple roots which are denoted by $\beta_1,\dots \beta_4$, where
\begin{equation}
\beta_1=-\alpha_0=-(\alpha_1+2\alpha_2+\alpha_3+\alpha_4)\,\,;\,\,\,\,\beta_2=\alpha_1\,\,;\,\,\,\,\beta_3=\alpha_3\,\,;\,\,\,\,
\beta_4=\alpha_4\,.
\end{equation}
We shall refer all the properties of the generators of
$\mathfrak{g}=\mathfrak{so}(4,4)$ to the corresponding matrices in the real fundamental
representation ${\bf 8}$, see Appendix \ref{mdstu2}. The effect of triality is to permute the roots $\beta_2,\,\beta_3,\,\beta_4$.\par

\section{The Issue of Nilpotent Orbits}\label{sec3}
As stressed in Sect. \ref{sec1}, a particularly useful mathematical notion for the study of stationary solutions in the model under consideration is that of $H^*$-orbits in ${\mathfrak{K}^*}$, which provides the appropriate tool for characterizing their physical properties. The orbits of regular Kerr solutions (which include the extremal Kerr solutions), were originally studied in \cite{Breitenlohner:1987dg}. They are characterized by a semisimple $Q_O$, $Q_{\psi,O}$ being $H^*$-conjugate to $Q_O$ \cite{Andrianopoli:2013kya,Andrianopoli:2012ee}. As pointed out earlier, the extremal solutions we shall be dealing with in the present paper are characterized by a nilpotent $Q_O$, $Q_{\psi,O}$ being nilpotent too but in a distinct $H^*$-orbit. The nilpotent $H^*$-orbits describing these solutions can be obtained as singular limits of the Kerr orbit, a general geometric prescription being given in \cite{Andrianopoli:2013kya,Andrianopoli:2012ee}.\par
Constructing and
classifying  $H^*$-adjoint orbits in  ${\mathfrak{K}^*}$, with
particular reference to the nilpotent ones, amounts to grouping the elements of
${\mathfrak{K}^*}$ in orbits $\mathcal{O}$ (or conjugacy classes) with
respect to the adjoint action of $H^*$:
\begin{equation}
k_1,\,k_2\,\in\mathcal{O}\subset
\mathfrak{K}\,\Leftrightarrow\,\,\,\exists h\in
H^*\,\,:\,\,\,k_2=h^{-1}\,k_1\,h\,.
\end{equation}
A valuable approach to this task makes use of the theory of adjoint
orbits within a real Lie algebra $\mathfrak{g}$ with respect to the action of the
Lie group $G$ it generates \cite{collingwood}. In this respect the
Kostant-Sekiguchi theorem \cite{collingwood} is of invaluable help
since it allows a complete classification of such orbits.\footnote{The Kostant-Sekiguchi theorem refers to the real orbits with respect to the action of the transformations in the identity sector of $G$.}
 This
is however not enough for our purposes, since we are interested in
the adjoint action of  $H^*$ on ${\mathfrak{K}^*}$ and a same $G$-orbit
will in general branch into several $H^*$-orbits. To understand this splitting
one may use $H^*$-invariant quantities which are not $G$-invariant,
such as $\gamma$-labels \cite{Kim:2010bf} or tensor classifiers
\cite{Fre:2011uy}. These, however, cannot guarantee by themselves a
complete classification. In \cite{Dietrich:2016ojx} we used two different approaches to
such a classification: One, which was originally devised in
\cite{Fre:2011ns} and a new one which generalizes the original Vinberg method \cite{vinberg2} and makes use of the notion of \emph{carrier algebras}. In this section we review the former method and discuss the results, referring to \cite{Dietrich:2016ojx} for the mathematical details.\par
We start from the notion of standard triple
associated with a nilpotent element $e$ of a real Lie algebra
$\mathfrak{g}$: According to the Jacobson-Morozov theorem
\cite{collingwood}, such element can be thought of as part of a
\emph{standard triple} of $\mathfrak{sl}(2,\mathbb{R})$-generators
$\{e,\,f,\,h\}$, satisfying the following commutation relations:
\begin{equation}
[h,e]=2\,e\,\,;\,\,\,[h,f]=-2\,f\,\,;\,\,\,[e,\,f]=h\,.\label{sttriple}
\end{equation}
If we were interested in
the orbits in the complexification $\mathfrak{g}^\mathbb{C}$ of
$\mathfrak{g}$ with respect to the adjoint action of the group
$G^\mathbb{C}$ it generates, different $G^\mathbb{C}$-nilpotent
orbits correspond to inequivalent embeddings of
$\mathfrak{sl}(2,\mathbb{R})={\rm Span}(e,\,f,\,h)$ inside
$\mathfrak{g}$, and these would correspond to different branchings
of a given representation of $G^\mathbb{C}$ with respect to the
${\rm SL}(2,\mathbb{R})$-subgroup. These different branchings are
uniquely characterized by the spectrum of the adjoint action of $h$
on $\mathfrak{g}^\mathbb{C}$. Such spectrum is conveniently
described by fixing a Cartan subalgebra $\mathfrak{h}$ of
$\mathfrak{g}^\mathbb{C}$, in which $h$, being a semisimple
generator, can be rotated by means of a
$G^\mathbb{C}$-transformation, and evaluating the values of the
simple roots $\alpha_i$ of $\mathfrak{g}^{\mathbb{C}}$, associated
with $\mathfrak{h}$, on $h$:
\begin{equation}
\mbox{$G^\mathbb{C}$-orbit of
$e$}\,\leftrightarrow\,\mbox{$G^\mathbb{C}$-orbits of
$h$}\,\leftrightarrow\,\{\alpha_i(h)\}\,,
\end{equation}
where the integers $\alpha_i(h)$ are conventionally evaluated after
$h$ is rotated in the fundamental domain and can only have values
$0,1,2$. They  are called $\alpha$-labels and provide a complete
classification of the nilpotent $G^\mathbb{C}$-adjoint orbits in
$\mathfrak{g}^\mathbb{C}$. \par
 We can always rotate, by means of $G$, the standard triple associated with a nilpotent element into a \emph{Cayley triple} characterized by the property that $h$ and $e+f$ are non-compact, $h, e+f\in \mathfrak{K}$, while $e-f$ is compact, $e-f \in \mathfrak{H}$, see footnote 6 for the definition of $\mathfrak{H}$ and $\mathfrak{K}$. Then the problem of classifying nilpotent orbits in the real Lie algebra $\mathfrak{g}$ with respect to the adjoint action of $G$ can also be reduced to that of classifying some characteristic semisimple
 generators. Such generator  associated with the triple of $e$,  is no longer $h$, but rather the non-compact generator
 $i\,(e-f)$. This is a consequence of the \emph{Kostant-Sekiguchi (KS) theorem} whose content we briefly recall below. Having denoted  by $H$ the maximal compact subgroup of $G$, generated by $\mathfrak{H}$,
we denote by  $H^{\mathbb{C}}$  its complexification, generated by the complexification $\mathfrak{H}^\mathbb{C}=\mathfrak{H}+i\,\mathfrak{H}$ of $\mathfrak{H}$.\footnote{Clearly the complexifications $\mathfrak{H}^\mathbb{C}=\mathfrak{H}+i\,\mathfrak{H}$ of $\mathfrak{H}$ and $\mathfrak{H}^{*\,\mathbb{C}}={\mathfrak{H}^*}+i\,{\mathfrak{H}^*}$ of ${\mathfrak{H}^*}$ are isomorphic in $\mathfrak{g}^\mathbb{C}$.}
The Kostant-Sekiguchi (KS) theorem defines a one-to-one
 correspondence between $G$-orbits of a nilpotent element $e$ of
 $\mathfrak{g}$, and the orbits under the adjoint action of
 $H^{\mathbb{C}}$ on $\mathfrak{K}^{\mathbb{C}}$,  where the latter is  the complexification of the space of non-compact $\mathfrak{g}$-generators $\mathfrak{K}$ defined by the Cartan decomposition (\ref{cdec}): $\mathfrak{K}^\mathbb{C}=\mathfrak{K}+i\,\mathfrak{K}$. These orbits are in turn in one-to-one correspondence with the $H^{\mathbb{C}}$-adjoint orbit of the element $(e-f)$ of $\mathfrak{H}$. Such orbits are completely defined by
 the (real) spectrum of the adjoint action of $i\,(e-f)$ over $\mathfrak{H}^{\mathbb{C}}$,
or, equivalently, by the embedding of the same semisimple element
within a suitable Cartan subalgebra $\mathfrak{h}_{\mathfrak{H}^{\mathbb{C}}}$ of
$\mathfrak{H}^{\mathbb{C}}$. If $\beta_k$ are the simple roots of
$\mathfrak{H}^{\mathbb{C}}$, referred to $\mathfrak{h}_{\mathfrak{H}^{\mathbb{C}}}$, such embedding is defined by the so
called $\beta$-labels, which are the values $\beta_k(i(e-f))$. In
summary the KS theorem states the following correspondence:
\begin{equation}
\left[\mbox{$G$-Orbit of
$e$}\right]\,\longleftrightarrow\,\left[\mbox{${H}^{\mathbb{C}}$-orbits of
$i(e-f)$ in $\mathfrak{H}^\mathbb{C}$}\right]\,\leftrightarrow\,\{\beta_k(i\,(e-f))\}\,=\,\mbox{$\beta$-labels}\,,
\end{equation}
where the labels $\beta_k(i\,(e-f))$ are conventionally evaluated once
$i\,(e-f)$ is rotated into the fundamental domain and are
non-negative integers.  The $\alpha,\beta$-labels are classified in
the mathematical literature, for all Lie groups
\cite{collingwood}.\par Let us now come back to our original
problem: What are the possible $H^*$-orbits of nilpotent elements
$e$ in $\mathfrak{K}^*$? We know that $e$ is part of a standard triple.
Since $e$ is in ${\mathfrak{K}^*}$, compatibility of (\ref{sttriple})
with (\ref{HKrels}) requires that $h\in {\mathfrak{H}^*}$ and $f\in
{\mathfrak{K}^*}$. In particular $h$ is a semisimple, non-compact
element of ${\mathfrak{H}^*}$ ($\tau(h)=-h^T=-h$, $\sigma(h)=h$), and thus can be
chosen (modulo $H^*$-transformations of the triple) within a given
maximally non-compact Cartan subalgebra $\mathfrak{h}_{{\mathfrak{H}^*}}$ of
${\mathfrak{H}^*}$. Clearly different $G^{\mathbb{C}}$ or $G$-orbits
(uniquely defined by $\alpha,\,\beta$-labels, respectively)
correspond to different $H^*$-orbits. A same $G$-orbit will
branch with respect to the action of $H^*$. In \cite{Kim:2010bf},
the case $G={\rm G}_{2(2)}$, $H^*={\rm SL}(2,\mathbb{R})^2$ was
studied in detail, and the so-called $\gamma$-labels were introduced
to distinguish between different $H^*$-orbits. The notion of
$\gamma$-labels is similar to that of $\beta$-labels. Let us denote by $\mathfrak{H}^{*\,\mathbb{C}}={\mathfrak{H}^*}+i\,{\mathfrak{H}^*}$ the complexification of ${\mathfrak{H}^*}$, generating the subgroup ${H}^{*\,\mathbb{C}}$ of $G^\mathbb{C}$. The $\gamma$-labels identify
the $H^{*\,\mathbb{C}}$-orbits of $e$ in $\mathfrak{K}^{*\,\mathbb{C}}$ or, equivalently, of $h$ within $\mathfrak{H}^{*\,\mathbb{C}}$ and can thus either be described in terms of the spectrum of the
adjoint action of $h$ on $\mathfrak{H}^{*\,\mathbb{C}}$, or in terms of the values
of the simple roots $\beta'_k$ of $\mathfrak{H}^{*\,\mathbb{C}}$ (referred now to the Cartan subalgebra
$\mathfrak{h}_{\mathfrak{H}^{*\,\mathbb{C}}}$ of $\mathfrak{H}^{*\,\mathbb{C}}$) on $h$, taken in the fundamental domain:
\begin{equation}
\left[\mbox{$H^{*\,\mathbb{C}}$-orbits of
$e$ in $\mathfrak{K}^{*\,\mathbb{C}}$}\right]\,\leftrightarrow \,\left[\mbox{$H^{*\,\mathbb{C}}$-orbits of
$h$ in $\mathfrak{H}^{*\,\mathbb{C}}$}\right]\,\leftrightarrow \,\{\beta'_k(h)\}\,=\,\mbox{$\gamma$-labels}\,\,.
\end{equation}
These quantities are clearly invariant with respect to the adjoint
action of $H^*$ (and in general of its complexification
$H^{*\,\mathbb{C}}$) on the whole triple and in particular on $h$, and
thus \emph{different $\gamma$-labels correspond to different $H^{*\,\mathbb{C}}$-orbits
of $e$ in $\mathfrak{K}^{\mathbb{C}}$.} Clearly the sets of all possible $\beta$- and
$\gamma$-labels coincide.\par Summarizing, a nilpotent element $e$ in ${\mathfrak{K}^*}$ can be simultaneously characterized as belonging to a certain $G$-orbit in $\mathfrak{g}$ and to an $H^{*\,\mathbb{C}}$-orbit in $\mathfrak{K}^{*\,\mathbb{C}}$ through its $\beta$ and $\gamma$-labels, respectively. \par In \cite{bossard2} a systematic study of black hole solutions to the STU model was done in
terms of the  $H^{*\,\mathbb{C}}$-orbits inside $\mathfrak{K}^{\mathbb{C}}$. It was shown that the form of the first order
system of equations governing the (composite) solutions only depends on this orbit, namely on the corresponding $\gamma$-label.
 There is no mathematical property
guaranteeing  that $\gamma$-labels, together with the $\alpha$ and
$\beta$ ones, provide a complete classification of the
$H^*$-nilpotent orbits in $\mathfrak{K}$.  And indeed there are counterexamples, as it is shown in \cite{Chemissany:2012nb} and in the present paper: Different $H^*$- orbits sharing the same
$\alpha,\,\beta,\,\gamma$-labels.\par Let us now review the
constructive procedure introduced in \cite{Fre:2011ns}. Given a
nilpotent element $e$ of $\mathfrak{K}^*$ one can prove, see \cite{Dietrich:2016ojx}, that there exists an element $e'$ in the same
$H^*$-orbit, whose triple $\{e',f',h'\}$ has the property that
$f'=e'^T$. We
shall then restrict to triples of this kind.\par The neutral element
$h$ of a  triple $\{e,f,h\}$, should fall in one of the $H^*$-orbits
uniquely defined by the $ \gamma$-labels. We then take a
representative $h$ of each of these orbits and solve the matrix
equations in the unknown $e$:
\begin{eqnarray}
[h,e]\,&=& 2e\, \label{eq1},\\
\small[e,e^T\small] &=& h\,.\label{eq2}
\end{eqnarray}
Using a MATHEMATICA code, for each $h$ we find a set of solutions to
(\ref{eq1}), (\ref{eq2}). We group these solutions under the action
of the compact part $Z_{H^*}^{comp.}[h]$ of the little group $Z_{H^*}[h]$ of $h$ in $H^*$. In
all cases we could find that solutions which were not connected by
the adjoint action of $Z_{H^*}^{comp.}[h]$, could be distinguished by
$H^*$-invariant quantities.  Instances of such quantities are the signatures of
certain symmetric covariant (or contravariant) $H^*$-tensors, called \emph{ tensor
classifiers}, defined in \cite{Dietrich:2016ojx,Fre:2011ns,Chemissany:2012nb}. In principle, if one is able to find $H^*$-invariant quantities capable of distinguishing between
solutions $e$ to (\ref{eq1}), (\ref{eq2}) which share the same $\beta$-label (i.e. fall in the same $G$-orbit) but are not related by $Z_{H^*}^{comp.}[h]$, the resulting classification of the $H^*$-orbits can be claimed to be complete. This is the case of our present analysis.
In \cite{hvl} a different strategy for listing the nilpotent orbits of a symmetric
pair was developed; this method however involves computational problems which make it difficult to be implemented by some practical algorithm \cite{Dietrich:2016ojx}.

The complex $G^\mathbb{C}$ orbit, besides the $\alpha$-label, is also described by the branching of the fundamental representation of ${\rm SO}(4,4)$ with respect to the ${\rm SL}(2,\mathbb{R})$ subgroup generated by the corresponding triple $\{e,\,f,\,h\}$.\par
Let us enter the details of the particular model we are considering. As discussed above, the nilpotent $H^*$-orbits in ${\mathfrak{K}^*}$ are characterized by their $\alpha$-label, $\gamma$ and $\beta$-labels. When these are not enough to identify the orbit we use additional labels $\delta^{(1)},\delta^{(2)},\,\dots$. We shall use the following notation: For each $\alpha$-label $\alpha^{(\ell)}$  we denote by $\beta^{(\ell;\,k)},\,\gamma^{(\ell;\,k)}$, the corresponding sets of $\gamma$ and $\beta$-label, the range of the index $k$ depending on the  $\alpha$-label. For the sake of notational simplicity we shall omit the reference to the $\alpha$-label in the suffix of the $\gamma$ and $\beta$-ones when there is no ambiguity.

In the model under consideration there are eleven $\alpha$-labels, each of which is described by a weighted Dynkin diagram:\begin{center}
$
\alpha=(n_1,n_2,n_3,n_4)\,\,\equiv\,\,$\scalebox{1}{\begin{picture}(70,17)
  \put(18,0){\circle{6}}
\put(15,4){{\tiny $n_1$}}
  \put(33,0){\circle{6}}
  \put(48,0){\circle{6}}
  \put(33,15){\circle{6}}
\put(45,4){{\tiny $n_4$}}
\put(28,-8){{\tiny $n_2$}}
\put(36,15){{\tiny $n_3$}}
  \put(21,0){\line(1,0){9}}
  \put(36,0){\line(1,0){9}}
  \put(33,3){\line(0,1){9}}
\end{picture}}\end{center} where $n_i=0,1,2$ and are listed in table.\par
The $\beta$ and $\gamma$ are described in terms of a weighted extended Dynkin diagram (referred to different Cartan subalgebras of $\mathfrak{H}^{*\,\mathbb{C}}$ inside $\mathfrak{g}$):
    \begin{center}
    $
\beta,\,\gamma=(n_1,n_2,n_3,n_4)\,\,\equiv\,\,$\scalebox{1}{\begin{picture}(70,17)
  \put(18,0){\circle{6}}
\put(15,5){{\tiny $n_2$}}
  \put(33,-15){\circle{6}}
  \put(48,0){\circle{6}}
  \put(33,15){\circle{6}}
\put(45,5){{\tiny $n_4$}}
\put(38,-15){{\tiny $n_1$}}
\put(38,15){{\tiny $n_3$}}
\end{picture}}\end{center}
\vskip 0.3cm
 The list of $\alpha$, $\beta$ and $\gamma$-labels is given in Appendix \ref{labels}.  The complete set of nilpotent orbits is illustrated in Sects. \ref{secdir1}-\ref{secdir2}, see Tables \ref{a1:1}-\ref{a11}, where, for each orbit, a representative is given: for the orbits from $\alpha^{(1)}$ to $\alpha^{(6)}$, the representative is described either in terms of the generating solution of single-center black holes or in a QUbit-basis, while the latter representation only is used to describe the orbits with higher $\alpha$-label.
The  QUbit-description of the elements in ${\mathfrak{K}^*}$, transforming in the representation ${\bf (2,2,2,2)}$ of $H^*$, is defined using the following convention:
\begin{equation}
(\pm,\pm,\pm,\pm)\equiv |\pm\rangle\otimes |\pm\rangle\otimes |\pm\rangle\otimes |\pm\rangle\,,
\end{equation}
$|\pm\rangle$ being a basis of the doublet representation of each ${\rm SL}(2,\mathbb{R})$ factor in $H^*={\rm SO}(2,2)^2\equiv {\rm SL}(2,\mathbb{R})\times_{\mathbb{Z}_2} {\rm SL}(2,\mathbb{R})\times {\rm SL}(2,\mathbb{R})\times_{\mathbb{Z}_2} {\rm SL}(2,\mathbb{R})$.\par

\subsection{A General Overview of the Orbits}\label{generaloverv}
Regular and small extremal static or under-rotating single-center solutions are characterized by a Noether charge $Q$ in the fundamental representation of $G={\rm SO}(4,4)$, which is a step-$k$ nilpotent matrix with $k\le 3$ \cite{Gaiotto:2007ag,Bossard:2009at}:
\begin{equation}
Q^k=0\,\,,\,\,\,\,k\le 3\,.
\end{equation}
The corresponding nilpotent orbits of $Q_O$ are defined by $\alpha$-labels from $\alpha^{(1)}$ to $\alpha^{(6)}$. These orbits can be described in terms of a \emph{generating solution}, which corresponds to a common $H^*$-frame in which the representative $Q_O$ is simplest \cite{Bergshoeff:2008be,Chemissany:2009hq}, and the solution depends on the least number of parameters. More specifically representatives of each of these orbits can be found in a characteristic subspace $\mathfrak{K}^{(N)}$ of $\mathfrak{K}$ of the form:
\begin{equation}
\mathfrak{K}^{(N)}=\bigoplus_{\ell=1}^4[\mathfrak{sl}(2,\mathbb{R})\ominus \mathfrak{sl}(1,1)]_\ell\,.\label{gN1}
\end{equation}
which is defined as follows.  In the case of the STU model, the maximal compact subgroup of $H^*$ is $H_c={\rm SO}(2)^4$ which is the product  of the ${\rm SO}(2)_E$ contained in the Ehlers group ${\rm SL}(2,\mathbb{R})_E$ and $H_4={\rm SO}(2)^3$.
A pointed out in (\ref{sec1}), the spaces $\mathfrak{H}^{*\,(R)}\subset {\mathfrak{H}^*}$ and $\mathfrak{K}^{*\,(R)}\subset {\mathfrak{K}^*}$
both support a representation ${\bf R}'$ of $H_c$ and, by means of transformations in this group, generic generators in these spaces can be rotated into subspaces $\mathcal{J}^{(N)},\,\mathcal{K}^{(N)}$, which are in fact maximal abelian subspaces of  $\mathfrak{H}^{*\,(R)}$ and $\mathfrak{K}^{*\,(R)}$, respectively. The space $\mathcal{J}^{(N)}$ defines the non-compact rank $p$ of $H^*/H_c$ so that
\begin{equation}
p={\rm dim}(\mathcal{J}^{(N)})={\rm dim}(\mathcal{K}^{(N)})={\rm rank}\left(\frac{H^*}{H_c}\right)\,.
\end{equation}
In our case, just as in the case of any $\mathcal{N}=2$ symmetric supergravity with a rank-3 spacial K\"ahler manifold or in the case of maximal and half-maximal supergravities,  $p=4$. The spaces $\mathcal{J}^{(N)},\,\mathcal{K}^{(N)}$ are defined by the \emph{normal form} of the representation ${\bf R}'$ with respect to  $H_c$ \cite{Bergshoeff:2008be,Chemissany:2009hq}.
 The generators $\mathcal{K}_\ell,\,\mathcal{J}_\ell$, $\ell=0,\dots, p-1$, of $\mathcal{K}^{(N)},\,\mathcal{J}^{(N)}$, respectively, have the following form:
 \begin{equation}
\mathcal{J}_\ell=\frac{1}{2}\,(\mathcal{T}_\ell+\mathcal{T}_\ell^T)\,\,;\,\,\,\mathcal{K}_\ell=
\frac{1}{2}\,(\mathcal{T}_\ell-\mathcal{T}_\ell^T)\,,
 \end{equation}
 where $\mathcal{T}_\ell$ are the $T_M$ generators corresponding to four $\gamma_M$ roots which, in the basis $\{H_0,\,\frac{1}{2}\,H_{\boldsymbol{\alpha}_i}\}$, are described by mutually orthogonal 4-vectors $\tilde{\gamma}_\ell$.
 The generators $\mathcal{K}_\ell,\,\mathcal{J}_\ell$ together with $\mathcal{H}_\ell\equiv
\frac{1}{2}[\mathcal{T}_\ell,\,\mathcal{T}_\ell^T]$, generate the $p=4$ commuting $\mathfrak{sl}(2)_\ell={\rm Span}(\mathcal{J}_\ell, \mathcal{H}_\ell,\,\mathcal{K}_\ell)$ algebras in (\ref{gN1}). They  indeed satisfy the following relations:
\begin{align}
[\mathcal{H}_\ell, \,\mathcal{J}_{\ell'}]&=\delta_{\ell\ell'}\,\mathcal{K}_{\ell'}\,\,,\,\,\,[\mathcal{H}_\ell, \,\mathcal{K}_{\ell'}]=\delta_{\ell\ell'}\,\mathcal{J}_{\ell'}\,\,,\,\,\,[\mathcal{J}_\ell, \,\mathcal{K}_{\ell'}]=-\delta_{\ell\ell'}\,\mathcal{H}_{\ell'}\,.
\end{align}
We see that there are two maximal sets of $p=4$ mutually orthogonal roots $\{\tilde{\gamma}_{\ell}\}=\{\gamma_1,\,\gamma_6,\,\gamma_7,\,\gamma_8\}$
  and $\{\tilde{\gamma}_{\ell'}\}=\{\gamma_2,\,\gamma_3,\,\gamma_4,\,\gamma_5\}$, $\ell,\,\ell'=0,\dots, 3$, corresponding to the normal forms of the charge vector
  with non-vanishing charges $\{q_0,\,p^i\}_{i=1,2,3}$ and $\{p^0,\,q_i\}_{i=1,2,3}$, respectively (see (\ref{gammaord})). We shall choose the former set.\par
\emph{Extremal static single-center solutions} were classified in orbits with respect to the $D=4$ global symmetry group $G_4$ in \cite{Bellucci:2006xz}. They are described by geodesics on ${\Scr M}_{scal}$. The affine parameter is $\tau=-1/r$ and runs from $\tau=0$, corresponding to radial infinity, to $\tau=-\infty$ corresponding to the horizon. The general solution $\phi^I(\tau)$, with boundary conditions $[\phi_0,\,Q]$, is derived from the matrix equation:
\begin{equation}
\mathcal{M}(\phi(\tau))=\mathcal{M}(\phi_0)e^{2Q\tau}\,.\label{gensol}
\end{equation}
By means of $G$, they can be mapped into geodesics on a smaller manifold (generating solution) \cite{Bergshoeff:2008be,Chemissany:2009hq}
  \begin{eqnarray}\label{genman}
{\Scr M}_N=\prod_{\ell=1}^p\frac{{\rm SL}(2,\mathbb{R})_\ell}{{\rm SO}(1,1)_\ell}\sim (dS_2)^p\subset \frac{G}{H^*}\,,\label{MN}
\end{eqnarray}
where ${\rm SL}(2,\mathbb{R})_\ell$ are generated by the  $\mathfrak{sl}(2)_\ell$ algebras defined above. If we
  choose $\phi_0=O$, the Noether charge $Q_O$ of the generating solution will have the general form
  \begin{equation}
Q_O=\sum_{\ell} k_\ell \,N_\ell^{\epsilon_\ell}=\sum_{\ell} k_\ell \,(\mathcal{H}_\ell-\epsilon_\ell\,\mathcal{K}_\ell )\,\,,\,\,\,\,\epsilon_\ell=\pm 1\,.\label{L0G}
\end{equation}
where $\{N_\ell^{(\pm)}\}$ is a basis of nilpotent generators in the coset space $\mathfrak{K}^{(N)}$ of ${\Scr M}_N$:
\begin{equation}
N_\ell^{\epsilon_\ell}=\mathcal{H}_\ell-\epsilon_\ell\,\mathcal{K}_\ell\,\,,\,\,\,[\mathcal{J}_\ell, N_\ell^{\epsilon_\ell}]=\epsilon_\ell\, N_\ell^{\epsilon_\ell}\,,\,,
\end{equation}
All these combinations have vanishing NUT charge: ${\rm Tr}(T_\bullet^T Q )=0$. The coefficients $k_\ell$ of $\mathcal{H}_\ell$ define the scalar charges and ADM mass, while the coefficients of  $\mathcal{K}_\ell $ define the electric and magnetic charges, which can be computed using  Eq. (\ref{allthat}) to be:
\begin{equation}
q_0=-\epsilon_0\,k_0/\sqrt{2}\,\,,\,\,\,p^i=-\epsilon_i\,k_i/\sqrt{2}\,\,,i=1,2,3\,.\label{chG}
\end{equation}
The ADM mass is computed, having chosen $\phi_0=O$, by tracing $Q$ with $T_0$, as in  Eq. (\ref{allthat}) and reads
\begin{equation}
M_{ADM}=\lim_{\tau\rightarrow 0^-}\dot{U}=\frac{1}{4}\sum_{\ell}
k_\ell\,.\label{ADMass}
\end{equation}
Solving (\ref{gensol})  we find the following solution \cite{Bergshoeff:2008be,Chemissany:2009hq}:
\begin{align}
e^{-2U}&=\sqrt{{\bf H}_0 {\bf H}_1 {\bf H}_2 {\bf H}_3}\,,\,\,e^{\varphi_1}=\sqrt{\frac{{\bf H}_0 {\bf H}_1}{{\bf H}_2 {\bf H}_3}}\,,\,\,e^{\varphi_2}=\sqrt{\frac{{\bf H}_0 {\bf H}_2}{{\bf H}_1 {\bf H}_3}}\,,\,\,e^{\varphi_3}=\sqrt{\frac{{\bf H}_0 {\bf H}_3}{{\bf H}_1 {\bf H}_2}}\,,\label{gsol1}\\
\mathcal{Z}^0&= \frac{q_0\,\tau}{{\bf H}_0}\,\,,\,\,\,\mathcal{Z}_k=-\frac{p^k\,\tau}{{\bf H}_k}\,,\,\,k=1,2,3\,,\label{gsol2}
\end{align}
where we have introduced the harmonic functions:
\begin{equation}
{\bf H}_0=1-k_0\,\tau=1+\sqrt{2}\,\epsilon_0\,q_0\,\tau\,\,;\,\,\,{\bf H}_i=1-k_i\,\tau=1+\sqrt{2}\,\epsilon_i\,p^i\,\tau\,\,,\,\,\,i=1,2,3\,.
\end{equation}
We see that, if one of the $k_\ell$ $(\ell=0,1,2,3)$ is negative, the corresponding ${\bf H}_\ell$ vanishes at finite $\tau=1/k_\ell<0$ and so does  $e^{-2U}$. At this distance the $D=4$ scalar curvature $\mathcal{R}$ blows up, signaling a true space-time singularity. \emph{ Therefore regular solutions, with non-vanishing horizon area, correspond to positive, non vanishing $k_\ell$}.
In this case the horizon area is given by:
\begin{equation}
A_H=4\pi\,\lim_{\tau\rightarrow -\infty} \frac{e^{-2U}}{\tau^2}=4\pi\,\sqrt{k_0 k_1 k_2 k_3}=4\pi\,\sqrt{
\epsilon I_4(p,q)}=4\pi\,\sqrt{
|I_4(p,q)|}\,,
\end{equation}
where $I_4(p,q)=4 q_0 p^1 p^2 p^3 $ is the quartic $G_4$-invariant function of the electric and magnetic charges  expressed in the charges of the solution, see Eq. (\ref{qinv}), and $\epsilon\equiv \prod_{\ell}\epsilon_\ell$.\par
When some of the $k_\ell$ vanish the solution is a \emph{small black hole}. Also in this case we distinguish solutions featuring a singularity only in correspondence to the vanishing horizon (regular small) from the others.\par
Once a solution is mapped into the generating one, we can still act on it by means of the ${\rm SO}(1,1)^4$ isotropy group of ${\Scr M}_N$ generated by $\mathcal{J}_\ell$, which consists of residual Harrison transformations. Its effect is to rescale the $k_\ell$ by a positive number and thus will not affect the regularity of the solution. It will however map a singular generating solution (in the $k_\ell$ are not all positive), into a solution with \emph{vanishing $M_{ADM}$ and NUT charge}. \par
Single center solutions with nilpotent $Q$ (and $Q_\psi$) are the extremal static solutions considered above, rotating BPS solutions (which are singular) and the under-rotating solutions \cite{Rasheed,Larsen,Astefanesei}. As opposed to the static solutions ($Q_\psi=0$), the rotating ones are not dual to a generating solution with values in the smaller target space ${\Scr M}_N$. Nevertheless $Q_O$ can always be mapped in the space $\mathfrak{K}^{(N)}$ and thus be expressed as combination of the nilpotent $N^\pm_\ell$ generators of this space.
In Sect. \ref{secdir1} we list the real orbits with  $\alpha$-label from $\alpha^{(1)}$ to $\alpha^{(6)}$ are listed.
We see that, just as for the model studied in \cite{Chemissany:2012nb}, the $\gamma$-labels are related to the gradings $\epsilon_\ell$ of the nilpotent generators while  the $\gamma$-labels depend on the signs of $k_\ell$. The former can be changed by means of compact transformations $e^{\pi\,\mathcal{K}_\ell}$ in $G/H^*$, generated by $\mathcal{K}_\ell$, the latter by complex Harrison transformations $e^{i\,\pi\,\mathcal{J}_\ell}$ in $H^{*\,\mathbb{C}}$, generated by $\mathcal{J}_\ell$, which are \emph{outer automorphisms} of $H^*$ \cite{Chemissany:2012nb}. The orbits defined by $\gamma^{6;5},\,\beta^{6;5}$ are somewhat special in this respect: The $\beta$-label is not affected by a change in the sign of two of the $k_\ell$, although the $H^*$-orbit is. This implies an orbit degeneracy with respect to the $\gamma$ and $\beta$-labels: There are four orbits with $\gamma^{6;5},\,\beta^{6;5}$, distinguished by the labels $\delta^{(1)},\dots,\delta^{(4)}$, which are in fact just two modulo the STU triality.  This feature shows that the $\gamma$ and $\beta$-labels are not enough to identify a real orbit. The orbit describing regular solutions (characterized by $I_4<0$) is the only one (labeled by $\delta^{(1)}$) with positive $k_\ell$.

The regular (small) single-center solutions are characterized by orbits with $\alpha$-label ranging from $\alpha^{(1)},\dots,\alpha^{(5)}$ and those with non-negative $k_\ell$ (i.e. with no singularities at finite $r$), are characterized by the coincidence of the $\gamma$ and $\beta$-labels, corresponding to the classification given in \cite{Borsten:2011ai}.  The regular (non-small) solutions are described by $Q_O$ in the $\alpha^{(6)}$ $G^\mathbb{C}$-orbit: \begin{itemize}
 \item[a)]{The $(\gamma^{(6;1)},\,\beta^{(6;1)})$ orbit describes the regular static BPS solution (which has $I_4>0$);}
 \item[b)]{$(\gamma^{(6;i)},\,\beta^{(6;i)})$, $i=2,3,4$ define the  orbits of the non-BPS solutions with $I_4>0$;}
     \item[c)]{ $(\gamma^{(6;5)},\,\beta^{(6;5)}, \delta^{(1)})$ corresponds to the regular non-BPS solution with $I_4<0$.}
\end{itemize}
The small but regular solutions are defined by the orbits:
\begin{itemize}
\item[d)]{$(\gamma^{(1;1)},\,\beta^{(1;1)})$ describes  the BPS \emph{doubly critical solutions};}
\item[e)]{$(\gamma^{(2;k)},\,\beta^{(2;k)})$, $(\gamma^{(3;k)},\,\beta^{(3;k)})$ and $(\gamma^{(4;k)},\,\beta^{(4;k)})$, $k=1,2$ , which are related by triality, describe  the \emph{critical solutions}. The first $k=1$ orbit of each series describes BPS solutions while the second non-BPS ones;}
\item[f)]{$(\gamma^{(5;k)},\,\beta^{(5;k)})$, $k=1,2,3,4$, describe  the \emph{light-like solutions}. The first $k=1$ describes BPS black holes, while the remaining three non-BPS solutions.}
\end{itemize}
In \cite{Andrianopoli:2013kya,Andrianopoli:2013jra} it is shown how, using singular Harrison transformations, one can connect the orbit of the regular Kerr solution to any of the orbits with $\alpha$-label from $\alpha^{(1)}$ to $\alpha^{(6)}$, which describe both singular and non-singular (rotating) single-center solutions.

The $H^*$-representations therefore provide a \emph{$G$-invariant characterization of regular single-center solutions}. Alternatively one can implement regularity conditions directly on the four-dimensional solution or use a characterization of regular solutions based on the notion of \emph{fake-superpotential} \cite{Bossard:2009we,Chemissany:2012nb}.\footnote{\label{fot}According to this characterization, regular BPS and non-BPS solutions, with finite horizon area, should satisfy a generalized Bogomol'nyi bound: Their
ADM mass should be larger than any of the fake superpotentials associated with the three classes of solutions a), b), c),
computed on the same charges at infinity.}\par
Let us end this section by showing that, as mentioned in the Introduction, the asymptotic behavior of the solution, near the centers and at spatial infinity, is not enough to guarantee the regularity of the whole solution. This clearly applies to the multi-centered case.
To illustrate this let us consider a solution to the three-dimensional effective theory whose electric-magnetic charges belong to the $I_4<0$ $G_4$-orbit.
\begin{figure}[h]
\begin{center}
\centerline{\includegraphics[width=0.9\textwidth]{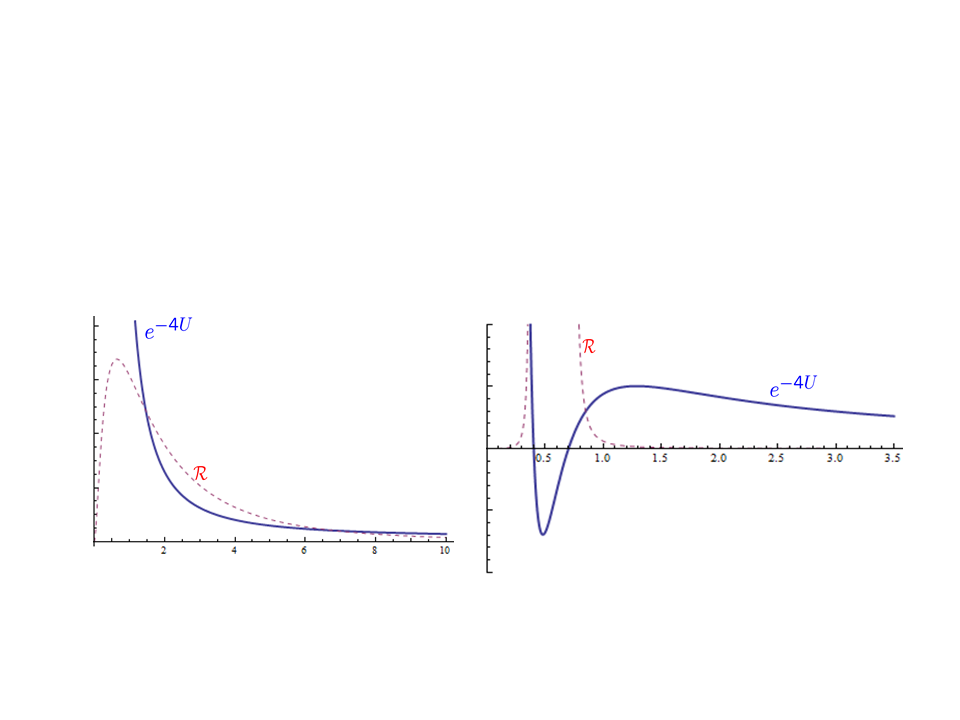}}
 \caption{\small Behavior of the warp function $e^{-4U}$ and of the Ricci scalar $\mathcal{R}$ against $r$ for representatives of the $\gamma^{(6;5)},\beta^{(6;5)},\,\delta^{(1)}$ (left) and  $\gamma^{(6;5)},\beta^{(6;5)},\,\delta^{(2)}$ (right). The latter shows two curvature singularities at finite $r$, although both solutions exhibit a regular asymptotic behavior. Clearly the values of $\mathcal{R}$ and of $e^{-4U}$ refer to different scales. They are plotted in the same graphs to illustrate the corresponding behaviors at the same values of $r$.}\label{fig1}
\end{center}
\end{figure}
For the sake of simplicity, we can take the generating solution with given $q_0,p^1,p^2,p^3$, say $q_0<0,\, p^i>0$. The solution with representative defined by $k_\ell>0$, $\epsilon_0=+1$ and $\epsilon_i=-1$ is regular and belongs to the orbit $\gamma^{(6;5)},\beta^{(6;5)},\,\delta^{(1)}$, see Fig. \ref{fig1} (left), while the one obtained from it by switching the sings of $k_2,k_3$ and of $\epsilon_2,\epsilon_3$, besides having the same $\gamma,\,\beta$-labels, electric-magnetic charges (for suitable choices of $|k_\ell|$), is singular and belongs to the orbit $\gamma^{(6;5)},\beta^{(6;5)},\,\delta^{(2)}$, see Fig. \ref{fig1} (right). The singularities of the latter, however, cannot be seen from an analysis of its asymptotic behavior near the horizon or at spatial infinity. The ADM mass of the singular solution, choosing $\phi_0=O$, is $(|k_0|+|k_1|-|k_2|-|k_3|)/4$ and is smaller than the one of the regular black hole with the same charges and boundary conditions, which is $(|k_0|+|k_1|+|k_2|+|k_3|)/4$. This shows that, if the latter saturates the generalized Bogomol'nyi bound, see footnote \ref{fot}, the former violates it. This simple argument extends to the most general representatives of the two orbits and has a bearing on the analysis of regularity of multi-center solutions of which one constituent is in the $I_4<0$ $G_4$-orbit: The singular behavior of the general solution cannot be avoided by requiring regularity near the centers of at spatial infinity, but either selecting the correct $H^*$-orbit associated with each constituent, or imposing regularity conditions directly on the explicit form of the solution.
\par
As mentioned earlier, orbits of the Noether charge matrix with $\alpha$-label between $\alpha^{(7)},\dots,\alpha^{(11)}$ can only contain regular multi-center solutions. In particular these orbits can be reached by combining two black holes of which one is  defined by an intrinsic matrix $Q^{(0)}_k$, see definition given in Sect. \ref{sec1}, in the orbit $\gamma^{(6;5)},\beta^{(6;5)},\,\delta^{(1)}$, describing a regular black hole with $I_4<0$.
 \subsection{Composition law of nilpotent $H^*$-orbits.}\label{compositionlaw}
 Once we fix $\phi_0=O$, we can define $Q_O^{(0)}\in \mathfrak{K}^*$ as the total Noether charge matrix corresponding to the non-interacting configuration, defined by sending $R_{ij}\rightarrow \infty$. It is reasonable to assume, as it is the case in all instances of solutions discussed here, the $H^*$-orbit of $Q_O$ not to depend on $R_{ij}$ (for $R_{ij}$ not too small, see discussion above) and thus to coincide with that of $Q_O^{(0)}$. The following composition rule holds:
 \begin{equation}
 Q_O^{(0)}=\sum_{k=1}^N Q_k^{(0)}\,.
 \end{equation}
Therefore a necessary condition in order for a multi-center solution to be regular is that the corresponding $Q_O^{(0)}$ be expressed as a combination of representatives $Q_k^{(0)}$ of $H^*$-orbits corresponding to regular single-center solutions.
We studied combinations of two representatives of the 16 orbits associated with regular single-center solutions, yielding nilpotent elements in the higher order orbits $\alpha^{(7)},\dots, \alpha^{(11)}$. The condition that given two nilpotent matrices $e,\,e'$ the sum $e+e'$ be nilpotent poses a strong restriction on the two representatives. We assumed, as a reasonable necessary condition for this, that the corresponding neutral elements $h,\,h'$ can be chosen to commute, so that they belong to a same Cartan subalgebra in the coset space of $H^*/H_c$. \footnote{As an illustrative example one can consider within $\mathfrak{sl}(2,\mathbb{R})$ the nilpositive elements $\sigma^+$ and $\sigma^{\prime\,+}$ with respect to the non-commuting Pauli matrices $\sigma_3$ and $\sigma_1$. No non-trivial combination of $\sigma^+$ and $\sigma^{\prime\,+}$ is nilpotent. In the general case $e,e',h,h'$ do not belong to a same $\mathfrak{sl}(2,\mathbb{R})$-algebra. However $e,e'$, in all the examples considered, turn out not to be orthogonal with respect to the Cartan-Killing metric, so that $e+e'$ is not nilpotent.}
Under this assumption we systematically worked out the composition rule using MATHEMATICA codes. The results are illustrated in Appendix \ref{sumrules}. This analysis singles out a number of  orbits for $Q_O^{(0)}$  which we may define as \emph{intrinsically singular} meaning by this that they cannot be reached by combining representatives of orbits associated with regular single-center solutions and thus, in light of the above necessary condition, do not contain regular solutions.
These orbits were are described in Fig. \ref{Figfigs} by the empty cells:
\begin{itemize}
\item{Orbits $\alpha^{(7)},\,\alpha^{(8)},\,\alpha^{(9)}$ and coinciding  $\gamma,\beta$-labels;}
\item{Orbits $\alpha^{(10)}$ and coinciding  $\gamma,\beta$-labels;}
\item{Orbits $\alpha^{(11)}$ and non-coinciding  $\gamma,\beta$-labels.}
 \end{itemize}
It is known that the information about the closure relation among nilpotent orbits is encoded in the \emph{Hasse diagram} \cite{collingwood}. The Hasse diagram associated with the identity sector ${\rm SO}_0(4,4)$ of ${\rm SO}(4,4)$ can be found in \cite{Bossard:2009we,Borsten:2010db,Djo}.
The corresponding diagram associated with the nilpotent orbits of $H^*$ in $\mathfrak{K}^*$ would be far more complicated and contain much more information than is relevant to our analysis. The problem we posed is of a different and more specific kind: Instead of determining the closure relations among the $H^*$-orbits, we determined which orbit can be obtained by combining two orbits associated with regular single-center solutions. Aside from determining the intrinsically singular orbits mentioned above, for each of the remaining orbits, with alpha-label ranging from $\alpha^{(7)}$ to $\alpha^{(11)}$, we determined the corresponding combinations of the 16 regular-single-center orbits. These combinations are given in the Appendix \ref{sumrules}.

\section{The non-BPS multi-center solutions}\label{rnoca}
In this section we discuss our results on the multi-centered solutions. The multi-center BPS black holes are characterized by a Noether charge matrix $Q_O$ in the orbits $(\gamma^{(6;1)},\beta^{(6;1)})$, $(\gamma^{(5;1)},\beta^{(5;1)})$, $(\gamma^{(3;1)},\beta^{(3;1)})$, $(\gamma^{(2;1)},\beta^{(2;1)})$, $(\gamma^{(1;1)},\beta^{(1;1)})$. They are described by as many harmonic functions as the electric and magnetic charges and are obtained from the single-center solutions by extending the number of poles of each harmonic function. They have been thoroughly studied in the literature and we shall not deal with them here.\par
 Below we shall analyze three classes of non-BPS composite solutions: the \emph{"Composite non-BPS"} \cite{bossard2}, which correspond to the orbits $\alpha^{(10)}$, the \emph{"Almost-BPS"} \cite{Goldstein:2008fq,Bena:2009ev}, which correspond to the orbits $\alpha^{(11)}$ and new classes of solutions described by the orbits with $\alpha$-labels $\alpha^{(7)},\,\alpha^{(8)},\,\alpha^{(9)}$. These latter orbits are connected by the $STU$ triality and are in the closure of the $\alpha^{(10)}$ and $\alpha^{(11)}$ ones. Although the corresponding solutions are much simpler that the generic \emph{"Composite non-BPS"} and the \emph{"Almost-BPS"} ones, they show characteristic features which are common to both of them.\footnote{While the $\alpha^{(10)}$ and $\alpha^{(11)}$ orbits contain composites of two regular non-small solutions, the $\alpha^{(7)},\,\alpha^{(8)},\,\alpha^{(9)}$ ones contain composites in which one of the centers is always a small black hole.} Moreover, in spite of their simplicity, these real orbits, when realized as composites of two representatives single-center ones, reveal an interesting pattern.\par
Let us briefly recall the description given in \cite{bossard2} of extremal solutions in terms of a system of graded, exactly solvable differential equations. With each $H^{*\,\mathbb{C}}$-orbit in $\mathfrak{K}^{*\,\mathbb{C}}$, defined by an $H^{*\,\mathbb{C}}$-orbit of the neutral element $h$ of the triplet in $\mathfrak{H}^{*\,\mathbb{C}}$, and thus by a $\gamma$-label, we associate a characteristic nilpotent subalgebra $\mathfrak{n}$ of $\mathfrak{g}$ consisting of the eigenspaces in $\mathfrak{g}$, with respect to the adjoint action of $h$, with positive grading:
\begin{equation}
\mathfrak{n}=\bigoplus_{k>0} \mathfrak{n}^{(k)}\,\,,\,\,\,\,[h,n]=k\,n\,,\,\,\,\forall n\in \mathfrak{n}^{(k)}\,.
\end{equation}
This nilpotent space  can be written in the form: $\mathfrak{n}=\mathfrak{n}_{\mathfrak{K}^*}\oplus \mathfrak{n}_{\mathfrak{H}^*}$, where the two subspaces are the intersections of $\mathfrak{n}$ with $\mathfrak{K}^*$ and $\mathfrak{H}^*$, respectively. For each of these $H^{*\,\mathbb{C}}$-orbits, in \cite{bossard2} the following ansatz for the scalar fields associated with extremal solutions was put forward:
\begin{equation}
\mathbb{L}(\phi^I(x^j))=\hat{\mathbb{L}}(x^i)\,{\tt h}(x^i)\,\,,\,\,\,\,\hat{\mathbb{L}}(x^i)=\exp(-\mathcal{Y}(x^i))\,\,\,,\,\,\,\,\,\mathcal{Y}(x^i)\in \mathfrak{n}_{\mathfrak{K}^*}\,,\label{defhatl}
\end{equation}
where ${\tt h}(x^i)$ is an element of $H^*$ and $\mathcal{Y}(x^i)$ is a matrix-valued function in $\mathfrak{n}_{\mathfrak{K}^*}$ in terms of whose components the scalar fields $\phi^I(x^i)$ in the solution are expressed.
 The graded structure of the algebra $\mathfrak{n}$ induces a graded structure of the scalar field equations (\ref{feqscal}), expressed in the components of $\mathcal{Y}(x^i)$, which makes them
exactly solvable, iteratively in the grading \cite{bossard2}. The scalar fields in the solution are then read off from the matrix $\mathcal{M}$:
\begin{equation}
\mathcal{M}(\phi^I(x^j))=\mathbb{L}(\phi^I(x^j))\eta \mathbb{L}(\phi^I(x^j))^\dagger=\hat{\mathbb{L}}(x^j)\eta \hat{\mathbb{L}}(x^j)^\dagger\,,\label{MLmatrix}
\end{equation}
where we have used the $H^*$-invariance of $\mathcal{M}$ and the definition (\ref{defhatl}) of $\hat{\mathbb{L}}$. The form of the solution $\phi^I(x^i)$ depends on the chosen neutral element $h$, which defines the nilpotent algebra $\mathfrak{n}$. Different embeddings of $\mathfrak{n}$ inside $\mathfrak{g}$ defining different ``duality frames'' for the solvable system, are related by  $H_c$ \cite{bossard2}. We refer the reader to Appendix \ref{mdstu2} for the explicit matrix forms of the relevant $\mathfrak{so}(4,4)$-generators.\par
In what follows we write for each $H^*$-orbit with higher $\alpha$-label from $\alpha^{(7)}$ to $\alpha^{(11)}$ admitting regular solutions, a regular double-center representative and illustrate its regularity property. \par
A solution contained in the corresponding $H^{*\,\mathbb{C}}$-orbit will be characterized by a Noether matrix at the origin $Q_O$ which is  the nilpositive element $e$ of the triple with neutral element $h$. It should therefore have grading two: $Q_O\in \mathfrak{n}^{(2)}_{\mathfrak{K}^*}$. When working out explicit solutions we shall further restrict $Q_O$ to belong to the $H^*$-orbit under consideration. \par
Let us emphasize once more that the orbit of $Q_O$ does not uniquely define a multi-center solution, whose features also depend on its constituents.  A same $H^*$-orbit of $Q_O$, for instance, may contain both regular and singular solutions. The intrinsically singular orbits of $Q_O$, on the other hand, only contain singular ones, see Sect. \ref{compositionlaw}.

\subsection{The orbit $\alpha^{(7)}\gamma^{(7;1)}\beta^{(i)}$}
Following the approach of \cite{bossard2}, let us start considering the graded decomposition of $\mathfrak{so}(4,4)$ respect to the neutral element $h$ associated with the ${H}^{*\,\mathbb{C}}$-orbit $\gamma^{(7;1)}$, see Appendix \ref{labels}. The nilpotent algebra defined by the positive gradings is
\begin{equation}{\mathfrak{n}^{(7;1)}\simeq(\textbf{1}+2\times \textbf{1})^{(1)}_{\mathfrak{K}^*}\oplus(2\times \textbf{1})^{(1)}_{\mathfrak{H}^*}\oplus(2\times \textbf{1})^{(2)}_{\mathfrak{K}^*}\oplus(2\times \textbf{1})^{(2)}_{\mathfrak{H}^*}\oplus\textbf{1}^{(3)}_{\mathfrak{K}^*}\,.}\label{n7_1}
\end{equation}
The suffix on $\mathfrak{n}$ refers to the $\gamma$-labels of the orbits under consideration.\par
Choosing an appropriate basis of this algebra, the only non-zero commutators are
\begin{equation}
\begin{split}
[\textit{\textbf{e}}_{0}^{(1)},\textit{\textbf{e}}^{(1),i}]=\textit{\textbf{f}}^{(2),i}\,,\,[\textit{\textbf{f}}^{(2),i},
\textit{\textbf{e}}^{(1),j}]=c^{ij}\,\textit{\textbf{e}}^{(3)}\,,\\
[\textit{\textbf{f}}^{(1),i},\textit{\textbf{e}}_{0}^{(1)}]=\textit{\textbf{e}}^{(2),i}\,,\,[\textit{\textbf{e}}^{(2),i},\textit{\textbf{f}}^{(1),j}]=
c^{ij}\,\textit{\textbf{e}}^{(3)}\,,\label{comrel1}
\end{split}
\end{equation}
having denoted by $\textit{\textbf{e}}$ the generators in ${\mathfrak{K}^*}$ and by $\textit{\textbf{f}}$ those in ${\mathfrak{H}^*}$, the number in superscript being the grading relative to the neutral element. The coefficients $c^{ij}$ are defined as: $c^{ij}=|\varepsilon^{ij}|$ .
It is interesting to note that the generators\, $\textit{\textbf{f}}^{(2),i}$, $\textit{\textbf{e}}^{(2),i}$, $\textit{\textbf{e}}^{(1),i}$, \, $\textit{\textbf{f}}^{(1),i}$ and $\textit{\textbf{e}}^{(3)}$  form a single Heisenberg subalgebra.\\
We now consider for the coset representative the Ansatz (\ref{defhatl}) with
\begin{equation}{\hat{\mathbb{L}}(x^i)\,=\,\exp(-\mathcal{Y}(x^i))=
\exp(-\tilde{V}\textit{\textbf{e}}_{0}^{(1)}-\tilde{L_{i}}\textit{\textbf{e}}^{(1),i}-\tilde{Z_{i}}\textit{\textbf{e}}^{(2),i}-
\tilde{M}\textit{\textbf{e}}^{(3)})}\label{ansazc}
\end{equation}
and write the graded field equations deduced from (\ref{feqscal}). With the following redefinitions
\begin{equation}
\begin{split}
\tilde{V}&\rightarrow{V}\,,\\
\tilde{L_{i}}&\rightarrow{2(2L_{i}-1)}\,,\\
\tilde{Z_{i}}&\rightarrow{4(Z_{i}-2)}\,,\\
\tilde{M}&\rightarrow{32M+\frac{4}{3}V\left[2(L_{1}+L_{2})+8L_{1}L_{2}-1\right]}\,,\\
\end{split}\label{redef1}
\end{equation}
we obtain the following equations of motion
\begin{equation}
\begin{split}
{d}\ast{d}{V}&=0\,,\\
{d}\ast{d}{L_{i}}&=0\,,\\
{d}\ast{d}{Z_{i}}&=0\,,\\
{d}\ast{d}{\left(M+\frac{1}{4}c^{ij}V L_{i}L_{j}\right)}&=\frac{1}{2}c^{ij}V {d}L_{i}\ast{d}L_{j}\,,\\\label{eqmot1}
\end{split}
\end{equation}
where, in writing the last equality, we have used the first three equations, namely that the functions $V$,\,$L_{i}$ and $Z_{i}$  are harmonic functions.\par
To work out a solution we choose the ``duality frame'' defined by the following choice of the neutral element $h$:
 \begin{align}
h&=\text{e}_{1,4}+\text{e}_{1,5}-2 \text{e}_{2,3}+2 \text{e}_{2,6}-2 \text{e}_{3,2}-2
   \text{e}_{3,7}+\text{e}_{4,1}-\text{e}_{4,8}+\text{e}_{5,1}-\text{e}_{5,8}+2 \text{e}_{6,2}+2 \text{e}_{6,7}-\nonumber\\&-2 \text{e}_{7,3}+2
   \text{e}_{7,6}-\text{e}_{8,4}-\text{e}_{8,5}\,.
\end{align}
From a solution to (\ref{eqmot1}) we can derive in this frame the expression of the scalars $\phi^I=(U,a,\mathcal{Z}^M, \varphi_i,\,\epsilon_i)$ by solving the matrix equation (\ref{MLmatrix}).
In particular, with the redefinitions (\ref{redef1}),\,$e^{-4\,U}$ has the usual form
\begin{equation}
e^{-4U}=L_{1}L_{2}Z_{1}Z_{2}-M^{2}\,,
\end{equation}
and the scalars $\mathcal{Z}^M$ and $a$ read
\begin{equation}
\begin{split}
\mathcal{Z}^0\,&=\,\frac{1}{\sqrt{2}}\,\left[\,1\,-\,\frac{2\,L_{1}\,L_{2}\,Z_{1}}{e^{-4U}}\,\right]\,,\\
\mathcal{Z}^1\,&=\,-\,\frac{(\,M\,+\,V\,L_{1}\,L_{2}\,)\,Z_{1}}{2\,\sqrt{2}\,e^{-4U}}\,,\\
\mathcal{Z}^2\,&=\,-\,\frac{\sqrt{2}\,M\,L_{2}}{e^{-4U}}\,,\\
\mathcal{Z}^3\,&=\,-\,\frac{\sqrt{2}\,M\,L_{1}}{e^{-4U}}\,,\\
\mathcal{Z}_0\,&=\,\frac{(\,M\,+\,V\,L_{1}\,L_{2}\,)\,Z_{2}}{2\,\sqrt{2}\,e^{-4U}}\,,\\
\mathcal{Z}_1\,&=\,\frac{1}{\sqrt{2}}\,\left[\,1\,-\,\frac{2\,L_{1}\,L_{2}\,Z_{2}}{e^{-4U}}\,\right]\,,\\
\mathcal{Z}_2\,&=\,\frac{1}{\sqrt{2}}\,\left[\,1\,-\,\frac{(M\,V\,+\,Z_{1}\,Z_{2})\,L_{1}}{2\,e^{-4U}}\,\right]\,,\\
\mathcal{Z}_3\,&=\,\frac{1}{\sqrt{2}}\,\left[\,1\,-\,\frac{(M\,V\,+\,Z_{1}\,Z_{2})\,L_{2}}{2\,e^{-4U}}\,\right]\,,\\
a\,&=\,-\,\frac{V\,L_{1}\,L_{2}\,(\,Z_{1}\,+\,Z_{2}\,)\,+\,M\,[\,4\,(\,L_{1}\,+\,L_{2}\,)\,+\,Z_{1}\,+\,Z_{2}\,-\,4\,]}{4\,e^{-4U}}\quad.\\
\label{chpq1}
\end{split}
\end{equation}
Referring to Appendix \ref{mdstu} for the choice of our duality frame, the four-dimensional scalar fields read
\begin{equation}
\begin{aligned}
\epsilon_{1}\,&=\,\frac{1}{\sqrt{4}}\,\left(\,\frac{M}{L_{1}\,L_{2}}\,+\,V\,\right)&,\quad&e^{-2\,\varphi_{1}}\,=&\,\frac{16\,L_{1}^{2}\,L_{2}^{2}}{e^{-4U}}\quad,\\
\epsilon_{2}\,&=\,\frac{M}{L_{1}\,Z_{1}}&,\quad&e^{-2\,\varphi_{2}}\,=&\,\frac{L_{1}^{2}\,Z_{1}^{2}}{e^{-4U}}\quad,\\
\epsilon_{3}\,&=\,\frac{M}{L_{2}\,Z_{1}}&,\quad&e^{-2\,\varphi_{3}}\,=&\,\frac{L_{2}^{2}\,Z_{1}^{2}}{e^{-4U}}\quad.
\label{scalar71}
\end{aligned}
\end{equation}
From the last of (\ref{eqmot1}), which can be rewritten as
\begin{equation}
{d}\ast{d}{M}=-\frac{1}{2}c^{ij}d\left(L_{i}L_{j}\ast{d}{V}\right)\,,
\end{equation}
we can explicitly compute the equations for $\omega$ and $\tilde{A}^M$ from (\ref{AF3}), (\ref{AF4}). We obtain
\begin{equation}
\ast{d}{\omega}=d{M}+L_{1}L_{2}{d}{V}\,\label{omega71}
\end{equation}
and
\begin{equation}
\begin{split}
\ast{d}{\tilde{A}^{0}}&=\frac{\ast{d}{\omega}}{\sqrt{2}}\quad,\quad\ast{d}{\tilde{A}^{1}}=\frac{{d}{Z_{1}}}{2\sqrt{2}}\quad,\quad\ast{d}{\tilde{A}^{2}}=\sqrt{2}\,{d}{L_{2}}\quad,
\quad\ast{d}{\tilde{A}^{3}}=\sqrt{2}\,{d}{L_{1}}\quad,\\
\ast{d}{\tilde{A}_0}&=-\frac{{d}{Z_{2}}}{2\sqrt{2}}\quad,\quad\ast{d}{\tilde{A}_1}=\frac{\ast{d}{\omega}}{\sqrt{2}}\quad,\quad\ast{d}{\tilde{A}_2}=\frac{\ast{d}{\omega}}{\sqrt{2}}+\frac{1}{2\sqrt{2}}\left(V{d}L_{1}-L_{1}{d}V\right)\quad,\\
\ast{d}{\tilde{A}_3}&=\frac{\ast{d}{\omega}}{\sqrt{2}}+\frac{1}{2\sqrt{2}}\left(V{d}L_{2}-L_{2}{d}V\right)\quad.\label{vect71}
\end{split}
\end{equation}
The electric-magnetic vectors $A^M$ in $D=4$ are then computed from Eqs. (\ref{chpq1}) and a solution to the above equations using (\ref{AF1}).
\subsubsection{The solution}\label{subsect1}
In what follows we shall focus on two-center solutions and work out the general regularity conditions. Eventually we shall fix the parameters so as to single out a regular representative of the orbit. In particular care has to be used, when choosing the values of the parameters, to keep the total Noether charge matrix $Q_O$ in the $H^*$-orbit under consideration.\par
In spherical coordinates ($r,\theta,\phi$), we can consider the first center, to be denoted by ``A'', at $r=0$ and the second one, denoted by ``B'', along the positive $z$-axis,\,\textit{i.e.} $\theta=0$, at a distance $R$ from the first. In the axisymmetric solution, the position of a point in space relative to the two centers is described by ($r,\theta$) and ($\Sigma,\theta_{\Sigma}$),\, respectively, where
\begin{equation}
\Sigma=\sqrt{R^2+r^2-2rR\cos\theta}\quad,\quad\cos\theta_{\Sigma}=\frac{r\cos\theta-R}{\Sigma}\,.\label{eq3}
\end{equation}
Being $V$,\,$L_{i}$ and $Z_{i}$ harmonic functions, we can write them in the following general form
\begin{equation}
V=h+\frac{Q_{6}}{r}+\frac{\tilde{Q}_{6}}{\Sigma}\quad,\quad\,L_{i}=l_{i}+\frac{Q_{i}}{r}+\frac{\tilde{Q}_{i}}{\Sigma}\,\quad,\quad\,Z_{i}=z_{i}+\frac{d_{i}}{r}+\frac{\tilde{d}_{i}}{\Sigma}\label{eqq1}\,.
\end{equation}
Substituting the above expressions in the last of equations (\ref{eqmot1}), we find
\begin{equation}
\begin{split}
{d}\ast{d}{\left[M+\frac{1}{4}c^{ij}\left(h+\frac{Q_{6}}{r}+\frac{\tilde{Q}_{6}}{\Sigma}\right)\left(l_{i}+\frac{Q_{i}}{r}+\frac{\tilde{Q}_{i}}{\Sigma}\right)\left(l_{j}+\frac{Q_{j}}{r}+\frac{\tilde{Q}_{j}}{\Sigma}\right)\right]}=\\
\quad=\frac{1}{2}c^{ij}\left(h+\frac{Q_{6}}{r}+\frac{\tilde{Q}_{6}}{\Sigma}\right){d}\left(l_{i}+\frac{Q_{i}}{r}+\frac{\tilde{Q}_{i}}{\Sigma}\right)\ast{d}\left(l_{j}+\frac{Q_{j}}{r}+\frac{\tilde{Q}_{i}}{\Sigma}\right)\,.
\end{split}\label{ddMa7}
\end{equation}
Using the relations listed in Appendix \ref{usef}, from the Eq. (\ref{ddMa7}) one obtains the following explicit expression for $M$:
\begin{equation}
\begin{split}
M=&m_0+\frac{m}{r}+\frac{\tilde{m}}{\Sigma}+\frac{\alpha\cos\theta}{r^2}+\frac{\tilde{\alpha}\cos\theta_\Sigma}{\Sigma^2}-\\&-
\frac{1}{2}\left(h+\frac{Q_{6}}{r}+\frac{\tilde{Q}_{6}}{\Sigma}\right)\left(l_{1}+\frac{Q_{1}}{r}+\frac{\tilde{Q}_{1}}{\Sigma}\right)
\left(l_{2}+\frac{Q_{2}}{r}+\frac{\tilde{Q}_{2}}{\Sigma}\right)+\\
&+\frac{hQ_{1}Q_{2}}{2r^2}+\frac{h\tilde{Q}_{1}\tilde{Q}_{2}}{2\Sigma^2}+\frac{(hR+Q_{6}\cos\theta-\tilde{Q}_{6}\cos\theta_\Sigma)(Q_{1}\tilde{Q}_{2}+
\tilde{Q}_{1}Q_{2})}{2rR\Sigma}+\\
&+\frac{rQ_{6}\tilde{Q}_{1}\tilde{Q}_{2}}{2R^2\Sigma^2}+\frac{\tilde{Q}_{6}Q_{1}Q_{2}\Sigma}{2R^2r^2}+
\frac{{Q}_{6}Q_{1}Q_{2}}{6r^3}+\frac{\tilde{Q}_{6}\tilde{Q}_{1}\tilde{Q}_{2}}{6\Sigma^3}\,.\label{eqq2}
\end{split}
\end{equation}
Making use of properties (\ref{Bom1})-(\ref{Bomfin}), we can also solve the Eq. (\ref{omega71})
\begin{equation}
\begin{split}
\omega&=\Bigg[k_{\omega}+\left(\,m+\frac{l_{1}l_{2}Q_{6}-h(l_{1}Q_{2}+l_{2}Q_{1})}{2}\right)\cos\theta+\left(\,\tilde{m}+\frac{l_{1}l_{2}\tilde{Q}_{6}-
h(l_{1}\tilde{Q}_{2}+l_{2}\tilde{Q}_{1})}{2}\right)\cos\theta_\Sigma+\\
&-\alpha\frac{\sin^{2}\theta}{r}-\tilde{\alpha}\frac{\sin^{2}\theta_{\Sigma}}{\Sigma}+
Q_{6}(l_{1}\tilde{Q}_{2}+l_{2}\tilde{Q}_{1})\frac{R\cos\theta-r}{2R\Sigma}+Q_{6}\tilde{Q}_{1}\tilde{Q}_{2}\frac{(r^{2}+R^{2})\cos\theta-2rR}{2R^{2}\Sigma^{2}}-\\
&-Q_{6}\left(Q_{1}\tilde{Q}_{2}+Q_{2}\tilde{Q}_{1}\right)
\frac{\sin^{2}\theta}{2R\Sigma}-\tilde{Q}_{6}(l_{1}Q_{2}+l_{2}Q_{1})\frac{R\cos\theta-r}{2R\Sigma}+\tilde{Q}_{6}Q_{1}Q_{2}\frac{R+r\cos\theta-2R\cos^{2}\theta}{2R^{2}\Sigma}+\\
&+\tilde{Q}_{6}\left(Q_{1}\tilde{Q}_{2}+Q_{2}\tilde{Q}_{1}\right)\frac{\sin^{2}\theta_{\Sigma}}{2Rr}\Bigg]{d}\phi\quad\,,\label{om71}
\end{split}
\end{equation}
and compute the explicit form of the vectors $\tilde{A}^M$ satisfying (\ref{vect71}), which are
\begin{equation}
\begin{split}
\tilde{A}^{0}&=\frac{\omega}{\sqrt{2}}\quad,\quad \tilde{A}^{1}=\frac{d_{1}\cos\theta+\tilde{d}_{1}\cos\theta_{\Sigma}}{2\sqrt{2}}d\varphi\quad,\quad \tilde{A}^{2}=\sqrt{2}\,(Q_{2}\cos\theta+\tilde{Q}_{2}\cos\theta_{\Sigma})d\varphi\quad,\\
\tilde{A}^{3}&=\sqrt{2}\,(Q_{1}\cos\theta+\tilde{Q}_{1}\cos\theta_{\Sigma})d\varphi\quad,\quad \tilde{A}_0=\frac{d_{2}\cos\theta+\tilde{d}_{2}\cos\theta_{\Sigma}}{2\sqrt{2}}d\varphi\quad,\quad \tilde{A}_1=\frac{\omega}{\sqrt{2}}\quad,\\
\tilde{A}_2&=\frac{\omega}{\sqrt{2}}+\frac{1}{2\sqrt{2}}\left[\left(hQ_{1}-l_{1}Q_{6}\right)\cos\theta+\left(h\tilde{Q}_{1}-l_{1}\tilde{Q}_{6}\right)
\cos\theta_{\Sigma}-\left(Q_{6}\tilde{Q}_{1}-\tilde{Q}_{6}Q_{1}\right)\frac{R\cos\theta-r}{R\Sigma}\right]d\varphi\quad,\\
\tilde{A}_3&=\frac{\omega}{\sqrt{2}}+\frac{1}{2\sqrt{2}}\left[\left(hQ_{2}-l_{2}Q_{6}\right)\cos\theta+\left(h\tilde{Q}_{2}-l_{2}\tilde{Q}_{6}\right)\cos\theta_{\Sigma}-\left(Q_{6}\tilde{Q}_{2}-\tilde{Q}_{6}Q_{2}\right)\frac{R\cos\theta-r}{R\Sigma}\right]d\varphi\quad.
\label{vects71}
\end{split}
\end{equation}
The electric-magnetic vectors $A^M$ in $D=4$ are then computed substituting (\ref{vects71})  and the expressions (\ref{chpq1}) for $\mathcal{Z}^M$ in Eq. (\ref{AF1}).\par
Although these expressions satisfy the equations of motion, these are not sufficient to ensure the regularity of the system, so further conditions are necessary.\\
\\
\textbf{Zero-NUT-charge condition.}\\
We require the absence of the NUT charge and Dirac-Misner singularities by imposing that each center is NUT charge - free and the $\omega$ is zero along the $Z$-axis when $\sin\theta=0$ , \emph{i.e.}
\begin{equation}
\begin{split}
\omega\stackrel{{\theta\rightarrow\pi}}=0\quad,\quad\omega\stackrel{\stackrel{\theta\rightarrow0}{r<R}}=0\quad,\quad\omega\stackrel{\stackrel{\theta\rightarrow0}{r>R}}=0\quad.\label{condnut7}
\end{split}
\end{equation}
This amounts to requiring Eq.s (\ref{nonut1}) and (\ref{nonut2}) as well as fixing the integration constant in the solution for $\omega_\varphi$.
The above equations imply three conditions on the integration constants $k_{\omega}$\,,\,$m$ and $\tilde{m}$
\begin{equation}
\begin{split}
k_{\omega}&=c^{ij}\frac{Q_{6}l_{i}\tilde{Q}_{j}-\tilde{Q}_{6}l_{i}Q_{j}}{2R}\quad,\\ m&=\frac{1}{2}\left[\left(h+\frac{\tilde{Q}_6}{R}\right)c^{ij}l_{i}Q_{j}-Q_{6}\prod^2_{i=1}(l_{i}+\frac{\tilde{Q}_{i}}{R})\right]\quad,\\
\tilde{m}&=\frac{1}{2}\left[\left(h+\frac{Q_6}{R}\right)c^{ij}l_{i}\tilde{Q}_{j}-\tilde{Q}_{6}\prod^2_{i=1}(l_{i}+\frac{Q_{i}}{R})\right]\,.\label{NUTcond7}
\end{split}
\end{equation}
The last two conditions encode the bubble equation relating the distance $R$ between the two centers to boundary conditions at spatial infinity. The above conditions are sufficient to exclude closed time-like curves (CTC) as they guarantee that the term $e^{-4\,U}\,r^{2}\sin^{2}\theta-\omega^2$ in the metric (\ref{ds2}) is positive everywhere.\\
\paragraph{Electric-magnetic charges}
Below, for the sake of completeness, we write the total electric-magnetic charge vector $\Gamma^M$ as well as those associated with the $A$ and $B$-centers (respectively denoted by $\Gamma^M_A,\,\Gamma^M_B$):
\begin{align}
\Gamma^M\,&=\,(\,p^\Lambda\,,\,q_\Lambda\,)\,=\,
\bigg(\,0\,,\,\frac{d_{1}+\tilde{d}_{1}}{2\sqrt{2}}\,,\,\sqrt{2}(Q_{2}+\tilde{Q}_{2})\,,\,\sqrt{2}(Q_{1}+\tilde{Q}_{1})\,,\,
-\frac{d_2+\tilde{d}_2}{2\sqrt{2}}\,,\,0\,,\\
&,\,\frac{l_{1}(Q_{6}+\tilde{Q}_{6})+h(Q_{1}+\tilde{Q}_{1})}{2\sqrt{2}}\,,\,\frac{l_{2}(Q_{6}+\tilde{Q}_{6})+h(Q_{2}+\tilde{Q}_{2})}{2\sqrt{2}}\,
\bigg)\,,\nonumber\\
{\Gamma^{M}}_{A}\,&=\,(\,{p^{\Lambda}}_{A}\,,\,{q_{\Lambda}}_{A}\,)\,=\,\bigg(\,0\,,\,\frac{d_1}{2\sqrt{2}}\,,\,\sqrt{2}Q_{2}\,,\,\sqrt{2}Q_{1}\,,\,
-\frac{d_2}{2\sqrt{2}}\,,\,0\,,\,\frac{-Q_{6}(\tilde{Q}_{1}+R{l}_{1})+Q_{1}(\tilde{Q}_{6}+h{R})}{2\sqrt{2}R}\,,\nonumber\\
&,\,\frac{-Q_{6}(\tilde{Q}_{2}+R{l}_{2})+Q_{2}(\tilde{Q}_{6}+h{R})}{2\sqrt{2}R}\,\bigg)\,,\nonumber\\
{\Gamma^{M}}_{B}\,&=\,(\,{p^{\Lambda}}_{B}\,,\,{q_{\Lambda}}_{B}\,)\,=\,\bigg(\,0\,,\,\frac{\tilde{d}_1}{2\sqrt{2}}\,,\,\sqrt{2}\tilde{Q}_2\,,\,
\sqrt{2}\tilde{Q}_1\,,\,-\frac{\tilde{d}_2}{2\sqrt{2}}\,,\,0\,,\,\frac{-\tilde{Q}_{6}(Q_{1}+R{l}_{1})+\tilde{Q}_{1}(Q_{6}+h{R})}{2\sqrt{2}R}\,,\\
&,\,\frac{-\tilde{Q}_{6}(Q_{2}+R{l}_{2})+\tilde{Q}_{2}(Q_{6}+h{R})}{2\sqrt{2}R}\,\bigg)\,.
\end{align}

\paragraph{Regularity near the centers and asymptotic flatness}\par
The horizon areas are expressed by the integral
\begin{equation}
A_H=\int\,d\theta\,d\varphi\sqrt{r^2(e^{-4U}r^2\sin^2\theta-\omega^2)}\label{HorArea}\,,
\end{equation}
computed about the two centers, namely in the limits $r\rightarrow 0$ and $\Sigma\rightarrow 0$. Since we allow small black holes, we only require this integral not to diverge at the two centers. This requires the $e^{-4U}$ go like $1/r^a$ and $1/\Sigma^a$, with $a\le 4$, in the two limits, respectively. In our case, we have for the first center
\begin{equation}
\begin{split}
e^{-4\,U}\stackrel{\stackrel{r\rightarrow0}{\theta=\pi}}{\cong}&\frac{1}{r^6}\left(-\frac{Q^2_{6}Q^2_{1}Q^2_{2}}{9}\right)+\\
&+\frac{1}{r^5}\left[Q_{6}Q_{1}Q_{2}\left(-\frac{2\alpha}{3}-\frac{Q_{6}c^{ij}Q_{i}(\tilde{Q}_{j}+Rl_{j})}{3R}\right)\right]+\mathcal{O}(r^{-4})
\end{split}
\end{equation}
and for the second one
\begin{equation}
\begin{split}
e^{-4\,U}\stackrel{\stackrel{r\rightarrow{R}}{\theta=0}}{\cong}&\frac{1}{(r-R)^6}\left(-\frac{\tilde{Q}^2_{6}\tilde{Q}^2_{1}\tilde{Q}^2_{2}}{9}\right)+\\
&+\frac{1}{(r-R)^5}
\left[\tilde{Q}_{6}\tilde{Q}_{1}\tilde{Q}_{2}\left(\frac{2\tilde{\alpha}}{3}-\frac{\tilde{Q}_{6}c^{ij}\tilde{Q}_{i}(Q_{j}+Rl_{j})}{3R}\right)\right]+
\mathcal{O}((r-R)^{-4})\,.
\end{split}
\end{equation}
Since we are interested in finding instances of solutions with total Noether charge in each orbit $\gamma^{(7;1)},\,\beta^{(7;k)}$, this feature is not affected by choosing
\begin{equation}
\tilde{Q}_6=0\quad\,,\,\quad\,Q_{1}=0\quad\,,\,\quad\,Q_{2}=0\,,
\end{equation}
in order to have a well-behaved $e^{-4\,U}$ near each center.
On the warp factor we further require the condition for asymptotic flatness, namely $\lim_{r \rightarrow +\infty}e^{-4\,U}=1$, which can be solved for $h$ to give:
\begin{equation}
h=\frac{2(m_0-\sqrt{l_{1}l_{2}l_{3}l_{4}-1})}{l_{1}l_{2}}\,.
\end{equation}
Since we are only interested in studying instances of solutions keeping $Q_O$ generic in a chosen orbit, as a  last condition, for the sake of simplicity, we shall set the values of the scalars at the origin,\,\textit{i.e.} at the infinity, requiring that $\lim_{r \rightarrow +\infty}\mathcal{M}^{(3)}_{scal}=\eta$ and obtaining
\begin{equation}
m_0=0\quad\,,\,\quad\,l_{1}=\frac{1}{2}\quad\,,\,\quad\,l_{2}=\frac{1}{2}\quad\,,\,\quad\,l_{3}=2\quad\,,\,\quad\,l_{4}=2\quad\,,\,\quad\,h=0\,.
\end{equation}
The advantage of this choice is that in this ``frame'' the $M_{ADM}$ and the topological mass, \textit{i.e.} the projection of the matrix $\mathcal{M}$ on $H_0$, have the same value.\\
With the above choice of the parameters the geometry near the first center goes like
\begin{equation}
\begin{split}
e^{-4\,U}\stackrel{\stackrel{r\rightarrow0}{\theta=\pi}}{\cong}&\frac{1}{r^4}\left[-\alpha^{2}\right]+\frac{1}{r^3}\left[\frac{-\alpha Q_{6}(R+2\tilde{Q}_{1})(R+2\tilde{Q}_{2})}{2R^{2}}\right]+\mathcal{O}(r^{-2})\quad,\label{conda7g1fc}
\end{split}
\end{equation}
while close to the second one we have
\begin{equation}
\begin{split}
e^{-4\,U}\stackrel{\stackrel{r\rightarrow{R}}{\theta=0}}{\cong}&\frac{1}{(r-R)^4}\left[-\tilde{\alpha}^{2}+\tilde{Q}_{1}\tilde{Q}_{2}\tilde{d}_{1}\tilde{d}_{2}\right]+\mathcal{O}((r-R)^{-3})\equiv\\
&\equiv\frac{1}{(r-R)^4}\left[-\tilde{\alpha}^{2}-I_{4}(p^{\Lambda}_{B},q_{\Lambda\,B})\right]+\mathcal{O}((r-R)^{-3})\quad,
\end{split}
\end{equation}
where $I_{4}(p^{\Lambda}_{B},q_{\Lambda\,B})$ is the quartic invariant associated with the second center (see Eq. (\ref{qinv}) for the explicit expression of this quantity). We note that, in order to have a well defined warp factor for the first center in (\ref{conda7g1fc}), the angular momentum term $\alpha$ has to be set to zero, revealing the singular nature of this center as a small black hole.
\paragraph{Charges, angular momentum and distance $R$}\par
After having derived the regularity conditions at the horizons, one can determine the total electromagnetic charges associated to the solution, making use of (\ref{elmagch}),  obtaining
\begin{equation}
\Gamma^M\,=\,(\,p^\Lambda\,,\,q_\Lambda\,)\,=\,\left(\,0\,,\,\frac{d_1+\tilde{d}_1}{2\sqrt{2}}\,,\,\sqrt{2}\tilde{Q}_2\,,\,\sqrt{2}\tilde{Q}_1\,,\,-\frac{d_2+\tilde{d}_2}{2\sqrt{2}}\,,\,0\,,\,-\frac{Q_6}{4\sqrt{2}}\,,\,-\frac{Q_6}{4\sqrt{2}}\,\right)\quad.
\end{equation}
By the same token, one can compute the electromagnetic charges associated with each black hole, considering a cycle about each center. These are given by
\begin{equation}
{\Gamma^{M}}_{A}\,=\,(\,{p^{\Lambda}}_{A}\,,\,{q_{\Lambda}}_{A}\,)\,=\,\left(\,0\,,\,\frac{d_1}{2\sqrt{2}}\,,\,0\,,\,0\,,\,-\frac{d_2}{2\sqrt{2}}\,,\,0\,,\,-\frac{Q_6(R+2\tilde{Q}_1)}{4\sqrt{2}R}\,,\,-\frac{Q_6(R+2\tilde{Q}_2)}{4\sqrt{2}R}\,\right)\quad
\end{equation}
for the center located in $r=0$ and
\begin{equation}
{\Gamma^{M}}_{B}\,=\,(\,{p^{\Lambda}}_{B}\,,\,{q_{\Lambda}}_{B}\,)\,=\,\left(\,0\,,\,\frac{\tilde{d}_1}{2\sqrt{2}}\,,\,\sqrt{2}\tilde{Q}_2\,,\,\sqrt{2}\tilde{Q}_1\,,\,-\frac{\tilde{d}_2}{2\sqrt{2}}\,,\,0\,,\,\frac{Q_6\,\tilde{Q}_1}{2\sqrt{2}R}\,,\,\frac{Q_6\,\tilde{Q}_2}{2\sqrt{2}R}\,\right)\quad
\end{equation}
for the center located in $r=R,\,\theta=0$ .\\
The angular momentum is computed through the use of $Q_\psi$ defined in (\ref{qupsi}) and the last equation of (\ref{allthat}). We obtain
\begin{equation}
\begin{split}
M_\varphi&=-\frac{1}{2}\tilde{\alpha}-\frac{Q_{6}[4\tilde{Q}_1\tilde{Q}_2+R(\tilde{Q}_1+\tilde{Q}_2)]}{8R}=\\
&=(M_{\varphi_A}=0)+M_{\varphi_B}-\frac{1}{2}({\Gamma^{M}_{A}})^{T}\,\mathbb{C}\,{\Gamma^{M}_{B}}\quad,
\end{split}
\end{equation}
where we have identified the angular momentum of the $B$-center $M_{\varphi_B}$ with $-\tilde{\alpha}/2$.
Therefore the total angular momentum is the sum of the angular momenta associated to each black hole plus the symplectic product of the electromagnetic charges.
In particular, one can consider the leading term for $\omega_\varphi$ in the expression (\ref{om71})
\begin{align}
\omega_\varphi=\frac{1}{r}\left[-\tilde{\alpha}-\frac{Q_6(\tilde{Q}_1+\tilde{Q}_2)}{4}-\frac{Q_6\tilde{Q}_1\tilde{Q}_2}{R}\right]\sin^2\theta+
\mathcal{O}((r)^{-2})\quad,
\end{align}
and verify that the following relation holds
\begin{align}
\omega_\varphi\,\cong\,\frac{2\,M_\varphi\,\sin^2\theta}{r}+\dots\quad.
\end{align}
We conclude this section, before entering into a detailed analysis of specific examples, by observing that there exists a lower bound for the distance $R$ between the two centers which adds to the constraints following from the zero-NUT charge condition. This minimum value of $R$ can be determined by requiring the positivity of the leading term in the warp function $e^{-4U}$\, near each center, in order to remove the presence of poles in this function. Obviously this condition is strictly connected to the real single center orbit which each component black hole belongs to, so we will derive this lower bound case by case.
\\
\subsubsection{The orbits $\mathcal{O}^{7;1,1}$ and $\mathcal{O}^{7;1,2}$}
So far, we have not made a choice for the real nilpotent orbit within the $\alpha^{(7)}$-class, only limiting our analysis to a specific $\gamma$-label (i.e. $H^{*\,\mathbb{C}}$-orbit). In order to further restrict ourselves to specific $H^*$-orbits, we have to fix the $\beta$-labels, choosing the Noether charges matrix $Q_O$ either in the orbit $\alpha^{(7)}\gamma^{(7;1)}\beta^{(7;1)}$ or in $\alpha^{(7)}\gamma^{(7;1)}\beta^{(7;2)}$.\par  One center, say $B$, can be always chosen to be described by the \emph{generating solution}, namely so that $Q_B^{(0)}$, defined in Eq. (\ref{intrinsicQ}) by sending the $A$-center to infinity,  has the form (\ref{L0G}). This makes it easier to fix its intrinsic orbit. The other center, in general, cannot be described in the same frame as the generating solution if we want the total charge $Q_O$ to be a generic element of the chosen orbit. Therefore we first choose the parameters of the $B$-black hole so that it belongs to a given $H^*$-orbit. Then we choose the parameters of the $A$-center so that the whole system is in the chosen real orbit. \\
\paragraph{The orbit $\mathcal{\textbf{O}}^{\textbf{7;1,1}}$.}\par
Let us start with $\alpha^{(7)}\gamma^{(7;1)}\beta^{(7;1)}$ by suitably choosing the parameters associated with each center of the two-center solution.
We find that, requiring the total solution to be in the orbit $\gamma^{(7;1)},\,\beta^{(7;1)}$, the two centers cannot be chosen to be both regular, consistently with the general result illustrated in Sect. \ref{compositionlaw}. For example  a singular representative of this class has a total Noether charge $Q_O^{(0)}$ associated with the non-interacting configuration ($R\rightarrow \infty$) of the form:
\begin{equation}
Q_O^{(0)}=Q_A^{(0)}+Q_B^{(0)}\,,
\end{equation}
where $Q_B^{(0)}$ defines a singular single-center solution in the orbit $\alpha^{(6)}\gamma^{(6;5)}\beta^{(4)}$ while $Q_A^{(0)}$ corresponds to a regular small black hole in the orbit $\alpha^{(2)}\gamma^{(2;2)}\beta^{(2)}$ (non-BPS critical solution). More specifically, for the center  $B$, we can set
\begin{equation}
\tilde{Q}_1=-\frac{1}{2}\quad\,,\,\quad\,\tilde{Q}_2=\frac{1}{2}\quad\,,\,\quad\,\tilde{d}_1=2\quad\,,\,\quad\,\tilde{d}_2=2\quad\,,
\label{par1}\end{equation}
 We end up with a singular non-BPS single center solution described by a matrix $Q_B^{(0)}$ in the generating frame (\ref{L0G}), of the form
\begin{equation}
Q_B^{(0)}=N_{1}^{+}+N_{2}^{-}+N_{3}^{-}-N_{4}^{-}\,.
\end{equation}
If we send $A$ to infinity and shift the origin to $B$, the solution has $M_{ADM}=1/2$ \, and warp function $e^{-4\,U}=(r-1)(r+1)^{3}/r^{4}$ with a single pole at $r=1$ which is in general not removed by the interaction terms. \\
As for the second center, we can choose
\begin{equation}
d_1=2\quad\,,\,\quad\,d_2=2\quad\,,\,\quad\,Q_6=2\quad\,.
\label{par2}\end{equation}
 The second center corresponds to a non-BPS critical solution which, in the limit $R\rightarrow \infty$, would have $M_{ADM}=1/2$ and  warp function $e^{-4\,U}=1+2/r+3/(4\,r^2)$. The Noether charges of the global solution can be computed using the formulas given earlier and one can verify that the Noether charge matrix is indeed in the  $\alpha^{(7)}\gamma^{(7;1)}\beta^{(1)}$ orbit.

Although this is  just one instance of solution, we emphasize that in any two-center solution with $Q_O^{(0)}$ in the same $H^*$-orbit one of the two black holes is singular.

\paragraph{The orbit $\mathcal{\textbf{O}}^{\textbf{7;1,2}}$}\par
As is in the previous case, we choose the parameters of the $B$ center so that its intrinsic charge-matrix $Q_B^{(0)}$ be in the  \emph{generating solution}-frame. In particular we have chosen
\begin{equation}
\tilde{Q}_1=1\quad\,,\,\quad\,\tilde{Q}_2=\frac{1}{2}\quad\,,\,\quad\,\tilde{d}_1=2\quad\,,\,\quad\,\tilde{d}_2=2\quad\,.
\label{par3}\end{equation}
Sending the center $A$ to infinity and placing the center $B$ in the origin, we end up with a regular single-center solution in the orbit $\alpha^{(6)}\gamma^{(6;5)}\beta^{(5)}\delta^{(1)}$, whose $M_{ADM}=5/4$, and  warp function $e^{-4\,U}=(r+1)^{3}(r+2)/r^{4}$ everywhere positive.
As for the $A$-center, we choose the $d_i$'s and $Q_6$ as in (\ref{par2}) so that it describes a non-BPS critical single-center black hole, as before.\\
The total Noether charge matrix $Q_O$ with the above chosen parameters belongs to the orbit $\alpha^{(7)}\gamma^{(7;1)}\beta^{(2)}$, and is the sum of two regular black holes, with total mass $M_{ADM}=7/4$\, and $(p,\,q)$ charges
\begin{equation}
\Gamma^M\,=\,(\,p^\Lambda\,,\,q_\Lambda\,)\,=\,\left(\,0\,,\,\sqrt{2}\,,\,\frac{1}{\sqrt{2}}\,,\,\sqrt{2}\,,\,-\sqrt{2}\,,\,0\,,\,-\frac{1}{2\sqrt{2}}\,,\,-\frac{1}{2\sqrt{2}}\,\right)\quad.
\end{equation}
Being both centers regular, we can expect to have a stable multi-center system.
Having fixed all parameters we can now study the lower bound of $R$. As mentioned before, this value is determined by studying the behavior of the total warp function near each center and imposing the positivity of the leading terms, in order to remove the existence of poles. We have
\begin{equation}
\begin{split}
e^{-4\,U}\stackrel{\stackrel{r\rightarrow0}{\theta=\pi}}{\cong}&\frac{1}{r^2}\left[-\frac{4+12R+5R^2-6R^3-3R^4}{4R^4}\right]+\mathcal{O}(r^{-1})\quad,\label{ex4Ua7g1b2}
\end{split}
\end{equation}
for the $A$ center and
\begin{equation}
\begin{split}
e^{-4\,U}\stackrel{\stackrel{r\rightarrow{R}}{\theta=0}}{\cong}&\frac{1}{(r-R)^4}\left[2-\tilde{\alpha}^{2}\right]+\mathcal{O}((r-R)^{-3})\equiv\\
&\equiv\frac{1}{(r-R)^4}\left[-I_{4}(p^{\Lambda}_{B},q_{B\,\Lambda})-\tilde{\alpha}^{2}\right]+\mathcal{O}((r-R)^{-3})\quad,
\end{split}
\end{equation}
for the $B$ center. As one can see, requiring positivity of $e^{-4\,U}$ in the (\ref{ex4Ua7g1b2}) near the $A$ center implies
\begin{equation}
\begin{split}
R>\frac{1}{6}\left(3+\sqrt{33}\right)\,.
\end{split}
\end{equation}
\\
Below this value of the distance $R$, the solution exhibits curvature singularities, although composed of two regular black holes. One can interpret this result as follows: When the two centers get below a minimum distance the interaction forces are too strong and destabilize the system.\\
In order to study this system of regular black holes, we fix  $R$ to a value, say $12$, above the bound; once the mutual distance is chosen, the interaction between the two centers is fixed and the warp function $e^{-4\,U}$\, and the curvature scalar $\mathcal{R}$\, only depend on the coordinates $r$ and $\theta$. In order to exclude the presence of closed time-like curves (CTC) in the solution, we have to make sure that $e^{-4\,U}$ and $e^{-4\,U}\,r^{2}\sin^{2}\theta-\omega^2$ are positive everywhere. \\
We have verified that $e^{-4\,U}>0$ and $e^{-4\,U}\,r^{2}\sin^{2}\theta-\omega^2>0$ for any choice of $r$ and $\theta$ at $R$ fixed, see Figs. a), b), c) below.\\
\begin{figure}[H]
\begin{center}
\subfloat[][\small\,The behavior of $e^{-4\,U}\,r^{2}\sin^{2}\theta-\omega^2$.]{\includegraphics[width=7cm]{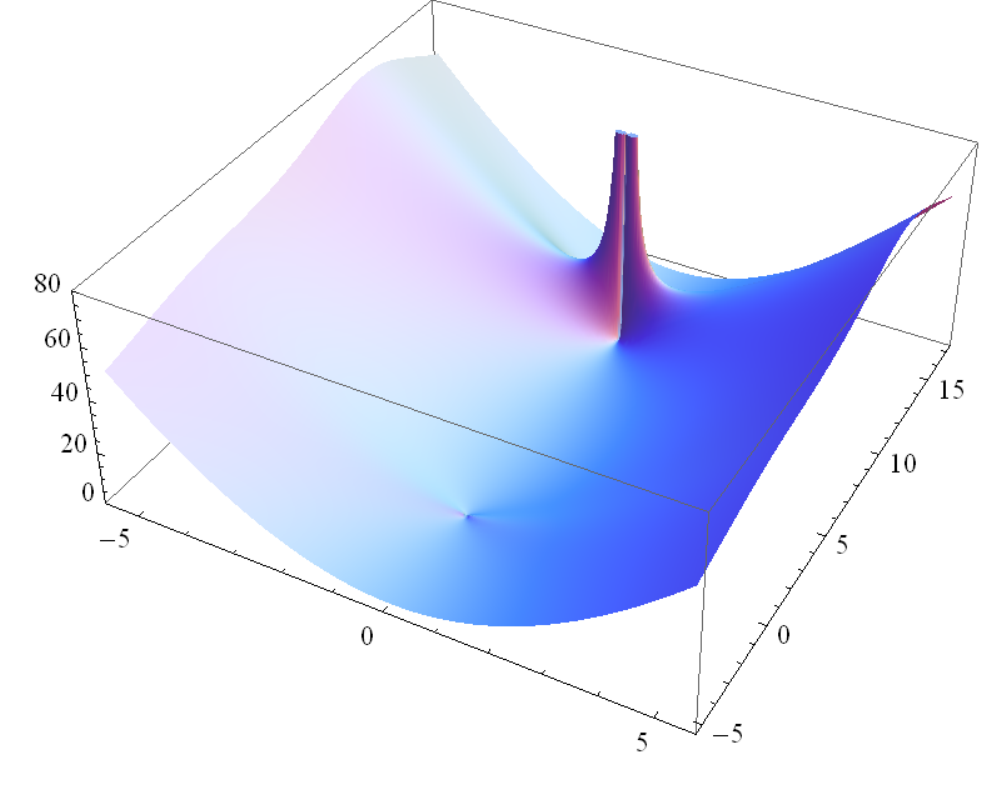}} \hfill
\subfloat[][\small\,The behavior of $e^{-4\,U}\,r^{2}\sin^{2}\theta-\omega^2$ near the $A$ center, corresponding to $\alpha^{(2)}\gamma^{(2;2)}\beta^{(2)}$.]{\includegraphics[width=7cm]{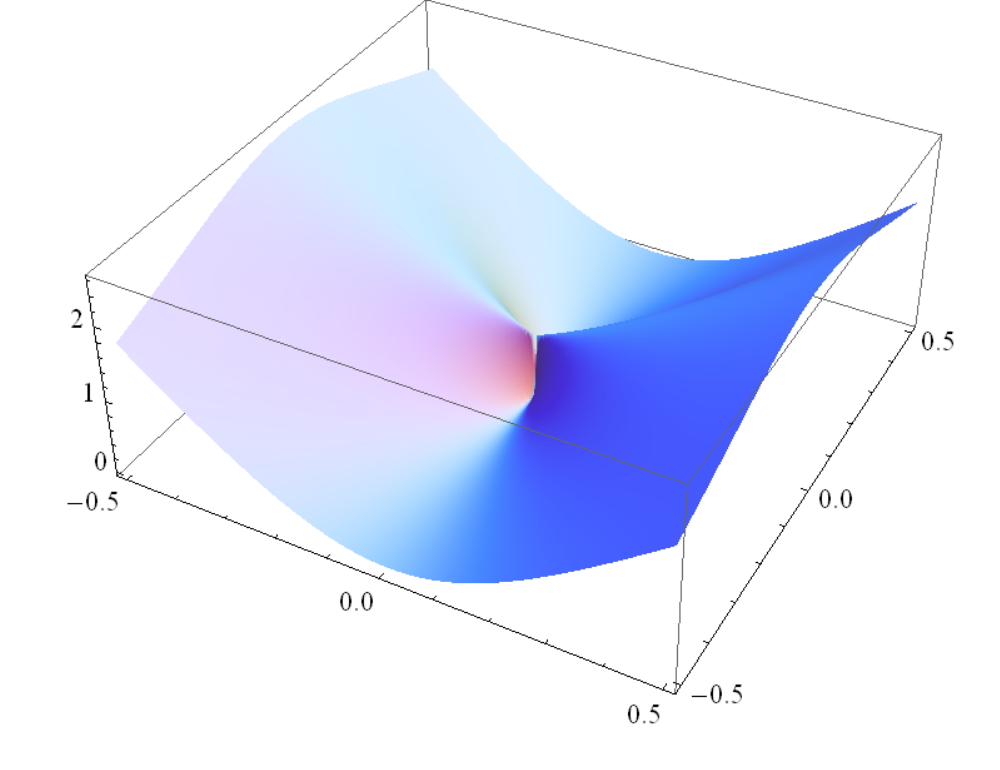}}\\
\subfloat[][\small\,The behavior of $e^{-4\,U}\,r^{2}\sin^{2}\theta-\omega^2$ near the $B$ center, corresponding to $\alpha^{(6)}\gamma^{(6;5)}\beta^{(5)}\delta^{(1)}$.]{\includegraphics[width=7cm]{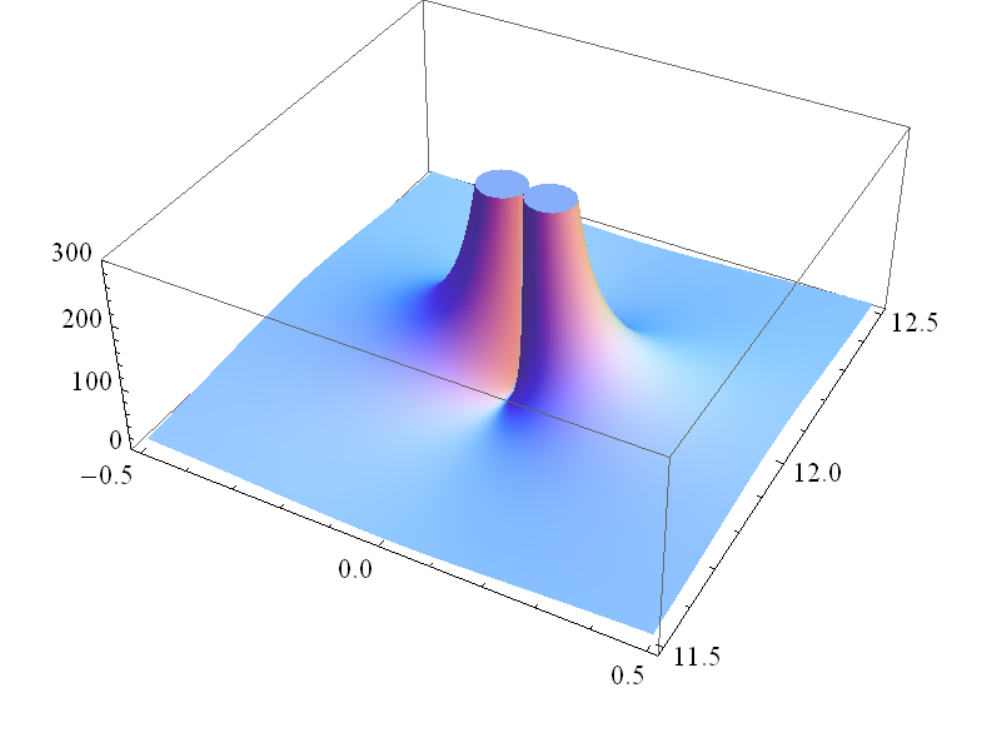}} \hfill
\end{center}
\end{figure}
The previous plots show that the presence of closed time-like curves can be excluded and this follows not only from the regularity of each center, that remains a necessary condition for this, but also from equations (\ref{NUTcond7}). The following plots show the behavior of the warp function $e^{-4U}$ and the curvature scalar $\mathcal{R}$ at the chosen value of the mutual distance $R$ between the two centers. The curvature scalar, which only depends on the scalar fields (see Eq. (\ref{eeq})) and is quadratic in their derivatives, shows for the $B$ center a characteristic ``volcano''-shape associated with the \emph{attractor mechanism} \cite{attractor} (see Figure (f)): it is non-vanishing about the center and zero at the horizon where the derivatives of the scalars vanish. The same quantity diverges at $r=0$ where the small black hole is located (see Figure (e)).
\begin{figure}[H]
\begin{center}
\subfloat[][The behavior of the warp function $e^{-4\,U}$.]{\includegraphics[width=7cm]{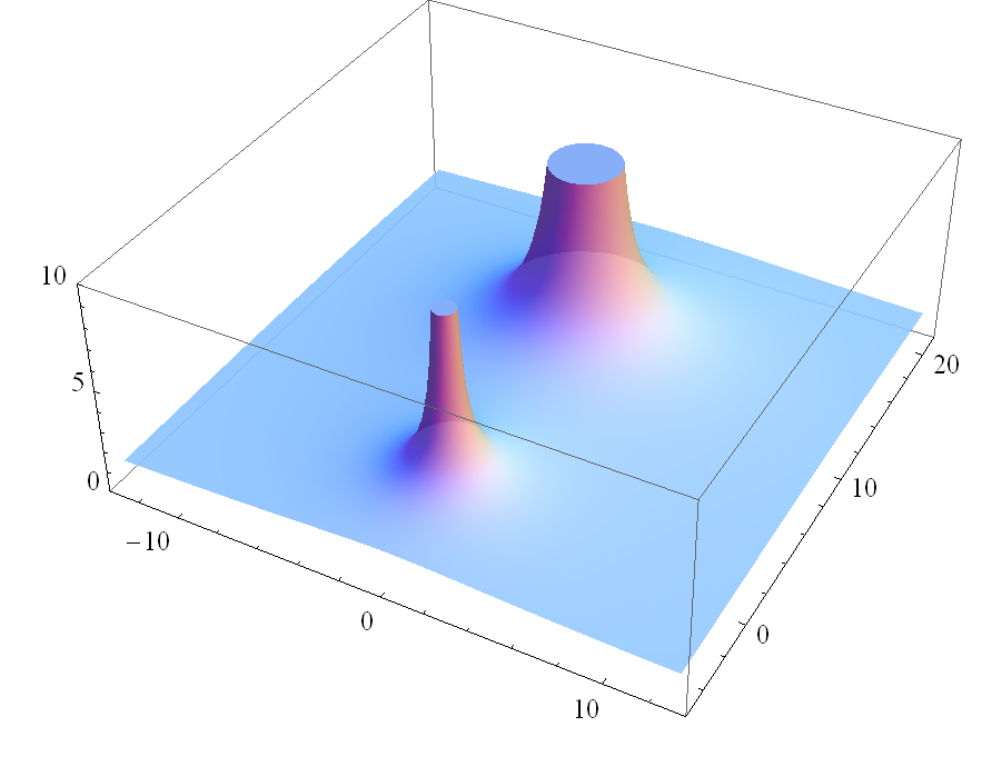}} \hfill
\subfloat[][The behavior of the curvature scalar $\mathcal{R}$.]{\includegraphics[width=7cm]{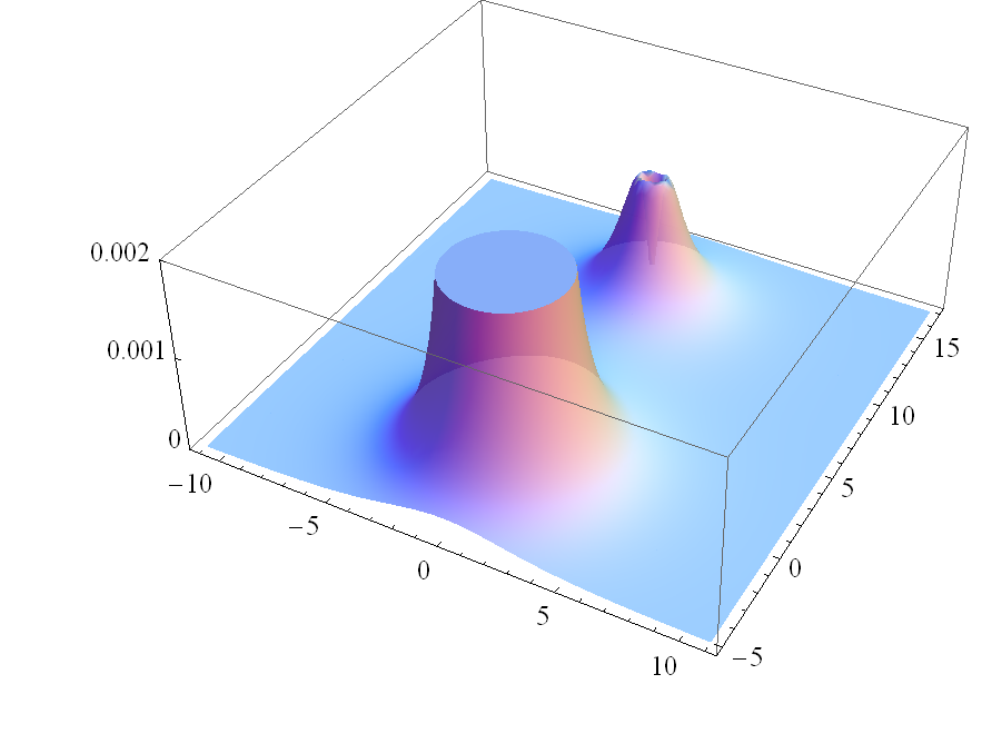}}\\
\subfloat[][Detail of the curvature scalar $\mathcal{R}$ for the regular center $\alpha^{(6)}\gamma^{(6;5)}\beta^{(5)}\delta^{(1)}$.]{\includegraphics[width=7cm]{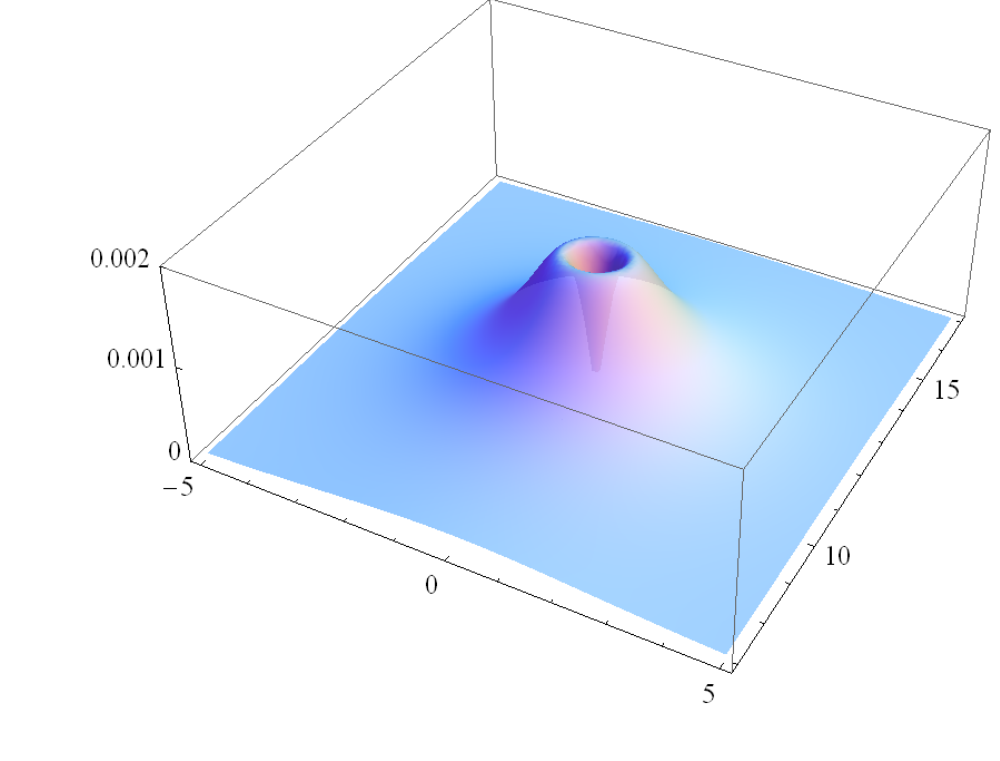}} \hfill
\end{center}
\end{figure}

\subsection{The orbit $\alpha^{(7)}\gamma^{(7;2)}\beta^{(i)}$}
Consider now the $H^{*\,\mathbb{C}}$-nilpotent orbit defined by $\gamma^{(7;2)}$.  The graded decomposition of $\mathfrak{so}(4,4)$ with respect to the corresponding neutral element $h$ yields a nilpotent algebra $\mathfrak{n}$ with the same structure (\ref{n7_1}),(\ref{comrel1}) as for the $\gamma^{(7;1)}$ case. We then consider the same Ansatz
 for the coset representative, with $\hat{\mathbb{L}}(x^i)$ given by (\ref{ansazc}) and obtain, after performing suitable redefinitions of the components of $\mathcal{Y}(x^i)$, the same equations of motion (\ref{eqmot1}).
This time we choose for the neutral element $h$ the following matrix:
 \begin{align}
h&=2 \text{e}_{1,4}+2 \text{e}_{1,5}-\text{e}_{2,3}+\text{e}_{2,6}-\text{e}_{3,2}-\text{e}_{3,7}+2 \text{e}_{4,1}-2 \text{e}_{4,8}+2
   \text{e}_{5,1}-2 \text{e}_{5,8}+\text{e}_{6,2}+\text{e}_{6,7}-\text{e}_{7,3}+\nonumber\\&+\text{e}_{7,6}-2 \text{e}_{8,4}-2 \text{e}_{8,5}\,.
\end{align}
  Being however the neutral element different with respect to the $\gamma^{(7;1)}$-case, the explicit from of the  solution $\phi^I(x^i)$ will also be different. The required redefinitions are
\begin{equation}
\begin{split}
\tilde{V}&\rightarrow{V}\,,\\
\tilde{L_{1}}&\rightarrow{2(2L_{1}-1)}\quad\,,\,\quad\tilde{L_{2}}\rightarrow{2(2L_{2}+1)}\,,\\
\tilde{Z_{1}}&\rightarrow{4(Z_{1}+2)}\quad\,\,\,,\,\,\quad\tilde{Z_{2}}\rightarrow{4(Z_{2}-2)}\,,\\
\tilde{M}&\rightarrow{32M+\frac{4}{3}V\left[8L_{1}L_{2}-2(L_{1}-L_{2})+1\right]}\,,\\\label{redef2}
\end{split}
\end{equation}
and yield for $e^{-4U}$ the same expression
\begin{equation}
e^{-4U}=L_{1}L_{2}Z_{1}Z_{2}-M^{2}\,,
\end{equation}
while for the  scalars $\mathcal{Z}^M$ and $a$ we find
\begin{equation}
\begin{split}
\mathcal{Z}^{0}\,&=\,\frac{1}{\sqrt{2}}\,\left[\,1\,+\,\frac{L_{1}\,(\,M\,V\,+\,Z_{1}\,Z_{2}\,)}{2\,e^{-4U}}\right]\,,\\
\mathcal{Z}^{1}\,&=\,-\frac{\,\sqrt{2}\,M\,L_{1}}{e^{-4U}}\,,\\
\mathcal{Z}^{2}\,&=\,-\frac{(\,M\,+\,V\,L_{1}\,L_{2}\,)\,Z_{2}}{2\,\sqrt{2}\,e^{-4U}}\,,\\
\mathcal{Z}^{3}\,&=\,\frac{(\,M\,+\,V\,L_{1}\,L_{2}\,)\,Z_{1}}{2\,\sqrt{2}\,e^{-4U}}\,,\\
\mathcal{Z}_{0}\,&=\,-\frac{\,\sqrt{2}\,M\,L_{2}}{e^{-4U}}\,,\\
\mathcal{Z}_{1}\,&=\,\frac{1}{\sqrt{2}}\,\left[\,1\,-\,\frac{L_{2}\,(\,M\,V\,+\,Z_{1}\,Z_{2}\,)}{2\,e^{-4U}}\right]\,,\\
\mathcal{Z}_{2}\,&=\,\frac{1}{\sqrt{2}}\,\left(\,1\,-\,\frac{2\,L_{1}\,L_{2}\,Z_{1}}{e^{-4U}}\right)\,,\\
\mathcal{Z}_{3}\,&=\,\frac{1}{\sqrt{2}}\,\left(\,1\,+\,\frac{2\,L_{1}\,L_{2}\,Z_{2}}{e^{-4U}}\right)\,,\\
a\,&=\,\frac{V\,L_{1}\,L_{2}(Z_{1}\,-\,Z_{2})\,-\,M\,[4(L_{1}\,-\,L_{2})\,-\,Z_{1}\,+\,Z_{2}\,-\,4]}{4\,e^{-4U}}\,.\\
\label{chpq2}
\end{split}
\end{equation}
The four-dimensional scalar fields read
\begin{equation}
\begin{aligned}
\epsilon_{1}\,&=\,-\,\frac{4\,(\,M\,+\,L_{1}\,L_{2}\,V\,)}{2\,M\,V\,+\,L_{1}\,L_{2}\,V^{2}\,+\,Z_{1}\,Z_{2}}&,\quad&e^{-2\,\varphi_{1}}\,&=&\,\frac{(\,2\,M\,V\,+\,L_{1}\,L_{2}\,V^{2}\,+\,Z_{1}\,Z_{2})^{2}}{16\,e^{-4U}}\quad,\\
\epsilon_{2}\,&=\,-\,\frac{M}{L_{1}\,Z_{1}}&,\quad&e^{-2\,\varphi_{2}}\,&=&\,\frac{L_{1}^{2}\,Z_{1}^{2}}{e^{-4U}}\quad,\\
\epsilon_{3}\,&=\,\frac{M}{L_{1}\,Z_{2}}&,\quad&e^{-2\,\varphi_{3}}\,&=&\,\frac{L_{1}^{2}\,Z_{2}^{2}}{e^{-4U}}\quad.
\label{scalar72}
\end{aligned}
\end{equation}
The equation for $\omega$ is the same as (\ref{omega71}) while  those for the $D=3$  vector fields $\tilde{A}^M$ read:\\
\begin{align}
\ast{d}{\tilde{A}^{0}}&=\frac{\ast{d}{\omega}}{\sqrt{2}}-\frac{1}{2\sqrt{2}}\left(V{d}L_{1}-L_{1}{d}V\right)\,,\,\,\ast{d}{\tilde{A}^{1}}=
\sqrt{2}\,{d}{L_{1}}\,,\,\,\ast{d}{\tilde{A}^{2}}=\frac{{d}{Z_{2}}}{2\sqrt{2}}\,,\,\,\ast{d}{\tilde{A}^{3}}=-\frac{{d}{Z_{1}}}{2\sqrt{2}}\quad,\nonumber\\
\ast{d}{\tilde{A}_{0}}&=\sqrt{2}\,{d}{L_{2}}\,,\,\,\ast{d}{\tilde{A}_{1}}=\frac{\ast{d}{\omega}}{\sqrt{2}}+\frac{1}{2\sqrt{2}}
\left(V{d}L_{2}-L_{2}{d}V\right)\,,\,\,\ast{d}{\tilde{A}_{2}}=\frac{\ast{d}{\omega}}{\sqrt{2}}\,,\,\,\ast{d}{\tilde{A}_{3}}=
\frac{\ast{d}{\omega}}{\sqrt{2}}\quad.
\label{vect72}
\end{align}

\subsubsection{The solution}
Referring to the equations (\ref{eqq1}) and (\ref{eqq2}) of \ref{subsect1}, which are independent of the orbit and continue to hold also in this case, we write below the conditions we have imposed to obtain our solution.
\paragraph{Regularity conditions and asymptotic flatness}\par
The absence of a NUT charge for each center implies the same conditions on the coefficients $m$ and $\tilde{m}$ as (\ref{NUTcond7}).
In order to have a well-behaved solution near each center, just as in the previous case, we need to require $e^{-4U}$ to go, like $1/r^a$ and $1/\Sigma^a$ near the $A$ and $B$ centers, respectively, with $a\le 4$. The resulting conditions are the same as as for the $\alpha^{(7)}\gamma^{(7;1)}\beta^{(i)}$, and we choose also in this case, the conditions $\tilde{Q}_6=0$,\,$Q_{1}=0$,\,$Q_{2}=0$ and $\alpha=0$ (the latter being required in order to have a well defined warp factor near the small black hole).
Asymptotic flatness further implies
\begin{equation}
h=\frac{2(m_0-\sqrt{l_{1}l_{2}l_{3}l_{4}-1})}{l_{1}l_{2}}\,.
\end{equation}
\paragraph{Electric-magnetic charges}
Below we write the total electric-magnetic charge vector $\Gamma^M$ as well as those associated with the $A$ and $B$-centers:
\begin{equation}
\begin{split}
\Gamma^M\,=\,(\,p^\Lambda\,,\,q_\Lambda\,)\,=\,\bigg(\,&\frac{l_{1}(Q_{6}+\tilde{Q}_{6})-h(Q_{1}+\tilde{Q}_{1})}{2\sqrt{2}}\,,\,\sqrt{2}(Q_{1}+\tilde{Q}_{1})\,,\,\frac{d_2+\tilde{d}_2}{2\sqrt{2}}\,,\,-\frac{d_1+\tilde{d}_1}{2\sqrt{2}}\,,\\
&,\,\sqrt{2}(Q_{2}+\tilde{Q}_{2})\,,\,-\frac{l_{2}(Q_{6}+\tilde{Q}_{6})-h(Q_{2}+\tilde{Q}_{2})}{2\sqrt{2}}\,,\,0\,,\,0\,\bigg)\quad,
\end{split}
\end{equation}\begin{equation}
\begin{split}
{\Gamma^{M}}_{A}\,=\,(\,{p^{\Lambda}}_{A}\,,\,{q_{\Lambda}}_{A}\,)\,=\,\bigg(\,&\frac{Q_{6}(\tilde{Q}_{1}+R{l}_{1})-Q_{1}(\tilde{Q}_{6}+h{R})}{2\sqrt{2}R}\,,\,\sqrt{2}Q_{1}\,,\,\frac{d_2}{2\sqrt{2}}\,,\,-\frac{d_1}{2\sqrt{2}}\,,\,\sqrt{2}Q_{2}\,,\\
&,\,\frac{-Q_{6}(\tilde{Q}_{2}+R{l}_{2})+Q_{2}(\tilde{Q}_{6}+h{R})}{2\sqrt{2}R}\,,\,0\,,\,0\,\bigg)\quad,
\end{split}
\end{equation}\begin{equation}
\begin{split}
{\Gamma^{M}}_{B}\,=\,(\,{p^{\Lambda}}_{B}\,,\,{q_{\Lambda}}_{B}\,)\,=\,\bigg(\,&\frac{\tilde{Q}_{6}(Q_{1}+R{l}_{1})-\tilde{Q}_{1}(Q_{6}+h{R})}{2\sqrt{2}R}\,,\,\sqrt{2}\tilde{Q}_1\,,\,\frac{\tilde{d}_2}{2\sqrt{2}}\,,\,-\frac{\tilde{d}_1}{2\sqrt{2}}\,,\,\sqrt{2}\tilde{Q}_2\,,\\
&,\,\frac{-\tilde{Q}_{6}(Q_{2}+R{l}_{2})+\tilde{Q}_{2}(Q_{6}+h{R})}{2\sqrt{2}R}\,,\,0\,,\,0\,\bigg)\quad.
\end{split}
\end{equation}

\subsubsection{The orbits $\mathcal{O}^{7;2,1}$ and $\mathcal{O}^{7;2,2}$}
Again, we have to distinguish between the real nilpotent orbits $\alpha^{(7)}\gamma^{(7;2)}\beta^{(i)}$ using the $\beta$-labels, requiring the Noether charge matrix $Q_O$ to be either in  $\alpha^{(7)}\gamma^{(7;2)}\beta^{(1)}$ or in $\alpha^{(7)}\gamma^{(7;2)}\beta^{(2)}$.\par
As a simplifying condition, without affecting the $H^*$-orbit of $Q_O$, we can set the values of the scalars at the origin,\,\textit{i.e.} at the infinity, requiring that $\lim_{r \rightarrow +\infty}\mathcal{M}^{(3)}_{scal}=\eta$ and find
\begin{equation}
m_0=0\quad\,,\,\quad\,l_{1}=\frac{1}{2}\quad\,,\,\quad\,l_{2}=-\frac{1}{2}\quad\,,\,\quad\,l_{3}=-2\quad\,,\,\quad\,l_{4}=2\quad\,,\,\quad\,h=0\,.
\end{equation}
\paragraph{The orbit $\mathcal{\textbf{O}}^{\textbf{7;2,1}}$}\par

We start with $\alpha^{(7)}\gamma^{(7;2)}\beta^{(1)}$, considering the system as composed by two single-center black holes: the $B$ described by the generating solution of the corresponding orbit. The orbit of the other center is determined by the tensor classifiers.\\
For the $B$ center, at the distance $R$ from the origin along the $z$-axes, we have chosen
\begin{equation}
\tilde{Q}_1=1\quad\,,\,\quad\,\tilde{Q}_2=-1\quad\,,\,\quad\,\tilde{d}_1=-2\quad\,,\,\quad\,\tilde{d}_2=2\quad\,.
\label{par3}\end{equation}
 We end up with a regular non-BPS single-center solution, with $Q_B^{(0)}$ of the form
\begin{equation}
Q_B^{(0)}=2N_{1}^{+}+2N_{2}^{-}+N_{3}^{-}+N_{4}^{-}\,,
\end{equation}
corresponding to the orbit $\alpha^{(6)}\gamma^{(6;5)}\beta^{(5)}\delta^{(1)}$.
For the $A$ center at the origin, we have chosen
\begin{equation}
d_1=-2\quad\,,\,\quad\,d_2=2\quad\,,\,\quad\,Q_6=2\quad\,,
\label{par4}\end{equation}
which defines a BPS critical small black hole in the orbit $\alpha^{(2)}\gamma^{(2;1)}\beta^{(1)}$. If isolated from the other center, this black hole would have  $M_{ADM}=1/2$, $e^{-4\,U}=1+2/r+3/(4\,r^2)$.
The Noether charge matrix with the parameters chosen as in (\ref{par3}) and (\ref{par4}), correspond to have a $\alpha^{(7)}\gamma^{(7;2)}\beta^{(1)}$ with total $M_{ADM}=2$ and $(p,\,q)$ charges
\begin{equation}
{\Gamma^{M}}\,=\,(\,{p^{\Lambda}}\,,\,{q_{\Lambda}}\,)\,=\,\left(\,\frac{1}{2\sqrt{2}}\,,\,\sqrt{2}\,,\,\sqrt{2}\,,\,\sqrt{2}\,,\,-\sqrt{2}\,,\,\frac{1}{2\sqrt{2}}\,,\,0\,,\,0\,\right)\quad.
\end{equation}
We can now proceed to the study of the behavior of the total warp function near each center and imposing the positivity of the leading terms, in order to remove the existence of poles and determine the lower bound of $R$. We have
\begin{equation}
\begin{split}
e^{-4\,U}\stackrel{\stackrel{r\rightarrow0}{\theta=\pi}}{\cong}&\frac{1}{r^2}\left[\frac{(R+2)^2(3R+2)(R-2)}{4R^4}\right]+\mathcal{O}(r^{-1})\quad,\label{ex4Ua7g2b1}
\end{split}
\end{equation}
for the $A$ center and
\begin{equation}
\begin{split}
e^{-4\,U}\stackrel{\stackrel{r\rightarrow{R}}{\theta=0}}{\cong}&\frac{1}{(r-R)^4}\left[4-\tilde{\alpha}^{2}\right]+\mathcal{O}((r-R)^{-3})\equiv\\
&\equiv\frac{1}{(r-R)^4}\left[-I_{4}(p^{\Lambda}_{B},q_{B\,\Lambda})-\tilde{\alpha}^{2}\right]+\mathcal{O}((r-R)^{-3})\quad,
\end{split}
\end{equation}
for the $B$ center. Requiring positivity of $e^{-4\,U}$ in the (\ref{ex4Ua7g2b1}) near the $A$ center implies the following lower bound for the distance $R$
\begin{equation}
\begin{split}
R>2\quad.\label{lbRa7g2b1}
\end{split}
\end{equation}
\\
Below this value of the distance $R$, the global warp function $e^{-4\,U}$ shows unphysical singularities and the system, although composed from two regular black holes, ceases to be globally regular.\\
We again exclude the presence of closed time-like curves  in the solution when Eq. (\ref{lbRa7g2b1}) holds; having fixed the value of $R$ above the lower bound, we can study the warp function $e^{-4\,U}$\, and the curvature scalar $\mathcal{R}$ , which are illustrated in the following plots.
\begin{figure}[H]
\centering
\subfloat[][The behavior of the warp function $e^{-4U}$.]{\includegraphics[width=7cm]{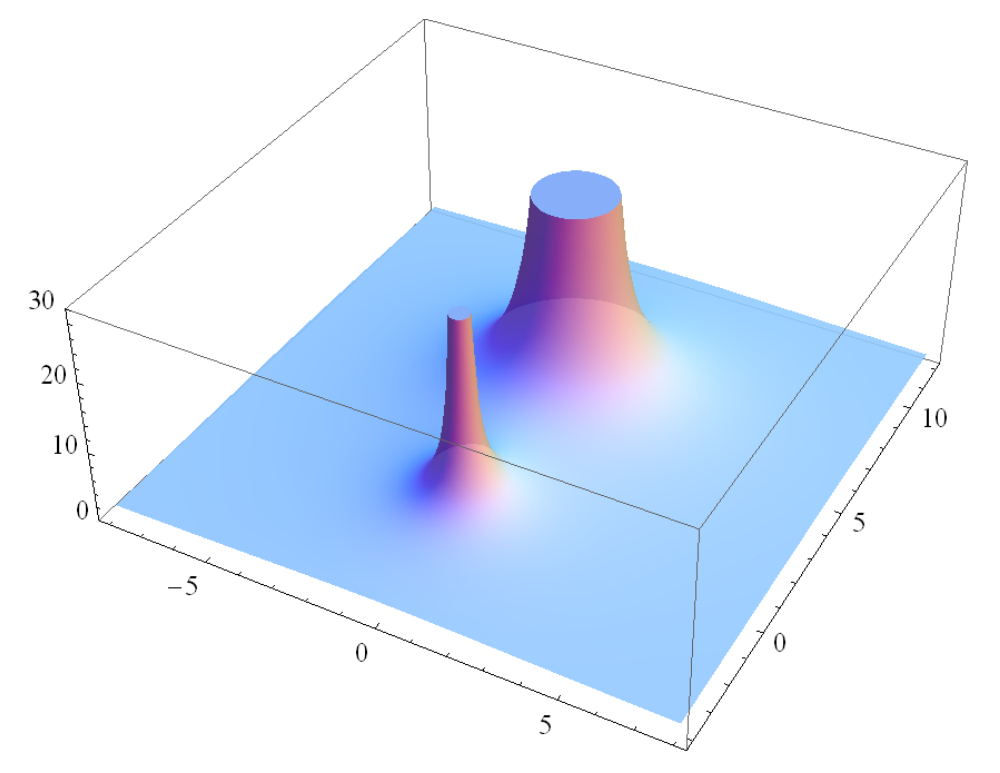}} \hfill
\subfloat[][The behavior of the curvature scalar $\mathcal{R}$.]{\includegraphics[width=7cm]{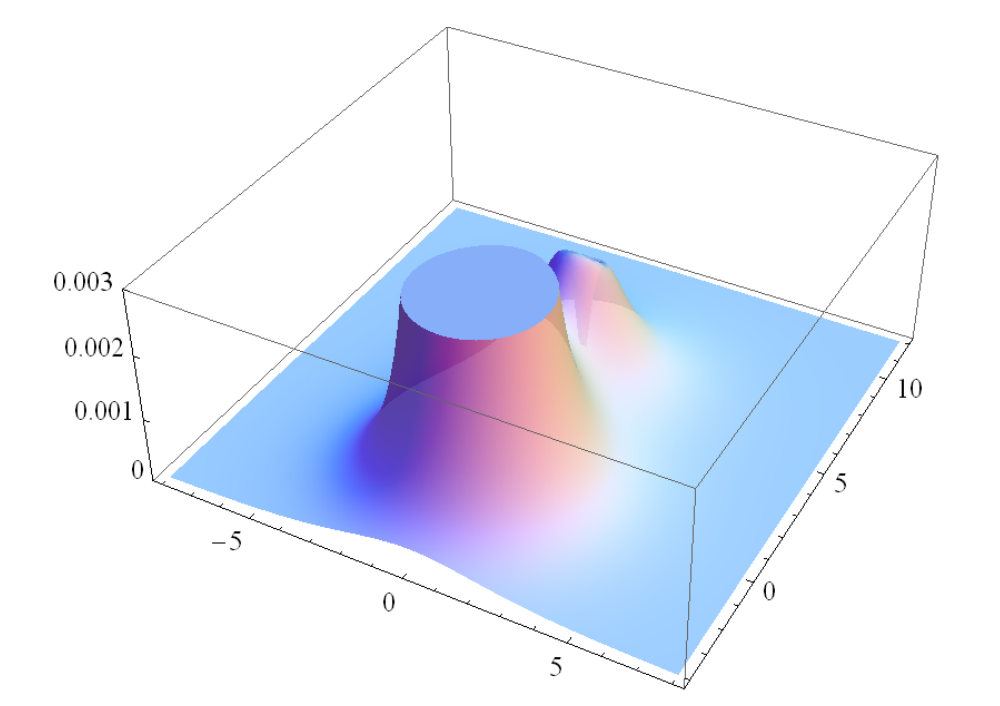}}\\
\subfloat[][Detail of the curvature scalar $\mathcal{R}$ near the regular $\alpha^{(6)}\gamma^{(6;5)}\beta^{(5)}\delta^{(1)}$.]{\includegraphics[width=7cm]{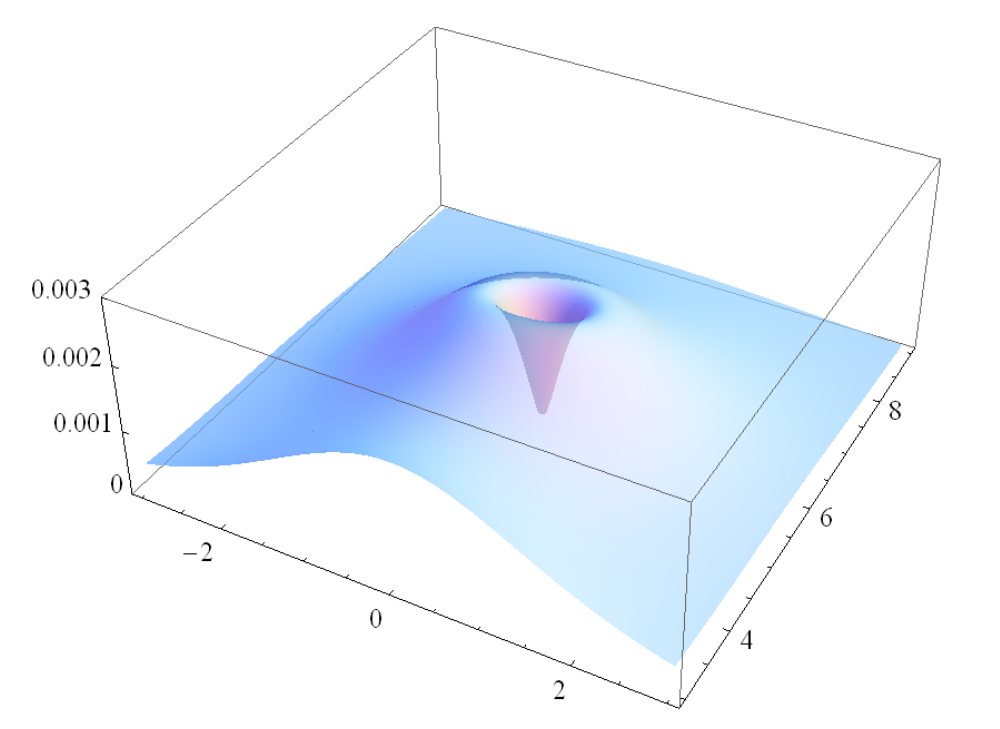}} \hfill
\end{figure}
Note that the curvature scalar $\mathcal{R}$, which diverges when approaching the small $A$-black hole (see Fig. h)), is finite about the center with finite horizon (see Fig. i)), and exhibits the characteristic shape related to the attractor mechanism.
\paragraph{The orbit $\mathcal{\textbf{O}}^{\textbf{7;2,2}}$}\par
Consistently with the general result illustrated in Sect. \ref{compositionlaw}, we could not find regular single center solutions yielding a total Noether charge matrix in this orbit. For instance we could produce a solution with $Q_O$ in $\mathcal{\textbf{O}}^{\textbf{7;2,2}}$ by choosing for the $B$-center
\begin{equation}
\tilde{Q}_1=-\frac{1}{2}\quad\,,\,\quad\,\tilde{Q}_2=-\frac{1}{2}\quad\,,\,\quad\,\tilde{d}_1=-2\quad\,,\,\quad\,\tilde{d}_2=2\quad\,,
\label{par5}\end{equation}
which defines  a singular single center solution corresponding to a $\alpha^{(6)}\gamma^{(6;5)}\beta^{(2)}$.\par
As for the $A$-center, we choose the $d_i$'s and $Q_6$ as in (\ref{par4}) fixing the other parameters to be zero, ending up in the same BPS critical single-center orbit $\alpha^{(2)}\gamma^{(2;1)}\beta^{(1)}$.\\
The Noether charge matrix $Q_O$, with the chosen parameters, is contained in the orbit  $\alpha^{(7)}\gamma^{(7;2)}\beta^{(2)}$, and the solution is the sum of a singular and a regular black hole, with total $M_{ADM}=1$ and $(p,\,q)$ charges
\begin{equation}
{\Gamma^{M}}\,=\,(\,{p^{\Lambda}}\,,\,{q_{\Lambda}}\,)\,=\,\left(\,\frac{1}{2\sqrt{2}}\,,\,-\frac{1}{\sqrt{2}}\,,\,\sqrt{2}\,,\,\sqrt{2}\,,\,-\frac{1}{\sqrt{2}}\,,\,\frac{1}{2\sqrt{2}}\,,\,0\,,\,0\,\right)\quad.
\end{equation}
As for $(\alpha^{(7)}\gamma^{(7;1)}\beta^{(1)})$ the presence of a singular black hole spoils the regularity of the whole system.
\subsection{Summary of representatives for\, $\alpha^{(7)}$,\,\,$\alpha^{(8)}$ and \,$\alpha^{(9)}$ }
Due to triality, the discussion for the orbits $\alpha^{(8)}$ and $\alpha^{(9)}$ follows the same lines as that for $\alpha^{(7)}$, so we omit it. We limit ourselves here to write, in the tables below, representatives of each of the corresponding $H^*$-orbits as sum of representatives of single-center orbits. For the $\alpha^{(7)}$-orbits these combinations correspond to the solutions discussed above.
\begin{table}[H]
\begin{center}
\begin{tabular}{|c|c|c|}\hline
\backslashbox{$\gamma$}{$\beta$}
&$\beta^{(1)}$&$\beta^{(2)}$ \\   \hline
$\gamma^{(1)}$&\multicolumn{1}{c|}{{\begin{tabular}{c}
\\ $(\alpha^{(6)}\gamma^{(6;5)}\beta^{(4)})\,+$\\
$(\alpha^{(2)}\gamma^{(2;2)}\beta^{(2)})$\\
\\
\end{tabular}}} & \multicolumn{1}{c|}{{\begin{tabular}{c}
\\ $(\alpha^{(6)}\gamma^{(6;5)}\beta^{(5)}\delta^{(1)})\,+$\\
$(\alpha^{(2)}\gamma^{(2;2)}\beta^{(2)})$\\
\\
\end{tabular}}}\\ \hline
$\gamma^{(2)}$&\multicolumn{1}{c|}{{\begin{tabular}{c}
\\ $(\alpha^{(6)}\gamma^{(6;5)}\beta^{(5)}\delta^{(1)})\,+$\\
$(\alpha^{(2)}\gamma^{(2;1)}\beta^{(1)})$\\
\\
\end{tabular}}}&\multicolumn{1}{c|}{{\begin{tabular}{c}
\\ $(\alpha^{(6)}\gamma^{(6;5)}\beta^{(2)})\,+$\\
$(\alpha^{(2)}\gamma^{(2;1)}\beta^{(1)})$\\
\\
\end{tabular}}}\\ \hline
\end{tabular}
\caption{\small Representatives of $\alpha^{(7)}$-orbit as combinations of representatives of single-center ones. }
\end{center}
\end{table}
\begin{table}[H]
\begin{center}
\begin{tabular}{|c|c|c|}\hline
\backslashbox{$\gamma$}{$\beta$}
&$\beta^{(1)}$&$\beta^{(2)}$ \\   \hline
$\gamma^{(1)}$&\multicolumn{1}{c|}{{\begin{tabular}{c}
\\ $(\alpha^{(6)}\gamma^{(6;5)}\beta^{(3)})\,+$\\
$(\alpha^{(3)}\gamma^{(3;2)}\beta^{(2)})$\\
\\
\end{tabular}}} & \multicolumn{1}{c|}{{\begin{tabular}{c}
\\ $(\alpha^{(6)}\gamma^{(6;5)}\beta^{(5)}\delta^{(1)})\,+$\\
$(\alpha^{(3)}\gamma^{(3;2)}\beta^{(2)})$\\
\\
\end{tabular}}}\\ \hline
$\gamma^{(2)}$&\multicolumn{1}{c|}{{\begin{tabular}{c}
\\ $(\alpha^{(6)}\gamma^{(6;5)}\beta^{(5)}\delta^{(1)})\,+$\\
$(\alpha^{(3)}\gamma^{(3;1)}\beta^{(1)})$\\
\\
\end{tabular}}}&\multicolumn{1}{c|}{{\begin{tabular}{c}
\\ $(\alpha^{(6)}\gamma^{(6;5)}\beta^{(1)})\,+$\\
$(\alpha^{(3)}\gamma^{(3;1)}\beta^{(1)})$\\
\\
\end{tabular}}}\\ \hline
\end{tabular}
\caption{\small Representatives of $\alpha^{(8)}$-orbit as combinations of representatives of single-center ones.. }
\end{center}
\end{table}
\begin{table}[H]
\begin{center}
\begin{tabular}{|c|c|c|}\hline
\backslashbox{$\gamma$}{$\beta$}
&$\beta^{(1)}$&$\beta^{(2)}$ \\   \hline
$\gamma^{(1)}$&\multicolumn{1}{c|}{{\begin{tabular}{c}
\\ $(\alpha^{(6)}\gamma^{(6;5)}\beta^{(4)})\,+$\\
$(\alpha^{(4)}\gamma^{(4;2)}\beta^{(2)})$\\
\\
\end{tabular}}} & \multicolumn{1}{c|}{{\begin{tabular}{c}
\\ $(\alpha^{(6)}\gamma^{(6;5)}\beta^{(5)}\delta^{(1)})\,+$\\
$(\alpha^{(4)}\gamma^{(4;2)}\beta^{(2)})$\\
\\
\end{tabular}}}\\ \hline
$\gamma^{(2)}$&\multicolumn{1}{c|}{{\begin{tabular}{c}
\\ $(\alpha^{(6)}\gamma^{(6;5)}\beta^{(5)}\delta^{(1)})\,+$\\
$(\alpha^{(4)}\gamma^{(4;1)}\beta^{(1)})$\\
\\
\end{tabular}}}&\multicolumn{1}{c|}{{\begin{tabular}{c}
\\ $(\alpha^{(6)}\gamma^{(6;5)}\beta^{(1)})\,+$\\
$(\alpha^{(4)}\gamma^{(4;1)}\beta^{(1)})$\\
\\
\end{tabular}}}\\ \hline
\end{tabular}
\caption{\small Representative of $\alpha^{(9)}$-orbit as combinations of representatives of single-center ones. }
\end{center}
\end{table}
The non-diagonal orbits, which contain regular solutions, can be obtained in a number of ways by combining regular-single-center orbits, as illustrated in Appendix \ref{sumrules}.
\subsection{The orbits $\alpha^{(10)}\gamma^{(10;(1,\dots,4))}\beta^{(1,\dots,4)}$}
Following \cite{bossard2}, we now consider now the orbits with give origin to the so-called \emph{composite non-BPS} solutions. For the construction of the general solution we review the analysis of \cite{bossard2}. We shall then discuss in detail, for three $H^*$ orbits, regular representative solutions. limiting ourselves to the detailed discussion of only three relevant instances of solutions, being the other solutions very similar. At the end of this section we report the complete results for each $\gamma$-$\beta$-label.
\subsubsection{The orbit $\alpha^{(10)}\gamma^{(10;1)}\beta^{(1,\dots,4)}$}
Let us start considering the graded decomposition of $\mathfrak{so}(4,4)$ respect to the weighted Dynkin diagram in \ref{labels}, which defines the orbits for $\alpha^{(10)}$. The nilpotent subalgebra $\mathfrak{n}$ has the following structure:
\begin{equation}{\mathfrak{n}^{(10;1,(1,\dots,4))}\cong(3\times\textbf{2})^{(1)}_{\mathfrak{K}^*}\oplus(3\times \textbf{1})^{(2)}_{\mathfrak{H}^*}\oplus\textbf{2}^{(3)}_{\mathfrak{K}^*}.}
\end{equation}
Choosing an appropriate basis of this algebra, the only non-zero commutators are \cite{bossard2}
\begin{equation}
\begin{split}
[\textit{\textbf{e}}_{\alpha}^{(1),i},\textit{\textbf{e}}_{\beta}^{(1),j}]&=\varepsilon_{\alpha\beta}c^{ijk}\textit{\textbf{f}}_{k}^{(2)}\,,\\
[\textit{\textbf{f}}_{i}^{(2)},\textit{\textbf{e}}_{\alpha}^{(1),j}]&=\delta^{j}_{i}\textit{\textbf{e}}_{\alpha}^{(3)}\,\,,\,\,\,\,i,j,k=1,2,3\,,\label{comrel2}
\end{split}
\end{equation}
having denoted by $\textit{\textbf{e}}$ the generators in ${\mathfrak{K}^*}$ and by $\textit{\textbf{f}}$ those in ${\mathfrak{H}^*}$, the number in superscript being the grading relative to the neutral element and $\alpha,\beta$ are doublet indices. The coefficients $c_{ijk}$ are defined as: $c_{ijk}=|\varepsilon_{ijk}|$ .\\
We now consider the Ansatz
\begin{equation}{\hat{\mathbb{L}}\,=\,\exp(-\tilde{K}_{i}^{\alpha}\textit{\textbf{e}}_{\alpha}^{(1),i}-\tilde{M}^{\alpha}\textit{\textbf{e}}_{\alpha}^{(3)})}
\end{equation}
and write in terms of the above functions the scalar field equations. With the following redefinitions
\begin{equation}
\begin{split}
\tilde{K}_{i}^{\alpha=1}&\rightarrow{2(L_{i}-1)}\,,\\
\tilde{K}_{i}^{\alpha=2}&\rightarrow{K_{i}}\,,\\
\tilde{M}^{\alpha=1}&\rightarrow{8M\,+\,\frac{4}{3}\,\left(\sum_{i}\,K_{i}\,-\,\sum_{k}\,c^{ijk}\,K_{i}\,L_{j}\,-\,c^{ijk}\,K_{i}\,L_{j}\,L_{k}\right)}\,,\\
\tilde{M}^{\alpha=2}&\rightarrow{2(V-1)\,+\,\frac{1}{3}\,\left(2\,\sum_{k}\,c^{ijk}\,K_{i}\,K_{j}\,+\,c^{ijk}\,K_{i}\,K_{j}\,L_{k}\right)}\,,\\
\end{split}\label{redef3}
\end{equation}
we obtain the following equations of motion \cite{bossard2}
\begin{equation}
\begin{split}
{d}\ast{d}{K_{i}}&\,=\,0\,,\\
{d}\ast{d}{L_{i}}&\,=\,0\,,\\
{d}\ast{d}{M}&\,=\,c^{ijk}\,L_{i}\,d\,L_{j}\ast{d}{K_{k}}\,,\\
{d}\ast{d}{V}&\,=\,c_{ijk}\,L_{i}\,dK_{j}\,\ast\,dK_{k}\,,\\\label{eqmot2}
\end{split}
\end{equation}
where, in writing the last equation, we have used the first two, namely the fact that \,$K_{i}$ and $L_{i}$ are harmonic functions.
With the redefinitions (\ref{redef3}),\,the warp function $e^{-4U}$ has the usual form
\begin{equation}
e^{-4\,U}\,=\,V\,L_{1}\,L_{2}\,L_{3}\,-\,M^{2}\,,
\end{equation}
and the scalars $\mathcal{Z}^M$ and $a$ read
\begin{equation}
\begin{split}
\mathcal{Z}^{0}&\,=\,\frac{1}{\sqrt{2}}\,\left[1\,-\,\frac{\prod_{i=1}^{3}\,L_{i}}{e^{-4\,U}}\right]\,,\\
\mathcal{Z}^{1}&\,=\,\frac{1}{\sqrt{2}}\,\left[\frac{L_{1}\,(\,M\,-\,K_{1}\,L_{2}\,L_{3}\,)}{e^{-4\,U}}\right]\,,\\
\mathcal{Z}^{2}&\,=\,\frac{1}{\sqrt{2}}\,\left[\frac{L_{2}\,(\,M\,-\,K_{2}\,L_{1}\,L_{3}\,)}{e^{-4\,U}}\right]\,,\\
\mathcal{Z}^{3}&\,=\,\frac{1}{\sqrt{2}}\,\left[\frac{L_{3}\,(\,M\,-\,K_{3}\,L_{1}\,L_{2}\,)}{e^{-4\,U}}\right]\,,\\
\mathcal{Z}_{0}&\,=\,\frac{1}{2\,\sqrt{2}}\,\left[\frac{2\,\prod_{i=1}^{3}\,K_{i}\,L_{i}\,-\,2\,M\,V\,+\,c^{ijk}\,(\,V\,L_{i}\,L_{j}\,K_{k}\,-\,M\,L_{i}\,K_{j}\,K_{k}\,)}{e^{-4\,U}}\right]\,,\\
\mathcal{Z}_{1}&\,=\,\frac{1}{\sqrt{2}}\,\left[\,1\,-\,\frac{2\,K_{2}\,K_{3}\,\prod_{i=1}^{3}\,L_{i}\,+\,c^{1jk}\,(\,V\,L_{j}\,L_{k}\,-\,2\,M\,K_{j}\,L_{k}\,)}{2\,e^{-4\,U}}\right]\,,\\
\mathcal{Z}_{2}&\,=\,\frac{1}{\sqrt{2}}\,\left[\,1\,-\,\frac{2\,K_{1}\,K_{3}\,\prod_{i=1}^{3}\,L_{i}\,+\,c^{2jk}\,(\,V\,L_{j}\,L_{k}\,-\,2\,M\,K_{j}\,L_{k}\,)}{2\,e^{-4\,U}}\right]\,,\\
\mathcal{Z}_{3}&\,=\,\frac{1}{\sqrt{2}}\,\left[\,1\,-\,\frac{2\,K_{1}\,K_{2}\,\prod_{i=1}^{3}\,L_{i}\,+\,c^{3jk}\,(\,V\,L_{j}\,L_{k}\,-\,2\,M\,K_{j}\,L_{k}\,)}{2\,e^{-4\,U}}\right]\,,\\
a&\,=\,\frac{\,2\,M\,(\,V\,-\,2\,+\,\sum_{i}\,L_{i}\,)\,+\,c^{ijk}\,(\,M\,K_{i}\,K_{j}\,L_{k}\,-\,V\,K_{i}\,L_{j}\,L_{k}\,)\,-\,2\,\prod_{i}\,L_{i}\,(\,\sum_{j}\,K_{j}\,+\,\prod_{j}\,K_{j}\,)}{4\,e^{-4\,U}}\,.\\
\label{chpq3}
\end{split}
\end{equation}
The four-dimensional scalar fields read
\begin{equation}
\begin{aligned}
\epsilon_{1}\,&=\,K_{1}\,-\,\frac{M}{L_{2}\,L_{3}}&,\quad&e^{-2\,\varphi_{1}}\,&=&\,\frac{L_{2}^{2}\,L_{3}^{2}}{e^{-4U}}\quad,\\
\epsilon_{2}\,&=\,K_{2}\,-\,\frac{M}{L_{1}\,L_{3}}&,\quad&e^{-2\,\varphi_{2}}\,&=&\,\frac{L_{1}^{2}\,L_{3}^{2}}{e^{-4U}}\quad,\\
\epsilon_{3}\,&=\,K_{3}\,-\,\frac{M}{L_{1}\,L_{2}}&,\quad&e^{-2\,\varphi_{3}}\,&=&\,\frac{L_{1}^{2}\,L_{2}^{2}}{e^{-4U}}\quad.
\label{scalar101}
\end{aligned}
\end{equation}
Now, it is useful rewrite the third of (\ref{eqmot2}) as
\begin{equation}
\begin{split}
{d}\left(\ast{d}{M}\right)\,=\,\frac{1}{2}\,c^{ijk}\,{d}\left(\,L_{i}\,L_{j}\ast{d}{K_{k}}\,\right)\,,\\
\end{split}
\end{equation}
so we can obtain the equation for $\omega$
\begin{equation}
\begin{split}
\ast{d}{\omega}\,=\,-{d}{M}\,+\,\frac{1}{2}\,c^{ijk}\,L_{i}\,L_{j}\,{d}{K_{k}}\label{om101}\,,
\end{split}
\end{equation}
and, from the Eq. (\ref{AF4}), those for $\tilde{A}^{M}$
\begin{equation}
\begin{split}
\ast{d}{\tilde{A}^{0}}&=\frac{\ast{d}{\omega}}{\sqrt{2}}\quad,\quad\ast{d}{\tilde{A}^{1}}=\frac{{d}L_{1}}{\sqrt{2}}\quad,\quad
\ast{d}{\tilde{A}^{2}}=\frac{{d}L_{2}}{\sqrt{2}}\quad,\quad\ast{d}{\tilde{A}^{3}}=\frac{{d}L_{3}}{\sqrt{2}}\quad,\\
\ast{d}{\tilde{A}_{0}}&=\frac{1}{\sqrt{2}}\left[-{d}V+\frac{1}{2}c^{ijk}\left[L_{i}{d}(K_{j}K_{k})-K_{j}K_{k}{d}L_{i}\right]\right]\quad,\\
\ast{d}{\tilde{A}_{1}}&=\frac{1}{\sqrt{2}}\left[\ast{d}{\omega}+c^{1jk}\left(K_{j}{d}L_{k}-L_{k}{d}K_{j}\right)\right]\quad,\\
\ast{d}{\tilde{A}_{2}}&=\frac{1}{\sqrt{2}}\left[\ast{d}{\omega}+c^{2jk}\left(K_{j}{d}L_{k}-L_{k}{d}K_{j}\right)\right]\quad,\\
\ast{d}{\tilde{A}_{3}}&=\frac{1}{\sqrt{2}}\left[\ast{d}{\omega}+c^{3jk}\left(K_{j}{d}L_{k}-L_{k}{d}K_{j}\right)\right]\quad.\label{vect101}\\
\end{split}
\end{equation}
\subsubsection{The solution}
Just as in Subsection \ref{subsect1}, we consider the first center ($A$) located at $r=0$ and the second one ($B$) at $r=R,\,\theta=0$, so Eq.s (\ref{eq3}) hold.
Being $V$,\,$L_{i}$ and $Z_{i}$ harmonic functions, we can write them in the following general form
\begin{equation}
L_{i}=l_{i}+\frac{Q_{i}}{r}+\frac{\tilde{Q}_{i}}{\Sigma}\,\quad,\quad\,K_{i}=k_{i}+\frac{d_{i}}{r}+\frac{\tilde{d}_{i}}{\Sigma}\,.
\end{equation}
Substituting the above expressions in the third and fourth of equations (\ref{eqmot2}), we find
\begin{equation}
\begin{split}
{d}\ast{d}{M}\,=\,c^{ijk}\left(l_{i}+\frac{Q_{i}}{r}+\frac{\tilde{Q}_{i}}{\Sigma}\right){d}\left(l_{j}+\frac{Q_{j}}{r}+\frac{\tilde{Q}_{j}}{\Sigma}\right)\ast{d}\left(k_{k}+\frac{d_{k}}{r}+\frac{\tilde{d}_{k}}{\Sigma}\right)\,
\end{split}\label{ddMa101}
\end{equation}
and
\begin{equation}
\begin{split}
{d}\ast{d}{V}\,=\,c^{ijk}\left(l_{i}+\frac{Q_{i}}{r}+\frac{\tilde{Q}_{i}}{\Sigma}\right){d}\left(k_{j}+\frac{d_{j}}{r}+\frac{\tilde{d}_{j}}{\Sigma}\right)\ast{d}\left(k_{k}+\frac{d_{k}}{r}+\frac{\tilde{d}_{k}}{\Sigma}\right)\,.
\end{split}\label{ddMa102}
\end{equation}
Using the properties in Appendix \ref{usef} as well as those obtained from them by interchanging the two centers $r\leftrightarrow\Sigma$,
 we can solve Eq.s (\ref{ddMa101}) by writing $M$ and $V$ as a suitable combination of terms, plus harmonic functions:
\begin{equation}
\begin{split}
M=&m_0+\frac{m}{r}+\frac{\tilde{m}}{\Sigma}+\frac{\alpha\cos\theta}{r^2}+\frac{\tilde{\alpha}\cos\theta_\Sigma}{\Sigma^2}+\frac{c^{ijk}}{2}\Biggl[\frac{l_{i}Q_{j}d_{k}}{r^2}+\frac{l_{i}\tilde{Q}_{j}\tilde{d}_{k}}{\Sigma^2}+\frac{l_{i}(Q_{j}\tilde{d}_{k}+\tilde{Q}_{j}d_{k})}{r\Sigma}+\\
&-\frac{\tilde{Q}_{i}\tilde{Q}_{j}d_{k}(r\cos\theta-R)}{rR\Sigma^2}-\frac{Q_{i}\tilde{Q}_{j}d_{k}(r\cos\theta-R)}{r^2R\Sigma}+\frac{Q_{i}Q_{j}\tilde{d}_{k}\cos\theta}{rR\Sigma}+\frac{Q_{i}\tilde{Q}_{j}\tilde{d}_{k}\cos\theta}{R\Sigma^2}+\\
&+Q_{i}Q_{j}d_{k}\left(\frac{R-r\cos\theta(1+\delta)}{3Rr^3}\right)+\tilde{Q}_{i}\tilde{Q}_{j}\tilde{d}_{k}\left(\frac{r\cos\theta(1+\tilde{\delta})-R\tilde{\delta}}{3R\Sigma^3}\right)\Biggr]\,,\label{eq5}
\end{split}
\end{equation}
\begin{equation}
\begin{split}
V=&h+\frac{Q_{6}}{r}+\frac{\tilde{Q}_6}{\Sigma}+\frac{\beta\cos\theta}{r^2}+\frac{\tilde{\beta}\cos\theta_\Sigma}{\Sigma^2}+\\
&+c^{ijk}\Biggl[\frac{1}{2}\left(l_{i}+\frac{Q_{i}}{r}+\frac{\tilde{Q}_{i}}{\Sigma}\right)\left(k_{j}+\frac{d_{j}}{r}+\frac{\tilde{d}_{j}}{\Sigma}\right)\left(k_{k}+\frac{d_{k}}{r}+\frac{\tilde{d}_{k}}{\Sigma}\right)-\frac{k_{i}Q_{j}d_{k}}{r^2}-\frac{k_{i}\tilde{Q}_{j}\tilde{d}_{k}}{\Sigma^2}+\\
&-\frac{k_{i}(Q_{j}\tilde{d}_{k}+\tilde{Q}_{j}d_{k})}{r\Sigma}+\frac{\tilde{d}_{i}\tilde{d}_{j}Q_{k}(r\cos\theta-R)}{rR\Sigma^2}+\frac{d_{i}\tilde{d}_{j}Q_{k}(r\cos\theta-R)}{r^2R\Sigma}-\frac{d_{i}d_{j}\tilde{Q}_{k}\cos\theta}{rR\Sigma}+\\
&-\frac{d_{i}\tilde{d}_{j}\tilde{Q}_{k}\cos\theta}{R\Sigma^2}-d_{i}d_{j}Q_{k}\left(\frac{R-r\cos\theta(1+\delta)}{3Rr^3}\right)-\tilde{d}_{i}\tilde{d}_{j}\tilde{Q}_{k}\left(\frac{r\cos\theta(1+\tilde{\delta})-R\tilde{\delta}}{3R\Sigma^3}\right)\Biggr]\,,\label{eq6}
\end{split}
\end{equation}
where the $\delta$ and $\tilde{\delta}$-parameters multiply the additional harmonic functions of the form $\cos(\theta)/r^2$ and $\cos(\theta_\Sigma)/\Sigma^2$ in the solutions for $M$ and $V$, when integrating equations of the form:
${d}\ast{d}f=\frac{1}{r}{d}\left(\frac{1}{r}\right)\ast{d}\left(\frac{1}{r}\right),\,
{d}\ast{d}f=\frac{1}{\Sigma}{d}\left(\frac{1}{\Sigma}\right)\ast{d}\left(\frac{1}{\Sigma}\right).$\par
Similarly, from (\ref{om101}) and (\ref{vect101}), we can obtain the expressions for $\omega$
\begin{equation}
\begin{split}
\omega&=\bigg[k_{\omega}+\alpha\frac{\sin^{2}\theta}{r}+\tilde{\alpha}\frac{\sin^{2}\theta_{\Sigma}}{\Sigma}-\left(m-c^{ijk}\frac{l_{i}l_{j}d_{k}}{2}\right)\cos\theta-\left(\tilde{m}-c^{ijk}\frac{l_{i}l_{j}\tilde{d}_{k}}{2}\right)\cos\theta_{\Sigma}+\\
&+c^{ijk}\big[-\frac{Q_{i}Q_{j}d_{k}(1+\delta)\sin^{2}\theta}{6rR}+\frac{\tilde{Q}_{i}\tilde{Q}_{j}\tilde{d}_{k}(1+\tilde{\delta})\sin^{2}\theta_{\Sigma}}{6R\Sigma}-\frac{l_{i}(\tilde{Q}_{j}d_{k}-Q_{j}\tilde{d}_{k})(r-R\cos\theta)}{2R\Sigma}+\\
&-\frac{Q_{i}(\tilde{Q}_{j}d_{k}-Q_{j}\tilde{d}_{k})\sin^{2}\theta}{2R\Sigma}-\frac{\tilde{Q}_{i}(\tilde{Q}_{j}d_{k}-Q_{j}\tilde{d}_{k})\sin^{2}\theta_{\Sigma}}{2Rr}\big]\bigg]d\varphi\quad,\\
\end{split}
\end{equation}
and for the $D=3$ vectors $\tilde{A}^{M}$
\begin{equation}
\begin{split}
\tilde{A}^{0}&=\frac{\omega}{\sqrt{2}}\quad,\quad\tilde{A}^{1}=\frac{Q_{1}\cos\theta+\tilde{Q}_{1}\cos\theta_{\Sigma}}{\sqrt{2}}d\varphi\quad,\quad\tilde{A}^{2}=\frac{Q_{2}\cos\theta+\tilde{Q}_{2}\cos\theta_{\Sigma}}{\sqrt{2}}d\varphi\quad,\\
\tilde{A}^{3}&=\frac{Q_{3}\cos\theta+\tilde{Q}_{3}\cos\theta_{\Sigma}}{\sqrt{2}}d\varphi\quad,\\
\tilde{A}_{0}&=\Bigg[-(Q_{6}+c^{ijk}k_{i}k_{j}Q_{k})\cos\theta-(\tilde{Q}_{6}+c^{ijk}k_{i}k_{j}\tilde{Q}_{k})\cos\theta_{\Sigma}+\beta\frac{\sin^{2}\theta}{r}+
\tilde{\beta}\frac{\sin^{2}\theta_{\Sigma}}{\Sigma}+\quad\\
&+c^{ijk}\left[k_{i}(d_{j}\tilde{Q}_{k}-\tilde{d}_{j}Q_{k})\frac{R\cos\theta-r}{R\Sigma}+d_{i}(d_{j}\tilde{Q}_{k}-\tilde{d}_{j}Q_{k})\frac{\sin^{2}\theta}{R\Sigma}+\tilde{d}_{i}(d_{j}\tilde{Q}_{k}-\tilde{d}_{j}Q_{k})\frac{\sin^{2}\theta_{\Sigma}}{Rr}\right]\Bigg]d\varphi\quad,\\
\tilde{A}_{1}&=\omega+c^{1jk}\left[(k_{j}Q_{k}-l_{j}d_{k})\cos\theta+(k_{j}\tilde{Q}_{k}-l_{j}\tilde{d}_{k})\cos\theta_{\Sigma}-(d_{j}\tilde{Q}_{k}-\tilde{d}_{j}Q_{k})\frac{R\cos\theta-r}{R\Sigma}\right]d\varphi\quad,\\
\tilde{A}_{2}&=\omega+c^{2jk}\left[(k_{j}Q_{k}-l_{j}d_{k})\cos\theta+(k_{j}\tilde{Q}_{k}-l_{j}\tilde{d}_{k})\cos\theta_{\Sigma}-(d_{j}\tilde{Q}_{k}-\tilde{d}_{j}Q_{k})\frac{R\cos\theta-r}{R\Sigma}\right]d\varphi\quad,\\
\tilde{A}_{3}&=\omega+c^{3jk}\left[(k_{j}Q_{k}-l_{j}d_{k})\cos\theta+(k_{j}\tilde{Q}_{k}-l_{j}\tilde{d}_{k})\cos\theta_{\Sigma}-(d_{j}\tilde{Q}_{k}-\tilde{d}_{j}Q_{k})\frac{R\cos\theta-r}{R\Sigma}\right]d\varphi\quad.\\
\end{split}
\end{equation}
Replacing the above expressions and the solution (\ref{chpq3}) for the $\mathcal{Z}^M$ scalars in (\ref{AF1}) one obtains the explicit form of the $D=4$ vector fields.\par
Below we proceed to impose the regularity condition on the solution.
\paragraph{Zero-NUT charge condition}\par
Imposing Eq.s  (\ref{condnut7}) on $\omega$ , in order to exclude the presence of Dirac-Misner singularities we find for $k_{\omega}$, $m$ and $\tilde{m}$ the following conditions
\begin{equation}
\begin{split}
k_{\omega}\,&=\,\frac{c^{ijk}}{2\,R}\,l_{i}\,\left(\tilde{Q}_{j}\,d_{k}\,-\,Q_{j}\,\tilde{d}_{k}\,\right)\quad,\\
m\,&=\,\frac{c^{ijk}}{2R}\,l_{i}\left(\,R\,l_{j}\,d_{k}\,+\,\tilde{Q}_{j}\,d_{k}\,-\,Q_{j}\,\tilde{d}_{k}\right)\quad,\\
\tilde{m}\,&=\,\frac{c^{ijk}}{2R}\,l_{i}\left(\,R\,l_{j}\,\tilde{d}_{k}\,+\,Q_{j}\,\tilde{d}_{k}\,-\,\tilde{Q}_{j}\,d_{k}\right)\,.\label{NUTcond2}
\end{split}
\end{equation}
The difference between last two conditions represents the bubble equation relating $R$ to the asymptotic data at spatial infinity.
\paragraph{Regularity near the centers and asymptotic flatness}\par
Let us now require each center to be either a small or a regular black hole with finite horizon area. This implies the absence of terms in the expansion of  $e^{-4U}$ near the $A$-center (and $B$-center) diverging faster that $1/r^4$ (and $1/\Sigma^4$, respectively). In our case, we have for the first center
\begin{equation}
\begin{split}
e^{-4\,U}\stackrel{\stackrel{r\rightarrow0}{\theta=\pi}}{\cong}&\frac{1}{r^6}\left[\frac{c^{ijk}}{18}(Q_{1}Q_{2}Q_{3}Q_{i}d_{j}d_{k}-Q^2_{i}Q^2_{j}d^2_{k})\right]+\mathcal{O}(r^{-5})
\end{split}
\end{equation}
and for the second one
\begin{equation}
\begin{split}
e^{-4\,U}\stackrel{\stackrel{r\rightarrow{R}}{\theta=0}}{\cong}&\frac{1}{(r-R)^6}\left[\frac{c^{ijk}}{18}(\tilde{Q}_{1}\tilde{Q}_{2}\tilde{Q}_{3}\tilde{Q}_{i}\tilde{d}_{j}\tilde{d}_{k}-\tilde{Q}^2_{i}\tilde{Q}^2_{j}\tilde{d}^2_{k})\right]+\mathcal{O}(r^{-5})\,.
\end{split}
\end{equation}
The above divergent terms can be removed, following \cite{bossard2},  by requiring
\begin{equation}
\gamma\,=\,\frac{d_{1}}{Q_{1}}\,=\,\frac{d_{2}}{Q_{2}}\,=\,\frac{d_{3}}{Q_{3}}\quad\,,\,\quad\tilde{\gamma}\,=\,\frac{\tilde{d}_{1}}{\tilde{Q}_{1}}\,=\,\frac{\tilde{d}_{2}}{\tilde{Q}_{2}}\,=\,\frac{\tilde{d}_{3}}{\tilde{Q}_{3}}\,\quad.\label{conda101}
\end{equation}
With this choice, we have
\begin{equation}
\begin{split}
e^{-4\,U}\stackrel{\stackrel{r\rightarrow0}{\theta=\pi}}{\cong}&\frac{1}{r^5}\left[-\frac{Q_{1}Q_{2}Q_{3}(R(\beta-2\alpha\gamma)+4\gamma^{2}(1+\delta)Q_{1}Q_{2}Q_{3})}{R}\right]+\mathcal{O}(r^{-4})\,,
\end{split}
\end{equation}
and
\begin{equation}
\begin{split}
e^{-4\,U}\stackrel{\stackrel{r\rightarrow{R}}{\theta=0}}{\cong}&\frac{1}{(r-R)^5}\left[\frac{\tilde{Q}_{1}\tilde{Q}_{2}\tilde{Q}_{3}[R(\tilde{\beta}-2\tilde{\alpha}\tilde{\gamma})-4\tilde{\gamma}^{2}(1+\tilde{\delta})\tilde{Q}_{1}\tilde{Q}_{2}\tilde{Q}_{3}]}{R}\right]+\mathcal{O}((r-R)^{-4})\,,
\end{split}
\end{equation}
so we can further impose, following \cite{bossard2}, the conditions
\begin{equation}
\beta\,=\,2\,\gamma\,\alpha\,\quad\,,\,\quad\,\tilde{\beta}\,=\,2\,\tilde{\gamma}\,\tilde{\alpha}\,\quad\,,\,\quad\,\delta\,=\,-1\,\quad\,,\,\quad\,\tilde{\delta}\,=\,-1\quad.\label{conda102}
\end{equation}
On the warp factor the asymptotic flatness condition $\lim_{r \rightarrow +\infty}e^{-4\,U}=1$ implies
\begin{equation}
h=\frac{1\,+\,m^2_0}{l_{1}\,l_{2}\,l_{3}}-\frac{c^{ijk}}{2}(\,l_{i}\,k_{j}\,k_{k}\,)\,.
\end{equation}
Being interested in analyzing a specific solution with $Q_O$ generic in a chosen $H^*$-orbit, we further restrict to parameters so that $\phi_0=O$. This amounts to  requiring that $\lim_{r \rightarrow +\infty}\mathcal{M}^{(3)}_{scal}=\eta$\, and implies
\begin{equation}
m_0\,=\,0\quad\,,\,\quad\,l_{i}\,=\,1\quad\,,\,\quad\,k_{i}\,=\,0\,.
\end{equation}
With these conditions, we can write down the expressions for the charges of each center, in a more compact form:
\begin{equation}
\begin{split}
{\Gamma^{M}_{A}}=({p^{\Lambda}_{A}},{q_{A\,\Lambda}})=\bigg(&0,\frac{Q_1}{\sqrt{2}},\frac{Q_2}{\sqrt{2}},\frac{Q_3}{\sqrt{2}},-\frac{Q_6}{\sqrt{2}},-\frac{R\gamma(Q_2+Q_3)+(\gamma-\tilde{\gamma})c^{1jk}Q_{j}\tilde{Q}_{k}}{\sqrt{2}R},\\
&-\frac{R\gamma(Q_1+Q_3)+(\gamma-\tilde{\gamma})c^{2jk}Q_{j}\tilde{Q}_{k}}{\sqrt{2}R},-\frac{R\gamma(Q_1+Q_2)+(\gamma-\tilde{\gamma})c^{3jk}Q_{j}\tilde{Q}_{k}}{\sqrt{2}}\bigg)\quad
\end{split}
\end{equation}
for the center $A$ and
\begin{equation}
\begin{split}
{\Gamma^{M}_{B}}=({p^{\Lambda}_{B}},{q_{B\,\Lambda}})=\bigg(&0,\frac{\tilde{Q}_1}{\sqrt{2}},\frac{\tilde{Q}_2}{\sqrt{2}},\frac{\tilde{Q}_3}{\sqrt{2}},-\frac{\tilde{Q}_6}{\sqrt{2}},-\frac{R\tilde{\gamma}(\tilde{Q}_2+\tilde{Q}_3)+(\tilde{\gamma}-\gamma)c^{1jk}Q_{j}\tilde{Q}_{k}}{\sqrt{2}R},\\
&-\frac{R\tilde{\gamma}(\tilde{Q}_1+\tilde{Q}_3)+(\tilde{\gamma}-\gamma)c^{2jk}Q_{j}\tilde{Q}_{k}}{\sqrt{2}R},-\frac{R\tilde{\gamma}(\tilde{Q}_1+{Q}_2)+(\tilde{\gamma}-\gamma)c^{3jk}Q_{j}\tilde{Q}_{k}}{\sqrt{2}}\bigg)\quad
\end{split}
\end{equation}
for the center $B$. The quartic invariant computed with the above charges read:
\begin{equation}
\begin{split}
I_{4}(p^{\Lambda}_{A},q_{A\,\Lambda})=Q_{1}Q_{2}Q_{3}\left[-Q_{6}+\gamma^{2}\sum_{i}Q_{i}+\frac{\gamma(\gamma-\tilde{\gamma})}{R}\sum_{i}c^{ijk}Q_{j}\tilde{Q}_{k}+\frac{(\gamma-\tilde{\gamma})^{2}}{2R^{2}}c^{ijk}Q_{i}\tilde{Q}_{j}\tilde{Q}_{k}\right]\quad
\end{split}
\end{equation}
and
\begin{equation}
\begin{split}
I_{4}(p^{\Lambda}_{B},q_{B\,\Lambda})=\tilde{Q}_{1}\tilde{Q}_{2}\tilde{Q}_{3}\left[-\tilde{Q}_{6}+\tilde{\gamma}^{2}\sum_{i}\tilde{Q}_{i}+\frac{\tilde{\gamma}(\tilde{\gamma}-\gamma)}{R}\sum_{i}c^{ijk}Q_{j}\tilde{Q}_{k}+\frac{(\gamma-\tilde{\gamma})^{2}}{2R^{2}}c^{ijk}Q_{i}Q_{j}\tilde{Q}_{k}\right]\quad.
\end{split}
\end{equation}
Now, one can see that the geometry near the first center behaves like
\begin{equation}
e^{-4\,U}\cong \frac{1}{r^4}\left[-I_{4}(p^{\Lambda}_{A},q_{A,\Lambda})-
\alpha^{2}\right]+\mathcal{O}((r)^{-3})\quad,\label{conda10g1fc}
\end{equation}
while near the second one as
\begin{equation}
e^{-4\,U}\cong \frac{1}{(r-R)^4}\left[-I_{4}(p^{\Lambda}_{B},q_{B,\Lambda})-\tilde{\alpha}^{2}\right]+
\mathcal{O}((r-R)^{-3})\quad.\label{condb10g1sc}
\end{equation}
Regularity them requires the quartic invariants $I_{4}(p^{\Lambda}_{A},q_{A,\Lambda}),\,I_{4}(p^{\Lambda}_{B},q_{B,\Lambda})$ to be negative \cite{bossard2} and, in our formalism, this restricts to two centers to belong to the $\alpha^{(6)},\,\gamma^{(6; 5)},\,\beta^{(6; 5)}, \delta^{(1)}$ orbit. Note that had one of the centers been in the  $\alpha^{(6)},\,\gamma^{(6; 5)},\,\beta^{(6; 5)}, \delta^{(2)}$, the asymptotic behavior of the solution described by (\ref{conda10g1fc}) and (\ref{condb10g1sc}) would be the same  but the solution would be singular (see the discussion at the end of Sect. \ref{generaloverv}).
\paragraph{Charges and angular momentum}\par
Having implemented the regularity conditions, one can determine the electromagnetic charges associated with the solution, making use of (\ref{elmagch}),  and with each center
\begin{equation}
\begin{split}
\Gamma^M=(p^\Lambda,q_\Lambda)=\bigg(&0,\frac{Q_1+\tilde{Q}_1}{\sqrt{2}},\frac{Q_2+\tilde{Q}_2}{\sqrt{2}},\frac{Q_3+\tilde{Q}_3}{\sqrt{2}},-\frac{Q_6+\tilde{Q}_6}{\sqrt{2}},-\frac{\gamma(Q_2+Q_3)+\tilde{\gamma}(\tilde{Q}_2+\tilde{Q}_3)}{\sqrt{2}},\\
&-\frac{\gamma(Q_1+Q_3)+\tilde{\gamma}(\tilde{Q}_1+\tilde{Q}_3)}{\sqrt{2}},-\frac{\gamma(Q_1+Q_2)+\tilde{\gamma}(\tilde{Q}_1+\tilde{Q}_2)}{\sqrt{2}}\bigg)\quad,
\end{split}
\end{equation}
\begin{equation}
\begin{split}
{\Gamma^{M}_{A}}=({p^{\Lambda}_{A}},{q_{A\,\Lambda}})=\bigg(&0,\frac{Q_1}{\sqrt{2}},\frac{Q_2}{\sqrt{2}},\frac{Q_3}{\sqrt{2}},-\frac{Q_6}{\sqrt{2}},-\frac{R\gamma(Q_2+Q_3)+(\gamma-\tilde{\gamma})c^{1jk}Q_{j}\tilde{Q}_{k}}{\sqrt{2}R},\\
&-\frac{R\gamma(Q_1+Q_3)+(\gamma-\tilde{\gamma})c^{2jk}Q_{j}\tilde{Q}_{k}}{\sqrt{2}R},-\frac{R\gamma(Q_1+Q_2)+(\gamma-\tilde{\gamma})c^{3jk}Q_{j}\tilde{Q}_{k}}{\sqrt{2}}\bigg)\quad,
\end{split}
\end{equation}\begin{equation}
\begin{split}
{\Gamma^{M}_{B}}=({p^{\Lambda}_{B}},{q_{B\,\Lambda}})=\bigg(&0,\frac{\tilde{Q}_1}{\sqrt{2}},\frac{\tilde{Q}_2}{\sqrt{2}},\frac{\tilde{Q}_3}{\sqrt{2}},-\frac{\tilde{Q}_6}{\sqrt{2}},-\frac{R\tilde{\gamma}(\tilde{Q}_2+\tilde{Q}_3)+(\tilde{\gamma}-\gamma)c^{1jk}Q_{j}\tilde{Q}_{k}}{\sqrt{2}R},\\
&-\frac{R\tilde{\gamma}(\tilde{Q}_1+\tilde{Q}_3)+(\tilde{\gamma}-\gamma)c^{2jk}Q_{j}\tilde{Q}_{k}}{\sqrt{2}R},-\frac{R\tilde{\gamma}(\tilde{Q}_1+{Q}_2)+(\tilde{\gamma}-\gamma)c^{3jk}Q_{j}\tilde{Q}_{k}}{\sqrt{2}}\bigg)\quad.
\end{split}
\end{equation}
The total angular momentum is computed to be
\begin{equation}
\begin{split}
M_\varphi&=\frac{1}{2}\left(\alpha+\tilde{\alpha}\right)-\frac{\gamma-\tilde{\gamma}}{4R}\left[R\sum_{i}c^{ijk}Q_{j}\tilde{Q}_{k}+c^{ijk}Q_{i}\tilde{Q}_{j}(Q_{k}+\tilde{Q}_{k})\right]=\\
&=M_{\varphi_A}+M_{\varphi_B}-\frac{1}{2}\,({\Gamma}_{A})^{T}\,\mathbb{C}\,{\Gamma}_{B}\quad,
\end{split}
\end{equation}
where $M_{\varphi_A}\equiv \alpha/2$ and $M_{\varphi_B}\equiv \tilde{\alpha}/2$.
Just as noticed by \cite{bossard2}, the total angular momentum is the sum of the angular momenta of the single centers plus the symplectic product of electromagnetic charges associated with each center. One can indeed verify that the above expression for $M_\varphi$ is precisely the one occurring in the asymptotic expansion of $\omega_\varphi$, according to the formula (\ref{omegaMphi}).

\subsubsection{The orbit $\mathcal{O}^{10;1,1}$}
Let us consider the double-center solutions with $Q_O$ in the orbit $\alpha^{(10)}\gamma^{(10;1)}\beta^{(1)}$. The peculiarity of this orbit (and in particular of all the $\alpha^{(10)}$ orbits with coinciding $\gamma$ and $\beta$ labels ) is that it can not be obtained as the sum of two orbits associated with regular single-center solutions (see discussion in Sect. \ref{compositionlaw}).\par
For instance, by choosing the charge parameters as follows:
\begin{align}
Q_6&=-1\,\,,\,\,\,Q_1=1\,\,,\,\,\,Q_2=-1\,\,,\,\,\,Q_3=1\,\,,\,\,\,\gamma=\frac{1}{2}\,\,,\nonumber\\
\tilde{Q}_6&=1\,\,,\,\,\,\tilde{Q}_1=-2\,\,,\,\,\,\tilde{Q}_2=-5\,\,,\,\,\,
\tilde{Q}_3=6\,\,,\,\,\,\tilde{\gamma}=1\quad\,,
\end{align} we can find a two-center solution in which the $A$-center is in the orbit $\alpha^{(6)}\gamma^{(6;5)}\beta^{(5)}\delta^{(4)}$ and the $B$-one in the orbit $\alpha^{(6)}\gamma^{(6;5)}\beta^{(5)}\delta^{(3)}$. Without entering into the details of the solution we see from the figures below that $e^{-4\,U}$ has poles at finite $r$ and  $e^{-4\,U}r^2\sin^2\theta-\omega^2$ becomes negative near the centers, revealing the presence of CTC.
\begin{figure}[H]
\centering
\subfloat[][The plot of the warp funtion $e^{-4\,U}$.]{\includegraphics[width=7cm]{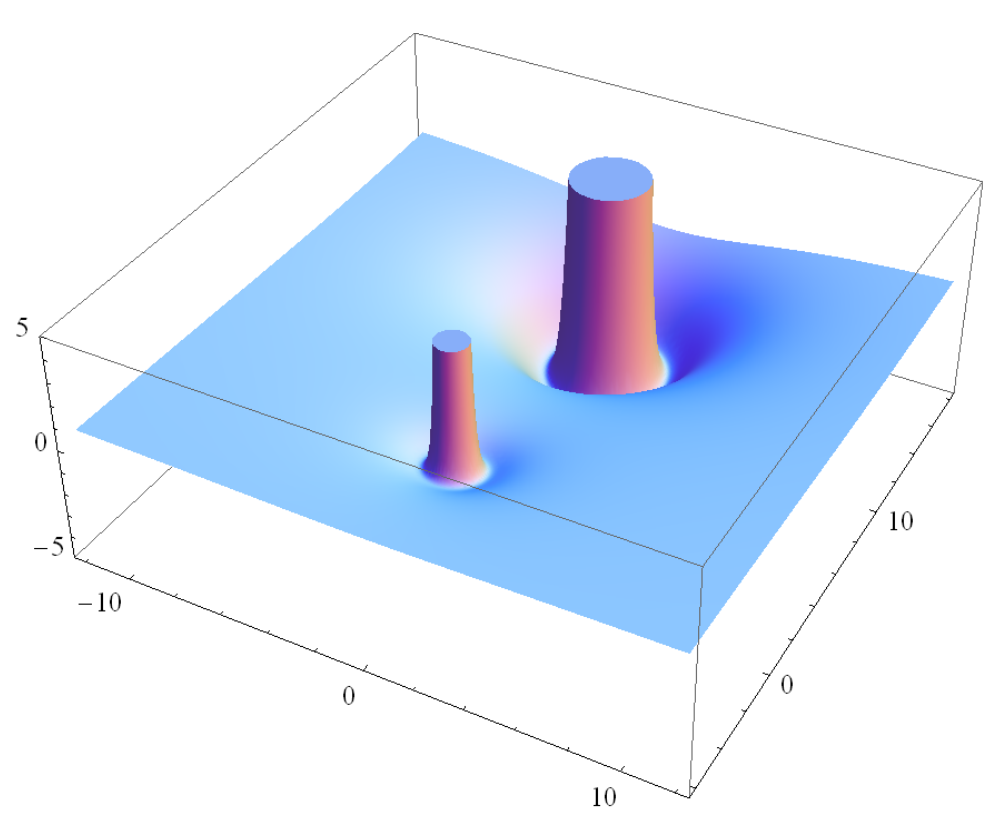}} \hfill
\subfloat[][The intersection of $e^{-4\,U}r^2\sin^2\theta-\omega^2$ with the plane $z=0$.]{\includegraphics[width=7cm]{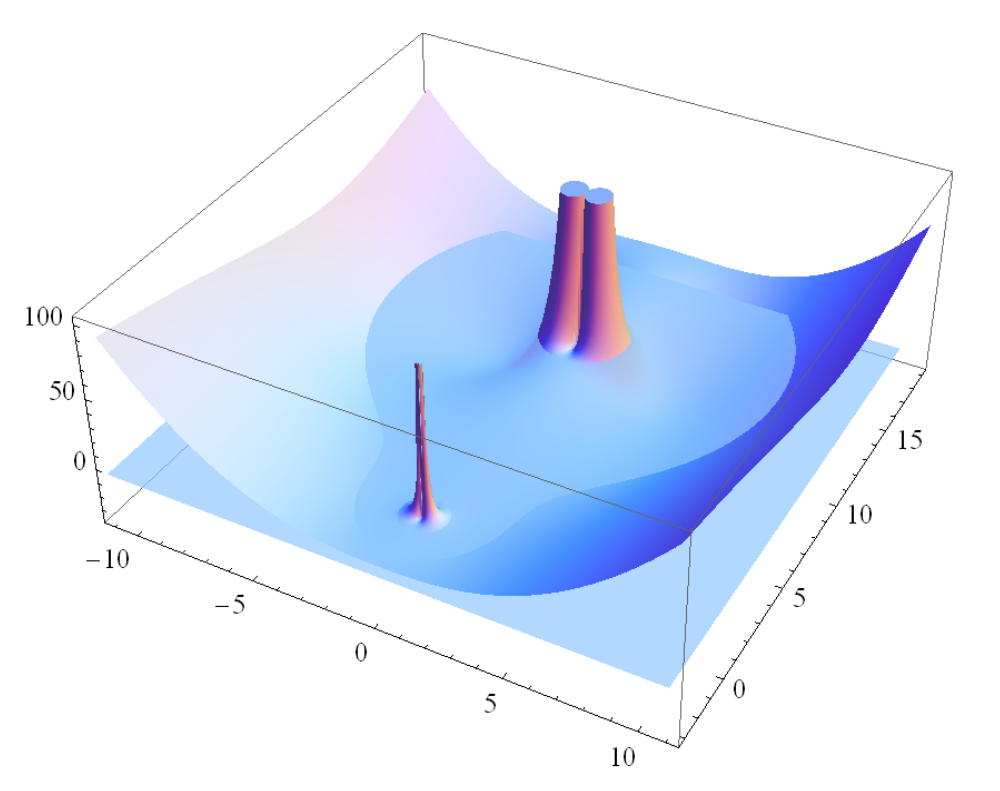}}
\end{figure}

\subsubsection{The orbits $\mathcal{O}^{10;1,2}$\, ( Singular solution )}
Let us consider now the $\alpha^{(10)}\gamma^{(10;1)}\beta^{(2)}$ orbit and illustrate in this example the general fact that a same orbit, which is not intrinsically singular,
can describe (depending on their composition) both singular and regular double-center solutions.

\paragraph{ Singular solution}\par
Choosing for instance the charge parameters as follows:
 \begin{align}
Q_6&=1\quad\,,\quad\,Q_1=1\quad\,,\quad\,Q_2=1\quad\,,\quad\,Q_3=1\quad\,,\quad\,\gamma=\frac{1}{2}\,\,,\nonumber\\
\tilde{Q}_6&=3\quad\,,\quad\,\tilde{Q}_1=1\quad\,,\quad\,\tilde{Q}_2=2\quad\,,\quad\,\tilde{Q}_3=-3\quad\,,\quad\,\tilde{\gamma}=1\quad\,,
\end{align}
we end up with an $A$-center in the $\alpha^{(6)}\gamma^{(6;5)}\beta^{(5)}\delta^{(1)}$-orbit describing a regular, large non-BPS black hole with $I_4<0$ and a $B$-center in the orbit $\alpha^{(6)}\gamma^{(6;5)}\beta^{(4)}$, describing a singular single-center solution. The whole solution is therefore singular as it can be ascertained from the figures below, showing the behavior of  $e^{-4\,U}$ and $e^{-4\,U}r^2\sin^2\theta-\omega^2$.
\begin{figure}[H]
\centering
\subfloat[][The plot of $e^{-4\,U}$.]{\includegraphics[width=7cm]{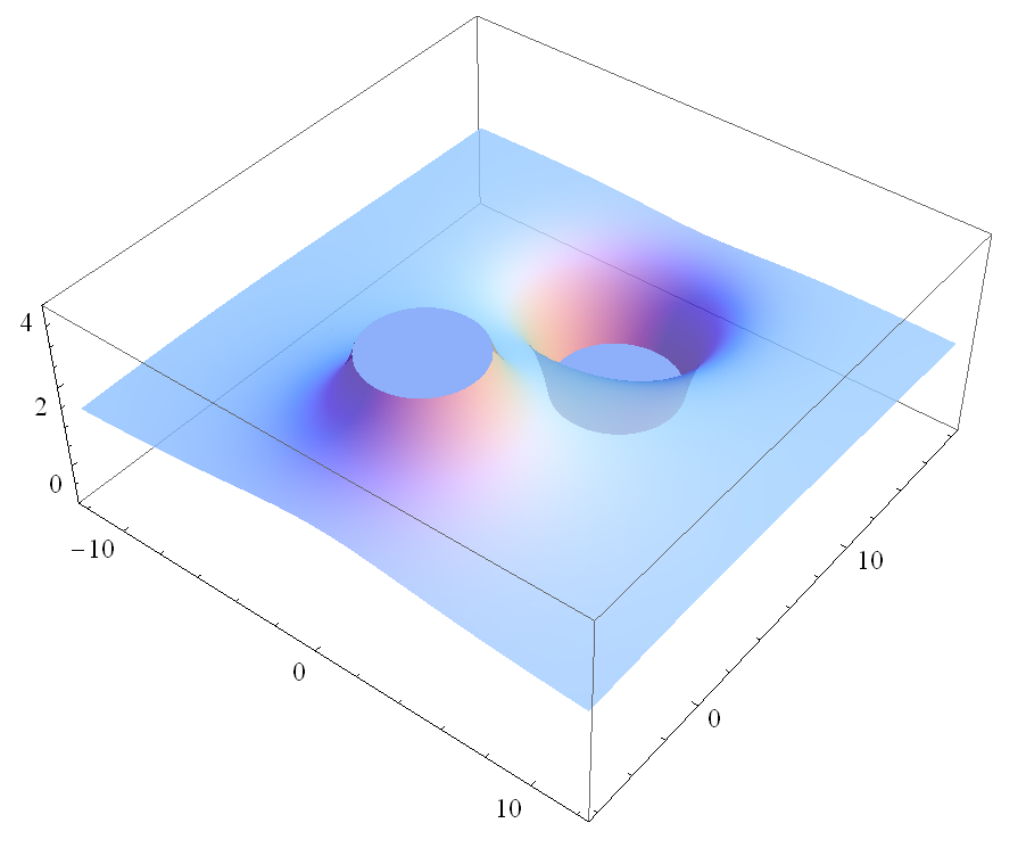}} \hfill
\subfloat[][The plot of $e^{-4\,U}r^2\sin^2\theta-\omega^2$.]{\includegraphics[width=7cm]{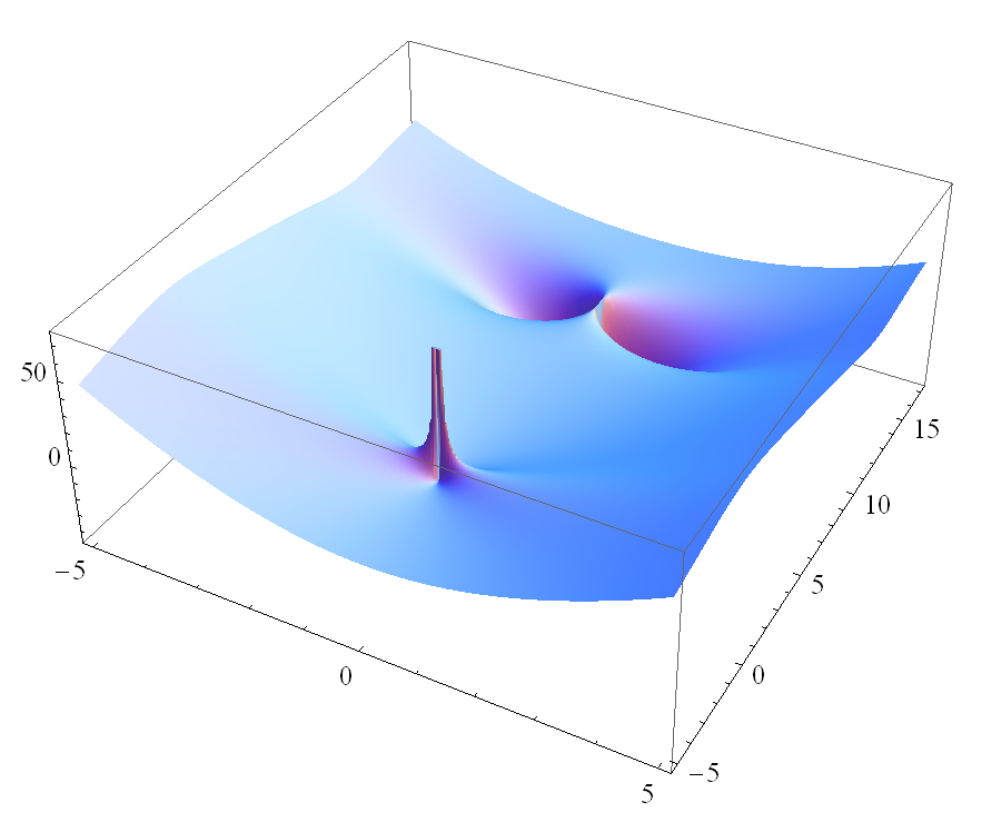}}\\
\caption{\small Fig. a) The plot of $e^{-4\,U}$ shows that near the second center the warp function is not positive, revealing a true singularity. In Fig. b) $e^{-4\,U}r^2\sin^2\theta-\omega^2$ is not positive near the singular center, so it is not possible exclude the presence of CTC.}
\end{figure}
\paragraph{ Regular solution}\par
We can keep $Q_O$ within $\alpha^{(10)}\gamma^{(10;1)}\beta^{(2)}$ and make the following choices for the charge parameters
 \begin{align}
Q_6&=3\quad\,,\quad\,Q_1=1\quad\,,\quad\,Q_2=1\quad\,,\quad\,Q_3=1\quad\,,\quad\,\gamma=\frac{1}{2}\,\,,\nonumber\\
\tilde{Q}_6&=10\quad\,,\quad\,\tilde{Q}_1=3\quad\,,\quad\,\tilde{Q}_2=4\quad\,,\quad\,\tilde{Q}_3=2\quad\,,\quad\,\tilde{\gamma}=1\,,
\end{align}
which imply both centers to be in the $\alpha^{(6)}\gamma^{(6;5)}\beta^{(5)}\delta^{(1)}$-orbit describing regular, large non-BPS black holes with $I_4<0$.
The electric-magnetic charges of the two centers read
\begin{equation}
{\Gamma^{M}_{A}}\,=\,(\,{p^{\Lambda}_{A}}\,,\,{q_{A\,\Lambda}}\,)\,=\,\left(\,0\,,\,\frac{1}{\sqrt{2}}\,,\,\frac{1}{\sqrt{2}}\,,\,\frac{1}{\sqrt{2}}\,,\,-\frac{3}{\sqrt{2}}\,,\,-\frac{1}{\sqrt{2}}\,,\,-\frac{1}{\sqrt{2}}\,,\,-\frac{1}{\sqrt{2}}\,\right)\quad,
\end{equation}
\begin{equation}
{\Gamma^{M}_{B}}\,=\,(\,{p^{\Lambda}_{B}}\,,\,{q_{B\,\Lambda}}\,)\,=\,\left(\,0\,,\,\frac{3}{\sqrt{2}}\,,\,2\sqrt{2}\,,\,\sqrt{2}\,,\,-5\sqrt{2}\,,\,-3\sqrt{2}\,,\,-\frac{5}{\sqrt{2}}\,,\,-\frac{7}{\sqrt{2}}\,\right)\quad.
\end{equation}
and the total charge vector is the sum of the two.\par
Having fixed all parameters in order to have two regular single center black holes and a stable system, we can now study the lower bound of $R$. We have
\begin{equation}
\begin{split}
e^{-4\,U}\stackrel{\stackrel{r\rightarrow0}{\theta=\pi}}{\cong}&\frac{1}{r^4}\left[\frac{-26+18R+R^2(9-4\alpha^{2})}{4R^2}\right]+\mathcal{O}(r^{-3})\quad,\label{ex4Ua10g1b2fc}
\end{split}
\end{equation}
for the $A$ center and
\begin{equation}
\begin{split}
e^{-4\,U}\stackrel{\stackrel{r\rightarrow{R}}{\theta=0}}{\cong}&\frac{1}{(r-R)^4}\left[24-\frac{54(1+4R)}{R^{2}}-\tilde{\alpha}^{2}\right]+\mathcal{O}((r-R)^{-3})\quad,\label{ex4Ua10g1b2sc}
\end{split}
\end{equation}
for the $B$ center. Sending to zero the terms associated with the angular momenta,$\alpha$ and $\tilde{\alpha}$, requiring the positivity of $e^{-4\,U}$ in the Eq.s (\ref{ex4Ua10g1b2fc})-(\ref{ex4Ua10g1b2sc}) near each center implies the following lower bound for the distance $R$
\begin{equation}
\begin{split}
R>\frac{3}{2}\left(3+\sqrt{10}\right)\quad\label{lba10g1b2}.
\end{split}
\end{equation}
\\
Below this value of the distance $R$, the global warp function $e^{-4\,U}$ shows unphysical singularities and the system, although composed of two regular black holes, ceases to be globally regular.\\
As before, we fix  $R$ at a value above the lower bound, say $12$. In the following plots we show the behavior for $e^{-4\,U}$ and $e^{-4\,U}\,r^{2}\sin^{2}\theta-\omega^2$, which are both positive everywhere. This excludes the presence of CTC in the solution when the lower bound (\ref{lba10g1b2}) holds.\\
\begin{figure}[H]
\centering
\subfloat[][\small The plot of $e^{-4\,U}\,r^{2}\sin^{2}\theta-\omega^2>0$.]{\includegraphics[width=7cm]{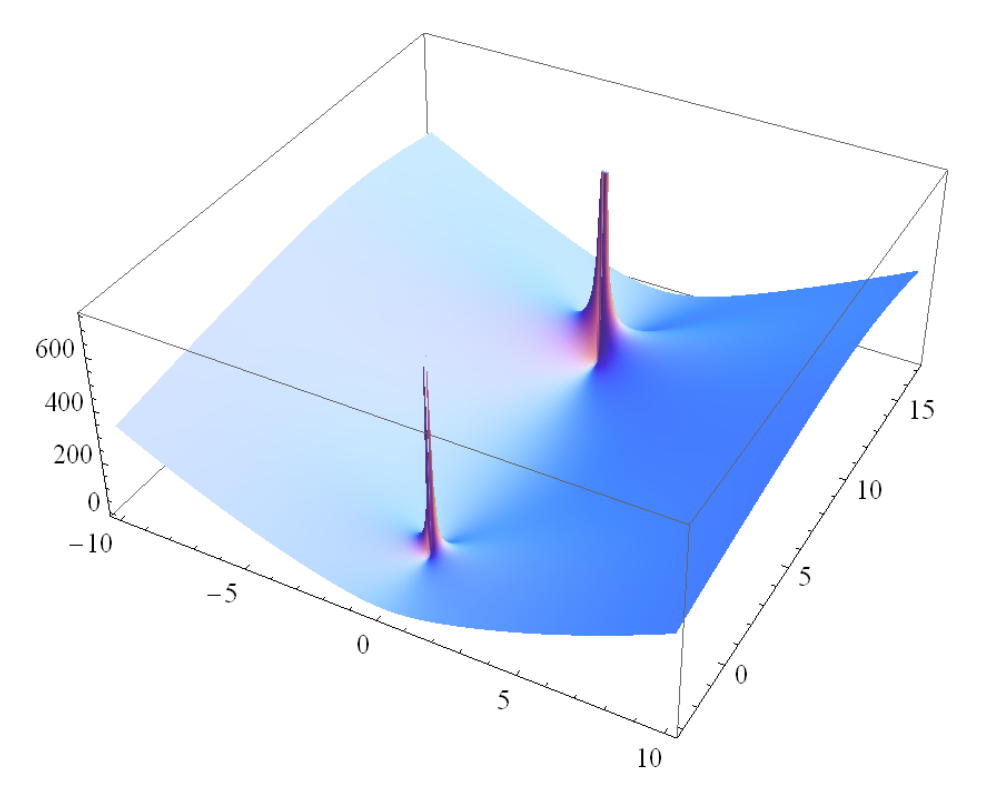}} \hfill
\subfloat[][The plot of $e^{-4\,U}$.]{\includegraphics[width=7cm]{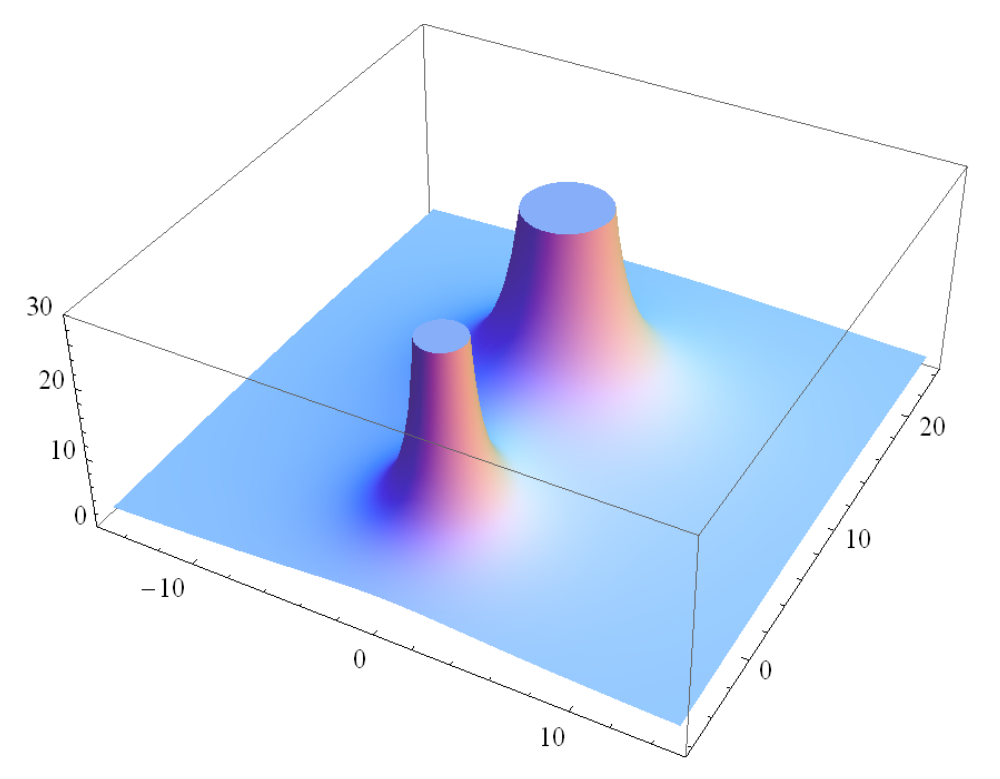}}\hfill
\end{figure}
The following plots illustrate the behavior of the  curvature scalar  $\mathcal{R}$ which shows for both centers the typical shape due to the  \emph{attractor mechanism}.\\
\begin{figure}[H]
\centering
\subfloat[][The plot of the curvature scalar $\mathcal{R}$.]{\includegraphics[width=7cm]{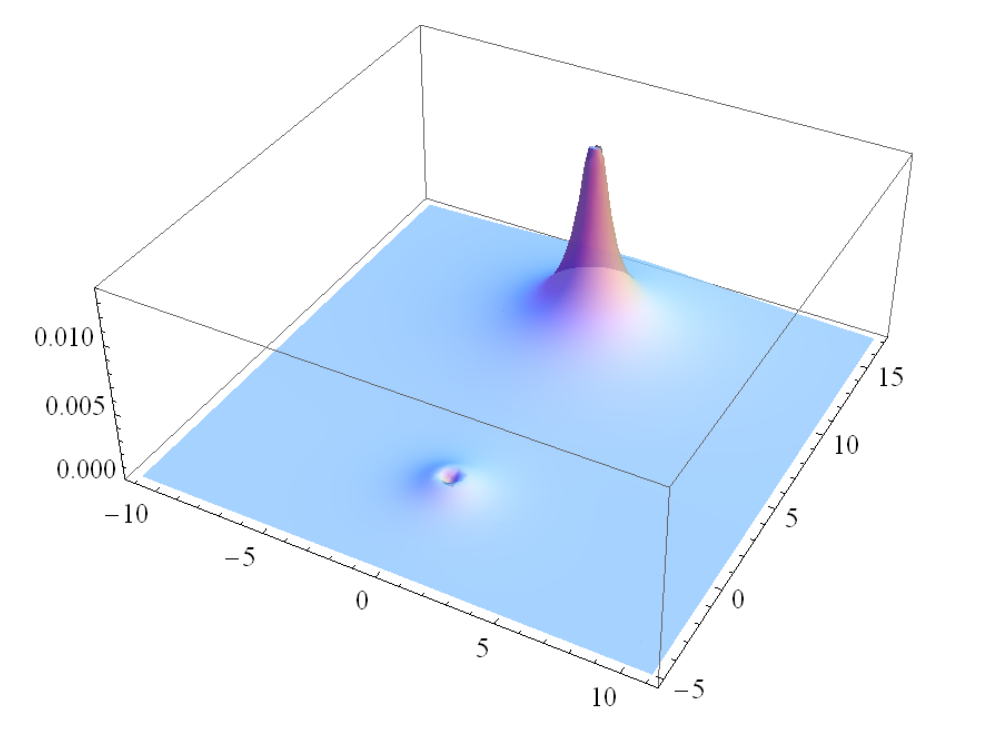}}\hfill
\subfloat[][Detail of the curvature scalar near the $B$ center.]{\includegraphics[width=7cm]{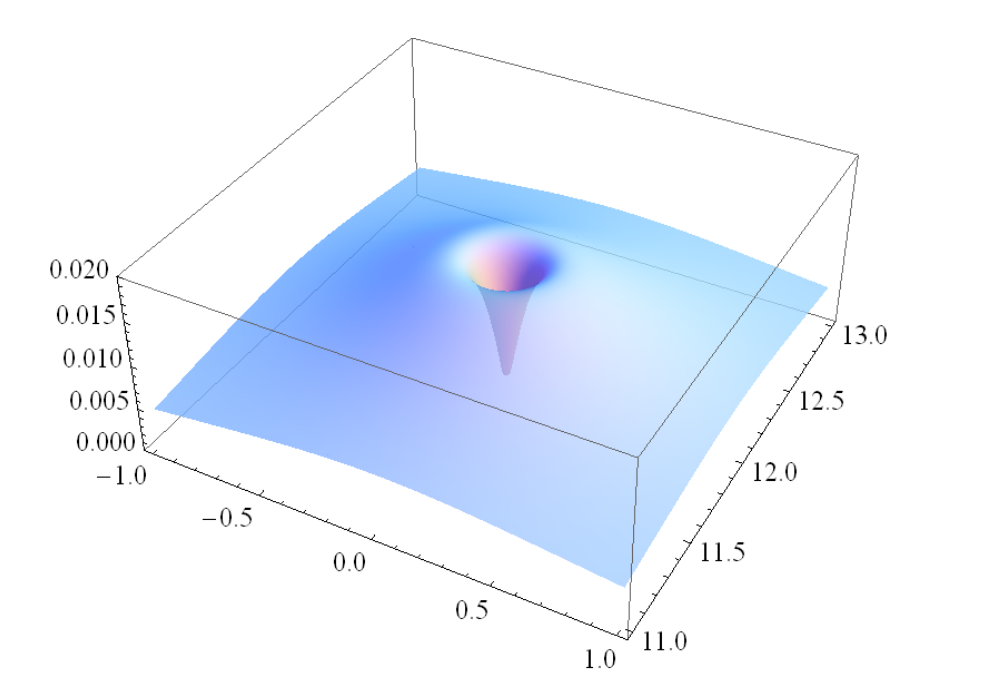}} \\
\subfloat[][Detail of the curvature scalar near the $A$ center.]{\includegraphics[width=7cm]{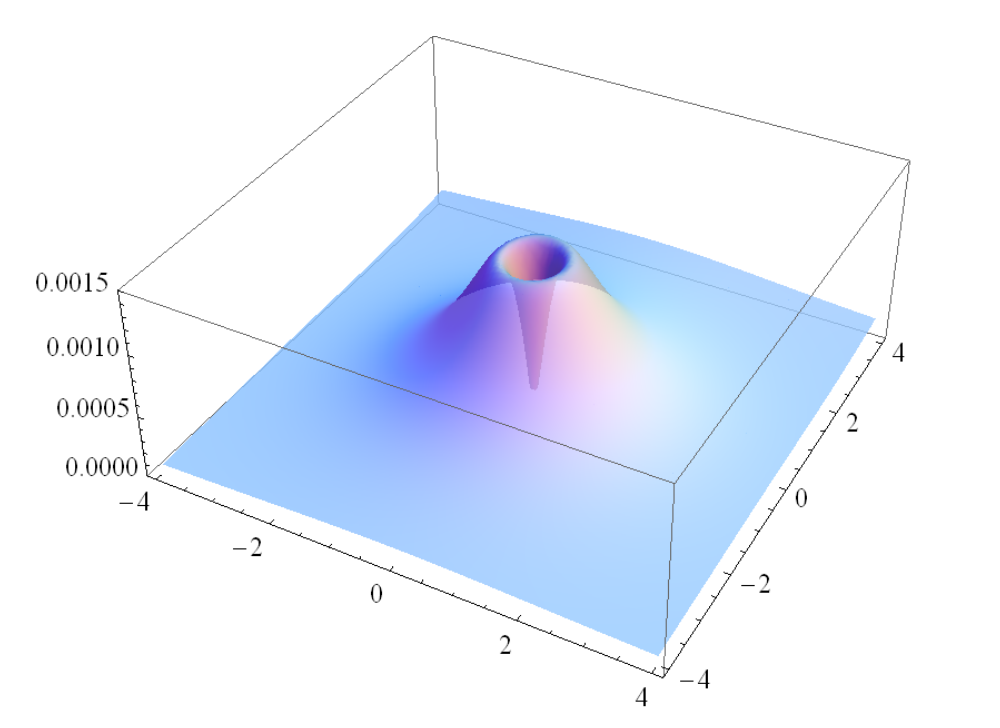}}\hfill
\end{figure}
\subsubsection{The other real $\alpha^{(10)}$-orbits}
As for the other $H^*$-orbits with $\alpha$-label $\alpha^{(10)}$, the corresponding solutions satisfy the same equations (\ref{eqmot2}). The ``diagonal'' orbits $\gamma^{(10;k)},\,\beta^{(10;k)}$ are intrinsically singular. As for the others, they can be obtained by combining regular-single-center orbits as illustrated in Appendix \ref{sumrules}. We did compute examples of them which are qualitatively analogous to the solution discussed above.
\subsection{Summary of representatives for the $\alpha^{(10)}$-orbits }
In the following table we give representatives of all the  $\alpha^{(10)}$-orbits as sum of representatives of single-center ones, keeping in mind that each orbit can have many representations in terms of (regular or non-regular) single center black holes. For the  diagonal, ``intrinsically singular'' orbits, we give a representation in terms of two constituents of which at least one is necessarily singular.
\begin{table}[H]
\begin{center}
\resizebox{\columnwidth}{!}{%
\begin{tabular}{|c|c|c|c|c|}\hline
\backslashbox{$\gamma$}{$\beta$}
&$\beta^{(1)}$&$\beta^{(2)}$&$\beta^{(3)}$&$\beta^{(4)}$ \\   \hline
$\gamma^{(1)}$&\multicolumn{1}{c|}{{\begin{tabular}{c}
\\ $\alpha^{(6)}\gamma^{(6;5)}\beta^{(5)}\delta^{(4)}\,+$\\
$\alpha^{(6)}\gamma^{(6;5)}\beta^{(5)}\delta^{(3)}$\\
\\
\end{tabular}}} & \multicolumn{1}{c|}{{\begin{tabular}{c}
\\ $\alpha^{(6)}\gamma^{(6;5)}\beta^{(5)}\delta^{(1)}\,+$\\
$\alpha^{(6)}\gamma^{(6;5)}\beta^{(5)}\delta^{(1)}$\\
\\
\end{tabular}}} & \multicolumn{1}{c|}{{\begin{tabular}{c}
\\ $\alpha^{(6)}\gamma^{(6;5)}\beta^{(5)}\delta^{(1)}\,+$\\
$\alpha^{(6)}\gamma^{(6;5)}\beta^{(5)}\delta^{(1)}$\\
\\
\end{tabular}}} & \multicolumn{1}{c|}{{\begin{tabular}{c}
\\ $\alpha^{(6)}\gamma^{(6;5)}\beta^{(5)}\delta^{(1)}\,+$\\
$\alpha^{(6)}\gamma^{(6;5)}\beta^{(5)}\delta^{(1)}$\\
\\
\end{tabular}}}\\ \hline
$\gamma^{(2)}$&\multicolumn{1}{c|}{{\begin{tabular}{c}
\\ $\alpha^{(6)}\gamma^{(6;5)}\beta^{(5)}\delta^{(1)}\,+$\\
$\alpha^{(6)}\gamma^{(6;5)}\beta^{(5)}\delta^{(1)}$\\
\\
\end{tabular}}}&\multicolumn{1}{c|}{{\begin{tabular}{c}
\\ $\alpha^{(6)}\gamma^{(6;5)}\beta^{(5)}\delta^{(2)}\,+$\\
$\alpha^{(6)}\gamma^{(6;5)}\beta^{(5)}\delta^{(3)}$\\
\\
\end{tabular}}}&\multicolumn{1}{c|}{{\begin{tabular}{c}
\\ $\alpha^{(6)}\gamma^{(6;5)}\beta^{(5)}\delta^{(1)}\,+$\\
$\alpha^{(6)}\gamma^{(6;5)}\beta^{(5)}\delta^{(1)}$\\
\\
\end{tabular}}}&\multicolumn{1}{c|}{{\begin{tabular}{c}
\\ $\alpha^{(6)}\gamma^{(6;5)}\beta^{(5)}\delta^{(1)}\,+$\\
$\alpha^{(6)}\gamma^{(6;5)}\beta^{(5)}\delta^{(1)}$\\
\\
\end{tabular}}}\\ \hline
$\gamma^{(3)}$&\multicolumn{1}{c|}{{\begin{tabular}{c}
\\ $\alpha^{(6)}\gamma^{(6;5)}\beta^{(5)}\delta^{(1)}\,+$\\
$\alpha^{(6)}\gamma^{(6;5)}\beta^{(5)}\delta^{(1)}$\\
\\
\end{tabular}}}&\multicolumn{1}{c|}{{\begin{tabular}{c}
\\ $\alpha^{(6)}\gamma^{(6;5)}\beta^{(5)}\delta^{(1)}\,+$\\
$\alpha^{(6)}\gamma^{(6;5)}\beta^{(5)}\delta^{(1)}$\\
\\
\end{tabular}}}&\multicolumn{1}{c|}{{\begin{tabular}{c}
\\ $\alpha^{(6)}\gamma^{(6;5)}\beta^{(5)}\delta^{(2)}\,+$\\
$\alpha^{(6)}\gamma^{(6;5)}\beta^{(5)}\delta^{(4)}$\\
\\
\end{tabular}}}&\multicolumn{1}{c|}{{\begin{tabular}{c}
\\ $\alpha^{(6)}\gamma^{(6;5)}\beta^{(5)}\delta^{(1)}\,+$\\
$\alpha^{(6)}\gamma^{(6;5)}\beta^{(5)}\delta^{(1)}$\\
\\
\end{tabular}}}\\ \hline
$\gamma^{(4)}$&\multicolumn{1}{c|}{{\begin{tabular}{c}
\\ $\alpha^{(6)}\gamma^{(6;5)}\beta^{(5)}\delta^{(1)}\,+$\\
$\alpha^{(6)}\gamma^{(6;5)}\beta^{(5)}\delta^{(1)}$\\
\\
\end{tabular}}}&\multicolumn{1}{c|}{{\begin{tabular}{c}
\\ $\alpha^{(6)}\gamma^{(6;5)}\beta^{(5)}\delta^{(1)}\,+$\\
$\alpha^{(6)}\gamma^{(6;5)}\beta^{(5)}\delta^{(1)}$\\
\\
\end{tabular}}}&\multicolumn{1}{c|}{{\begin{tabular}{c}
\\ $\alpha^{(6)}\gamma^{(6;5)}\beta^{(5)}\delta^{(1)}\,+$\\
$\alpha^{(6)}\gamma^{(6;5)}\beta^{(5)}\delta^{(1)}$\\
\\
\end{tabular}}}&\multicolumn{1}{c|}{{\begin{tabular}{c}
\\ $\alpha^{(6)}\gamma^{(6;5)}\beta^{(5)}\delta^{(3)}\,+$\\
$\alpha^{(6)}\gamma^{(6;5)}\beta^{(5)}\delta^{(3)}$\\
\\
\end{tabular}}}\\ \hline
\end{tabular}}
\caption{\small The representative of $\alpha^{(10)}$-orbit in term of the two black holes. }
\end{center}
\end{table}
\subsection{The orbit $\alpha^{(11)}\gamma^{(11;(1,\dots,4))}\beta^{(1,\dots,4)}$}
Following \cite{Bena:2009ev}, let us start considering the graded decomposition of $\mathfrak{so}(4,4)$ respect to the weighted Dynkin diagram in \ref{labels}, which defines $\alpha^{(11)}$. The nilpotent algebra $\mathfrak{n}$ algebra has the structure \cite{bossard2}:
\begin{equation}{\mathfrak{n}^{(11;1,(1,\dots,4))}\cong(\textbf{1}\oplus3\times\textbf{1})^{(1)}_{\mathfrak{K}^*}\oplus(3\times \textbf{1})^{(2)}_{\mathfrak{H}^*}\oplus(3\times \textbf{1})^{(3)}_{\mathfrak{K}^*}\oplus\textbf{1}^{(4)}_{\mathfrak{H}^*}\oplus\textbf{1}^{(5)}_{\mathfrak{K}^*}}\,.
\end{equation}
Choosing an appropriate basis, the only non-zero commutators are
\begin{equation}
\begin{split}
[\,\textit{\textbf{e}}_{0}^{(1)}\,,\,\textit{\textbf{e}}_{i}^{(1)}\,]&\,=\,\textit{\textbf{f}}_{i}^{(2)}\,,\,[\,\textit{\textbf{f}}_{i}^{(2)}\,,\,\textit{\textbf{e}}_{j}^{(1)}\,]\,=\,c_{ijk}\,\textit{\textbf{e}}^{(3)\,k}\,,\,[\,\textit{\textbf{e}}^{(3)\,i}\,,\,\textit{\textbf{e}}_{j}^{(1)}\,]\,=\,\delta^{i}_{j}\,\textit{\textbf{f}}^{(4)}\,,\\
[\,\textit{\textbf{f}}^{(4)}\,,\,\textit{\textbf{e}}_{0}^{(1)}\,]&\,=\,\textit{\textbf{e}}^{5}\,,\,[\,\textit{\textbf{f}}_{i}^{(2)}\,,\,\textit{\textbf{e}}^{(3)\,j}\,]\,=\,\delta_{i}^{j}\,\textit{\textbf{e}}^{(5)}\label{comrel3}\,,
\end{split}
\end{equation}
having denoted by $\textit{\textbf{e}}$ the generators in ${\mathfrak{K}^*}$ and by $\textit{\textbf{f}}$ those in ${\mathfrak{H}^*}$, the number in superscript being the grading relative to the neutral element. The coefficients $c_{ijk}$ are defined as: $c_{ijk}=|\varepsilon_{ijk}|$ .\\
We now consider the Ansatz (\ref{defhatl}) with
\begin{equation}{\hat{\mathbb{L}}\,=\,exp(\,-\,\tilde{V}\,\textit{\textbf{e}}_{0}^{(1)}\,-\,\tilde{K}^{i}\,\textit{\textbf{e}}_{i}^{(1)}\,-\,\tilde{Z}_{i}\,
\textit{\textbf{e}}^{(3)\,i}\,-\,\tilde{M}\,\textit{\textbf{e}}^{(5)})}\,
\end{equation}
and write the graded field equations deduced from (\ref{feqscal}). With the following redefinitions
\begin{equation}
\begin{split}
\tilde{V}&\rightarrow{\,a\,V\,-\,l_{0}}\,,\\
\tilde{K}^{i}&\rightarrow{\,K_{i}\,}\,,\\
\tilde{Z}_{i}&\rightarrow{\,a\,Z_{i}\,-\,\frac{1}{3}\,c_{ijk}\,K_{j}\,K_{k}\,\left(\,a\,V\,+\,2\,l_{0}\right)\,-\,l_{i}}\,,\\
\tilde{M}&\rightarrow{\,2\,a^{2}\,V\,\mu\,}-\,\frac{1}{3}\,(\,a\,V\,-\,l_{0}\,)\,\sum_{i}\,l_{i}\,K_{i}\,-\,\frac{1}{3}\,a\,(\,2\,a\,V\,+\,l_{0}\,)\,\sum_{i}\,K_{i}\,Z_{i}\,\\
&-\,\frac{1}{15}\,\left[\,4\,(\,a\,V\,+\,l_{0}\,)^{2}\,-\,a\,l_{0}\,V\,\right]\,\prod_{i}\,K_{i}\,,
\end{split}\label{redef4}
\end{equation}
we obtain the following equations of motion  \cite{Bena:2009ev}
\begin{equation}
\begin{split}
{d}\ast{d}{V}&\,=\,0\,,\\
{d}\ast{d}{K_{i}}&\,=\,0\,,\\
{d}\ast{d}{Z_{i}}&\,=\,\frac{1}{2}\,c_{ijk}\,V\,d\,\ast\,{d}{(\,K_{j}\,K_{k}\,)}\,,\\
{d}\ast{d}{(\,\mu\,V\,)}&\,=\,{d}{(\,V\,Z_{i}\,)}\ast\,{d}{K_{i}}\,,\\\label{eqmot3}
\end{split}
\end{equation}
where, in writing the last two equations, we have used the property that $V$ and $K^{i}$  are harmonic functions.
With the redefinitions (\ref{redef4}) and choosing\,$a=4\sqrt{2}$\,\,,\,\,$l_{0}=-2$\,,\,$l_{1}=l_{2}=8$\,,\,$l_{3}=-8$\,,\,\,$e^{-4\,U}$ has the form
\begin{equation}
e^{-4\,U}\,=\,V\,Z_{1}\,Z_{2}\,Z_{3}\,-\,(\,\mu\,V\,)^{2}\,,
\end{equation}
and the electromagnetic fields and $a$ read
\begin{equation}
\begin{split}
\mathcal{Z}^{0}&\,=\,-\,\frac{1}{\sqrt{2}}\,-\,\frac{\mu\,V^2\,K_{1}\,-\,V\,Z_{2}\,Z_{3}}{e^{-4\,U}}\,,\\
\mathcal{Z}^{1}&\,=\,-\,\frac{2\,\mu\,V^{2}}{e^{-4\,U}}\,,\\
\mathcal{Z}^{2}&\,=\,\frac{\mu\,V^{2}\,K_{1}\,K_{2}\,+\,(\,\mu\,-\,K_{1}\,Z_{1}\,-\,K_{2}\,Z_{2})\,V\,Z_{3}}{2\,e^{-4\,U}}\,,\\
\mathcal{Z}^{3}&\,=\,-\frac{\mu\,V^{2}\,K_{1}\,K_{3}\,+\,(\,\mu\,-\,K_{1}\,Z_{1}\,-\,K_{3}\,Z_{3})\,V\,Z_{2}}{2\,e^{-4\,U}}\,,\\
\mathcal{Z}_{0}&\,=\,-\frac{\mu\,V^{2}\,K_{2}\,K_{3}\,+\,(\,\mu\,-\,K_{2}\,Z_{2}\,-\,K_{3}\,Z_{3})\,V\,Z_{1}}{2\,e^{-4\,U}}\,,\\
\mathcal{Z}_{1}&\,=\,-\,\frac{1}{\sqrt{2}}\,+\,\frac{2\,\mu\,V\,(\,V\,\prod_{i}\,K_{i}\,+\,\prod_{i}\,K_{i}\,Z_{i}\,)\,-\,2\,\prod_{i}\,Z_{i}\,-\,\sum_{k}\,c^{ijk}\,K_{i}\,Z_{i}\,K_{j}\,Z_{j}}{8\,e^{-4\,U}}\,,\\
\mathcal{Z}_{2}&\,=\,\frac{1}{\sqrt{2}}\,-\,\frac{\mu\,V^2\,K_{3}\,-\,V\,Z_{1}\,Z_{2}}{e^{-4\,U}}\,,\\
\mathcal{Z}_{3}&\,=\,\frac{1}{\sqrt{2}}\,+\,\frac{\mu\,V^2\,K_{2}\,-\,V\,Z_{1}\,Z_{3}}{e^{-4\,U}}\,,\\
a&\,=\,\frac{1}{4\,e^{-4\,U}}\,\Biggl[\mu V[4+\sqrt{2}V(4+K_{1}K_{2}-K_{1}K_{3}-K_{2}K_{3})-\sqrt{2}(Z_{1}+Z_{2}-Z_{3})]+\\
&\,+\sqrt{2}V[(K_{1}+K_{2})Z_{1}Z_{2}-(K_{1}-K_{3})Z_{1}Z_{3}-(K_{2}-K_{3})Z_{2}Z_{3}]\Biggr]\,.\\
\label{chpq2}
\end{split}
\end{equation}
The four-dimensional scalar fields read
\begin{equation}
\begin{aligned}
\epsilon_{1}\,&=\,\frac{2\,V\,(\,\mu\,-\,K_{1}\,Z_{1}\,)}{V\,K_{1}\,(\,2\,\mu\,-\,K_{1}\,Z_{1}\,)\,-\,Z_{2}\,Z_{3}}&,\quad&e^{-2\,\varphi_{1}}\,&=&\,\frac{(\,V\,K_{1}\,(\,2\,\mu\,-\,K_{1}\,Z_{1}\,)\,-\,Z_{2}\,Z_{3}\,)^{2}}{4\,e^{-4U}}\quad,\\
\epsilon_{2}\,&=\,-\,\frac{1}{2}\,\left(\,K_{2}\,-\,\frac{\mu}{Z_{2}}\,\right)&,\quad&e^{-2\,\varphi_{2}}\,&=&\,\frac{4\,V^{2}\,Z_{2}^{2}}{e^{-4U}}\quad,\\
\epsilon_{3}\,&=\,\frac{1}{2}\,\left(\,K_{3}\,-\,\frac{\mu}{Z_{3}}\,\right)&,\quad&e^{-2\,\varphi_{3}}\,&=&\,\frac{4\,V^{2}\,Z_{3}^{2}}{e^{-4U}}\quad.
\label{scalar111}
\end{aligned}
\end{equation}
Now, it is useful rewrite the last equation of (\ref{eqmot3}) as
\begin{equation}
\begin{split}
{d}\left(\ast{d}{\,\mu\,V\,}\right)&\,=\,{d}\left(\,V\,Z_{i}\,\ast\,{d}{K_{i}}\right)\,,\\
\end{split}
\end{equation}
so we can obtain the explicit form for $\omega$
\begin{equation}
\begin{split}
\ast{d}{\omega}\,=\,{d}{(\mu\,V)}\,-\,\,V\,Z_{i}\,{d}{K^{i}}\label{om111}\,,
\end{split}
\end{equation}
and for the $D=3$ vectors $\tilde{A}^{M}$
\begin{equation}
\begin{split}
\ast{d}{\tilde{A}^{0}}&=-\frac{\ast{d}{\omega}}{\sqrt{2}}-V{d}{K_{1}}+K_{1}{d}{V}\quad,\quad\ast{d}{\tilde{A}^{1}}=2{d}{V}\quad,\\
\ast{d}{\tilde{A}^{2}}&=-\frac{1}{2}\left({d}{Z_{3}}-V{K_{1}}{d}{K_{2}}-V{K_{2}}{d}{K_{1}}+{K_{1}}{K_{2}}{d}{V}\right)\quad,\\
\ast{d}{\tilde{A}^{3}}&=\frac{1}{2}\left({d}{Z_{2}}-V{K_{1}}{d}{K_{3}}-V{K_{3}}{d}{K_{1}}+{K_{1}}{K_{3}}{d}{V}\right)\quad,\\
\ast{d}{\tilde{A}_{0}}&=\frac{1}{2}\left({d}{Z_{1}}-V{K_{2}}{d}{K_{3}}-V{K_{3}}{d}{K_{2}}+{K_{2}}{K_{3}}{d}{V}\right)\quad,\\
\ast{d}{\tilde{A}_{1}}&=-\frac{\ast{d}{\omega}}{\sqrt{2}}-\frac{1}{4}\left[\sum_{i}{K_{i}}{d}{Z_{i}}+{K_{1}}{K_{2}}{K_{3}}{d}{V}-{V}{d}\left({{K_{1}}{K_{2}}{K_{3}}}\right)-\sum_{i}{Z_{i}}{d}{K_{i}}\right]\quad,\\
\ast{d}{\tilde{A}_{2}}&=\frac{\ast{d}{\omega}}{\sqrt{2}}-V{d}{K_{3}}+K_{3}{d}{V}\quad,\quad\ast{d}{\tilde{A}_{3}}=\frac{\ast{d}{\omega}}{\sqrt{2}}+V{d}{K_{2}}-K_{2}{d}{V}\quad.\label{vect111}\\
\end{split}
\end{equation}
\subsubsection{The solution}\label{subsect2}
We follow the procedure of \cite{Bena:2009ev} for the derivation of a two-center solution to the equations of motion. The general solution we find will differ from that in \cite{Bena:2009ev} by a duality transformation. We shall find examples of regular solutions in which one center is a large non-BPS black hole with $I_4<0$ and the other is a large black hole (BPS or non-BPS) with $I_4>0$.\par
As usual we choose the $A$-center to be located at $r=0$ and the $B$-one at $\theta=0,\,r=R$.
From the first two equations in (\ref{eqmot3}) follow that $V$ and $K_{i}$ are harmonic functions, so we can write them in the form
\begin{equation}
V\,=\,h\,+\,\frac{Q_{6}}{r}\,\quad,\quad\,K_{i}\,=\,\frac{d_{i}}{\Sigma}\,.
\end{equation}
Substituting the above expressions in the third equation of (\ref{eqmot3}), we find
\begin{equation}
\begin{split}
{d}\ast{d}{Z_{i}}\,=\,c^{ijk}\left(\,h\,+\,\frac{Q_{6}}{r}\right)\,{d}\left(\,\frac{d_{j}}{\Sigma}\,\right)\ast{d}\left(\,\frac{d_{k}}{\Sigma}\,\right)\,,
\end{split}\label{ddZ111}
\end{equation}
which can be easily solved by using the following properties
\begin{equation}
\begin{split}
{d}\left(\frac{1}{\Sigma}\right)\ast{d}\left(\frac{1}{\Sigma}\right)={d}\ast{d}\left(\frac{1}{2\Sigma^{2}}\right)\quad,\quad\frac{1}{r}\,{d}\left(\frac{1}{\Sigma}\right)\ast{d}\left(\frac{1}{\Sigma}\right)={d}\ast{d}\left(\frac{r}{2R^{2}\Sigma^{2}}\right)\quad.
\end{split}
\end{equation}
Adding the harmonic contribution, we obtain for $Z_{i}$
\begin{equation}
Z_{i}=l_{i}+\frac{Q_{i}}{\Sigma}+\frac{\tilde{Q}_{i}}{r}+\frac{c_{ijk}d_{j}d_{k}}{2\Sigma^{2}}\left(h+\frac{Q_{6}r}{R^{2}}\right)\,.
\end{equation}
Now, we have to solve the fourth equation of (\ref{eqmot3})
\begin{equation}
\begin{split}
{d}\ast{d}{(\mu\,V)}\,=\,{d}\left[\left(\,h\,+\,\frac{Q_{6}}{r}\right)\,\left(l_{i}+\frac{Q_{i}}{\Sigma}+\frac{\tilde{Q}_{i}}{r}+\frac{c_{ijk}d_{j}d_{k}}{2\Sigma^{2}}\left(h+\frac{Q_{6}r}{R^{2}}\right)\right)\right]\ast{d}\left(\,\frac{d_{i}}{\Sigma}\,\right)\,,\\
\end{split}\label{ddZ112}
\end{equation}
and we do it term by term:
\begin{equation}
\begin{split}
{d}\ast{d}{(\mu_{(1)}\,V)}\,&=\,{d}\left[\left(\,h\,+\,\frac{Q_{6}}{r}\right)\,l_{i}\right]\ast{d}\left(\,\frac{d_{i}}{\Sigma}\,\right)={d}\ast{d}\left[\frac{Vl_{i}d_{i}}{2\Sigma}\right]\,\\
&\Rightarrow\,\mu_{(1)}=\frac{l_{i}d_{i}}{2\Sigma}\,,\\
\\
{d}\ast{d}{(\mu_{(2)}\,V)}\,&=\,{d}\left[\left(\,h\,+\,\frac{Q_{6}}{r}\right)\,\frac{Q_{i}}{\Sigma}\right]\ast{d}\left(\,\frac{d_{i}}{\Sigma}\,\right)={d}\ast{d}\left[\frac{Q_{i}d_{i}}{2\Sigma^{2}}\left(h+\frac{Q_{6}\cos\theta}{R}\right)\right]\,\\
&\Rightarrow\,\mu_{(2)}=\frac{Q_{i}d_{i}}{2V\Sigma^{2}}\left(h+\frac{Q_{6}\cos\theta}{R}\right)\,,\\
\\
{d}\ast{d}{(\mu_{(3)}\,V)}\,&=\,{d}\left[\left(\,h\,+\,\frac{Q_{6}}{r}\right)\,\frac{\tilde{Q}_{i}}{r}\right]\ast{d}\left(\,\frac{d_{i}}{\Sigma}\,\right)={d}\ast{d}\left[\frac{\tilde{Q}_{i}d_{i}}{r\Sigma}\left(\frac{h}{2}+\frac{Q_{6}\cos\theta}{R}\right)\right]\,\\
&\Rightarrow\,\mu_{(3)}=\frac{\tilde{Q}_{i}d_{i}}{rV\Sigma}\left(\frac{h}{2}+\frac{Q_{6}\cos\theta}{R}\right)\,,\\
\\
{d}\ast{d}{(\mu_{(4)}\,V)}\,&=\,{d}\left[\frac{hQ_{6}c_{ijk}d_{j}d_{k}}{2\Sigma^2}\left(\frac{1}{r}+\frac{r}{R^2}\right)\right]\ast{d}\left(\,\frac{d_{i}}{\Sigma}\,\right)={d}\ast{d}\left[\frac{hQ_{6}c_{ijk}d_{i}d_{j}d_{k}}{6}\,\frac{3r^{2}+R^{2}}{2rR^{2}\Sigma^{3}}\right]\,\\
&\Rightarrow\,\mu_{(4)}=\frac{hQ_{6}c_{ijk}d_{i}d_{j}d_{k}}{6}\,\frac{3r^{2}+R^{2}}{2rVR^{2}\Sigma^{3}}\,.\\
\end{split}
\end{equation}
It remain to solve the term
\begin{equation}
\begin{split}
{d}\ast{d}{(\mu_{(5)}\,V)}\,=\,{d}\left[\frac{c_{ijk}d_{j}d_{k}}{2\Sigma^2}\left(h^{2}+\frac{Q_{6}^{2}}{R^2}\right)\right]\ast{d}\left(\,\frac{d_{i}}{\Sigma}\,\right)\,,\\
\end{split}
\end{equation}
which admits two possible solutions
\begin{equation}
\begin{split}
{d}\ast{d}{(\mu_{(5)}\,V)}\,=\,{d}\left[\frac{c_{ijk}d_{j}d_{k}}{2\Sigma^2}\left(h^{2}+\frac{Q_{6}^{2}}{R^2}\right)\right]\ast{d}\left(\,\frac{d_{i}}{\Sigma}\,\right)&=_{1}{d}\ast{d}\left[\frac{c_{ijk}d_{i}d_{j}d_{k}}{6\Sigma^{3}}\,\left(h^{2}+\frac{Q_{6}^{2}}{R^2}\right)\right]\,\\
&=_{2}{d}\ast{d}\left[\frac{c_{ijk}d_{i}d_{j}d_{k}}{6}\,\frac{r\cos\theta}{R\Sigma^{3}}\,\left(h^{2}+\frac{Q_{6}^{2}}{R^2}\right)\right]\,\\
\Rightarrow\,\mu_{(5,1)}=\frac{c_{ijk}d_{i}d_{j}d_{k}}{6V\Sigma^{3}}\,\left(h^{2}+\frac{Q_{6}^{2}}{R^2}\right)\,,\\
\Rightarrow\,\mu_{(5,2)}=\frac{c_{ijk}d_{i}d_{j}d_{k}}{6}\,\frac{r\cos\theta}{RV\Sigma^{3}}\,\left(h^{2}+\frac{Q_{6}^{2}}{R^2}\right)\,,
\end{split}
\end{equation}
so we can choose the linear combination of these terms: $\mu_{(5)}=\mu_{(5,2)}+\delta(\mu_{(5,2)}-\mu_{(5,1)})$.
Finally, adding the harmonic contribution, we obtain
\begin{equation}
\begin{split}
\mu&=\frac{m_{0}}{V}+\frac{m}{V\Sigma}+\frac{\tilde{m}}{Vr}+\frac{\beta\cos\theta}{Vr^{2}}+\frac{l_{i}d_{i}}{2\Sigma}+\frac{hQ_{i}d_{i}}{2V\Sigma^{2}}+\frac{Q_{6}Q_{i}d_{i}\cos\theta}{2RV\Sigma^{2}}+\frac{h\tilde{Q}_{i}d_{i}}{2Vr\Sigma}+\frac{Q_{6}\tilde{Q}_{i}d_{i}\cos\theta}{RVr\Sigma}+\\
&+\frac{c_{ijk}}{6}d_{i}d_{j}d_{k}\left[\left(h^{2}+\frac{Q_{6}^{2}}{R^2}\right)\left(\frac{r\cos\theta}{RV\Sigma^{3}}+\delta\frac{\cos\theta_{\Sigma}}{RV\Sigma^{2}}\right)+hQ_{6}\frac{3r^{2}+R^{2}}{2rVR^{2}\Sigma^{3}}\right]\,.
\end{split}
\end{equation}
Similarly, from (\ref{om111}) and (\ref{vect111}), we obtain the expressions for $\omega$
\begin{equation}
\begin{split}
\omega&=\bigg[k_{\omega}-\beta\frac{\sin^{2}\theta}{r}+\left(m-\frac{{h}l_{i}d_{i}}{2}\right)\cos\theta_{\Sigma}+\tilde{m}\cos\theta+\left(Q_{6}l_{i}d_{i}+h\tilde{Q}_{i}d_{i}\right)\frac{R\cos\theta-r}{2R\Sigma}+\\
&-Q_{6}Q_{i}d_{i}\frac{\sin^{2}\theta_{\Sigma}}{2rR}-Q_{6}\tilde{Q}_{i}d_{i}\frac{\sin^{2}\theta}{r\Sigma}-\frac{c^{ijk}d_{i}d_{j}d_{k}}{6}\big[\left(h^{2}+\frac{Q_{6}^2}{R^{2}}\right)(1+\delta)\frac{\sin^{2}\theta_{\Sigma}}{R\Sigma}+\\
&+hQ_{6}\frac{r(3R^{2}+r^{2})-R(3r^{2}+R^{2})\cos\theta}{2R^{3}\Sigma^{3}}\big]\bigg]d\varphi\quad,\label{omega111}\\
\end{split}
\end{equation}
and for the $D=3$ vectors $\tilde{A}^{M}$
\begin{equation}
\begin{split}
\tilde{A}^{0}&=-\frac{\omega}{\sqrt{2}}-\left[h{d}_{1}\cos\theta_{\Sigma}-Q_{6}{d}_{1}\frac{R\cos\theta-r}{R\Sigma}\right]d\varphi\quad,\quad{\tilde{A}}^{1}=2Q_{6}\cos\theta{d}\phi\quad,\\
{\tilde{A}}^{2}&=-\frac{1}{2}\left[Q_{3}\cos\theta_{\Sigma}+\tilde{Q}_{3}\cos\theta+Q_{6}{d}_{1}{d}_{2}\frac{(r^{2}+R^{2})\cos\theta-2rR}{R^{2}\Sigma^{2}}\right]d\varphi\quad,\\
{\tilde{A}}^{3}&=\frac{1}{2}\left[Q_{2}\cos\theta_{\Sigma}+\tilde{Q}_{2}\cos\theta+Q_{6}{d}_{1}{d}_{3}\frac{(r^{2}+R^{2})\cos\theta-2rR}{R^{2}\Sigma^{2}}\right]d\varphi\quad,\\
{\tilde{A}}_{0}&=\frac{1}{2}\left[Q_{1}\cos\theta_{\Sigma}+\tilde{Q}_{1}\cos\theta+Q_{6}{d}_{2}{d}_{3}\frac{(r^{2}+R^{2})\cos\theta-2rR}{R^{2}\Sigma^{2}}\right]d\varphi\quad,\\
\tilde{A}_{1}&=-\frac{\omega}{\sqrt{2}}+\frac{1}{4}\sum_{i}\bigg[l_{i}d_{i}\cos\theta_{\Sigma}+d_{i}\tilde{Q}_{i}\frac{r-R\cos\theta}{R\Sigma}+Q_{6}d_{1}d_{2}d_{3}\frac{r(3R^{2}+r^{2})-R(3r^{2}+R^{2})\cos\theta}{R^{3}\Sigma^{3}}\bigg]d\varphi\quad,\\
\tilde{A}_{2}&=\frac{\omega}{\sqrt{2}}-\left[h{d}_{3}\cos\theta_{\Sigma}-Q_{6}{d}_{3}\frac{R\cos\theta-r}{R\Sigma}\right]d\varphi\quad,\\
\tilde{A}_{3}&=\frac{\omega}{\sqrt{2}}+\left[h{d}_{2}\cos\theta_{\Sigma}-Q_{6}{d}_{2}\frac{R\cos\theta-r}{R\Sigma}\right]d\varphi\quad.\\
\end{split}
\end{equation}
Although these expressions satisfy the equations of motion, these are not sufficient to ensure the regularity of the system, so further conditions are necessary.
\paragraph{Zero-NUT charge condition}\par
We require the absence of the NUT charge for all centers and impose the conditions (\ref{condnut7}) on $\omega$ , in order to exclude the presence of Dirac-Misner singularities. This implies three conditions on the coefficients $k_{\omega}$, $m$ and $\tilde{m}$
\begin{equation}
\begin{split}
k_{\omega}\,&=\,\frac{(\,Q_6\,l_{i}\,+\,h\,\tilde{Q}_{i})\,d_{i}}{2\,R}\,+\,\frac{h\,Q_{6}\,d_{1}\,d_{2}\,d_{3}}{2\,R^{3}}\quad,\\
m\,&=\,\left(\,h\,+\,\frac{Q_{6}}{R}\,\right)\,\frac{l_{i}\,d_{i}}{2}\,+\,\frac{h\,Q_{6}}{2\,R^{3}}\,\frac{c_{ijk}\,d_{i}\,d_{j}\,d_{k}}{6}\,+\,\frac{h\,\tilde{Q}_{i}\,d_{i}}{2\,R}\,,\\
\tilde{m}\,&=\,-\,\frac{Q_{6}}{2\,R}\,\left(\,l_{i}\,d_{i}\,+\,\frac{h}{R^{2}}\,\frac{c_{ijk}\,d_{i}\,d_{j}\,d_{k}}{6}\,\right)\,-\,\frac{h\,\tilde{Q}_{i}\,d_{i}}{2\,R}\,.\label{NUTcond32}
\end{split}
\end{equation}
Note that only the difference between the last two equations depends on $R$ and provides the bubble equation for the solution.

\paragraph{Horizon regularity condition and asymptotic flatness}\par
Just as we did for the other orbits, we shall now impose regularity constraints on the behavior of the warp factor near the centers.
Near the $A$-center we have:
\begin{equation}
\begin{split}
e^{-4\,U}\stackrel{\stackrel{r\rightarrow0}{\theta=\pi}}{\cong}&\frac{1}{r^4}\,\left(\,Q_{6}\,\tilde{Q}_{1}\,\tilde{Q}_{2}\,\tilde{Q}_{3}\,-\,\beta^{2}\,\right)+\mathcal{O}(r^{-3})\cong\\
&\cong\frac{1}{r^4}\,\left(\,-I_{4}(p^{\Lambda}_{A},q_{A,\Lambda})\,-\,\beta^{2}\,\right)+\mathcal{O}(r^{-3})\quad;
\end{split}
\end{equation}
and near the second one:
\begin{equation}
\begin{split}
e^{-4\,U}\stackrel{\stackrel{r\rightarrow{R}}{\theta=0}}{\cong}&\frac{1}{(r-R)^5}\,\left[\,-\,\frac{2\,(\,Q_{6}\,+\,h\,R\,)^{2}\,(\,\delta\,Q_{6}^{2}\,+
\,h^{2}\,R^{2}\,(\,1\,+\,\delta\,)\,)\,d_{1}^{2}\,d_{2}^{2}\,d_{3}^{2}}{R^{5}}\right]+\mathcal{O}((\,r\,-\,R\,)^{-4})\,.
\end{split}
\end{equation}
In order to set the above leading term to zero we can choose \cite{Bena:2009ev}
\begin{equation}
\delta\,=\,-\,\frac{h^{2}\,R^{2}}{Q_{6}^{2}\,+\,h^{2}\,R^{2}}\,\quad.\label{conda111}
\end{equation}
On the warp factor we ask, also, the asymptotic flatness, requiring $\lim_{r \rightarrow +\infty}e^{-4\,U}=1$ and solving for $h$, which is
\begin{equation}
h=\frac{1\,+\,m_{0}^{2}}{l_{1}\,l_{2}\,l_{3}}\,.
\end{equation}
To work out an explicit solution with $Q_O$ generic in the chosen $\alpha^{(11)}$-orbit, we set the values of the scalars at the origin,\,\textit{i.e.} at the infinity, requiring that $\lim_{r \rightarrow +\infty}\mathcal{M}^{(3)}_{scal}=\eta$\, and obtaining
\begin{equation}
m_0\,=\,0\quad\,,\,\quad\,l_{1}\,=\,l_{2}\,=\,\sqrt{2}\quad\,,\,\quad\,l_{3}\,=\,-\,\sqrt{2}\,.
\end{equation}
With these conditions, the expression for the quartic invariant near the $B$ center is
\begin{equation}
\begin{split}
I_{4}(p^{\Lambda}_{B},q_{B\,\Lambda})&=\left(\frac{Q_{6}}{R}-\frac{1}{2\sqrt{2}}\right)^{3}\bigg[\left(\frac{Q_{6}}{R}-\frac{1}{2\sqrt{2}}\right)^{-1}\frac{\sum_{k}(c_{ijk}d_{i}Q_{i}d_{j}Q_{j}-d_{k}^{2}Q_{k}^{2})}{4}+\\
&-\frac{\prod_{i}d_{i}\sum_{j}d_{j}\tilde{Q}_{j}}{R}-\frac{Q_{6}\prod_{i}d_{i}^{2}}{R^{3}}\bigg]\quad.
\end{split}
\end{equation}
Comparing the above quartic invariant with the expression which describes the geometry near the $B$ center, one note that
\begin{equation}
\begin{split}
e^{-4\,U}\stackrel{\stackrel{r\rightarrow{R}}{\theta=0}}{\cong}&\frac{1}{(r-R)^4}\left[I_{4}(p^{\Lambda}_{B},q_{B,\Lambda})+\left[\left(\frac{Q_{6}}{R}-\frac{1}{2\sqrt{2}}\right)\frac{Q_{6}d_{1}d_{2}d_{3}}{R^{2}}\right]^{2}\right]+\mathcal{O}((r-R)^{-3})\quad,\label{conda11g1sc}
\end{split}
\end{equation}
and it is tempting to interpret the squared term (despite the ``+'' sign in front of the brackets) as an \emph{angular momentum induced} by the interaction, vanishing in the non-interacting limit $R\rightarrow 0$. This conjecture is supported, see discussion below, by the appearance of the same term in the general formula of the total angular momentum as a contribution from the $B$-center.
\paragraph{ADM-Mass, charges and angular momentum}\par
Let us give by using the Eq. in (\ref{allthat}), the $M_{ADM}$ of the system
\begin{equation}
M_{ADM}=\frac{-4Q_{6}+Q_{1}+Q_{2}-Q_{3}+\tilde{Q}_{1}+\tilde{Q}_{2}-\tilde{Q}_{3}}{4\sqrt{2}}+\frac{Q_{6}(-d_{1}d_{2}+d_{1}d_{3}+d_{2}d_{3})}{4\sqrt{2}R^{2}}=M^A+M^B\quad,
\end{equation}
which is the sum of contributions from the $A$ and $B$ centers, computed through the corresponding Noether charge matrices
\begin{equation}
M^{(A)}=\frac{-4Q_{6}+\tilde{Q}_{1}+\tilde{Q}_{2}-\tilde{Q}_{3}}{4\sqrt{2}}+\frac{Q_{6}(-d_{1}d_{2}+d_{1}d_{3}+d_{2}d_{3})}{4\sqrt{2}R^{2}}\quad,
\end{equation}
and
\begin{equation}
M^{(B)}=\frac{Q_{1}+Q_{2}-Q_{3}}{4\sqrt{2}}\quad.
\end{equation}
Let us now compute the electromagnetic charges associated with the solution, making use of (\ref{elmagch}):
\begin{equation}
\begin{split}
\Gamma^M=(p^\Lambda,q_\Lambda)=\bigg(&\frac{d_{1}}{2\sqrt{2}},2Q_{6},-\frac{Q_{6}d_{1}d_{2}+R^{2}(Q_{3}+\tilde{Q}_{3})}{2R^2},\frac{Q_{6}d_{1}d_{3}+R^{2}(Q_{2}+\tilde{Q}_{2})}{2R^2},\\
&\frac{Q_{6}d_{2}d_{3}+R^{2}(Q_{1}+\tilde{Q}_{1})}{2R^2},\frac{d_{1}+d_{2}-d_{3}}{2\sqrt{2}},\frac{d_{3}}{2\sqrt{2}},-\frac{d_{2}}{2\sqrt{2}}\bigg)\,,
\end{split}
\end{equation}
\begin{equation}
\begin{split}
{\Gamma^{M}_{A}}=({p^{\Lambda}_{A}},{q_{A\,\Lambda}})=\bigg(&\frac{Q_{6}d_{1}}{R},2Q_{6},-\frac{Q_{6}d_{1}d_{2}+R^{2}\tilde{Q}_{3}}{2R^2},\frac{Q_{6}d_{1}d_{3}+R^{2}\tilde{Q}_{2}}{2R^2},\frac{Q_{6}d_{2}d_{3}+R^{2}\tilde{Q}_{1}}{2R^2},\\
&-\frac{Q_{6}d_{1}d_{2}d_{3}+R^{2}\sum_{i}d_{i}\tilde{Q}_{i}}{4R^{3}},\frac{Q_{6}d_{3}}{R},-\frac{Q_{6}d_{2}}{R}\bigg)\quad,
\end{split}
\end{equation}
\begin{equation}
\begin{split}
{\Gamma^{M}_{B}}=({p^{\Lambda}_{B}},{q_{B\,\Lambda}})=\bigg(&-\left(\frac{Q_{6}}{R}-\frac{1}{2\sqrt{2}}\right)d_{1},0,-\frac{Q_{3}}{2},\frac{Q_{2}}{2},\frac{Q_{1}}{2},\frac{d_{1}+d_{2}-d_{3}}{2\sqrt{2}}+\frac{\sum_{i}d_{i}\tilde{Q}_{i}}{4R}+\frac{Q_{6}d_{1}d_{2}d_{3}}{4R^{3}},\\
&-\left(\frac{Q_{6}}{R}-\frac{1}{2\sqrt{2}}\right)d_{3},\left(\frac{Q_{6}}{R}-\frac{1}{2\sqrt{2}}\right)d_{2}\bigg)\quad.
\end{split}
\end{equation}
The  angular momentum is easily computed through the use of $Q_\psi$ defined in (\ref{qupsi}) and the last equation of (\ref{allthat}):
\begin{equation}
\begin{split}
M_\varphi&=-\frac{\beta}{2}-\frac{-4Q_{6}(d_{1}+d_{2}-d_{3})+\sum_{i}d_{i}\tilde{Q}_{i}}{8\sqrt{2}}-\frac{Q_{6}+\sum_{i}d_{i}(Q_{i}+2\tilde{Q}_{i})}{4R}+\\
&-\left(\frac{Q_{6}}{R}-\frac{1}{4\sqrt{2}}\right)\frac{Q_{6}d_{1}d_{2}d_{3}}{2R^{2}}=M_{\varphi_A}+M_{\varphi_B}-\frac{1}{2}\,
({\Gamma_{A}})^{T}\,\mathbb{C}\,{\Gamma_{B}}\label{momanga11g1}\quad,
\end{split}
\end{equation}
where last expression holds provided we identify $M_{\varphi_A}$ with $-\beta/2$ and define
\begin{equation}
M_{\varphi_B}\equiv-\left(\frac{Q_{6}}{R}-\frac{1}{2\sqrt{2}}\right)\frac{Q_{6}d_{1}d_{2}d_{3}}{R^{2}}\,,\label{betatilde}
\end{equation}
which motivates the interpretation given earlier of last term in (\ref{conda11g1sc}) as originating from an induced angular momentum. One can verify that $M_\varphi$ is precisely the expression given by the asymptotic expansion (\ref{omegaMphi}) of $\omega_\varphi$.
\subsubsection{The orbits $\mathcal{O}^{11;1,1}$}
In this section we discuss representative double-center solutions for the $\alpha^{(11)}$-orbits. We need to specialize $Q_O$ to the various $\gamma$ and $\beta$ labels. The field equations can always be brought to the form (\ref{eqmot3}). We find that, as opposed to the $\alpha^{(7)},\dots, \alpha^{(10)}$-cases, the intrinsically singular orbits are the off-diagonal ones, characterized by non-coinciding $\gamma$ and $\beta$ labels. As for the diagonal ones, we can find representatives consisting of a large non-BPS center ($\alpha^{(6)}\gamma^{(6;5)}\beta^{(6;5)}\delta^{(1)}$-orbit) and a large BPS or non-BPS black hole with $I_4>0$ ($\alpha^{(6)}\gamma^{(6;k)}\beta^{(6;k)}$-orbit, $k=1,2,3,4$), the $\gamma$ label of the $\alpha^{(11)}$-orbit being in one-to-one correspondence with that of the second center. We shall discuss in detail only a solution with $Q_O$ in the first orbit, being the representatives in the other orbits analogous. Representatives of each $\alpha^{(11)}$-orbit as sum of two single-center ones is given in Table \ref{tablea11}.

As an example of double-center solution with $Q_O$ generic in the $\alpha^{(11)}\gamma^{(11;1)}\beta^{(11;1)}$-orbit, we fix the charge parameters as follows
\begin{align}
\tilde{Q}_6\,&=\,-\sqrt{2}\quad\,,\,\quad\,\tilde{Q}_{1}\,=\,\tilde{Q}_{2}\,=\,\sqrt{2}\quad\,,\,\quad\,\tilde{Q}_{3}=\,-\,\sqrt{2}\,,\label{par9}\\
d_{1}\,=\,d_{2}\,&=\,2\quad\,,\,\quad\,d_{3}\,=\,-2\quad\,,\,\quad\,Q_{1}\,=\,Q_{2}\,=\,4\sqrt{2}\quad\,,\,\quad\,Q_{3}\,=\,-4\sqrt{2}\,.\label{par111}
\end{align}
This corresponds to choosing the $A$-center to be a  regular, large non-BPS single-center solution with $I_4<0$ ($\alpha^{(6)}\gamma^{(6;5)}\beta^{(6;5)}\delta^{(1)}$-orbit) and  an intrinsic charge matrix of the form
\begin{equation}
Q_A^{(0)}=N_{1}^{-}+4N_{2}^{+}+N_{3}^{-}+N_{4}^{-}\,,
\end{equation}
and the $(p,\,q)$ charges read
\begin{equation}
{\Gamma^{M}_{A}}\,=\,(\,{p^{\Lambda}_{A}}\,,\,{q_{A\,\Lambda}}\,)\,=\,\left(\,0\,,\,-2\sqrt{2}\,,\,\frac{1}{\sqrt{2}}\,,\,\frac{1}{\sqrt{2}}\,,\,\frac{1}{\sqrt{2}}\,,\,0\,,\,0\,,\,0\,\right)\quad.
\end{equation}
If isolated from the other center and warp function would be $e^{-4\,U}=(r+4)(r+1)^{3}/r^{4}$ and have no zeros. \par
As for the $B$-center, the chosen charges
  define a regular BPS solution ($\alpha^{(6)}\gamma^{(6;1)}\beta^{(1)}$-orbit) with warp function $e^{-4\,U}=(r^{4}+12r^{3}+42r^{2}+52r+9)/r^{4})$ if isolated from the $A$-center\, and $(p,\,q)$ charges
\begin{equation}
{\Gamma^{M}_{B}}\,=\,(\,{p^{\Lambda}_{B}}\,,\,{q_{B\,\Lambda}}\,)\,=\,\left(\,\frac{1}{\sqrt{2}}\,,\,0\,,\,2\sqrt{2}\,,\,2\sqrt{2}\,,\,2\sqrt{2}\,,\,\frac{3}{\sqrt{2}}\,,\,-\frac{1}{\sqrt{2}}\,,\,-\frac{1}{\sqrt{2}}\,\right)\quad.
\end{equation}
The total electric-magnetic charge vector is
\begin{equation}
\Gamma^M\,=\,(\,p^\Lambda\,,\,q_\Lambda\,)\,=\,\Gamma^M_A+\Gamma^M_B\,=\,\left(\,\frac{1}{\sqrt{2}}\,,\,-2\sqrt{2}\,,\,\frac{4+5R^{2}}{\sqrt{2}R^{2}}\,,\,\frac{4+5R^{2}}{\sqrt{2}R^{2}}\,,\,\frac{4+5R^{2}}{\sqrt{2}R^{2}}\,,\,\frac{3}{\sqrt{2}}\,,\,-\frac{1}{\sqrt{2}}\,,\,-\frac{1}{\sqrt{2}}\,\right)\quad.
\end{equation}
Let us now study the behavior of warp function near each center and impose the positivity of its leading terms:
We have
\begin{equation}
\begin{split}
e^{-4\,U}\stackrel{\stackrel{r\rightarrow0}{\theta=\pi}}{\cong}&\frac{1}{r^4}\left(4-\beta^{2}\right)+\mathcal{O}(r^{-3})\quad,
\end{split}
\end{equation}
for the $A$-center and
\begin{equation}
\begin{split}
e^{-4\,U}\stackrel{\stackrel{r\rightarrow{R}}{\theta=0}}{\cong}&\frac{1}{(r-R)^4}\left[\frac{(R+4)^{2}(9R^{3}-15R^{2}-12R-4)}{R^5}\right]+
\mathcal{O}((r-R)^{-3})\quad\label{ex4Ua11g1b1},
\end{split}
\end{equation}
for the $B$-one. Requiring positivity of the leading term in (\ref{ex4Ua11g1b1})  immediately implies the following lower bound for the distance $R$:
\begin{equation}
\begin{split}
R>2,323\quad\label{lba11g1b1}.
\end{split}
\end{equation}
\\
Below this value of the distance $R$, the global warp function $e^{-4\,U}$ shows unphysical singularities and the system, although composed from two regular black holes, ceases to be globally regular, as showed in the following figure, where the behavior of the warp function when the distance between the two centers is  lower than this bound, is illustrated.
\begin{figure}[H]
\centering
\includegraphics[width=7cm]{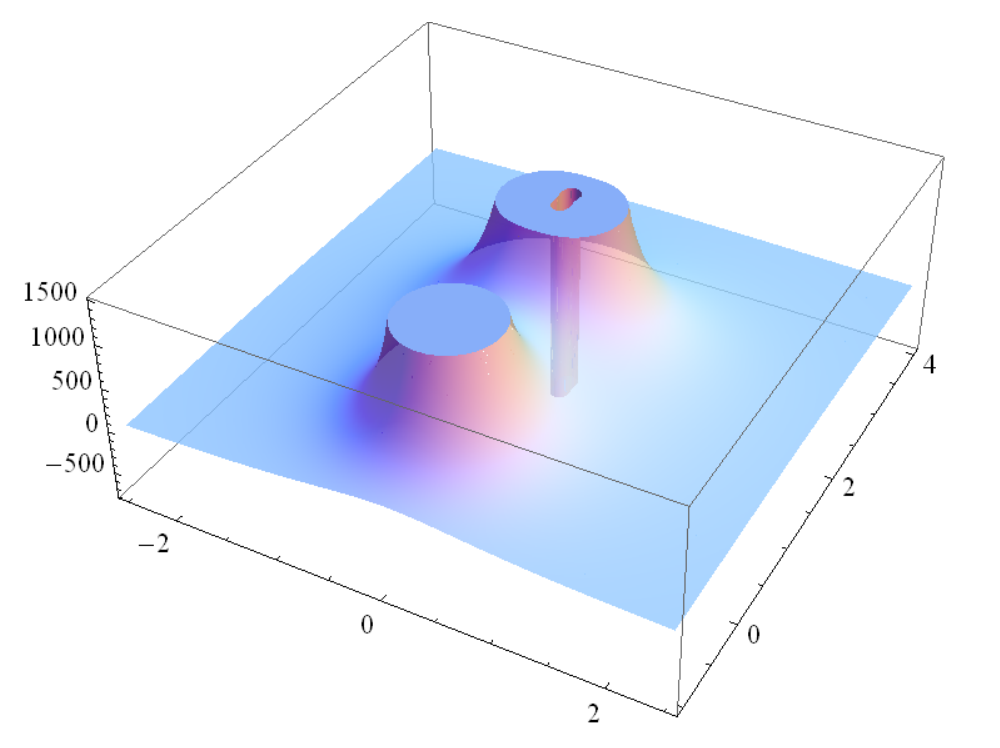}
\caption{The warp function when the mutual distance between the two centers is less than the lower bound in (\ref{lba11g1b1}). }
\end{figure}
One can interpret this result thinking that when the two centers approach one another the interaction forces are too strong and destabilize the system.\\
By setting the value of $R$ to be larger than the lower bound, say $12$, we exclude the presence of  CTC in the solution, as the behavior of  the function $e^{-4\,U}\,r^{2}\sin^{2}\theta-\omega^2$ shown in the figure below. \\
\begin{figure}[H]
\centering
\subfloat[][\small\,The behavior of $e^{-4\,U}\,r^{2}\sin^{2}\theta-\omega^2$.]{\includegraphics[width=7cm]{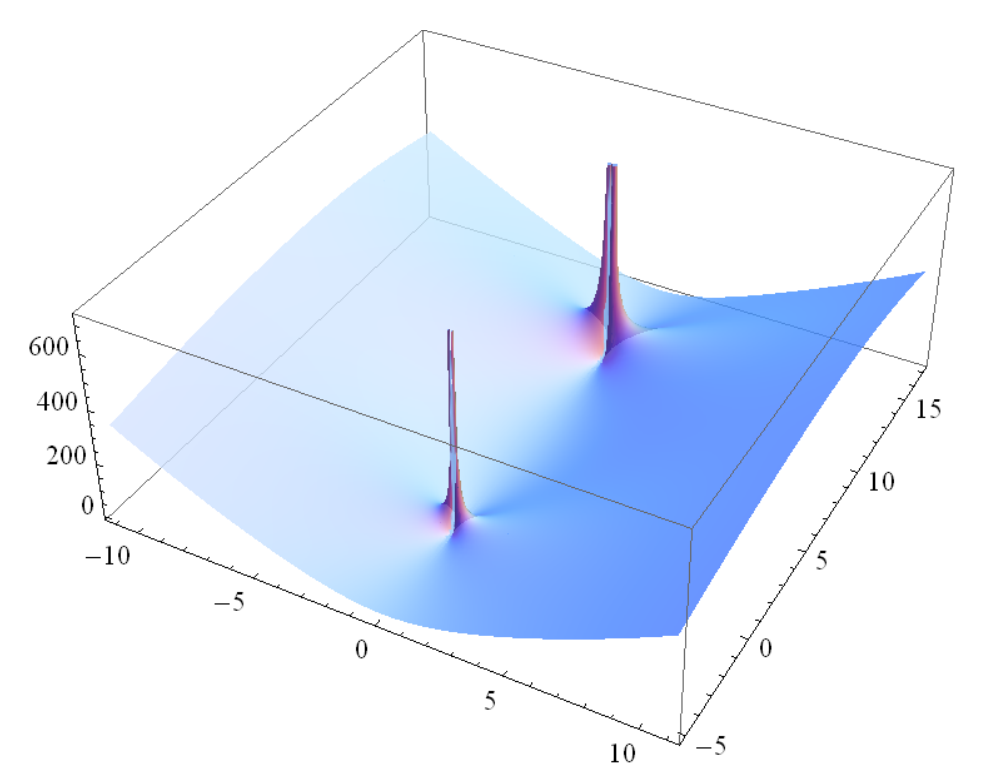}} \hfill
\subfloat[][\small\,The behavior of $e^{-4\,U}\,r^{2}\sin^{2}\theta-\omega^2$ near the $A$ center, corresponding to a regular non-BPS $\alpha^{(6)}\gamma^{(6;5)}\beta^{(5)}\delta^{(1)}$.]{\includegraphics[width=7cm]{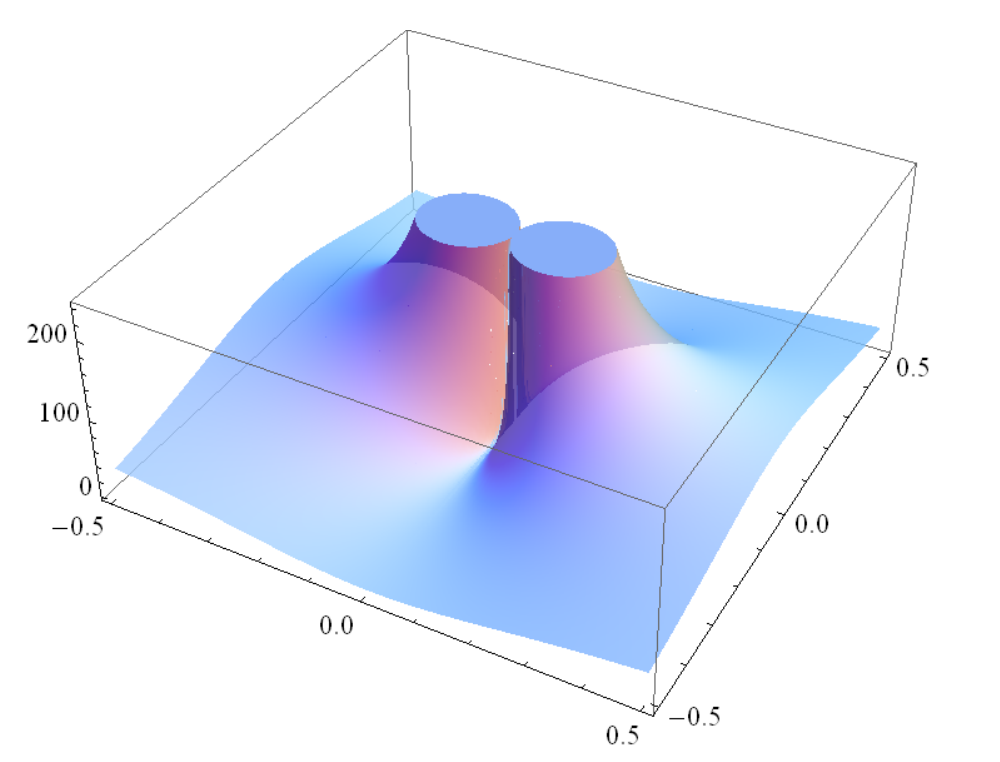}}\\
\subfloat[][\small\,The behavior of $e^{-4\,U}\,r^{2}\sin^{2}\theta-\omega^2$ near the $B$ center, corresponding to a regular BPS $\alpha^{(6)}\gamma^{(6;1)}\beta^{(1)}$.]{\includegraphics[width=7cm]{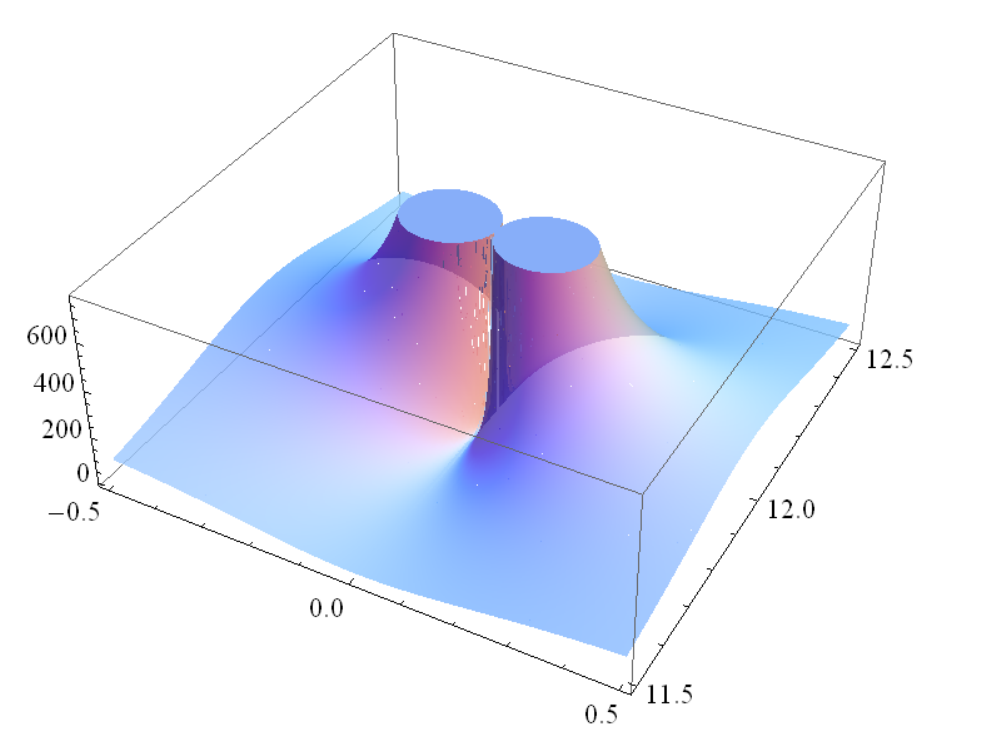}} \hfill
\end{figure}
In the following plots, the function $e^{-4U}$ and the curvature scalar $\mathcal{R}$ are plotted at the same value of the distance $R$. The curvature scalar shows for both the centers the volcano shape associated with the \emph{attractor mechanism}.
\begin{figure}[H]
\centering
\subfloat[][The behavior of the warp function $e^{-4\,U}$.]{\includegraphics[width=7cm]{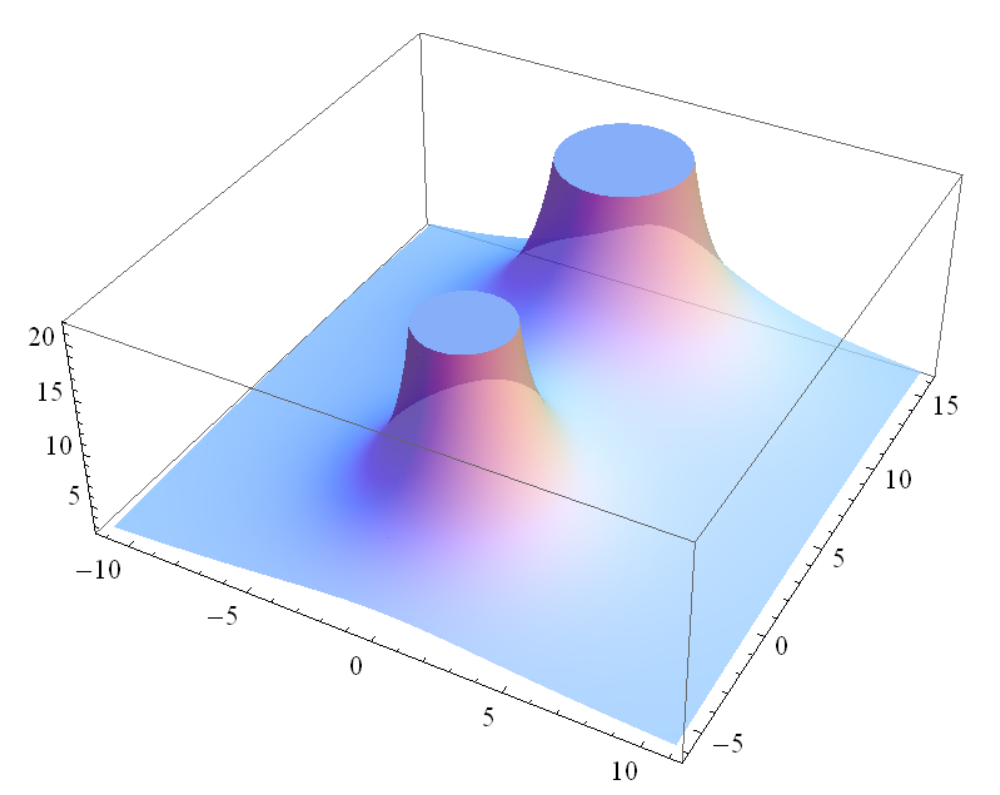}} \hfill
\subfloat[][The behavior of the curvature scalar $\mathcal{R}$.]{\includegraphics[width=7cm]{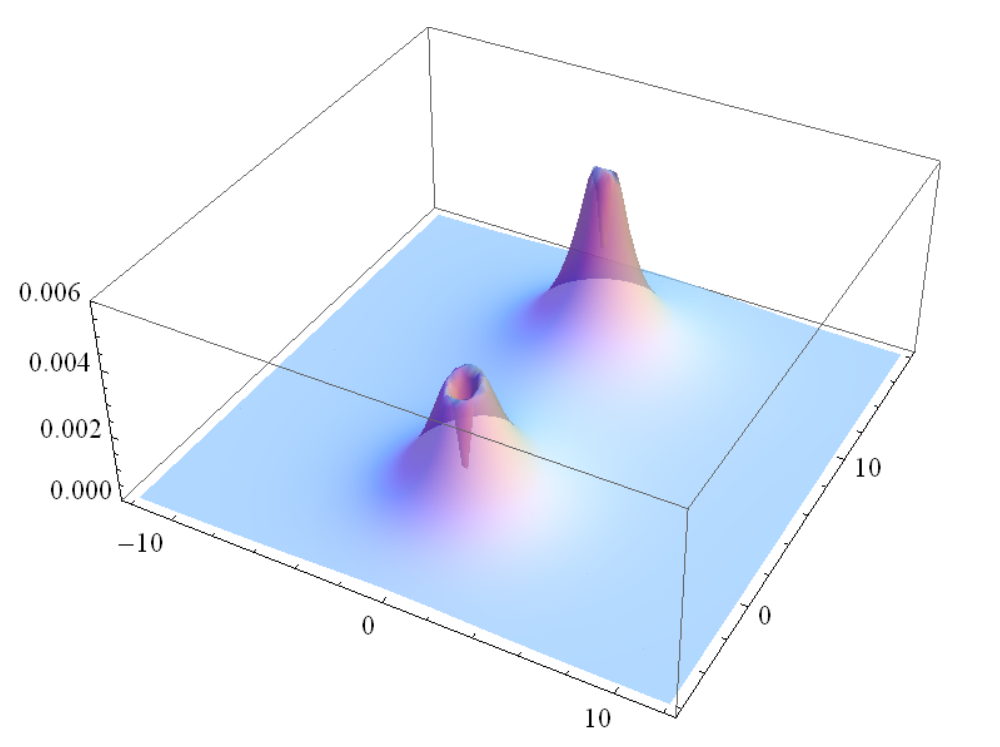}}\\
\subfloat[][Detail of the curvature scalar $\mathcal{R}$ for the regular non-BPS center $\alpha^{(6)}\gamma^{(6;5)}\beta^{(5)}\delta^{(1)}$ at the origin.]{\includegraphics[width=7cm]{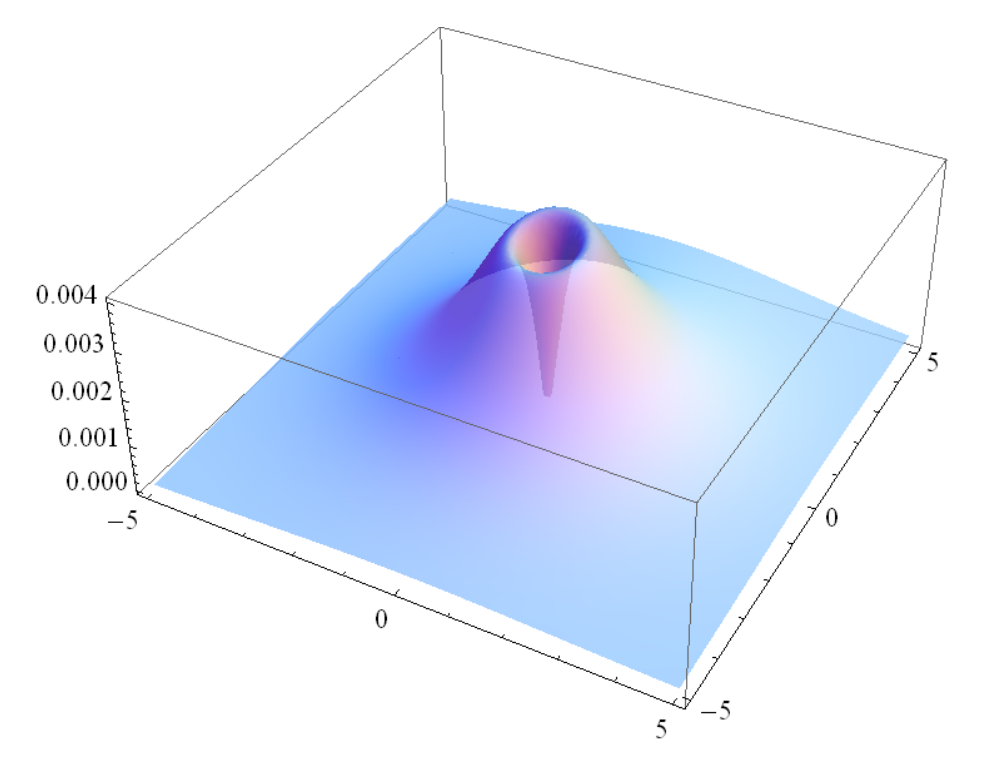}} \hfill
\subfloat[][Detail of the curvature scalar $\mathcal{R}$ for the regular BPS center $\alpha^{(6)}\gamma^{(6;1)}\beta^{(1)}$ at the distance $R=12$ from the origin.]{\includegraphics[width=7cm]{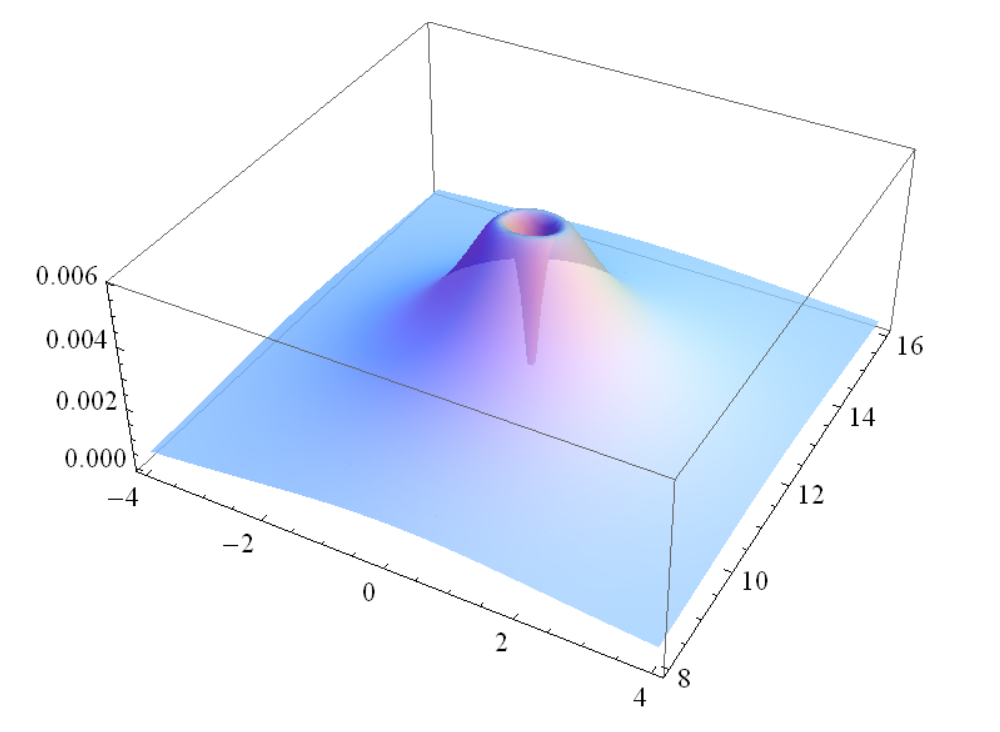}}\\
\end{figure}
\subsection{Summary of representatives for\, $\alpha^{(11)}$ }
In the following table are reported all orbits of $\alpha^{(11)}$ in terms of single-center constituent.
\begin{table}[H]
\begin{center}
\resizebox{\columnwidth}{!}{%
\begin{tabular}{|c|c|c|c|c|}\hline
\backslashbox{$\gamma$}{$\beta$}
&$\beta^{(1)}$&$\beta^{(2)}$&$\beta^{(3)}$&$\beta^{(4)}$ \\   \hline
$\gamma^{(1)}$&\multicolumn{1}{c|}{{\begin{tabular}{c}
\\ $\alpha^{(6)}\gamma^{(6;5)}\beta^{(5)}\delta^{(1)}\,+$\\
$\alpha^{(6)}\gamma^{(6;1)}\beta^{(1)}$\\
\\
\end{tabular}}} & \multicolumn{1}{c|}{{\begin{tabular}{c}
\\ $\alpha^{(6)}\gamma^{(6;5)}\beta^{(5)}\delta^{(1)}\,+$\\
$\alpha^{(6)}\gamma^{(6;1)}\beta^{(2)}$\\
\\
\end{tabular}}} & \multicolumn{1}{c|}{{\begin{tabular}{c}
\\ $\alpha^{(6)}\gamma^{(6;5)}\beta^{(5)}\delta^{(1)}\,+$\\
$\alpha^{(6)}\gamma^{(6;1)}\beta^{(3)}$\\
\\
\end{tabular}}} & \multicolumn{1}{c|}{{\begin{tabular}{c}
\\ $\alpha^{(6)}\gamma^{(6;5)}\beta^{(5)}\delta^{(1)}\,+$\\
$\alpha^{(6)}\gamma^{(6;1)}\beta^{(4)}$\\
\\
\end{tabular}}}\\ \hline
$\gamma^{(2)}$&\multicolumn{1}{c|}{{\begin{tabular}{c}
\\ $\alpha^{(6)}\gamma^{(6;5)}\beta^{(5)}\delta^{(1)}\,+$\\
$\alpha^{(6)}\gamma^{(6;2)}\beta^{(1)}$\\
\\
\end{tabular}}}&\multicolumn{1}{c|}{{\begin{tabular}{c}
\\ $\alpha^{(6)}\gamma^{(6;5)}\beta^{(5)}\delta^{(1)}\,+$\\
$\alpha^{(6)}\gamma^{(6;2)}\beta^{(2)}$\\
\\
\end{tabular}}}&\multicolumn{1}{c|}{{\begin{tabular}{c}
\\ $\alpha^{(6)}\gamma^{(6;5)}\beta^{(5)}\delta^{(1)}\,+$\\
$\alpha^{(6)}\gamma^{(6;2)}\beta^{(3)}$\\
\\
\end{tabular}}}&\multicolumn{1}{c|}{{\begin{tabular}{c}
\\ $\alpha^{(6)}\gamma^{(6;5)}\beta^{(5)}\delta^{(1)}\,+$\\
$\alpha^{(6)}\gamma^{(6;2)}\beta^{(4)}$\\
\\
\end{tabular}}}\\ \hline
$\gamma^{(3)}$&\multicolumn{1}{c|}{{\begin{tabular}{c}
\\ $\alpha^{(6)}\gamma^{(6;5)}\beta^{(5)}\delta^{(1)}\,+$\\
$\alpha^{(6)}\gamma^{(6;3)}\beta^{(1)}$\\
\\
\end{tabular}}}&\multicolumn{1}{c|}{{\begin{tabular}{c}
\\ $\alpha^{(6)}\gamma^{(6;5)}\beta^{(5)}\delta^{(1)}\,+$\\
$\alpha^{(6)}\gamma^{(6;3)}\beta^{(2)}$\\
\\
\end{tabular}}}&\multicolumn{1}{c|}{{\begin{tabular}{c}
\\ $\alpha^{(6)}\gamma^{(6;5)}\beta^{(5)}\delta^{(1)}\,+$\\
$\alpha^{(6)}\gamma^{(6;3)}\beta^{(3)}$\\
\\
\end{tabular}}}&\multicolumn{1}{c|}{{\begin{tabular}{c}
\\ $\alpha^{(6)}\gamma^{(6;5)}\beta^{(5)}\delta^{(1)}\,+$\\
$\alpha^{(6)}\gamma^{(6;3)}\beta^{(4)}$\\
\\
\end{tabular}}}\\ \hline
$\gamma^{(4)}$&\multicolumn{1}{c|}{{\begin{tabular}{c}
\\ $\alpha^{(6)}\gamma^{(6;5)}\beta^{(5)}\delta^{(1)}\,+$\\
$\alpha^{(6)}\gamma^{(6;4)}\beta^{(1)}$\\
\\
\end{tabular}}}&\multicolumn{1}{c|}{{\begin{tabular}{c}
\\ $\alpha^{(6)}\gamma^{(6;5)}\beta^{(5)}\delta^{(1)}\,+$\\
$\alpha^{(6)}\gamma^{(6;4)}\beta^{(2)}$\\
\\
\end{tabular}}}&\multicolumn{1}{c|}{{\begin{tabular}{c}
\\ $\alpha^{(6)}\gamma^{(6;5)}\beta^{(5)}\delta^{(1)}\,+$\\
$\alpha^{(6)}\gamma^{(6;4)}\beta^{(3)}$\\
\\
\end{tabular}}}&\multicolumn{1}{c|}{{\begin{tabular}{c}
\\ $\alpha^{(6)}\gamma^{(6;5)}\beta^{(5)}\delta^{(1)}\,+$\\
$\alpha^{(6)}\gamma^{(6;4)}\beta^{(4)}$\\
\\
\end{tabular}}}\\ \hline
\end{tabular}}
\caption{\small The single-center representatives of $\alpha^{(11)}$-orbit. }\label{tablea11}
\end{center}
\end{table}

\section{Conclusions and outlook}
In the present work we applied the mathematical tool of the real nilpotent orbits to a systematic study of single and multi-center solutions in the STU model.
 The real orbits allow to completely characterize in a $G$-invariant way, through the \emph{regular-single-center orbits}, the regular (large or small) black holes and formulate stringent regularity conditions for multi-center solutions. The $H^*$-orbits associated with single-center solutions may also describe multi-center ones. We restricted ourselves to axisymmetric two-black hole systems although this approach applies to the most general multi-center in the model. As an outcome of this analysis we defined composition rules, which allow to express representatives of the non-BPS multi-center orbits $\alpha^{(7)},\dots,\alpha^{(11)}$ as sum of two nilpotent elements in regular-single-center ones. There are orbits which are never obtained in this way (the empty boxes in Fig. \ref{Figfigs}) and we interpreted them as being \emph{intrinsically singular} in that, we claim, they do not contain regular multi-center representatives. As for the remaining orbits with $\alpha$-label from $\alpha^{(7)}$ to $\alpha^{(11)}$, they may contain both singular and regular solutions, depending on their inner composition. This is not the case for the single-center orbits from $\alpha^{(1)}$ to $\alpha^{(6)}$, each of which either contain only singular solutions or only regular ones.\par
 As far as the $\alpha^{(10)}$ and $\alpha^{(11)}$-orbits are concerned, we recover the general results of \cite{bossard2} and \cite{Bena:2009ev}, related to the composite non-BPS and the almost-BPS solutions, and show through worked out examples, that if each constituent in taken in regular-single-center orbits, the total solution can be made regular. Example of regular double-center solutions in the  $\alpha^{(10)}$-orbits consist of two non-BPS large black holes with $I_4<0$, while examples in the   $\alpha^{(11)}$-orbits describe two large black holes, one with $I_4<0$ and the other with $I_4>0$.\par
 We also discuss the  $\alpha^{(7)},\dots ,\alpha^{(9)}$ orbits, which are in the closure of the  $\alpha^{(10)}$ and $\alpha^{(11)}$-ones. They are associated with characteristic graded-field equations which we solved.\par
The composition rules in Appendix \ref{sumrules} reveal a variety of possible internal structures of multi-center solutions in each of the $\alpha^{(10)},\dots, \alpha^{(11)}$-orbits. We expect not all of these combinations to yield a regular solution. It would be interesting to study, in light of these rules, possible new multi-center systems in these orbits and their properties, like the presence of  walls of marginal stability \cite{Bossard:2013nwa,BPSmc}. It would also be interesting to apply the entropy analysis of \cite{Crichigno:2016lac} to the composite non-BPS bound and almost-BPS bound states.
 \section{Acknowledgements}
 We are very grateful to G. Bossard for helpful discussions. We also wish to thank the Galileo Galilei Institute for Theoretical Physics for the hospitality during the last stage of preparation of this work.

\appendix
\section{The Invariant Labels}\label{labels}
\begin{table}[H]
\begin{center}\resizebox{0.70\columnwidth}{!}{%
\begin{tabular}{c|c|c}
{\bf $\pmb{\alpha}$-label} &{\bf$\pmb{\beta}$-$\pmb{\gamma}$-labels} & {\bf$\pmb{\delta}$-label (if present)} \\
\hline
$\alpha^{(1)}=(0,\,1,\,0,\,0)$&$\gamma^{(1;1)}=\beta^{(1;1)}=(1,\,1,\,1,\,1)$&\\
\hline
$\alpha^{(2)}=(2,\,0,\,0,\,0)$&$\gamma^{(2;1)}=\beta^{(2;1)}=(2,\,2,\,0,\,0)$&\\
&$\gamma^{(2;2)}=\beta^{(2;2)}=(0,\,0,\,2,\,2)$&\\
\hline
$\alpha^{(3)}=(0,\,0,\,2,\,0)$&$\gamma^{(3;1)}=\beta^{(3;1)}=(2,\,0,\,0,\,2)$&\\
&$\gamma^{(3;2)}=\beta^{(3;2)}=(0,\,2,\,2,\,0)$&\\
\hline
$\alpha^{(4)}=(0,\,0,\,0,\,2)$&$\gamma^{(4;1)}=\beta^{(4;1)}=(2,\,0,\,2,\,0)$&\\
&$\gamma^{(4;2)}=\beta^{(4;2)}=(0,\,2,\,0,\,2)$&\\
\hline
$\alpha^{(5)}=(1,\,0,\,1,\,1)$&$\gamma^{(5;1)}=\beta^{(5;1)}=(3,\,1,\,1,\,1)$&\\
&$\gamma^{(5;2)}=\beta^{(5;2)}=(1,\,3,\,1,\,1)$&\\
&$\gamma^{(5;3)}=\beta^{(5;3)}=(1,\,1,\,3,\,1)$&\\
&$\gamma^{(5;4)}=\beta^{(5;4)}=(1,\,1,\,1,\,3)$&\\
\hline
$\alpha^{(6)}=(0,\,2,\,0,\,0)$&$\gamma^{(6;1)}=\beta^{(6;1)}=(4,\,0,\,0,\,0)$&\\
&$\gamma^{(6;2)}=\beta^{(6;2)}=(0,\,4,\,0,\,0)$&\\
&$\gamma^{(6;3)}=\beta^{(6;3)}=(0,\,0,\,4,\,0)$&\\
&$\gamma^{(6;4)}=\beta^{(6;4)}=(0,\,0,\,0,\,4)$&\\
&$\gamma^{(6;5)}=\beta^{(6;5)}=(2,\,2,\,2,\,2)$&$\delta^{(1,\dots,4)}$\\
\hline
$\alpha^{(7)}=(2,\,2,\,0,\,0)$&$\gamma^{(7;1)}=\beta^{(7;1)}=(2,\,2,\,4,\,4)$&\\
&$\gamma^{(7;2)}=\beta^{(7;2)}=(4,\,4,\,2,\,2)$&\\
\hline
$\alpha^{(8)}=(0,\,2,\,2,\,0)$&$\gamma^{(8;1)}=\beta^{(8;1)}=(2,\,4,\,4,\,2)$&\\
&$\gamma^{(8;2)}=\beta^{(8;2)}=(4,\,2,\,2,\,3)$&\\
\hline
$\alpha^{(9)}=(0,\,2,\,0,\,2)$&$\gamma^{(9;1)}=\beta^{(9;1)}=(2,\,4,\,2,\,4)$&\\
&$\gamma^{(9;2)}=\beta^{(9;2)}=(4,\,2,\,4,\,2)$&\\
\hline
$\alpha^{(10)}=(2,\,0,\,2,\,2)$&$\gamma^{(10;1)}=\beta^{(10;1)}=(0,\,4,\,4,\,4)$&\\
&$\gamma^{(10;2)}=\beta^{(10;2)}=(4,\,0,\,4,\,4)$&\\
&$\gamma^{(10;3)}=\beta^{(10;3)}=(4,\,4,\,0,\,4)$&\\
&$\gamma^{(10;4)}=\beta^{(10;4)}=(4,\,4,\,4,\,0)$&\\
\hline
$\alpha^{(11)}=(2,\,2,\,2,\,2)$&$\gamma^{(11;1)}=\beta^{(11;1)}=(8,\,4,\,4,\,4)$&\\
&$\gamma^{(11;2)}=\beta^{(11;2)}=(4,\,8,\,4,\,4)$&\\
&$\gamma^{(11;3)}=\beta^{(11;3)}=(4,\,4,\,8,\,4)$&\\
&$\gamma^{(11;4)}=\beta^{(11;4)}=(4,\,4,\,4,\,8)$&
\end{tabular}}
\end{center}
\end{table}

\section{Useful formulae}\label{usef}
These properties are used in the calculation of equations of motion:
\begin{align}
&{d}\left(\frac{1}{r}\right)\ast{d}\left(\frac{1}{r}\right)={d}\ast{d}\left(\frac{1}{2r^{2}}\right)\quad,\\
&{d}\left(\frac{1}{\Sigma}\right)\ast{d}\left(\frac{1}{\Sigma}\right)={d}\ast{d}\left(\frac{1}{2\Sigma^{2}}\right)\quad,\\
&{d}\left(\frac{1}{r}\right)\ast{d}\left(\frac{1}{\Sigma}\right)={d}\ast{d}\left(\frac{1}{2r\Sigma}\right)\quad,\\
&\frac{1}{r}{d}\left(\frac{1}{r}\right)\ast{d}\left(\frac{1}{\Sigma}\right)={d}\ast{d}\left(\frac{\cos\theta}{2rR\Sigma}\right)\quad,\\
&\frac{1}{\Sigma}{d}\left(\frac{1}{r}\right)\ast{d}\left(\frac{1}{\Sigma}\right)=-{d}\ast{d}\left(\frac{\cos\theta_{\Sigma}}{2rR\Sigma}\right)\quad,\\
&\frac{1}{r}{d}\left(\frac{1}{\Sigma}\right)\ast{d}\left(\frac{1}{\Sigma}\right)={d}\ast{d}\left(\frac{r}{2R^{2}\Sigma^{2}}\right)\quad,\\
&\frac{1}{r}{d}\left(\frac{1}{r}\right)\ast{d}\left(\frac{1}{r}\right)=_{1}{d}\ast{d}\left(\frac{1}{6r^{3}}\right)\quad,\\
&\frac{1}{r}{d}\left(\frac{1}{r}\right)\ast{d}\left(\frac{1}{r}\right)=_{2}{d}\ast{d}\left(-\frac{\Sigma\cos\theta_{\Sigma}}{6Rr^{3}}\right)\quad,\\
&\frac{1}{\Sigma}{d}\left(\frac{1}{\Sigma}\right)\ast{d}\left(\frac{1}{\Sigma}\right)=_{1}{d}\ast{d}\left(\frac{1}{6\Sigma^{3}}\right)\quad,\\
&\frac{1}{\Sigma}{d}\left(\frac{1}{\Sigma}\right)\ast{d}\left(\frac{1}{\Sigma}\right)=_{2}{d}\ast{d}\left(\frac{r\cos\theta}{6R\Sigma^{3}}\right)\quad.
\end{align}
The following relations are useful in the computation of \,$\omega$:
\begin{align}
&{d}\left(\frac{1}{r}\right)=\ast{d}\left(\cos\theta{d}\phi\right)\quad,\label{Bom1}\\
&{d}\left(\frac{1}{\Sigma}\right)=\ast{d}\left(\cos\theta_{\Sigma}{d}\phi\right)\quad,\\
&{d}\left(\frac{\cos\theta}{r^{2}}\right)=\ast{d}\left(-\frac{\sin^{2}\theta}{r}{d}\phi\right)\quad,\\
&{d}\left(\frac{\cos\theta_{\Sigma}}{\Sigma^{2}}\right)=\ast{d}\left(-\frac{\sin^{2}\theta_{\Sigma}}{\Sigma}{d}\phi\right)\quad,\\
&{d}\left(\frac{1}{r\Sigma}\right)-\frac{2}{r}{d}\left(\frac{1}{\Sigma}\right)=\ast{d}\left(\frac{R\cos\theta-r}{R\Sigma}{d}\phi\right)\quad,\\
&{d}\left(\frac{1}{r\Sigma}\right)-\frac{2}{\Sigma}{d}\left(\frac{1}{r}\right)=\ast{d}\left(-\frac{R\cos\theta-r}{R\Sigma}{d}\phi\right)\quad,\\
&{d}\left(\frac{\cos\theta}{rR\Sigma}\right)-\frac{1}{r^{2}}{d}\left(\frac{1}{\Sigma}\right)=\ast{d}\left(-\frac{\sin^{2}\theta}{R\Sigma}{d}\phi\right)\quad,\\
&{d}\left(\frac{\cos\theta_{\Sigma}}{rR\Sigma}\right)+\frac{1}{\Sigma^{2}}{d}\left(\frac{1}{r}\right)=\ast{d}\left(-\frac{\sin^{2}\theta_{\Sigma}}{Rr}{d}\phi\right)\quad,\\
&{d}\left(\frac{\cos\theta_{\Sigma}}{Rr^{2}}\right)+\frac{2}{r\Sigma}{d}\left(\frac{1}{r}\right)=\ast{d}\left(-\frac{\sin^{2}\theta}{R\Sigma}{d}\phi\right)\quad,\\
&{d}\left(\frac{\cos\theta}{R\Sigma^{2}}\right)-\frac{2}{r\Sigma}{d}\left(\frac{1}{\Sigma}\right)=\ast{d}\left(-\frac{\sin^{2}\theta_{\Sigma}}{Rr}{d}\phi\right)\quad,\\
&\frac{1}{r^{2}}{d}\left(\frac{1}{\Sigma}\right)+{d}\left(\frac{\Sigma^{2}-R^{2}}{2r^{2}R^{2}\Sigma}\right)=\ast{d}\left(\frac{R+r\cos\theta-2R\cos^{2}\theta}{2R^{2}\Sigma}{d}\phi\right)\quad,\\
&\frac{1}{\Sigma^{2}}{d}\left(\frac{1}{r}\right)+{d}\left(\frac{r^{2}-R^{2}}{2rR^{2}\Sigma^{2}}\right)=\ast{d}\left(\frac{(r^{2}+R^{2})\cos\theta-2rR}{2R^{2}\Sigma^{2}}{d}\phi\right)\quad.\label{Bomfin}
\end{align}
\section{The nilpotent orbits in detail}

\subsubsection{The single-center orbits}\label{secdir1}
\subsubsection{$\alpha^{(1)}=\,(0,\,1,\,0,\,0)$   ($O[{2^2,1^4}]$)}
These orbits correspond to \emph{small doubly-critical black holes}.
By setting three charges of the generating solution to
zero we end up in the orbit $O[{2^2,1^4}]$ .
We can take as representative the matrices $Q_O=N_\ell^\pm\,$\, which have order of nilpotency $2$.
\begin{table}[H]
\begin{center}
\begin{tabular}{|c|c|c|c|}\hline
\backslashbox{$\gamma$}{$\beta$}
&$\beta^{(1)}$=(1,\,1,\,1,\,1)& \\ \hline
$\gamma^{(1)}$=(1,\,1,\,1,\,1)&
\multicolumn{1}{c|}{{\begin{tabular}{c}
\\$
\renewcommand{\arraystretch}{1}
\begin{array}{c}
N_\ell^\pm\end{array}  $ \\
\\
\end{tabular}}} &  BPS \\ \hline
\end{tabular}

\end{center}
\caption{\small The representative of $\alpha^{(1)}$-orbit in terms of the generating solution.}\label{a1:1}
\end{table}
We express the same generators in terms of QUbit representation.
\begin{table}[H]
\begin{center}
\begin{tabular}{|c|c|c|}\hline
\backslashbox{$\gamma$}{$\beta$}
&$\beta^{(1)}$\,=\,(1,\,1,\,1,\,1) \\ \hline
$\gamma^{(1)}$\,=\,(1,\,1,\,1,\,1)&\multicolumn{1}{c|}{{\begin{tabular}{c}
\\
$(+,\,+,\,-,\,-)$ \\
\\
\end{tabular}}}\\ \hline
\end{tabular}
\caption{\small The representative of $\alpha^{(1)}$-orbit in QUbit representation.}\label{a1:2}
\end{center}
\end{table}
\subsubsection{$\alpha^{(2)}=(2,\,0,\,0,\,0)$   ($O[{3,1^5}]$)}
The $\alpha^{(2)}$\,,\,$\alpha^{(3)}$\,,\,$\alpha^{(4)}$-orbits in the following sections are related by triality. This is apparent from the form of the corresponding labels.

They are associated with \emph{small critical black holes}.
In the normal form defined by the generating solution, we can take as nilpositive representatives the matrices \,$Q_O=N_1^\pm \pm N_2^\pm,$\, which have order of nilpotency $3$.
\begin{table}[H]
\begin{center}
\begin{tabular}{|c|c|c|c|}\hline
\backslashbox{$\gamma$}{$\beta$}
&$\beta^{(1)}$=(2,\,2,\,0,\,0)&$\beta^{(2)}$=(0,\,0,\,2,\,2) & \\   \hline
$\gamma^{(1)}$=(2,\,2,\,0,\,0)&\multicolumn{1}{c|}{{\begin{tabular}{c}
\\ $N_1^+ + N_2^+$\\
\\
\end{tabular}}} & $N_1^+ - N_2^+$ & BPS\\ \hline
$\gamma^{(2)}$=(0,\,0,\,2,\,2)&\multicolumn{1}{c|}{{\begin{tabular}{c}
\\ $N_1^- - N_2^+$ \\
\\
\end{tabular}}}&  $N_1^- + N_2^+$ & non BPS\\ \hline
\end{tabular}
\caption{\small The representative of $\alpha^{(2)}$-orbit in terms of the generating solution.}\label{a2:1}
\end{center}
\end{table}
In the QUbit basis we have:
\begin{table}[H]
\begin{center}
\begin{tabular}{|c|c|c|}\hline
\backslashbox{$\gamma$}{$\beta$}
&$\beta^{(1)}$=(2,\,2,\,0,\,0)&$\beta^{(2)}$=(0,\,0,\,2,\,2)  \\   \hline
$\gamma^{(1)}$=(2,\,2,\,0,\,0)&\multicolumn{1}{c|}{{\begin{tabular}{c}
\\ $(+,\,+,\,-,\,+)+(+,\,+,\,+,\,-)$ \\
\\
\end{tabular}}}& $(+,\,+,\,-,\,+)-(+,\,+,\,+,\,-)$ \\ \hline
$\gamma^{(2)}$=(0,\,0,\,2,\,2)& \multicolumn{1}{c|}{{\begin{tabular}{c}
\\ $(+,\,+,\,-,\,-)+(-,\,-,\,-,\,-)$ \\
\\
\end{tabular}}}&  $(+,\,+,\,-,\,-)-(-,\,-,\,-,\,-)$ \\ \hline
\end{tabular}
\caption{\small The representative of $\alpha^{(2)}$-orbit in the QUbit representation.}\label{a2:2}
\end{center}
\end{table}

\subsubsection{$\alpha^{(3)}=(0,\,0,\,2,\,0)$     ($O[{2^4}]^{II}$)}

In the normal form defined by the generating solution, we can take as nilpositive representatives the matrices $Q_O=N_1^\pm \pm N_4^\pm,$\, which have order of nilpotency $2$.
\begin{table}[H]
\begin{center}
\begin{tabular}{|c|c|c|c|}\hline
\backslashbox{$\gamma$}{$\beta$}
&$\beta^{(1)}$=(2,\,0,\,0,\,2)&$\beta^{(2)}$=(0,\,2,\,2,\,0) & \\ \hline
$\gamma^{(1)}$=(2,\,0,\,0,\,2)& \multicolumn{1}{c|}{{\begin{tabular}{c}
\\
$N_1^+ + N_4^+$ \\
\\
\end{tabular}}} & $N_1^+ - N_4^+$ & BPS \\ \hline
$\gamma^{(2)}$=(0,\,2,\,2,\,0)& \multicolumn{1}{c|}{{\begin{tabular}{c}
\\$N_1^- - N_4^+$ \\
\\
\end{tabular}}}&  $N_1^- + N_4^+$ & non BPS\\ \hline
\end{tabular}
\caption{\small The representative of $\alpha^{(3)}$-orbit in terms of the generating solution.}\label{a3:1}
\end{center}
\end{table}
In the QUbit basis we have:
\begin{table}[H]
\begin{center}
\begin{tabular}{|c|c|c|}\hline
\backslashbox{$\gamma$}{$\beta$}
&$\beta^{(1)}$=(2,\,0,\,0,\,2)&$\beta^{(2)}$=(0,\,2,\,2,\,0)  \\   \hline
$\gamma^{(1)}$=(2,\,0,\,0,\,2)& \multicolumn{1}{c|}{{\begin{tabular}{c}
\\$(+,\,+,\,-,\,-)-(+,\,-,\,+,\,-)$\\
\\
\end{tabular}}} & $(+,\,+,\,-,\,-)+(+,\,-,\,+,\,-)$ \\ \hline
$\gamma^{(2)}$=(0,\,2,\,2,\,0)&\multicolumn{1}{c|}{{\begin{tabular}{c}
\\ $(+,\,+,\,-,\,+)-(-,\,+,\,-,\,-)$\\
\\
\end{tabular}}} &  $(+,\,+,\,-,\,+)+(-,\,+,\,-,\,-)$ \\ \hline
\end{tabular}
\caption{\small The representative of $\alpha^{(3)}$-orbit in the QUbit representation.}\label{a3:2}
\end{center}
\end{table}
\subsubsection{$\alpha^{(4)}\,=\,(0,\,0,\,0,\,2)$   ($O[{2^4}]^{I}$)}
We can take as representatives the matrices $Q_O=N_1^\pm \pm N_3^\pm,$ which have order of nilpotency $2$.
\begin{table}[H]
\begin{center}
\begin{tabular}{|c|c|c|c|}\hline
\backslashbox{$\gamma$}{$\beta$}
&$\beta^{(1)}$=(2,\,0,\,2,\,0)&$\beta^{(2)}$=(0,\,2,\,0,\,2) & \\ \hline
$\gamma^{(1)}$=(2,\,0,\,2,\,0)&\multicolumn{1}{c|}{{\begin{tabular}{c}
\\ $N_1^+ + N_3^+$ \\
\\
\end{tabular}}}& $N_1^+ - N_3^+$ & BPS\\ \hline
$\gamma^{(2)}$=(0,\,2,\,0,\,2)& \multicolumn{1}{c|}{{\begin{tabular}{c}
\\$N_1^- - N_3^+$ \\
\\
\end{tabular}}}&  $N_1^- + N_3^+$ & non BPS\\ \hline
\end{tabular}
\caption{\small The representative of $\alpha^{(4)}$-orbit in terms of the generating solution.}\label{a4:1}
\end{center}
\end{table}
In the QUbit basis we have:
\begin{table}[H]
\begin{center}
\begin{tabular}{|c|c|c|}\hline
\backslashbox{$\gamma$}{$\beta$}
&$\beta^{(1)}$=(2,\,0,\,2,\,0)&$\beta^{(2)}$=(0,\,2,\,0,\,2)  \\   \hline
$\gamma^{(1)}$=(2,\,0,\,2,\,0)& \multicolumn{1}{c|}{{\begin{tabular}{c}
\\$(+,\,+,\,-,\,-)-(+,\,-,\,-,\,+)$\\
\\
\end{tabular}}} & $(+,\,+,\,-,\,-)+(+,\,-,\,-,\,+)$ \\ \hline
$\gamma^{(2)}$=(0,\,2,\,0,\,2)&\multicolumn{1}{c|}{{\begin{tabular}{c}
\\ $(+,\,+,\,+,\,-)-(-,\,+,\,-,\,-)$\\
\\
\end{tabular}}} &  $(+,\,+,\,+,\,-)+(-,\,+,\,-,\,-)$ \\ \hline
\end{tabular}
\caption{\small The representative of $\alpha^{(4)}$-orbit in the QUbit representation.}\label{a4:2}
\end{center}
\end{table}
\subsubsection{$\alpha^{(5)}\,=\,(1,\,0,\,1,\,1)$   ($O[{3,2^2,1}]$)}
These orbits describe \emph{small light-like black holes}.
We can take representatives of the form $Q_O=N_1^\pm \pm N_3^\pm \pm N_4^\pm,$ which have order of nilpotency $3$.
\begin{table}[H]
\begin{center}
\resizebox{\columnwidth}{!}{%
\begin{tabular}{|c|c|c|c|c|c|}\hline
\backslashbox{$\gamma$}{$\beta$}
& $\beta^{(1)}$=(3,\,1,\,1,\,1) & $\beta^{(2)}$=(1,\,3,\,1,\,1) & $\beta^{(3)}$=(1,\,1,\,3,\,1) & $\beta^{(4)}$=(1,\,1,\,1,\,3) & \\ \hline
$\gamma^{(1)}$=(3,\,1,\,1,\,1)& \multicolumn{1}{c|}{{\begin{tabular}{c}
\\
$N_1^+ + N_3^+ + N_4^+$\\
\\
\end{tabular}}} & $-N_1^+ + N_3^+ + N_4^+$ & $N_1^+ + N_3^+ - N_4^+$ & $N_1^+ - N_3^+ + N_4^+$ & BPS \\ \hline
$\gamma^{(2)}$=(1,\,3,\,1,\,1)&\multicolumn{1}{c|}{{\begin{tabular}{c}
\\ $N_1^- - N_3^+ - N_4^+$
\\
\\
\end{tabular}}} & $N_1^- + N_3^+ + N_4^+$ & $N_1^- - N_3^+ + N_4^+$ & $N_1^- + N_3^+ - N_4^+$ & non BPS\\ \hline
$\gamma^{(3)}$=(1,\,1,\,3,\,1) &\multicolumn{1}{c|}{{\begin{tabular}{c}
\\ $N_1^+ + N_3^+ - N_4^-$ \\
\\
\end{tabular}}}& $N_1^+ - N_3^+ + N_4^-$ & $N_1^+ + N_3^+ + N_4^-$ & $ N_1^+ - N_3^+ - N_4^- $ & non BPS\\ \hline
$\gamma^{(4)}$=(1,\,1,\,1,\,3)&\multicolumn{1}{c|}{{\begin{tabular}{c}
\\ $N_1^+ - N_3^- + N_4^+$ \\
\\
\end{tabular}}}& $N_1^+ + N_3^- - N_4^+$ & $N_1^+ - N_3^- - N_4^+$ & $N_1^+ + N_3^- + N_4^+$ & non BPS\\ \hline
\end{tabular}}
\caption{\small The representative of $\alpha^{(5)}$-orbit in terms of the generating solution.}\label{a5:1}
\end{center}
\end{table}
In the QUbit basis we have:
\begin{table}[H]
\begin{center}
\resizebox{\columnwidth}{!}{%
\begin{tabular}{|c|c|c|c|c|}\hline
\backslashbox{$\gamma$}{$\beta$}
& $\beta^{(1)}$=(3,\,1,\,1,\,1) & $\beta^{(2)}$=(1,\,3,\,1,\,1) & $\beta^{(3)}$=(1,\,1,\,3,\,1) & $\beta^{(4)}$=(1,\,1,\,1,\,3) \\ \hline
$\gamma^{(1)}$=(3,\,1,\,1,\,1)&\multicolumn{1}{c|}{{\begin{tabular}{c}
\\
$(+,\,+,\,-,\,+)+(+,\,+,\,+,\,-)$\\$+(+,\,-,\,-,\,-)$
\\
\\
\end{tabular}}}
 &
\multicolumn{1}{c|}{{\begin{tabular}{c}
\\
$(+,\,+,\,-,\,+)+(+,\,+,\,+,\,-)$\\$-(+,\,-,\,-,\,-)$
\\
\\
\end{tabular}}}
 &
\multicolumn{1}{c|}{{\begin{tabular}{c}
\\
$(+,\,+,\,-,\,+)-(+,\,+,\,+,\,-)$\\$+(+,\,-,\,-,\,-)$
\\
\\
\end{tabular}}}
 &
\multicolumn{1}{c|}{{\begin{tabular}{c}
\\
$(+,\,+,\,-,\,+)-(+,\,+,\,+,\,-)$\\$-(+,\,-,\,-,\,-)$
\\
\\
\end{tabular}}} \\ \hline
$\gamma^{(2)}$=(1,\,3,\,1,\,1)& \multicolumn{1}{c|}{{\begin{tabular}{c}
\\
$(+,\,+,\,-,\,+)+(+,\,+,\,+,\,-)$\\$-(-,\,+,\,-,\,-)$
\\
\\
\end{tabular}}} & \multicolumn{1}{c|}{{\begin{tabular}{c}
\\
$(+,\,+,\,-,\,+)+(+,\,+,\,+,\,-)$\\$+(-,\,+,\,-,\,-)$
\\
\\
\end{tabular}}} & \multicolumn{1}{c|}{{\begin{tabular}{c}
\\
$(+,\,+,\,-,\,+)-(+,\,+,\,+,\,-)$\\$+(-,\,+,\,-,\,-)$
\\
\\
\end{tabular}}} & \multicolumn{1}{c|}{{\begin{tabular}{c}
\\
$(+,\,+,\,-,\,+)-(+,\,+,\,+,\,-)$\\$-(-,\,+,\,-,\,-)$
\\
\\
\end{tabular}}}\\ \hline
$\gamma^{(3)}$=(1,\,1,\,3,\,1) & \multicolumn{1}{c|}{{\begin{tabular}{c}
\\
$(+,\,+,\,-,\,+)-(-,\,+,\,-,\,-)$\\$+(+,\,-,\,-,\,-)$
\\
\\
\end{tabular}}} & \multicolumn{1}{c|}{{\begin{tabular}{c}
\\
$(+,\,+,\,-,\,+)+(-,\,+,\,-,\,-)$\\$-(+,\,-,\,-,\,-)$
\\
\\
\end{tabular}}} & \multicolumn{1}{c|}{{\begin{tabular}{c}
\\
$(+,\,+,\,-,\,+)+(-,\,+,\,-,\,-)$\\$+(+,\,-,\,-,\,-)$
\\
\\
\end{tabular}}} & \multicolumn{1}{c|}{{\begin{tabular}{c}
\\
$(+,\,+,\,-,\,+)-(-,\,+,\,-,\,-)$\\$-(+,\,-,\,-,\,-)$
\\
\\
\end{tabular}}}\\ \hline
$\gamma^{(4)}$=(1,\,1,\,1,\,3)& \multicolumn{1}{c|}{{\begin{tabular}{c}
\\
$(+,\,+,\,+,\,-)-(-,\,+,\,-,\,-)$\\$+(+,\,-,\,-,\,-)$
\\
\\
\end{tabular}}} & \multicolumn{1}{c|}{{\begin{tabular}{c}
\\
$(+,\,+,\,+,\,-)+(-,\,+,\,-,\,-)$\\$-(+,\,-,\,-,\,-)$
\\
\\
\end{tabular}}} & \multicolumn{1}{c|}{{\begin{tabular}{c}
\\
$(+,\,+,\,+,\,-)-(-,\,+,\,-,\,-)$\\$-(+,\,-,\,-,\,-)$
\\
\\
\end{tabular}}} & \multicolumn{1}{c|}{{\begin{tabular}{c}
\\
$(+,\,+,\,+,\,-)+(-,\,+,\,-,\,-)$\\$+(+,\,-,\,-,\,-)$
\\
\\
\end{tabular}}}\\ \hline
\end{tabular}}
\caption{\small The representative of $\alpha^{(5)}$-orbit in the QUbit representation. }\label{a5:2}
\end{center}
\end{table}
\subsubsection{$\alpha^{(6)}\,=\,(0,\,2,\,0,\,0)$   ($O[{3^2,1^2}]$)}

These orbits describe \emph{large black holes}.
We can take as representatives the matrices $Q_O=N_1^\pm \pm N_2^\pm \pm N_3^\pm \pm N_4^\pm,$ which have order of nilpotency $3$.
\begin{table}[H]
\begin{center}
\resizebox{\columnwidth}{!}{%
\begin{tabular}{|c|c|c|c|c|c|c|}\hline
\backslashbox{$\gamma$}{$\beta$}
& $\beta^{(1)}$=(4,\,0,\,0,\,0) & $\beta^{(2)}$=(0,\,4,\,0,\,0) & $\beta^{(3)}$=(0,\,0,\,4,\,0) & $\beta^{(4)}$=(0,\,0,\,0,\,4) & $\beta^{(5)}$=(2,\,2,\,2,\,2) & \\ \hline
$\gamma^{(1)}$=(4,\,0,\,0,\,0)&\multicolumn{1}{c|}{{\begin{tabular}{c}
\\ $N_1^+ + N_2^+ + N_3^+ + N_4^+$ \\
\\
\end{tabular}}}& \multicolumn{1}{c|}{$N_1^+ + N_2^+ - N_3^+ - N_4^+$} & \multicolumn{1}{c|}{$N_1^+ - N_2^+ + N_3^+ - N_4^+$} & \multicolumn{1}{c|}{$N_1^+ - N_2^+ - N_3^+ + N_4^+$}  & $N_1^+ + N_2^+ + N_3^+ - N_4^+$ & BPS\\ \hline
$\gamma^{(2)}$=(0,\,4,\,0,\,0)& \multicolumn{1}{c|}{{\begin{tabular}{c}
\\$N_1^+ + N_2^+ - N_3^- - N_4^-$\\
\\
\end{tabular}}} & \multicolumn{1}{c|}{$N_1^+ + N_2^+ + N_3^- + N_4^-$} & \multicolumn{1}{c|}{$N_1^+ - N_2^+ - N_3^- + N_4^-$} & \multicolumn{1}{c|}{$N_1^+ - N_2^+ + N_3^- - N_4^-$} & \multicolumn{1}{c|}{$N_1^+ + N_2^+ + N_3^- - N_4^-$} & non BPS\\ \hline
$\gamma^{(3)}$=(0,\,0,\,4,\,0)& \multicolumn{1}{c|}{{\begin{tabular}{c}
\\$N_1^- - N_2^+ + N_3^- - N_4^+$\\
\\
\end{tabular}}} & \multicolumn{1}{c|}{$N_1^- - N_2^+ - N_3^- + N_4^+$} & \multicolumn{1}{c|}{$N_1^- + N_2^+ + N_3^- + N_4^+$} & \multicolumn{1}{c|}{$N_1^- + N_2^+ - N_3^- - N_4^+$} & \multicolumn{1}{c|}{$N_1^- + N_2^+ - N_3^- + N_4^+$} & non BPS\\ \hline
$\gamma^{(4)}$=(0,\,0,\,0,\,4)& \multicolumn{1}{c|}{{\begin{tabular}{c}\\$N_1^- - N_2^+ - N_3^+ + N_4^-$\\
\\
\end{tabular}}} & \multicolumn{1}{c|}{$N_1^- - N_2^+ + N_3^+ - N_4^-$} & \multicolumn{1}{c|}{$N_1^- + N_2^+ - N_3^+ - N_4^-$} & \multicolumn{1}{c|}{$N_1^- + N_2^+ + N_3^+ + N_4^-$} & \multicolumn{1}{c|}{$N_1^- + N_2^+ + N_3^+ - N_4^-$} & non BPS\\ \hline
$\gamma^{(5)}$=(2,\,2,\,2,\,2)& $N_1^- + N_2^- + N_3^- - N_4^+$ & \multicolumn{1}{c|}{$N_1^- + N_2^- - N_3^- + N_4^+$} & \multicolumn{1}{c|}{$N_1^- - N_2^- + N_3^- + N_4^+$} & \multicolumn{1}{c|}{$-N_1^- + N_2^- + N_3^- + N_4^+$} &
{\begin{tabular}{l | r}
$N_1^- + N_2^- + N_3^- + N_4^+$ & $\delta^{(1)}$\\ \hline
\multicolumn{1}{c|}{$N_1^- + N_2^- - N_3^- - N_4^+$} & $\delta^{(2)}$\\ \hline
\multicolumn{1}{c|}{$N_1^- - N_2^- - N_3^- + N_4^+$} & $\delta^{(3)}$\\ \hline
\multicolumn{1}{c|}{$N_1^- - N_2^- + N_3^- - N_4^+$} & $\delta^{(4)}$\\
\end{tabular}} & non BPS\\ \hline
\end{tabular}}
\caption{\small The representative of $\alpha^{(6)}$-orbit in terms of the generating solution.  }\label{a6:1}
\end{center}
\end{table}
In the QUbit basis we have:
\begin{table}[H]
\begin{center}
\resizebox{\columnwidth}{!}{%
\begin{tabular}{|c|c|c|c|c|c|}\hline
\backslashbox{$\gamma$}{$\beta$}
& $\beta^{(1)}$=(4,\,0,\,0,\,0) & $\beta^{(2)}$=(0,\,4,\,0,\,0) & $\beta^{(3)}$=(0,\,0,\,4,\,0) & $\beta^{(4)}$=(0,\,0,\,0,\,4) & $\beta^{(5)}$=(2,\,2,\,2,\,2) \\ \hline
$\gamma^{(1)}$=(4,\,0,\,0,\,0)& {\begin{tabular}{c}
\\
$(+,\,+,\,+,\,+)-(+,\,+,\,-,\,-)$\\$+(+,\,-,\,+,\,-)+(+,\,-,\,-,\,+)$
\\
\\
\end{tabular}} & \multicolumn{1}{c|}{{\begin{tabular}{c}
\\
$-(+,\,+,\,+,\,+)+(+,\,+,\,-,\,-)$\\$+(+,\,-,\,+,\,-)+(+,\,-,\,-,\,+)$
\\
\\
\end{tabular}}} & \multicolumn{1}{c|}{{\begin{tabular}{c}
\\
$(+,\,+,\,+,\,+)+(+,\,+,\,-,\,-)$\\$+(+,\,-,\,+,\,-)-(+,\,-,\,-,\,+)$
\\
\\
\end{tabular}}} & \multicolumn{1}{c|}{{\begin{tabular}{c}
\\
$(+,\,+,\,+,\,+)+(+,\,+,\,-,\,-)$\\$-(+,\,-,\,+,\,-)+(+,\,-,\,-,\,+)$
\\
\\
\end{tabular}}}  & {\begin{tabular}{c}
\\
$(+,\,+,\,+,\,+)+(+,\,+,\,-,\,-)$\\$+(+,\,-,\,+,\,-)+(+,\,-,\,-,\,+)$
\\
\\
\end{tabular}}\\ \hline
$\gamma^{(2)}$=(0,\,4,\,0,\,0)& \multicolumn{1}{c|}{{\begin{tabular}{c}
\\
$(-,\,+,\,+,\,+)+(+,\,+,\,-,\,+)$\\$+(+,\,+,\,+,\,-)-(-,\,+,\,-,\,-)$
\\
\\
\end{tabular}}} & \multicolumn{1}{c|}{{\begin{tabular}{c}
\\
$-(-,\,+,\,+,\,+)+(+,\,+,\,-,\,+)$\\$+(+,\,+,\,+,\,-)+(-,\,+,\,-,\,-)$
\\
\\
\end{tabular}}} & \multicolumn{1}{c|}{{\begin{tabular}{c}
\\
$(-,\,+,\,+,\,+)+(+,\,+,\,-,\,+)$\\$-(+,\,+,\,+,\,-)+(-,\,+,\,-,\,-)$
\\
\\
\end{tabular}}} & \multicolumn{1}{c|}{{\begin{tabular}{c}
\\
$(-,\,+,\,+,\,+)-(+,\,+,\,-,\,+)$\\$+(+,\,+,\,+,\,-)+(-,\,+,\,-,\,-)$
\\
\\
\end{tabular}}} & \multicolumn{1}{c|}{{\begin{tabular}{c}
\\
$(-,\,+,\,+,\,+)+(+,\,+,\,-,\,+)$\\$+(+,\,+,\,+,\,-)+(-,\,+,\,-,\,-)$
\\
\\
\end{tabular}}}\\ \hline
$\gamma^{(3)}$=(0,\,0,\,4,\,0)& \multicolumn{1}{c|}{{\begin{tabular}{c}
\\
$(-,\,+,\,-,\,+)+(+,\,+,\,-,\,-)$\\$-(+,\,-,\,-,\,+)+(-,\,-,\,-,\,-)$
\\
\\
\end{tabular}}} & \multicolumn{1}{c|}{{\begin{tabular}{c}
\\
$-(-,\,+,\,-,\,+)+(+,\,+,\,-,\,-)$\\$+(+,\,-,\,-,\,+)+(-,\,-,\,-,\,-)$
\\
\\
\end{tabular}}} & \multicolumn{1}{c|}{{\begin{tabular}{c}
\\
$(-,\,+,\,-,\,+)-(+,\,+,\,-,\,-)$\\$+(+,\,-,\,-,\,+)+(-,\,-,\,-,\,-)$
\\
\\
\end{tabular}}} & \multicolumn{1}{c|}{{\begin{tabular}{c}
\\
$(-,\,+,\,-,\,+)+(+,\,+,\,-,\,-)$\\$+(+,\,-,\,-,\,+)-(-,\,-,\,-,\,-)$
\\
\\
\end{tabular}}} & \multicolumn{1}{c|}{{\begin{tabular}{c}
\\
$(-,\,+,\,-,\,+)+(+,\,+,\,-,\,-)$\\$+(+,\,-,\,-,\,+)+(-,\,-,\,-,\,-)$
\\
\\
\end{tabular}}}\\ \hline
$\gamma^{(4)}$=(0,\,0,\,0,\,4)& \multicolumn{1}{c|}{{\begin{tabular}{c}
\\
$(-,\,+,\,+,\,-)+(+,\,+,\,-,\,-)$\\$-(+,\,-,\,+,\,-)+(-,\,-,\,-,\,-)$
\\
\\
\end{tabular}}} & \multicolumn{1}{c|}{{\begin{tabular}{c}
\\
$-(-,\,+,\,+,\,-)+(+,\,+,\,-,\,-)$\\$+(+,\,-,\,+,\,-)+(-,\,-,\,-,\,-)$
\\
\\
\end{tabular}}} & \multicolumn{1}{c|}{{\begin{tabular}{c}
\\
$(-,\,+,\,+,\,-)+(+,\,+,\,-,\,-)$\\$+(+,\,-,\,+,\,-)-(-,\,-,\,-,\,-)$
\\
\\
\end{tabular}}} & \multicolumn{1}{c|}{{\begin{tabular}{c}
\\
$(-,\,+,\,+,\,-)-(+,\,+,\,-,\,-)$\\$+(+,\,-,\,+,\,-)+(-,\,-,\,-,\,-)$
\\
\\
\end{tabular}}} & \multicolumn{1}{c|}{{\begin{tabular}{c}
\\
$(-,\,+,\,+,\,-)+(+,\,+,\,-,\,-)$\\$+(+,\,-,\,+,\,-)+(-,\,-,\,-,\,-)$
\\
\\
\end{tabular}}}\\ \hline
$\gamma^{(5)}$=(2,\,2,\,2,\,2)& {\begin{tabular}{c}
\\
$(+,\,+,\,-,\,+)+(+,\,+,\,+,\,-)$\\$-(-,\,+,\,-,\,-)+(+,\,-,\,-,\,-)$
\\
\\
\end{tabular}} & \multicolumn{1}{c|}{{\begin{tabular}{c}
\\
$(+,\,+,\,-,\,+)+(+,\,+,\,+,\,-)$\\$+(-,\,+,\,-,\,-)-(+,\,-,\,-,\,-)$
\\
\\
\end{tabular}}} & \multicolumn{1}{c|}{{\begin{tabular}{c}
\\
$(+,\,+,\,-,\,+)-(+,\,+,\,+,\,-)$\\$+(-,\,+,\,-,\,-)-(+,\,-,\,-,\,-)$
\\
\\
\end{tabular}}} & \multicolumn{1}{c|}{{\begin{tabular}{c}
\\
$(+,\,+,\,-,\,+)-(+,\,+,\,+,\,-)$\\$-(-,\,+,\,-,\,-)-(+,\,-,\,-,\,-)$
\\
\\
\end{tabular}}} &
{\begin{tabular}{l | r}
{\begin{tabular}{c}
\\
$(+,\,+,\,-,\,+)+(+,\,+,\,+,\,-)$\\$+(-,\,+,\,-,\,-)+(+,\,-,\,-,\,-)$
\\
\\
\end{tabular}} & $\delta^{(1)}$\\ \hline
\multicolumn{1}{c|}{{\begin{tabular}{c}
\\
$(+,\,+,\,-,\,+)+(+,\,+,\,+,\,-)$\\$-(-,\,+,\,-,\,-)-(+,\,-,\,-,\,-)$
\\
\\
\end{tabular}}} & $\delta^{(2)}$\\ \hline
\multicolumn{1}{c|}{{\begin{tabular}{c}
\\
$(+,\,+,\,-,\,+)-(+,\,+,\,+,\,-)$\\$+(-,\,+,\,-,\,-)-(+,\,-,\,-,\,-)$
\\
\\
\end{tabular}}} & $\delta^{(3)}$\\ \hline
\multicolumn{1}{c|}{{\begin{tabular}{c}
\\
$(+,\,+,\,-,\,+)-(+,\,+,\,+,\,-)$\\$-(-,\,+,\,-,\,-)+(+,\,-,\,-,\,-)$
\\
\\
\end{tabular}}} & $\delta^{(4)}$\\
\end{tabular}}\\ \hline
\end{tabular}}
\caption{\small The representative of $\alpha^{(6)}$-orbit in terms of QUbit basis. }\label{a6:2}
\end{center}
\end{table}
\subsection{The multi-center orbits}\label{secdir2}

The following orbits do not contain regular single-center solutions, but only multi-center ones. Their representatives have order of nilpotency greater than 3 and can not be written in the normal form defined by the generating solution. All solutions in these orbits are \emph{non-BPS}. The orbits with $\alpha$-labels $\alpha^{(7)},\alpha^{(8)},\alpha^{(9)}$ are related by triality and contain, as representative solutions, systems consisting of a large and a small black holes.
\subsubsection{$\alpha^{(7)}\,=\,(2,\,2,\,0,\,0)$    ($O[{5,1^3}]$)}
The order of nilpotency is $5$. Representatives of these orbits can be written in  the QUbit basis as follows:
\begin{table}[H]
\begin{center}
\resizebox{\columnwidth}{!}{%
\begin{tabular}{|c|c|c|}\hline
\backslashbox{$\gamma$}{$\beta$}
& $\beta^{(1)}$=(2,\,2,\,4,\,4) & $\beta^{(2)}$=(4,\,4,\,2,\,2) \\ \hline
$\gamma^{(1)}$=(2,\,2,\,4,\,4)& {\begin{tabular}{c}
\\
$(+, \, +, \, -, \,+)- (+, \, +, \, +, \, -) $\\$ + (-, \, -, \, -, \, -)$
\\
\\
\end{tabular}} & {\begin{tabular}{c}
\\
$(+, \, +, \, - \, +)+ (+, \, +, \, +, \, -) $\\$ + (-, \, -, \, -, \, -)$
\\
\\
\end{tabular}}\\
\hline
$\gamma^{(2)}$=(4,\,4,\,2,\,2)& {\begin{tabular}{c}
\\
$(+, \, +, \, +, \, +) + (-, \, +, \, -, \, -)$\\$ + (+, \, -, \, -, \, -)$
\\
\\
\end{tabular}} & {\begin{tabular}{c}
\\
$(+, \, +, \, +, \, +) + (-, \, +, \, -, \, -)$\\$ - (+, \, -, \, -, \, -)$
\\
\\
\end{tabular}} \\ \hline
\end{tabular}}
\caption{\small The representative of $\alpha^{(7)}$-orbit in the QUbit basis.}\label{a7}
\end{center}
\end{table}
\subsubsection{$\alpha^{(8)}\,=\,(0,\,2,\,2,\,0)$    ($O[{4^2}]^{II}$)}

The order of nilpotency is $4$ and representatives of the orbits in the  QUbit basis are:
\begin{table}[H]
\begin{center}
\resizebox{\columnwidth}{!}{%
\begin{tabular}{|c|c|c|}\hline
\backslashbox{$\gamma$}{$\beta$}
& $\beta^{(1)}$=(2,\,4,\,4,\,2) & $\beta^{(2)}$=(4,\,2,\,2,\,4) \\ \hline
$\gamma^{(1)}$=(2,\,4,\,4,\,2)& {\begin{tabular}{c}
\\
$(-,\,+,\,-,\,+)-(+,\,+,\,+,\,-)$\\$+(+,\,-,\,-,\,-) $
\\
\\
\end{tabular}}&{\begin{tabular}{c}
\\
$(-,\,+,\,-,\,+)-(+,\,+,\,+,\,-)$\\$-(+,\,-,\,-,\,-) $
\\
\\
\end{tabular}}\\
\hline
$\gamma^{(2)}$=(4,\,2,\,2,\,4)& {\begin{tabular}{c}
\\
$(+,\,+,\,-,\,+)+(-,\,+,\,-,\,-)$\\$-(+,\,-,\,+,\,-) $
\\
\\
\end{tabular}} & {\begin{tabular}{c}
\\
$(+,\,+,\,-,\,+)-(-,\,+,\,-,\,-)$\\$-(+,\,-,\,+,\,-) $
\\
\\
\end{tabular}} \\ \hline
\end{tabular}}
\caption{\small The representative of $\alpha^{(8)}$-orbit in the QUbit basis . }\label{a8}
\end{center}
\end{table}
\subsubsection{$\alpha^{(9)}\,=\,(0,\,2,\,0,\,2)$    ($O[{4^2}]^{I}$)}
The order of nilpotency is $4$ and representatives of the orbits in the  QUbit basis are:
\begin{table}[H]
\begin{center}
\resizebox{\columnwidth}{!}{%
\begin{tabular}{|c|c|c|}\hline
\backslashbox{$\gamma$}{$\beta$}
& $\beta^{(1)}$=(2,\,4,\,2,\,4) & $\beta^{(2)}$=(4,\,2,\,4,\,2) \\ \hline
$\gamma^{(1)}$=(2,\,4,\,2,\,4)& {\begin{tabular}{c}
\\
$ (-,\,+,\,+,\,-)-(+,\,+,\,-,\,+)$\\$+(+,\,-,\,-,\,-) $
\\
\\
\end{tabular}}&{\begin{tabular}{c}
\\
$ (-,\,+,\,+,\,-)-(+,\,+,\,-,\,+)$\\$-(+,\,-,\,-,\,-) $
\\
\\
\end{tabular}}\\
\hline
$\gamma^{(2)}$=(4,\,2,\,4,\,2)& {\begin{tabular}{c}
\\
$ (+,\,+,\,+,\,-)+(-,\,+,\,-,\,-)$\\$-(+,\,-,\,-,\,+) $
\\
\\
\end{tabular}} & {\begin{tabular}{c}
\\
$ (+,\,+,\,+,\,-)-(-,\,+,\,-,\,-)$\\$-(+,\,-,\,-,\,+) $
\\
\\
\end{tabular}} \\ \hline
\end{tabular}}
\caption{\small The representative of $\alpha^{(9)}$-orbit in the QUbit basis . }\label{a9}
\end{center}
\end{table}
\subsubsection{$\alpha^{(10)}\,=\,(2,\,0,\,2,\,2)$    ($O[{5,3}]$)}
These orbits contain multi-center \emph{composite non-BPS black holes} \cite{bossard2}.
The order of nilpotency is $5$ and representatives of the orbits in the  QUbit basis are:
\begin{table}[H]
\begin{center}
\rotatebox{90}{
\resizebox{\columnwidth}{!}{%
\begin{tabular}{|c|c|c|c|c|}\hline
\backslashbox{$\gamma$}{$\beta$}
& $\beta^{(1)}$=(0,\,4,\,4,\,4) & $\beta^{(2)}$=(4,\,0,\,4,\,4) & $\beta^{(3)}$=(4,\,4,\,0,\,4) & $\beta^{(4)}$=(4,\,4,\,4,\,0)\\ \hline
$\gamma^{(1)}$=(0,\,4,\,4,\,4) & {\begin{tabular}{c}
\\
$(-,\,+,\,-,\,+)+(+,\,+,\,+,\,-)$\\$-(+,\,-,\,-,\,-)-(-,\,-,\,-,\,-) $
\\
\\
\end{tabular}} & {\begin{tabular}{c}
\\
$(-,\,+,\,-,\,+)+(+,\,+,\,+,\,-)$\\$+(+,\,-,\,-,\,-)+(-,\,-,\,-,\,-) $
\\
\\
\end{tabular}} & {\begin{tabular}{c}
\\
$(-,\,+,\,-,\,+)+(+,\,+,\,+,\,-)$\\$+(+,\,-,\,-,\,-)-(-,\,-,\,-,\,-) $
\\
\\
\end{tabular}} & {\begin{tabular}{c}
\\
$(-,\,+,\,-,\,+)+(+,\,+,\,+,\,-)$\\$-(+,\,-,\,-,\,-)+(-,\,-,\,-,\,-) $
\\
\\
\end{tabular}}\\
\hline
$\gamma^{(2)}$=(4,\,0,\,4,\,4) & {\begin{tabular}{c}
\\
$(+,\,+,\,-,\,+)+(-,\,+,\,-,\,-)$\\$-(-,\,-,\,-,\,-)-(+,\,-,\,+,\,-) $
\\
\\
\end{tabular}} & {\begin{tabular}{c}
\\
$(+,\,+,\,-,\,+)-(-,\,+,\,-,\,-)$\\$-(-,\,-,\,-,\,-)+(+,\,-,\,+,\,-) $
\\
\\
\end{tabular}} & {\begin{tabular}{c}
\\
$(+,\,+,\,-,\,+)-(-,\,+,\,-,\,-)$\\$+(-,\,-,\,-,\,-)+(+,\,-,\,+,\,-) $
\\
\\
\end{tabular}} & {\begin{tabular}{c}
\\
$(+,\,+,\,-,\,+)+(-,\,+,\,-,\,-)$\\$+(-,\,-,\,-,\,-)-(+,\,-,\,+,\,-) $
\\
\\
\end{tabular}}\\
\hline
$\gamma^{(3)}$=(4,\,4,\,0,\,4) & {\begin{tabular}{c}
\\
$(+,\,+,\,+,\,+)+(-,\,+,\,-,\,-)$\\$+(+,\,-,\,-,\,-)-(+,\,-,\,+,\,-) $
\\
\\
\end{tabular}} & {\begin{tabular}{c}
\\
$(+,\,+,\,+,\,+)-(-,\,+,\,-,\,-)$\\$-(+,\,-,\,-,\,-)+(+,\,-,\,+,\,-) $
\\
\\
\end{tabular}} & {\begin{tabular}{c}
\\
$(+,\,+,\,+,\,+)+(-,\,+,\,-,\,-)$\\$-(+,\,-,\,-,\,-)-(+,\,-,\,+,\,-) $
\\
\\
\end{tabular}} & {\begin{tabular}{c}
\\
$(+,\,+,\,+,\,+)+(-,\,+,\,-,\,-)$\\$-(+,\,-,\,-,\,-)+(+,\,-,\,+,\,-) $
\\
\\
\end{tabular}}\\
\hline
$\gamma^{(4)}$=(4,\,4,\,4,\,0) & {\begin{tabular}{c}
\\
$(+,\,+,\,+,\,+)+(-,\,+,\,-,\,-)$\\$+(+,\,-,\,-,\,-)-(+,\,-,\,-,\,+)$
\\
\\
\end{tabular}} & {\begin{tabular}{c}
\\
$(+,\,+,\,+,\,+)+(-,\,+,\,-,\,-)$\\$+(+,\,-,\,-,\,-)+(+,\,-,\,-,\,+)$
\\
\\
\end{tabular}} & {\begin{tabular}{c}
\\
$(+,\,+,\,+,\,+)+(-,\,+,\,-,\,-)$\\$-(+,\,-,\,-,\,-)+(+,\,-,\,-,\,+)$
\\
\\
\end{tabular}} & {\begin{tabular}{c}
\\
$(+,\,+,\,+,\,+)+(-,\,+,\,-,\,-)$\\$-(+,\,-,\,-,\,-)-(+,\,-,\,-,\,+)$
\\
\\
\end{tabular}} \\
\hline
\end{tabular}}}
\caption{\small The representative of $\alpha^{(10)}$-orbit in the QUbit basis.}\label{a10}
\end{center}
\end{table}
\subsubsection{$\alpha^{(11)}\,=\,(2,\,2,\,2,\,2))$    ($O[{7,1}]$)}

These orbits contain multi-center \emph{almost-BPS black holes} \cite{Goldstein:2008fq,Bena:2009ev}.
The order of nilpotency is $7$ and representatives of the orbits in the  QUbit basis are:
\begin{table}[H]
\begin{center}
\rotatebox{90}{
\resizebox{\columnwidth}{!}{%
\begin{tabular}{|c|c|c|c|c|}\hline
\backslashbox{$\gamma$}{$\beta$}
& $\beta^{(1)}$=(8,\,4,\,4,\,4) & $\beta^{(2)}$=(4,\,8,\,4,\,4) & $\beta^{(3)}$=(4,\,4,\,8,\,4) & $\beta^{(2)}$=(4,\,4,\,4,\,8)\\ \hline
$\gamma^{(1)}$=(8,\,4,\,4,\,4) & {\begin{tabular}{c}
\\
$(+,\,+,\,+,\,+)-(-,\,+,\,-,\,-)$\\$+(+,\,-,\,+,\,-)+(+,\,-,\,-,\,+)$
\\
\\
\end{tabular}} & {\begin{tabular}{c}
\\
$(+,\,+,\,+,\,+)+(-,\,+,\,-,\,-)$\\$-(+,\,-,\,+,\,-)-(+,\,-,\,-,\,+)$
\\
\\
\end{tabular}} & {\begin{tabular}{c}
\\
$(+,\,+,\,+,\,+)+(-,\,+,\,-,\,-)$\\$+(+,\,-,\,+,\,-)-(+,\,-,\,-,\,+)$
\\
\\
\end{tabular}} & {\begin{tabular}{c}
\\
$(+,\,+,\,+,\,+)+(-,\,+,\,-,\,-)$\\$-(+,\,-,\,+,\,-)+(+,\,-,\,-,\,+)$
\\
\\
\end{tabular}}\\
\hline
$\gamma^{(2)}$=(4,\,8,\,4,\,4) & {\begin{tabular}{c}
\\
$(+,\,+,\,+,\,+)-(-,\,+,\,-,\,+)$\\$-(-,\,+,\,+,\,-)+(+,\,-,\,-,\,-)$
\\
\\
\end{tabular}} & {\begin{tabular}{c}
\\
$(+,\,+,\,+,\,+)+(-,\,+,\,-,\,+)$\\$+(-,\,+,\,+,\,-)-(+,\,-,\,-,\,-)$
\\
\\
\end{tabular}} & {\begin{tabular}{c}
\\
$(+,\,+,\,+,\,+)-(-,\,+,\,-,\,+)$\\$+(-,\,+,\,+,\,-)+(+,\,-,\,-,\,-)$
\\
\\
\end{tabular}} & {\begin{tabular}{c}
\\
$(+,\,+,\,+,\,+)+(-,\,+,\,-,\,+)$\\$-(-,\,+,\,+,\,-)+(+,\,-,\,-,\,-)$
\\
\\
\end{tabular}}\\
\hline
$\gamma^{(3)}$=(4,\,4,\,8,\,4) & {\begin{tabular}{c}
\\
$(-,\,+,\,-,\,+)+(+,\,+,\,+,\,-)$\\$+(-,\,-,\,-,\,-)-(+,\,-,\,-,\,+)$
\\
\\
\end{tabular}} & {\begin{tabular}{c}
\\
$(-,\,+,\,-,\,+)-(+,\,+,\,+,\,-)$\\$-(-,\,-,\,-,\,-)-(+,\,-,\,-,\,+)$
\\
\\
\end{tabular}} & {\begin{tabular}{c}
\\
$(-,\,+,\,-,\,+)-(+,\,+,\,+,\,-)$\\$+(-,\,-,\,-,\,-)+(+,\,-,\,-,\,+)$
\\
\\
\end{tabular}} & {\begin{tabular}{c}
\\
$(-,\,+,\,-,\,+)-(+,\,+,\,+,\,-)$\\$-(-,\,-,\,-,\,-)+(+,\,-,\,-,\,+)$
\\
\\
\end{tabular}}\\
\hline
$\gamma^{(4)}$=(4,\,4,\,4,\,8) & {\begin{tabular}{c}
\\
$(-,\,+,\,+,\,-)+(+,\,+,\,-,\,+)$\\$+(-,\,-,\,-,\,-)-(+,\,-,\,+,\,-)$
\\
\\
\end{tabular}} & {\begin{tabular}{c}
\\
$(-,\,+,\,+,\,-)-(+,\,+,\,-,\,+)$\\$-(-,\,-,\,-,\,-)-(+,\,-,\,+,\,-)$
\\
\\
\end{tabular}} & {\begin{tabular}{c}
\\
$(-,\,+,\,+,\,-)-(+,\,+,\,-,\,+)$\\$-(-,\,-,\,-,\,-)+(+,\,-,\,+,\,-)$
\\
\\
\end{tabular}} & {\begin{tabular}{c}
\\
$(-,\,+,\,+,\,-)-(+,\,+,\,-,\,+)$\\$+(-,\,-,\,-,\,-)+(+,\,-,\,+,\,-)$
\\
\\
\end{tabular}} \\
\hline
\end{tabular}}}
\caption{\small The representative of $\alpha^{(11)}$-orbit in the QUbit basis.}\label{a11}
\end{center}
\end{table}
\subsection{Sum rules}\label{sumrules}
Here we give the combinations of regular-single-center orbits which yield $H^*$-orbits with $\alpha$-label between $\alpha^{(7)}, \dots, \alpha^{(11)}$.
This is a purely mathematical result and does not imply that each combination actually correspond to a regular solution, but does exclude the intrinsically ``singular'' $H^*$-orbits (not in the list below), which can not be obtained in this way and thus we claim not to contain regular solutions.
\subsubsection{The orbit $\alpha^{(7)}\gamma^{(7;1)}\beta^{(2)}$}
\begin{center}
\resizebox{\textwidth}{!}{\begin{tabular}{|c|c|c|}
                            $\alpha^{(6)}\gamma^{(6;5)}\beta^{(5)}\delta^{(1)}+(\alpha^{(6)}\gamma^{(6;5)}\beta^{(5)}\delta^{(1)})'$ &   &   \\
                            $\alpha^{(6)}\gamma^{(6;5)}\beta^{(5)}\delta^{(1)}+(\alpha^{(5)}\gamma^{(5;2)}\beta^{(2)})'$ & $\alpha^{(5)}\gamma^{(5;2)}\beta^{(2)}+(\alpha^{(5)}\gamma^{(5;2)}\beta^{(2)})'$ &  \\
                            $\alpha^{(6)}\gamma^{(6;5)}\beta^{(5)}\delta^{(1)}+(\alpha^{(5)}\gamma^{(5;1)}\beta^{(1)})'$ & $\alpha^{(5)}\gamma^{(5;2)}\beta^{(2)}+(\alpha^{(5)}\gamma^{(5;1)}\beta^{(1)})'$ &  \\
                            $\alpha^{(6)}\gamma^{(6;5)}\beta^{(5)}\delta^{(1)}+(\alpha^{(2)}\gamma^{(2;2)}\beta^{(2)})'$ & $\alpha^{(5)}\gamma^{(5;2)}\beta^{(2)}+(\alpha^{(2)}\gamma^{(2;2)}\beta^{(2)})'$ &  \\
                            $\alpha^{(6)}\gamma^{(6;5)}\beta^{(5)}\delta^{(1)}+(\alpha^{(2)}\gamma^{(2;1)}\beta^{(1)})'$ & $\alpha^{(5)}\gamma^{(5;2)}\beta^{(2)}+(\alpha^{(2)}\gamma^{(2;1)}\beta^{(1)})'$ &  \\
                            $\alpha^{(6)}\gamma^{(6;5)}\beta^{(5)}\delta^{(1)}+(\alpha^{(1)}\gamma^{(1;1)}\beta^{(1)})'$  & $\alpha^{(5)}\gamma^{(5;2)}\beta^{(2)}+(\alpha^{(1)}\gamma^{(1;1)}\beta^{(1)})'$ &  \\
                              & $\alpha^{(5)}\gamma^{(5;1)}\beta^{(1)}+(\alpha^{(5)}\gamma^{(5;1)}\beta^{(1)})'$ &  \\
                              & $\alpha^{(5)}\gamma^{(5;1)}\beta^{(1)}+(\alpha^{(2)}\gamma^{(2;2)}\beta^{(2)})'$ &  \\
                              & $\alpha^{(5)}\gamma^{(5;1)}\beta^{(1)}+(\alpha^{(2)}\gamma^{(2;1)}\beta^{(1)})'$ & $\alpha^{(2)}\gamma^{(2;1)}\beta^{(1)}+(\alpha^{(2)}\gamma^{(2;2)}\beta^{(2)})'$ \\
                              & $\alpha^{(5)}\gamma^{(5;1)}\beta^{(1)}+(\alpha^{(1)}\gamma^{(1;1)}\beta^{(1)})'$ & $\alpha^{(2)}\gamma^{(2;1)}\beta^{(1)}+(\alpha^{(1)}\gamma^{(1;1)}\beta^{(1)})'$ \\
                          \end{tabular}
}
\end{center}
\subsubsection{The orbit $\alpha^{(7)}\gamma^{(7;2)}\beta^{(1)}$}
\begin{center}
\resizebox{\textwidth}{!}{\begin{tabular}{|c|c|c|}
                            $\alpha^{(6)}\gamma^{(6;5)}\beta^{(5)}\delta^{(1)}+(\alpha^{(6)}\gamma^{(6;5)}\beta^{(5)}\delta^{(1)})'$ &   &   \\
                            $\alpha^{(6)}\gamma^{(6;5)}\beta^{(5)}\delta^{(1)}+(\alpha^{(5)}\gamma^{(5;4)}\beta^{(4)})'$ & $\alpha^{(5)}\gamma^{(5;4)}\beta^{(4)}+(\alpha^{(5)}\gamma^{(5;4)}\beta^{(4)})'$ &  \\
                            $\alpha^{(6)}\gamma^{(6;5)}\beta^{(5)}\delta^{(1)}+(\alpha^{(5)}\gamma^{(5;3)}\beta^{(3)})'$ & $\alpha^{(5)}\gamma^{(5;4)}\beta^{(4)}+(\alpha^{(5)}\gamma^{(5;3)}\beta^{(3)})'$ &  \\
                            $\alpha^{(6)}\gamma^{(6;5)}\beta^{(5)}\delta^{(1)}+(\alpha^{(2)}\gamma^{(2;2)}\beta^{(2)})'$ & $\alpha^{(5)}\gamma^{(5;4)}\beta^{(4)}+(\alpha^{(2)}\gamma^{(2;2)}\beta^{(2)})'$ &  \\
                            $\alpha^{(6)}\gamma^{(6;5)}\beta^{(5)}\delta^{(1)}+(\alpha^{(2)}\gamma^{(2;1)}\beta^{(1)})'$ & $\alpha^{(5)}\gamma^{(5;4)}\beta^{(4)}+(\alpha^{(2)}\gamma^{(2;1)}\beta^{(1)})'$ &  \\
                            $\alpha^{(6)}\gamma^{(6;5)}\beta^{(5)}\delta^{(1)}+(\alpha^{(1)}\gamma^{(1;1)}\beta^{(1)})'$  & $\alpha^{(5)}\gamma^{(5;4)}\beta^{(4)}+(\alpha^{(1)}\gamma^{(1;1)}\beta^{(1)})'$ &  \\
                              & $\alpha^{(5)}\gamma^{(5;3)}\beta^{(3)}+(\alpha^{(5)}\gamma^{(5;3)}\beta^{(3)})'$ &  \\
                              & $\alpha^{(5)}\gamma^{(5;3)}\beta^{(3)}+(\alpha^{(2)}\gamma^{(2;2)}\beta^{(2)})'$ &  \\
                              & $\alpha^{(5)}\gamma^{(5;3)}\beta^{(3)}+(\alpha^{(2)}\gamma^{(2;1)}\beta^{(1)})'$ & $\alpha^{(2)}\gamma^{(2;2)}\beta^{(2)}+(\alpha^{(2)}\gamma^{(2;1)}\beta^{(1)})'$ \\
                              & $\alpha^{(5)}\gamma^{(5;3)}\beta^{(3)}+(\alpha^{(1)}\gamma^{(1;1)}\beta^{(1)})'$ & $\alpha^{(2)}\gamma^{(2;2)}\beta^{(2)}+(\alpha^{(1)}\gamma^{(1;1)}\beta^{(1)})'$ \\
                          \end{tabular}}
\end{center}
\subsubsection{The orbit $\alpha^{(8)}\gamma^{(8;1)}\beta^{(2)}$}
\begin{center}
\resizebox{\textwidth}{!}{\begin{tabular}{|c|c|c|}
                            $\alpha^{(6)}\gamma^{(6;5)}\beta^{(5)}\delta^{(1)}+(\alpha^{(6)}\gamma^{(6;5)}\beta^{(5)}\delta^{(1)})'$ &   &   \\
                            $\alpha^{(6)}\gamma^{(6;5)}\beta^{(5)}\delta^{(1)}+(\alpha^{(5)}\gamma^{(5;4)}\beta^{(4)})'$ & $\alpha^{(5)}\gamma^{(5;4)}\beta^{(4)}+(\alpha^{(5)}\gamma^{(5;4)}\beta^{(4)})'$ &  \\
                            $\alpha^{(6)}\gamma^{(6;5)}\beta^{(5)}\delta^{(1)}+(\alpha^{(5)}\gamma^{(5;1)}\beta^{(1)})'$ & $\alpha^{(5)}\gamma^{(5;4)}\beta^{(4)}+(\alpha^{(5)}\gamma^{(5;1)}\beta^{(1)})'$ &  \\
                            $\alpha^{(6)}\gamma^{(6;5)}\beta^{(5)}\delta^{(1)}+(\alpha^{(3)}\gamma^{(3;2)}\beta^{(2)})'$ & $\alpha^{(5)}\gamma^{(5;4)}\beta^{(4)}+(\alpha^{(3)}\gamma^{(3;2)}\beta^{(2)})'$ &  \\
                            $\alpha^{(6)}\gamma^{(6;5)}\beta^{(5)}\delta^{(1)}+(\alpha^{(3)}\gamma^{(3;1)}\beta^{(1)})'$ & $\alpha^{(5)}\gamma^{(5;4)}\beta^{(4)}+(\alpha^{(3)}\gamma^{(3;1)}\beta^{(1)})'$ &  \\
                            $\alpha^{(6)}\gamma^{(6;5)}\beta^{(5)}\delta^{(1)}+(\alpha^{(1)}\gamma^{(1;1)}\beta^{(1)})'$  & $\alpha^{(5)}\gamma^{(5;4)}\beta^{(4)}+(\alpha^{(1)}\gamma^{(1;1)}\beta^{(1)})'$ &  \\
                              & $\alpha^{(5)}\gamma^{(5;1)}\beta^{(1)}+(\alpha^{(5)}\gamma^{(5;1)}\beta^{(1)})'$ &  \\
                              & $\alpha^{(5)}\gamma^{(5;1)}\beta^{(1)}+(\alpha^{(3)}\gamma^{(3;2)}\beta^{(2)})'$ &  \\
                              & $\alpha^{(5)}\gamma^{(5;1)}\beta^{(1)}+(\alpha^{(3)}\gamma^{(3;1)}\beta^{(1)})'$ & $\alpha^{(3)}\gamma^{(3;1)}\beta^{(1)}+(\alpha^{(3)}\gamma^{(3;2)}\beta^{(2)})'$ \\
                              & $\alpha^{(5)}\gamma^{(5;1)}\beta^{(1)}+(\alpha^{(1)}\gamma^{(1;1)}\beta^{(1)})'$ & $\alpha^{(3)}\gamma^{(3;1)}\beta^{(1)}+(\alpha^{(1)}\gamma^{(1;1)}\beta^{(1)})'$ \\
                          \end{tabular}}
\end{center}
\subsubsection{The orbit $\alpha^{(8)}\gamma^{(8;2)}\beta^{(1)}$}
\begin{center}
\resizebox{\textwidth}{!}{\begin{tabular}{|c|c|c|}
                            $\alpha^{(6)}\gamma^{(6;5)}\beta^{(5)}\delta^{(1)}+(\alpha^{(6)}\gamma^{(6;5)}\beta^{(5)}\delta^{(1)})'$ &   &   \\
                            $\alpha^{(6)}\gamma^{(6;5)}\beta^{(5)}\delta^{(1)}+(\alpha^{(5)}\gamma^{(5;3)}\beta^{(3)})'$ & $\alpha^{(5)}\gamma^{(5;3)}\beta^{(3)}+(\alpha^{(5)}\gamma^{(5;3)}\beta^{(3)})'$ &  \\
                            $\alpha^{(6)}\gamma^{(6;5)}\beta^{(5)}\delta^{(1)}+(\alpha^{(5)}\gamma^{(5;2)}\beta^{(2)})'$ & $\alpha^{(5)}\gamma^{(5;3)}\beta^{(3)}+(\alpha^{(5)}\gamma^{(5;2)}\beta^{(2)})'$ &  \\
                            $\alpha^{(6)}\gamma^{(6;5)}\beta^{(5)}\delta^{(1)}+(\alpha^{(3)}\gamma^{(3;2)}\beta^{(2)})'$ & $\alpha^{(5)}\gamma^{(5;3)}\beta^{(3)}+(\alpha^{(3)}\gamma^{(3;2)}\beta^{(2)})'$ &  \\
                            $\alpha^{(6)}\gamma^{(6;5)}\beta^{(5)}\delta^{(1)}+(\alpha^{(3)}\gamma^{(3;1)}\beta^{(1)})'$ & $\alpha^{(5)}\gamma^{(5;3)}\beta^{(3)}+(\alpha^{(3)}\gamma^{(3;1)}\beta^{(1)})'$ &  \\
                            $\alpha^{(6)}\gamma^{(6;5)}\beta^{(5)}\delta^{(1)}+(\alpha^{(1)}\gamma^{(1;1)}\beta^{(1)})'$  & $\alpha^{(5)}\gamma^{(5;3)}\beta^{(3)}+(\alpha^{(1)}\gamma^{(1;1)}\beta^{(1)})'$ &  \\
                              & $\alpha^{(5)}\gamma^{(5;2)}\beta^{(2)}+(\alpha^{(5)}\gamma^{(5;2)}\beta^{(2)})'$ &  \\
                              & $\alpha^{(5)}\gamma^{(5;2)}\beta^{(2)}+(\alpha^{(3)}\gamma^{(3;2)}\beta^{(2)})'$ &  \\
                              & $\alpha^{(5)}\gamma^{(5;2)}\beta^{(2)}+(\alpha^{(3)}\gamma^{(3;1)}\beta^{(1)})'$ & $\alpha^{(3)}\gamma^{(3;2)}\beta^{(2)}+(\alpha^{(3)}\gamma^{(3;1)}\beta^{(1)})'$ \\
                              & $\alpha^{(5)}\gamma^{(5;2)}\beta^{(2)}+(\alpha^{(1)}\gamma^{(1;1)}\beta^{(1)})'$ & $\alpha^{(3)}\gamma^{(3;2)}\beta^{(2)}+(\alpha^{(1)}\gamma^{(1;1)}\beta^{(1)})'$ \\
                          \end{tabular}}
\end{center}
\subsubsection{The orbit $\alpha^{(9)}\gamma^{(9;1)}\beta^{(2)}$}
\begin{center}
\resizebox{\textwidth}{!}{\begin{tabular}{|c|c|c|}
                            $\alpha^{(6)}\gamma^{(6;5)}\beta^{(5)}\delta^{(1)}+(\alpha^{(6)}\gamma^{(6;5)}\beta^{(5)}\delta^{(1)})'$ &   &   \\
                            $\alpha^{(6)}\gamma^{(6;5)}\beta^{(5)}\delta^{(1)}+(\alpha^{(5)}\gamma^{(5;3)}\beta^{(3)})'$ & $\alpha^{(5)}\gamma^{(5;3)}\beta^{(3)}+(\alpha^{(5)}\gamma^{(5;3)}\beta^{(3)})'$ &  \\
                            $\alpha^{(6)}\gamma^{(6;5)}\beta^{(5)}\delta^{(1)}+(\alpha^{(5)}\gamma^{(5;1)}\beta^{(1)})'$ & $\alpha^{(5)}\gamma^{(5;3)}\beta^{(3)}+(\alpha^{(5)}\gamma^{(5;1)}\beta^{(1)})'$ &  \\
                            $\alpha^{(6)}\gamma^{(6;5)}\beta^{(5)}\delta^{(1)}+(\alpha^{(4)}\gamma^{(4;2)}\beta^{(2)})'$ & $\alpha^{(5)}\gamma^{(5;3)}\beta^{(3)}+(\alpha^{(4)}\gamma^{(4;2)}\beta^{(2)})'$ &  \\
                            $\alpha^{(6)}\gamma^{(6;5)}\beta^{(5)}\delta^{(1)}+(\alpha^{(4)}\gamma^{(4;1)}\beta^{(1)})'$ & $\alpha^{(5)}\gamma^{(5;3)}\beta^{(3)}+(\alpha^{(4)}\gamma^{(4;1)}\beta^{(1)})'$ &  \\
                            $\alpha^{(6)}\gamma^{(6;5)}\beta^{(5)}\delta^{(1)}+(\alpha^{(1)}\gamma^{(1;1)}\beta^{(1)})'$  & $\alpha^{(5)}\gamma^{(5;3)}\beta^{(3)}+(\alpha^{(1)}\gamma^{(1;1)}\beta^{(1)})'$ &  \\
                              & $\alpha^{(5)}\gamma^{(5;1)}\beta^{(1)}+(\alpha^{(5)}\gamma^{(5;1)}\beta^{(1)})'$ &  \\
                              & $\alpha^{(5)}\gamma^{(5;1)}\beta^{(1)}+(\alpha^{(4)}\gamma^{(4;2)}\beta^{(2)})'$ &  \\
                              & $\alpha^{(5)}\gamma^{(5;1)}\beta^{(1)}+(\alpha^{(4)}\gamma^{(4;1)}\beta^{(1)})'$ & $\alpha^{(4)}\gamma^{(4;1)}\beta^{(1)}+(\alpha^{(4)}\gamma^{(4;2)}\beta^{(2)})'$ \\
                              & $\alpha^{(5)}\gamma^{(5;1)}\beta^{(1)}+(\alpha^{(1)}\gamma^{(1;1)}\beta^{(1)})'$ & $\alpha^{(4)}\gamma^{(4;1)}\beta^{(1)}+(\alpha^{(1)}\gamma^{(1;1)}\beta^{(1)})'$ \\
                          \end{tabular}}
\end{center}
\subsubsection{The orbit $\alpha^{(9)}\gamma^{(9;2)}\beta^{(1)}$}
\begin{center}
\resizebox{\textwidth}{!}{\begin{tabular}{|c|c|c|}
                            $\alpha^{(6)}\gamma^{(6;5)}\beta^{(5)}\delta^{(1)}+(\alpha^{(6)}\gamma^{(6;5)}\beta^{(5)}\delta^{(1)})'$ &   &   \\
                            $\alpha^{(6)}\gamma^{(6;5)}\beta^{(5)}\delta^{(1)}+(\alpha^{(5)}\gamma^{(5;4)}\beta^{(4)})'$ & $\alpha^{(5)}\gamma^{(5;4)}\beta^{(4)}+(\alpha^{(5)}\gamma^{(5;4)}\beta^{(4)})'$ &  \\
                            $\alpha^{(6)}\gamma^{(6;5)}\beta^{(5)}\delta^{(1)}+(\alpha^{(5)}\gamma^{(5;2)}\beta^{(2)})'$ & $\alpha^{(5)}\gamma^{(5;4)}\beta^{(4)}+(\alpha^{(5)}\gamma^{(5;2)}\beta^{(2)})'$ &  \\
                            $\alpha^{(6)}\gamma^{(6;5)}\beta^{(5)}\delta^{(1)}+(\alpha^{(4)}\gamma^{(4;2)}\beta^{(2)})'$ & $\alpha^{(5)}\gamma^{(5;4)}\beta^{(4)}+(\alpha^{(4)}\gamma^{(4;2)}\beta^{(2)})'$ &  \\
                            $\alpha^{(6)}\gamma^{(6;5)}\beta^{(5)}\delta^{(1)}+(\alpha^{(4)}\gamma^{(4;1)}\beta^{(1)})'$ & $\alpha^{(5)}\gamma^{(5;4)}\beta^{(4)}+(\alpha^{(4)}\gamma^{(4;1)}\beta^{(1)})'$ &  \\
                            $\alpha^{(6)}\gamma^{(6;5)}\beta^{(5)}\delta^{(1)}+(\alpha^{(1)}\gamma^{(1;1)}\beta^{(1)})'$  & $\alpha^{(5)}\gamma^{(5;4)}\beta^{(4)}+(\alpha^{(1)}\gamma^{(1;1)}\beta^{(1)})'$ &  \\
                              & $\alpha^{(5)}\gamma^{(5;2)}\beta^{(2)}+(\alpha^{(5)}\gamma^{(5;2)}\beta^{(2)})'$ &  \\
                              & $\alpha^{(5)}\gamma^{(5;2)}\beta^{(2)}+(\alpha^{(4)}\gamma^{(4;2)}\beta^{(2)})'$ &  \\
                              & $\alpha^{(5)}\gamma^{(5;2)}\beta^{(2)}+(\alpha^{(4)}\gamma^{(4;1)}\beta^{(1)})'$ & $\alpha^{(4)}\gamma^{(4;2)}\beta^{(2)}+(\alpha^{(4)}\gamma^{(4;1)}\beta^{(1)})'$ \\
                              & $\alpha^{(5)}\gamma^{(5;2)}\beta^{(2)}+(\alpha^{(1)}\gamma^{(1;1)}\beta^{(1)})'$ & $\alpha^{(4)}\gamma^{(4;2)}\beta^{(2)}+(\alpha^{(1)}\gamma^{(1;1)}\beta^{(1)})'$ \\
                          \end{tabular}}
\end{center}
\subsubsection{The orbit $\alpha^{(10)}\gamma^{(10;1)}\beta^{(2)}$}
\begin{center}
\resizebox{\textwidth}{!}{\begin{tabular}{|c|c|c|}
                            $\alpha^{(6)}\gamma^{(6;5)}\beta^{(5)}\delta^{(1)}+(\alpha^{(6)}\gamma^{(6;5)}\beta^{(5)}\delta^{(1)})'$ &  &  \\
                            $\alpha^{(6)}\gamma^{(6;5)}\beta^{(5)}\delta^{(1)}+(\alpha^{(5)}\gamma^{(5;4)}\beta^{(4)})'$ & $\alpha^{(5)}\gamma^{(5;4)}\beta^{(4)}+(\alpha^{(5)}\gamma^{(5;3)}\beta^{(3)})'$ &  \\
                            $\alpha^{(6)}\gamma^{(6;5)}\beta^{(5)}\delta^{(1)}+(\alpha^{(5)}\gamma^{(5;3)}\beta^{(3)})'$ & $\alpha^{(5)}\gamma^{(5;4)}\beta^{(4)}+(\alpha^{(5)}\gamma^{(5;1)}\beta^{(1)})'$ &  \\
                            $\alpha^{(6)}\gamma^{(6;5)}\beta^{(5)}\delta^{(1)}+(\alpha^{(5)}\gamma^{(5;1)}\beta^{(1)})'$ & $\alpha^{(5)}\gamma^{(5;4)}\beta^{(4)}+(\alpha^{(4)}\gamma^{(4;1)}\beta^{(1)})'$ &  \\
                            $\alpha^{(6)}\gamma^{(6;5)}\beta^{(5)}\delta^{(1)}+(\alpha^{(4)}\gamma^{(4;1)}\beta^{(1)})'$ & $\alpha^{(5)}\gamma^{(5;3)}\beta^{(3)}+(\alpha^{(5)}\gamma^{(5;1)}\beta^{(1)})'$ &  \\
                            $\alpha^{(6)}\gamma^{(6;5)}\beta^{(5)}\delta^{(1)}+(\alpha^{(3)}\gamma^{(3;1)}\beta^{(1)})'$ & $\alpha^{(5)}\gamma^{(5;3)}\beta^{(3)}+(\alpha^{(3)}\gamma^{(3;1)}\beta^{(1)})'$ &  \\
                            & $\alpha^{(5)}\gamma^{(5;1)}\beta^{(1)}+(\alpha^{(5)}\gamma^{(5;1)}\beta^{(1)})'$ &  \\
                             & $\alpha^{(5)}\gamma^{(5;1)}\beta^{(1)}+(\alpha^{(4)}\gamma^{(4;1)}\beta^{(1)})'$ &  \\
                             & $\alpha^{(5)}\gamma^{(5;1)}\beta^{(1)}+(\alpha^{(3)}\gamma^{(3;1)}\beta^{(1)})'$ & $\alpha^{(4)}\gamma^{(4;1)}\beta^{(1)}+(\alpha^{(3)}\gamma^{(3;1)}\beta^{(1)})'$ \\
                          \end{tabular}
}
\end{center}
\subsubsection{The orbit $\alpha^{(10)}\gamma^{(10;1)}\beta^{(3)}$}
\begin{center}
\resizebox{\textwidth}{!}{\begin{tabular}{|c|c|c|}
                            $\alpha^{(6)}\gamma^{(6;5)}\beta^{(5)}\delta^{(1)}+(\alpha^{(6)}\gamma^{(6;5)}\beta^{(5)}\delta^{(1)})'$ &  &  \\
                            $\alpha^{(6)}\gamma^{(6;5)}\beta^{(5)}\delta^{(1)}+(\alpha^{(5)}\gamma^{(5;4)}\beta^{(4)})'$ & $\alpha^{(5)}\gamma^{(5;4)}\beta^{(4)}+(\alpha^{(5)}\gamma^{(5;2)}\beta^{(2)})'$ &  \\
                            $\alpha^{(6)}\gamma^{(6;5)}\beta^{(5)}\delta^{(1)}+(\alpha^{(5)}\gamma^{(5;2)}\beta^{(2)})'$ & $\alpha^{(5)}\gamma^{(5;4)}\beta^{(4)}+(\alpha^{(5)}\gamma^{(5;1)}\beta^{(1)})'$ &  \\
                            $\alpha^{(6)}\gamma^{(6;5)}\beta^{(5)}\delta^{(1)}+(\alpha^{(5)}\gamma^{(5;1)}\beta^{(1)})'$ & $\alpha^{(5)}\gamma^{(5;4)}\beta^{(4)}+(\alpha^{(2)}\gamma^{(2;1)}\beta^{(1)})'$ &  \\
                            $\alpha^{(6)}\gamma^{(6;5)}\beta^{(5)}\delta^{(1)}+(\alpha^{(3)}\gamma^{(3;1)}\beta^{(1)})'$ & $\alpha^{(5)}\gamma^{(5;2)}\beta^{(2)}+(\alpha^{(5)}\gamma^{(5;1)}\beta^{(1)})'$ &  \\
                            $\alpha^{(6)}\gamma^{(6;5)}\beta^{(5)}\delta^{(1)}+(\alpha^{(2)}\gamma^{(2;1)}\beta^{(1)})'$ & $\alpha^{(5)}\gamma^{(5;2)}\beta^{(2)}+(\alpha^{(3)}\gamma^{(3;1)}\beta^{(1)})'$ &  \\
                            & $\alpha^{(5)}\gamma^{(5;1)}\beta^{(1)}+(\alpha^{(5)}\gamma^{(5;1)}\beta^{(1)})'$ &  \\
                             & $\alpha^{(5)}\gamma^{(5;1)}\beta^{(1)}+(\alpha^{(3)}\gamma^{(3;1)}\beta^{(1)})'$ &  \\
                             & $\alpha^{(5)}\gamma^{(5;1)}\beta^{(1)}+(\alpha^{(2)}\gamma^{(2;1)}\beta^{(1)})'$ & $\alpha^{(3)}\gamma^{(3;1)}\beta^{(1)}+(\alpha^{(2)}\gamma^{(2;1)}\beta^{(1)})'$ \\
                          \end{tabular}
}
\end{center}
\subsubsection{The orbit $\alpha^{(10)}\gamma^{(10;1)}\beta^{(4)}$}
\begin{center}
\resizebox{\textwidth}{!}{\begin{tabular}{|c|c|c|}
                            $\alpha^{(6)}\gamma^{(6;5)}\beta^{(5)}\delta^{(1)}+(\alpha^{(6)}\gamma^{(6;5)}\beta^{(5)}\delta^{(1)})'$ &  &  \\
                            $\alpha^{(6)}\gamma^{(6;5)}\beta^{(5)}\delta^{(1)}+(\alpha^{(5)}\gamma^{(5;3)}\beta^{(3)})'$ & $\alpha^{(5)}\gamma^{(5;3)}\beta^{(3)}+(\alpha^{(5)}\gamma^{(5;2)}\beta^{(2)})'$ &  \\
                            $\alpha^{(6)}\gamma^{(6;5)}\beta^{(5)}\delta^{(1)}+(\alpha^{(5)}\gamma^{(5;2)}\beta^{(2)})'$ & $\alpha^{(5)}\gamma^{(5;3)}\beta^{(3)}+(\alpha^{(5)}\gamma^{(5;1)}\beta^{(1)})'$ &  \\
                            $\alpha^{(6)}\gamma^{(6;5)}\beta^{(5)}\delta^{(1)}+(\alpha^{(5)}\gamma^{(5;1)}\beta^{(1)})'$ & $\alpha^{(5)}\gamma^{(5;3)}\beta^{(3)}+(\alpha^{(2)}\gamma^{(2;1)}\beta^{(1)})'$ &  \\
                            $\alpha^{(6)}\gamma^{(6;5)}\beta^{(5)}\delta^{(1)}+(\alpha^{(4)}\gamma^{(4;1)}\beta^{(1)})'$ & $\alpha^{(5)}\gamma^{(5;2)}\beta^{(2)}+(\alpha^{(5)}\gamma^{(5;1)}\beta^{(1)})'$ &  \\
                            $\alpha^{(6)}\gamma^{(6;5)}\beta^{(5)}\delta^{(1)}+(\alpha^{(2)}\gamma^{(2;1)}\beta^{(1)})'$ & $\alpha^{(5)}\gamma^{(5;2)}\beta^{(2)}+(\alpha^{(4)}\gamma^{(4;1)}\beta^{(1)})'$ &  \\
                             & $\alpha^{(5)}\gamma^{(5;1)}\beta^{(1)}+(\alpha^{(5)}\gamma^{(5;1)}\beta^{(1)})'$ &  \\
                             & $\alpha^{(5)}\gamma^{(5;1)}\beta^{(1)}+(\alpha^{(4)}\gamma^{(4;1)}\beta^{(1)})'$ & \\
                             & $\alpha^{(5)}\gamma^{(5;1)}\beta^{(1)}+(\alpha^{(2)}\gamma^{(2;1)}\beta^{(1)})'$ & $\alpha^{(4)}\gamma^{(4;1)}\beta^{(1)}+(\alpha^{(2)}\gamma^{(2;1)}\beta^{(1)})'$ \\
                          \end{tabular}
}
\end{center}
\subsubsection{The orbit $\alpha^{(10)}\gamma^{(10;2)}\beta^{(1)}$}
\begin{center}
\resizebox{\textwidth}{!}{\begin{tabular}{|c|c|c|}
                            $\alpha^{(6)}\gamma^{(6;5)}\beta^{(5)}\delta^{(1)}+(\alpha^{(6)}\gamma^{(6;5)}\beta^{(5)}\delta^{(1)})'$ &  &  \\
                            $\alpha^{(6)}\gamma^{(6;5)}\beta^{(5)}\delta^{(1)}+(\alpha^{(5)}\gamma^{(5;4)}\beta^{(4)})'$ & $\alpha^{(5)}\gamma^{(5;4)}\beta^{(4)}+(\alpha^{(5)}\gamma^{(5;3)}\beta^{(3)})'$ &  \\
                            $\alpha^{(6)}\gamma^{(6;5)}\beta^{(5)}\delta^{(1)}+(\alpha^{(5)}\gamma^{(5;3)}\beta^{(3)})'$ & $\alpha^{(5)}\gamma^{(5;4)}\beta^{(4)}+(\alpha^{(5)}\gamma^{(5;2)}\beta^{(2)})'$ &  \\
                            $\alpha^{(6)}\gamma^{(6;5)}\beta^{(5)}\delta^{(1)}+(\alpha^{(5)}\gamma^{(5;2)}\beta^{(2)})'$ & $\alpha^{(5)}\gamma^{(5;4)}\beta^{(4)}+(\alpha^{(3)}\gamma^{(3;2)}\beta^{(2)})'$ &  \\
                            $\alpha^{(6)}\gamma^{(6;5)}\beta^{(5)}\delta^{(1)}+(\alpha^{(4)}\gamma^{(4;2)}\beta^{(2)})'$ & $\alpha^{(5)}\gamma^{(5;3)}\beta^{(3)}+(\alpha^{(5)}\gamma^{(5;2)}\beta^{(2)})'$ &  \\
                            $\alpha^{(6)}\gamma^{(6;5)}\beta^{(5)}\delta^{(1)}+(\alpha^{(3)}\gamma^{(3;2)}\beta^{(2)})'$ & $\alpha^{(5)}\gamma^{(5;3)}\beta^{(3)}+(\alpha^{(4)}\gamma^{(4;2)}\beta^{(2)})'$ &  \\
                             & $\alpha^{(5)}\gamma^{(5;2)}\beta^{(2)}+(\alpha^{(5)}\gamma^{(5;2)}\beta^{(2)})'$ &  \\
                             & $\alpha^{(5)}\gamma^{(5;2)}\beta^{(2)}+(\alpha^{(4)}\gamma^{(4;2)}\beta^{(2)})'$ &  \\
                             & $\alpha^{(5)}\gamma^{(5;2)}\beta^{(2)}+(\alpha^{(3)}\gamma^{(3;2)}\beta^{(2)})'$ & $\alpha^{(4)}\gamma^{(4;2)}\beta^{(2)}+(\alpha^{(3)}\gamma^{(3;2)}\beta^{(2)})'$ \\
                          \end{tabular}
}
\end{center}
\subsubsection{The orbit $\alpha^{(10)}\gamma^{(10;2)}\beta^{(3)}$}
\begin{center}
\resizebox{\textwidth}{!}{\begin{tabular}{|c|c|c|}
                            $\alpha^{(6)}\gamma^{(6;5)}\beta^{(5)}\delta^{(1)}+(\alpha^{(6)}\gamma^{(6;5)}\beta^{(5)}\delta^{(1)})'$ &  &  \\
                            $\alpha^{(6)}\gamma^{(6;5)}\beta^{(5)}\delta^{(1)}+(\alpha^{(5)}\gamma^{(5;4)}\beta^{(4)})'$ & $\alpha^{(5)}\gamma^{(5;4)}\beta^{(4)}+(\alpha^{(5)}\gamma^{(5;2)}\beta^{(2)})'$ &  \\
                            $\alpha^{(6)}\gamma^{(6;5)}\beta^{(5)}\delta^{(1)}+(\alpha^{(5)}\gamma^{(5;2)}\beta^{(2)})'$ & $\alpha^{(5)}\gamma^{(5;4)}\beta^{(4)}+(\alpha^{(5)}\gamma^{(5;1)}\beta^{(1)})'$ &  \\
                            $\alpha^{(6)}\gamma^{(6;5)}\beta^{(5)}\delta^{(1)}+(\alpha^{(5)}\gamma^{(5;1)}\beta^{(1)})'$ & $\alpha^{(5)}\gamma^{(5;4)}\beta^{(4)}+(\alpha^{(2)}\gamma^{(2;1)}\beta^{(1)})'$ &  \\
                            $\alpha^{(6)}\gamma^{(6;5)}\beta^{(5)}\delta^{(1)}+(\alpha^{(4)}\gamma^{(4;2)}\beta^{(2)})'$ & $\alpha^{(5)}\gamma^{(5;2)}\beta^{(2)}+(\alpha^{(5)}\gamma^{(5;2)}\beta^{(2)})'$ &  \\
                            $\alpha^{(6)}\gamma^{(6;5)}\beta^{(5)}\delta^{(1)}+(\alpha^{(2)}\gamma^{(2;1)}\beta^{(1)})'$ & $\alpha^{(5)}\gamma^{(5;2)}\beta^{(2)}+(\alpha^{(5)}\gamma^{(5;1)}\beta^{(1)})'$ &  \\
                             & $\alpha^{(5)}\gamma^{(5;2)}\beta^{(2)}+(\alpha^{(4)}\gamma^{(4;2)}\beta^{(2)})'$ &  \\
                             & $\alpha^{(5)}\gamma^{(5;2)}\beta^{(2)}+(\alpha^{(2)}\gamma^{(2;1)}\beta^{(1)})'$ &  \\
                             & $\alpha^{(5)}\gamma^{(5;1)}\beta^{(1)}+(\alpha^{(4)}\gamma^{(4;2)}\beta^{(2)})'$ & $\alpha^{(4)}\gamma^{(4;2)}\beta^{(2)}+(\alpha^{(2)}\gamma^{(2;1)}\beta^{(1)})'$ \\
                          \end{tabular}
}
\end{center}
\subsubsection{The orbit $\alpha^{(10)}\gamma^{(10;2)}\beta^{(4)}$}
\begin{center}
\resizebox{\textwidth}{!}{\begin{tabular}{|c|c|c|}
                            $\alpha^{(6)}\gamma^{(6;5)}\beta^{(5)}\delta^{(1)}+(\alpha^{(6)}\gamma^{(6;5)}\beta^{(5)}\delta^{(1)})'$ &  &  \\
                            $\alpha^{(6)}\gamma^{(6;5)}\beta^{(5)}\delta^{(1)}+(\alpha^{(5)}\gamma^{(5;3)}\beta^{(3)})'$ & $\alpha^{(5)}\gamma^{(5;3)}\beta^{(3)}+(\alpha^{(5)}\gamma^{(5;2)}\beta^{(2)})'$ &  \\
                            $\alpha^{(6)}\gamma^{(6;5)}\beta^{(5)}\delta^{(1)}+(\alpha^{(5)}\gamma^{(5;2)}\beta^{(2)})'$ & $\alpha^{(5)}\gamma^{(5;3)}\beta^{(3)}+(\alpha^{(5)}\gamma^{(5;1)}\beta^{(1)})'$ &  \\
                            $\alpha^{(6)}\gamma^{(6;5)}\beta^{(5)}\delta^{(1)}+(\alpha^{(5)}\gamma^{(5;1)}\beta^{(1)})'$ & $\alpha^{(5)}\gamma^{(5;3)}\beta^{(3)}+(\alpha^{(2)}\gamma^{(2;1)}\beta^{(1)})'$ &  \\
                            $\alpha^{(6)}\gamma^{(6;5)}\beta^{(5)}\delta^{(1)}+(\alpha^{(3)}\gamma^{(3;2)}\beta^{(2)})'$ & $\alpha^{(5)}\gamma^{(5;2)}\beta^{(2)}+(\alpha^{(5)}\gamma^{(5;1)}\beta^{(1)})'$ &  \\
                            $\alpha^{(6)}\gamma^{(6;5)}\beta^{(5)}\delta^{(1)}+(\alpha^{(2)}\gamma^{(2;1)}\beta^{(1)})'$ & $\alpha^{(5)}\gamma^{(5;2)}\beta^{(2)}+(\alpha^{(3)}\gamma^{(3;2)}\beta^{(2)})'$ &  \\
                            & $\alpha^{(5)}\gamma^{(5;2)}\beta^{(2)}+(\alpha^{(5)}\gamma^{(5;2)}\beta^{(2)})'$ &  \\
                             & $\alpha^{(5)}\gamma^{(5;2)}\beta^{(2)}+(\alpha^{(2)}\gamma^{(2;1)}\beta^{(1)})'$ &  \\
                             & $\alpha^{(5)}\gamma^{(5;1)}\beta^{(1)}+(\alpha^{(3)}\gamma^{(3;2)}\beta^{(2)})'$ & $\alpha^{(3)}\gamma^{(3;2)}\beta^{(2)}+(\alpha^{(2)}\gamma^{(2;1)}\beta^{(1)})'$ \\
                          \end{tabular}
}
\end{center}
\subsubsection{The orbit $\alpha^{(10)}\gamma^{(10;3)}\beta^{(1)}$}
\begin{center}
\resizebox{\textwidth}{!}{\begin{tabular}{|c|c|c|}
                            $\alpha^{(6)}\gamma^{(6;5)}\beta^{(5)}\delta^{(1)}+(\alpha^{(6)}\gamma^{(6;5)}\beta^{(5)}\delta^{(1)})'$ &  &  \\
                            $\alpha^{(6)}\gamma^{(6;5)}\beta^{(5)}\delta^{(1)}+(\alpha^{(5)}\gamma^{(5;4)}\beta^{(4)})'$ & $\alpha^{(5)}\gamma^{(5;4)}\beta^{(4)}+(\alpha^{(5)}\gamma^{(5;3)}\beta^{(3)})'$ &  \\
                            $\alpha^{(6)}\gamma^{(6;5)}\beta^{(5)}\delta^{(1)}+(\alpha^{(5)}\gamma^{(5;3)}\beta^{(3)})'$ & $\alpha^{(5)}\gamma^{(5;4)}\beta^{(4)}+(\alpha^{(5)}\gamma^{(5;2)}\beta^{(2)})'$ &  \\
                            $\alpha^{(6)}\gamma^{(6;5)}\beta^{(5)}\delta^{(1)}+(\alpha^{(5)}\gamma^{(5;2)}\beta^{(2)})'$ & $\alpha^{(5)}\gamma^{(5;4)}\beta^{(4)}+(\alpha^{(3)}\gamma^{(3;2)}\beta^{(2)})'$ &  \\
                            $\alpha^{(6)}\gamma^{(6;5)}\beta^{(5)}\delta^{(1)}+(\alpha^{(3)}\gamma^{(3;2)}\beta^{(2)})'$ & $\alpha^{(5)}\gamma^{(5;3)}\beta^{(3)}+(\alpha^{(5)}\gamma^{(5;3)}\beta^{(3)})'$ &  \\
                            $\alpha^{(6)}\gamma^{(6;5)}\beta^{(5)}\delta^{(1)}+(\alpha^{(2)}\gamma^{(2;2)}\beta^{(2)})'$ & $\alpha^{(5)}\gamma^{(5;3)}\beta^{(3)}+(\alpha^{(5)}\gamma^{(5;2)}\beta^{(2)})'$ &  \\
                             & $\alpha^{(5)}\gamma^{(5;3)}\beta^{(3)}+(\alpha^{(3)}\gamma^{(3;2)}\beta^{(2)})'$ &  \\
                             & $\alpha^{(5)}\gamma^{(5;3)}\beta^{(3)}+(\alpha^{(2)}\gamma^{(2;2)}\beta^{(2)})'$ &  \\
                             & $\alpha^{(5)}\gamma^{(5;2)}\beta^{(2)}+(\alpha^{(2)}\gamma^{(2;2)}\beta^{(2)})'$ & $\alpha^{(3)}\gamma^{(3;2)}\beta^{(2)}+(\alpha^{(2)}\gamma^{(2;2)}\beta^{(2)})'$ \\
                          \end{tabular}
}
\end{center}
\subsubsection{The orbit $\alpha^{(10)}\gamma^{(10;3)}\beta^{(2)}$}
\begin{center}
\resizebox{\textwidth}{!}{\begin{tabular}{|c|c|c|}
                            $\alpha^{(6)}\gamma^{(6;5)}\beta^{(5)}\delta^{(1)}+(\alpha^{(6)}\gamma^{(6;5)}\beta^{(5)}\delta^{(1)})'$ &  &  \\
                            $\alpha^{(6)}\gamma^{(6;5)}\beta^{(5)}\delta^{(1)}+(\alpha^{(5)}\gamma^{(5;4)}\beta^{(4)})'$ & $\alpha^{(5)}\gamma^{(5;4)}\beta^{(4)}+(\alpha^{(5)}\gamma^{(5;3)}\beta^{(3)})'$ &  \\
                            $\alpha^{(6)}\gamma^{(6;5)}\beta^{(5)}\delta^{(1)}+(\alpha^{(5)}\gamma^{(5;3)}\beta^{(3)})'$ & $\alpha^{(5)}\gamma^{(5;4)}\beta^{(4)}+(\alpha^{(5)}\gamma^{(5;1)}\beta^{(1)})'$ &  \\
                            $\alpha^{(6)}\gamma^{(6;5)}\beta^{(5)}\delta^{(1)}+(\alpha^{(5)}\gamma^{(5;1)}\beta^{(1)})'$ & $\alpha^{(5)}\gamma^{(5;4)}\beta^{(4)}+(\alpha^{(4)}\gamma^{(4;1)}\beta^{(1)})'$ &  \\
                            $\alpha^{(6)}\gamma^{(6;5)}\beta^{(5)}\delta^{(1)}+(\alpha^{(4)}\gamma^{(4;1)}\beta^{(1)})'$ & $\alpha^{(5)}\gamma^{(5;3)}\beta^{(3)}+(\alpha^{(5)}\gamma^{(5;3)}\beta^{(3)})'$ &  \\
                            $\alpha^{(6)}\gamma^{(6;5)}\beta^{(5)}\delta^{(1)}+(\alpha^{(2)}\gamma^{(2;2)}\beta^{(2)})'$ & $\alpha^{(5)}\gamma^{(5;3)}\beta^{(3)}+(\alpha^{(5)}\gamma^{(5;1)}\beta^{(1)})'$ &  \\
                             & $\alpha^{(5)}\gamma^{(5;3)}\beta^{(3)}+(\alpha^{(4)}\gamma^{(4;1)}\beta^{(1)})'$ &  \\
                             & $\alpha^{(5)}\gamma^{(5;3)}\beta^{(3)}+(\alpha^{(2)}\gamma^{(2;2)}\beta^{(2)})'$ &  \\
                             & $\alpha^{(5)}\gamma^{(5;1)}\beta^{(1)}+(\alpha^{(2)}\gamma^{(2;2)}\beta^{(2)})'$ & $\alpha^{(4)}\gamma^{(4;1)}\beta^{(1)}+(\alpha^{(2)}\gamma^{(2;2)}\beta^{(2)})'$ \\
                          \end{tabular}
}
\end{center}
\subsubsection{The orbit $\alpha^{(10)}\gamma^{(10;3)}\beta^{(4)}$}
\begin{center}
\resizebox{\textwidth}{!}{\begin{tabular}{|c|c|c|}
                            $\alpha^{(6)}\gamma^{(6;5)}\beta^{(5)}\delta^{(1)}+(\alpha^{(6)}\gamma^{(6;5)}\beta^{(5)}\delta^{(1)})'$ &  &  \\
                            $\alpha^{(6)}\gamma^{(6;5)}\beta^{(5)}\delta^{(1)}+(\alpha^{(5)}\gamma^{(5;3)}\beta^{(3)})'$ & $\alpha^{(5)}\gamma^{(5;3)}\beta^{(3)}+(\alpha^{(5)}\gamma^{(5;3)}\beta^{(3)})'$ &  \\
                            $\alpha^{(6)}\gamma^{(6;5)}\beta^{(5)}\delta^{(1)}+(\alpha^{(5)}\gamma^{(5;2)}\beta^{(2)})'$ & $\alpha^{(5)}\gamma^{(5;3)}\beta^{(3)}+(\alpha^{(5)}\gamma^{(5;2)}\beta^{(2)})'$ &  \\
                            $\alpha^{(6)}\gamma^{(6;5)}\beta^{(5)}\delta^{(1)}+(\alpha^{(5)}\gamma^{(5;1)}\beta^{(1)})'$ & $\alpha^{(5)}\gamma^{(5;3)}\beta^{(3)}+(\alpha^{(5)}\gamma^{(5;1)}\beta^{(1)})'$ &  \\
                            $\alpha^{(6)}\gamma^{(6;5)}\beta^{(5)}\delta^{(1)}+(\alpha^{(4)}\gamma^{(4;1)}\beta^{(1)})'$ & $\alpha^{(5)}\gamma^{(5;3)}\beta^{(3)}+(\alpha^{(4)}\gamma^{(4;1)}\beta^{(1)})'$ &  \\
                            $\alpha^{(6)}\gamma^{(6;5)}\beta^{(5)}\delta^{(1)}+(\alpha^{(3)}\gamma^{(3;2)}\beta^{(2)})'$ & $\alpha^{(5)}\gamma^{(5;3)}\beta^{(3)}+(\alpha^{(3)}\gamma^{(3;2)}\beta^{(2)})'$ &  \\
                            & $\alpha^{(5)}\gamma^{(5;2)}\beta^{(2)}+(\alpha^{(5)}\gamma^{(5;1)}\beta^{(1)})'$ &  \\
                             & $\alpha^{(5)}\gamma^{(5;2)}\beta^{(2)}+(\alpha^{(4)}\gamma^{(4;1)}\beta^{(1)})'$ &  \\
                             & $\alpha^{(5)}\gamma^{(5;1)}\beta^{(1)}+(\alpha^{(3)}\gamma^{(3;2)}\beta^{(2)})'$ & $\alpha^{(4)}\gamma^{(4;1)}\beta^{(1)}+(\alpha^{(3)}\gamma^{(3;2)}\beta^{(2)})'$ \\
                          \end{tabular}
}
\end{center}
\subsubsection{The orbit $\alpha^{(10)}\gamma^{(10;4)}\beta^{(1)}$}
\begin{center}
\resizebox{\textwidth}{!}{\begin{tabular}{|c|c|c|}
                            $\alpha^{(6)}\gamma^{(6;5)}\beta^{(5)}\delta^{(1)}+(\alpha^{(6)}\gamma^{(6;5)}\beta^{(5)}\delta^{(1)})'$ &  &  \\
                            $\alpha^{(6)}\gamma^{(6;5)}\beta^{(5)}\delta^{(1)}+(\alpha^{(5)}\gamma^{(5;4)}\beta^{(4)})'$ & $\alpha^{(5)}\gamma^{(5;4)}\beta^{(4)}+(\alpha^{(5)}\gamma^{(5;4)}\beta^{(4)})'$ &  \\
                            $\alpha^{(6)}\gamma^{(6;5)}\beta^{(5)}\delta^{(1)}+(\alpha^{(5)}\gamma^{(5;3)}\beta^{(3)})'$ & $\alpha^{(5)}\gamma^{(5;4)}\beta^{(4)}+(\alpha^{(5)}\gamma^{(5;3)}\beta^{(3)})'$ &  \\
                            $\alpha^{(6)}\gamma^{(6;5)}\beta^{(5)}\delta^{(1)}+(\alpha^{(5)}\gamma^{(5;2)}\beta^{(2)})'$ & $\alpha^{(5)}\gamma^{(5;4)}\beta^{(4)}+(\alpha^{(5)}\gamma^{(5;2)}\beta^{(2)})'$ &  \\
                            $\alpha^{(6)}\gamma^{(6;5)}\beta^{(5)}\delta^{(1)}+(\alpha^{(4)}\gamma^{(4;2)}\beta^{(2)})'$ & $\alpha^{(5)}\gamma^{(5;4)}\beta^{(4)}+(\alpha^{(4)}\gamma^{(4;2)}\beta^{(2)})'$ &  \\
                            $\alpha^{(6)}\gamma^{(6;5)}\beta^{(5)}\delta^{(1)}+(\alpha^{(2)}\gamma^{(2;2)}\beta^{(2)})'$ & $\alpha^{(5)}\gamma^{(5;4)}\beta^{(4)}+(\alpha^{(2)}\gamma^{(2;2)}\beta^{(2)})'$ &  \\
                             & $\alpha^{(5)}\gamma^{(5;3)}\beta^{(3)}+(\alpha^{(5)}\gamma^{(5;2)}\beta^{(2)})'$ &  \\
                             & $\alpha^{(5)}\gamma^{(5;3)}\beta^{(3)}+(\alpha^{(4)}\gamma^{(4;2)}\beta^{(2)})'$ &  \\
                             & $\alpha^{(5)}\gamma^{(5;2)}\beta^{(2)}+(\alpha^{(2)}\gamma^{(2;2)}\beta^{(2)})'$ & $\alpha^{(4)}\gamma^{(4;2)}\beta^{(2)}+(\alpha^{(2)}\gamma^{(2;2)}\beta^{(2)})'$ \\
                          \end{tabular}
}
\end{center}
\subsubsection{The orbit $\alpha^{(10)}\gamma^{(10;4)}\beta^{(2)}$}
\begin{center}
\resizebox{\textwidth}{!}{\begin{tabular}{|c|c|c|}
                            $\alpha^{(6)}\gamma^{(6;5)}\beta^{(5)}\delta^{(1)}+(\alpha^{(6)}\gamma^{(6;5)}\beta^{(5)}\delta^{(1)})'$ &  &  \\
                            $\alpha^{(6)}\gamma^{(6;5)}\beta^{(5)}\delta^{(1)}+(\alpha^{(5)}\gamma^{(5;4)}\beta^{(4)})'$ & $\alpha^{(5)}\gamma^{(5;4)}\beta^{(4)}+(\alpha^{(5)}\gamma^{(5;4)}\beta^{(4)})'$ &  \\
                            $\alpha^{(6)}\gamma^{(6;5)}\beta^{(5)}\delta^{(1)}+(\alpha^{(5)}\gamma^{(5;3)}\beta^{(3)})'$ & $\alpha^{(5)}\gamma^{(5;4)}\beta^{(4)}+(\alpha^{(5)}\gamma^{(5;3)}\beta^{(3)})'$ &  \\
                            $\alpha^{(6)}\gamma^{(6;5)}\beta^{(5)}\delta^{(1)}+(\alpha^{(5)}\gamma^{(5;1)}\beta^{(1)})'$ & $\alpha^{(5)}\gamma^{(5;4)}\beta^{(4)}+(\alpha^{(5)}\gamma^{(5;1)}\beta^{(1)})'$ &  \\
                            $\alpha^{(6)}\gamma^{(6;5)}\beta^{(5)}\delta^{(1)}+(\alpha^{(3)}\gamma^{(3;1)}\beta^{(1)})'$ & $\alpha^{(5)}\gamma^{(5;4)}\beta^{(4)}+(\alpha^{(3)}\gamma^{(3;1)}\beta^{(1)})'$ &  \\
                            $\alpha^{(6)}\gamma^{(6;5)}\beta^{(5)}\delta^{(1)}+(\alpha^{(2)}\gamma^{(2;2)}\beta^{(2)})'$ & $\alpha^{(5)}\gamma^{(5;4)}\beta^{(4)}+(\alpha^{(2)}\gamma^{(2;2)}\beta^{(2)})'$ &  \\
                             & $\alpha^{(5)}\gamma^{(5;3)}\beta^{(3)}+(\alpha^{(5)}\gamma^{(5;1)}\beta^{(1)})'$ &  \\
                             & $\alpha^{(5)}\gamma^{(5;3)}\beta^{(3)}+(\alpha^{(3)}\gamma^{(3;1)}\beta^{(1)})'$ &  \\
                             & $\alpha^{(5)}\gamma^{(5;1)}\beta^{(1)}+(\alpha^{(2)}\gamma^{(2;2)}\beta^{(2)})'$ & $\alpha^{(3)}\gamma^{(3;1)}\beta^{(1)}+(\alpha^{(2)}\gamma^{(2;2)}\beta^{(2)})'$ \\
                          \end{tabular}
}
\end{center}
\subsubsection{The orbit $\alpha^{(10)}\gamma^{(10;4)}\beta^{(3)}$}
\begin{center}
\resizebox{\textwidth}{!}{\begin{tabular}{|c|c|c|}
                            $\alpha^{(6)}\gamma^{(6;5)}\beta^{(5)}\delta^{(1)}+(\alpha^{(6)}\gamma^{(6;5)}\beta^{(5)}\delta^{(1)})'$ &  &  \\
                            $\alpha^{(6)}\gamma^{(6;5)}\beta^{(5)}\delta^{(1)}+(\alpha^{(5)}\gamma^{(5;4)}\beta^{(4)})'$ & $\alpha^{(5)}\gamma^{(5;4)}\beta^{(4)}+(\alpha^{(5)}\gamma^{(5;4)}\beta^{(4)})'$ &  \\
                            $\alpha^{(6)}\gamma^{(6;5)}\beta^{(5)}\delta^{(1)}+(\alpha^{(5)}\gamma^{(5;2)}\beta^{(2)})'$ & $\alpha^{(5)}\gamma^{(5;4)}\beta^{(4)}+(\alpha^{(5)}\gamma^{(5;2)}\beta^{(2)})'$ &  \\
                            $\alpha^{(6)}\gamma^{(6;5)}\beta^{(5)}\delta^{(1)}+(\alpha^{(5)}\gamma^{(5;1)}\beta^{(1)})'$ & $\alpha^{(5)}\gamma^{(5;4)}\beta^{(4)}+(\alpha^{(5)}\gamma^{(5;1)}\beta^{(1)})'$ &  \\
                            $\alpha^{(6)}\gamma^{(6;5)}\beta^{(5)}\delta^{(1)}+(\alpha^{(4)}\gamma^{(4;2)}\beta^{(2)})'$ & $\alpha^{(5)}\gamma^{(5;4)}\beta^{(4)}+(\alpha^{(4)}\gamma^{(4;2)}\beta^{(2)})'$ &  \\
                            $\alpha^{(6)}\gamma^{(6;5)}\beta^{(5)}\delta^{(1)}+(\alpha^{(3)}\gamma^{(3;1)}\beta^{(1)})'$ & $\alpha^{(5)}\gamma^{(5;4)}\beta^{(4)}+(\alpha^{(3)}\gamma^{(3;1)}\beta^{(1)})'$ &  \\
                              & $\alpha^{(5)}\gamma^{(5;2)}\beta^{(2)}+(\alpha^{(5)}\gamma^{(5;1)}\beta^{(1)})'$ &  \\
                             & $\alpha^{(5)}\gamma^{(5;2)}\beta^{(2)}+(\alpha^{(3)}\gamma^{(3;1)}\beta^{(1)})'$ &  \\
                             & $\alpha^{(5)}\gamma^{(5;1)}\beta^{(1)}+(\alpha^{(4)}\gamma^{(4;2)}\beta^{(2)})'$ & $\alpha^{(4)}\gamma^{(4;2)}\beta^{(2)}+(\alpha^{(3)}\gamma^{(3;1)}\beta^{(1)})'$ \\
                          \end{tabular}
}
\end{center}
\subsubsection{The orbit $\alpha^{(11)}\gamma^{(11;1)}\beta^{(1)}$}
\begin{center}
\resizebox{\textwidth}{!}{\begin{tabular}{|c|c|c|}
                            $\alpha^{(6)}\gamma^{(6;5)}\beta^{(5)}\delta^{(1)}+(\alpha^{(6)}\gamma^{(6;1)}\beta^{(1)})'$ & $\alpha^{(6)}\gamma^{(6;1)}\beta^{(1)}+(\alpha^{(6)}\gamma^{(6;1)}\beta^{(1)})'$ & \\
                            $\alpha^{(6)}\gamma^{(6;5)}\beta^{(5)}\delta^{(1)}+(\alpha^{(5)}\gamma^{(5;1)}\beta^{(1)})'$ & $\alpha^{(6)}\gamma^{(6;1)}\beta^{(1)}+(\alpha^{(5)}\gamma^{(5;4)}\beta^{(4)})'$ &  \\
                             & $\alpha^{(6)}\gamma^{(6;1)}\beta^{(1)}+(\alpha^{(5)}\gamma^{(5;3)}\beta^{(3)})'$ & $\alpha^{(5)}\gamma^{(5;1)}\beta^{(1)}+(\alpha^{(5)}\gamma^{(5;4)}\beta^{(4)})'$\\
                             & $\alpha^{(6)}\gamma^{(6;1)}\beta^{(1)}+(\alpha^{(5)}\gamma^{(5;2)}\beta^{(2)})'$ & $\alpha^{(5)}\gamma^{(5;1)}\beta^{(1)}+(\alpha^{(5)}\gamma^{(5;3)}\beta^{(3)})'$\\
                             & $\alpha^{(6)}\gamma^{(6;1)}\beta^{(1)}+(\alpha^{(5)}\gamma^{(5;1)}\beta^{(1)})'$ & $\alpha^{(5)}\gamma^{(5;1)}\beta^{(1)}+(\alpha^{(5)}\gamma^{(5;2)}\beta^{(2)})'$\\
                             & $\alpha^{(6)}\gamma^{(6;1)}\beta^{(1)}+(\alpha^{(4)}\gamma^{(4;2)}\beta^{(2)})'$ & $\alpha^{(5)}\gamma^{(5;1)}\beta^{(1)}+(\alpha^{(4)}\gamma^{(4;2)}\beta^{(2)})'$ \\
                             & $\alpha^{(6)}\gamma^{(6;1)}\beta^{(1)}+(\alpha^{(3)}\gamma^{(3;2)}\beta^{(2)})'$ & $\alpha^{(5)}\gamma^{(5;1)}\beta^{(1)}+(\alpha^{(3)}\gamma^{(3;2)}\beta^{(2)})'$ \\
                            & $\alpha^{(6)}\gamma^{(6;1)}\beta^{(1)}+(\alpha^{(2)}\gamma^{(2;2)}\beta^{(2)})'$ & $\alpha^{(5)}\gamma^{(5;1)}\beta^{(1)}+(\alpha^{(2)}\gamma^{(2;2)}\beta^{(2)})'$  \\
                            & $\alpha^{(6)}\gamma^{(6;1)}\beta^{(1)}+(\alpha^{(1)}\gamma^{(1;1)}\beta^{(1)})'$ & $\alpha^{(5)}\gamma^{(5;1)}\beta^{(1)}+(\alpha^{(1)}\gamma^{(1;1)}\beta^{(1)})'$ \\
                          \end{tabular}
}
\end{center}
\subsubsection{The orbit $\alpha^{(11)}\gamma^{(11;2)}\beta^{(2)}$}
\begin{center}
\resizebox{\textwidth}{!}{\begin{tabular}{|c|c|c|}
                            $\alpha^{(6)}\gamma^{(6;5)}\beta^{(5)}\delta^{(1)}+(\alpha^{(6)}\gamma^{(6;2)}\beta^{(2)})'$  & $\alpha^{(6)}\gamma^{(6;2)}\beta^{(2)}+(\alpha^{(6)}\gamma^{(6;2)}\beta^{(2)})'$ &  \\
                            $\alpha^{(6)}\gamma^{(6;5)}\beta^{(5)}\delta^{(1)}+(\alpha^{(5)}\gamma^{(5;2)}\beta^{(2)})'$ & $\alpha^{(6)}\gamma^{(6;2)}\beta^{(2)}+(\alpha^{(5)}\gamma^{(5;4)}\beta^{(4)})'$ &  \\
                             & $\alpha^{(6)}\gamma^{(6;2)}\beta^{(2)}+(\alpha^{(5)}\gamma^{(5;3)}\beta^{(3)})'$ & $\alpha^{(5)}\gamma^{(5;2)}\beta^{(2)}+(\alpha^{(5)}\gamma^{(5;4)}\beta^{(4)})'$\\
                             & $\alpha^{(6)}\gamma^{(6;2)}\beta^{(2)}+(\alpha^{(5)}\gamma^{(5;2)}\beta^{(2)})'$ & $\alpha^{(5)}\gamma^{(5;2)}\beta^{(2)}+(\alpha^{(5)}\gamma^{(5;3)}\beta^{(3)})'$\\
                             & $\alpha^{(6)}\gamma^{(6;2)}\beta^{(2)}+(\alpha^{(5)}\gamma^{(5;1)}\beta^{(1)})'$ & $\alpha^{(5)}\gamma^{(5;2)}\beta^{(2)}+(\alpha^{(5)}\gamma^{(5;1)}\beta^{(1)})'$\\
                             & $\alpha^{(6)}\gamma^{(6;2)}\beta^{(2)}+(\alpha^{(4)}\gamma^{(4;1)}\beta^{(1)})'$ & $\alpha^{(5)}\gamma^{(5;2)}\beta^{(2)}+(\alpha^{(4)}\gamma^{(4;1)}\beta^{(1)})'$ \\
                             & $\alpha^{(6)}\gamma^{(6;2)}\beta^{(2)}+(\alpha^{(3)}\gamma^{(3;1)}\beta^{(1)})'$ & $\alpha^{(5)}\gamma^{(5;2)}\beta^{(2)}+(\alpha^{(3)}\gamma^{(3;1)}\beta^{(1)})'$ \\
                            & $\alpha^{(6)}\gamma^{(6;2)}\beta^{(2)}+(\alpha^{(2)}\gamma^{(2;2)}\beta^{(2)})'$ & $\alpha^{(5)}\gamma^{(5;2)}\beta^{(2)}+(\alpha^{(2)}\gamma^{(2;2)}\beta^{(2)})'$  \\
                            & $\alpha^{(6)}\gamma^{(6;2)}\beta^{(2)}+(\alpha^{(1)}\gamma^{(1;1)}\beta^{(1)})'$ & $\alpha^{(5)}\gamma^{(5;2)}\beta^{(2)}+(\alpha^{(1)}\gamma^{(1;1)}\beta^{(1)})'$ \\
                          \end{tabular}}
\end{center}
\subsubsection{The orbit $\alpha^{(11)}\gamma^{(11;3)}\beta^{(3)}$}
\begin{center}
\resizebox{\textwidth}{!}{\begin{tabular}{|c|c|c|}
                             $\alpha^{(6)}\gamma^{(6;5)}\beta^{(5)}\delta^{(1)}+(\alpha^{(6)}\gamma^{(6;3)}\beta^{(3)})'$ &$\alpha^{(6)}\gamma^{(6;3)}\beta^{(3)}+(\alpha^{(6)}\gamma^{(6;3)}\beta^{(3)})'$ &  \\
                            $\alpha^{(6)}\gamma^{(6;5)}\beta^{(5)}\delta^{(1)}+(\alpha^{(5)}\gamma^{(5;3)}\beta^{(3)})'$ & $\alpha^{(6)}\gamma^{(6;3)}\beta^{(3)}+(\alpha^{(5)}\gamma^{(5;4)}\beta^{(4)})'$ &  \\
                             & $\alpha^{(6)}\gamma^{(6;3)}\beta^{(3)}+(\alpha^{(5)}\gamma^{(5;3)}\beta^{(3)})'$ & $\alpha^{(5)}\gamma^{(5;3)}\beta^{(3)}+(\alpha^{(5)}\gamma^{(5;4)}\beta^{(4)})'$\\
                             & $\alpha^{(6)}\gamma^{(6;3)}\beta^{(3)}+(\alpha^{(5)}\gamma^{(5;2)}\beta^{(2)})'$ & $\alpha^{(5)}\gamma^{(5;3)}\beta^{(3)}+(\alpha^{(5)}\gamma^{(5;2)}\beta^{(2)})'$\\
                             & $\alpha^{(6)}\gamma^{(6;3)}\beta^{(3)}+(\alpha^{(5)}\gamma^{(5;1)}\beta^{(1)})'$ & $\alpha^{(5)}\gamma^{(5;3)}\beta^{(3)}+(\alpha^{(5)}\gamma^{(5;1)}\beta^{(1)})'$\\
                             & $\alpha^{(6)}\gamma^{(6;3)}\beta^{(3)}+(\alpha^{(4)}\gamma^{(4;2)}\beta^{(2)})'$ & $\alpha^{(5)}\gamma^{(5;3)}\beta^{(3)}+(\alpha^{(4)}\gamma^{(4;2)}\beta^{(2)})'$ \\
                             & $\alpha^{(6)}\gamma^{(6;3)}\beta^{(3)}+(\alpha^{(3)}\gamma^{(3;1)}\beta^{(1)})'$ & $\alpha^{(5)}\gamma^{(5;3)}\beta^{(3)}+(\alpha^{(3)}\gamma^{(3;1)}\beta^{(1)})'$ \\
                            & $\alpha^{(6)}\gamma^{(6;3)}\beta^{(3)}+(\alpha^{(2)}\gamma^{(2;1)}\beta^{(1)})'$ & $\alpha^{(5)}\gamma^{(5;3)}\beta^{(3)}+(\alpha^{(2)}\gamma^{(2;1)}\beta^{(1)})'$  \\
                            & $\alpha^{(6)}\gamma^{(6;3)}\beta^{(3)}+(\alpha^{(1)}\gamma^{(1;1)}\beta^{(1)})'$ & $\alpha^{(5)}\gamma^{(5;3)}\beta^{(3)}+(\alpha^{(1)}\gamma^{(1;1)}\beta^{(1)})'$ \\
                          \end{tabular}}
\end{center}
\subsubsection{The orbit $\alpha^{(11)}\gamma^{(11;4)}\beta^{(4)}$}
\begin{center}
\resizebox{\textwidth}{!}{\begin{tabular}{|c|c|c|}
                             $\alpha^{(6)}\gamma^{(6;5)}\beta^{(5)}\delta^{(1)}+(\alpha^{(6)}\gamma^{(6;4)}\beta^{(4)})'$ & $\alpha^{(6)}\gamma^{(6;4)}\beta^{(4)}+(\alpha^{(6)}\gamma^{(6;4)}\beta^{(4)})'$ &  \\
                            $\alpha^{(6)}\gamma^{(6;5)}\beta^{(5)}\delta^{(1)}+(\alpha^{(5)}\gamma^{(5;4)}\beta^{(4)})'$ & $\alpha^{(6)}\gamma^{(6;4)}\beta^{(4)}+(\alpha^{(5)}\gamma^{(5;4)}\beta^{(4)})'$ &  \\
                             & $\alpha^{(6)}\gamma^{(6;4)}\beta^{(4)}+(\alpha^{(5)}\gamma^{(5;3)}\beta^{(3)})'$ & $\alpha^{(5)}\gamma^{(5;4)}\beta^{(4)}+(\alpha^{(5)}\gamma^{(5;3)}\beta^{(3)})'$\\
                             & $\alpha^{(6)}\gamma^{(6;4)}\beta^{(4)}+(\alpha^{(5)}\gamma^{(5;2)}\beta^{(2)})'$ & $\alpha^{(5)}\gamma^{(5;4)}\beta^{(4)}+(\alpha^{(5)}\gamma^{(5;2)}\beta^{(2)})'$\\
                             & $\alpha^{(6)}\gamma^{(6;4)}\beta^{(4)}+(\alpha^{(5)}\gamma^{(5;1)}\beta^{(1)})'$ & $\alpha^{(5)}\gamma^{(5;4)}\beta^{(4)}+(\alpha^{(5)}\gamma^{(5;1)}\beta^{(1)})'$\\
                             & $\alpha^{(6)}\gamma^{(6;4)}\beta^{(4)}+(\alpha^{(4)}\gamma^{(4;1)}\beta^{(1)})'$ & $\alpha^{(5)}\gamma^{(5;4)}\beta^{(4)}+(\alpha^{(4)}\gamma^{(4;1)}\beta^{(1)})'$ \\
                             & $\alpha^{(6)}\gamma^{(6;4)}\beta^{(4)}+(\alpha^{(3)}\gamma^{(3;2)}\beta^{(2)})'$ & $\alpha^{(5)}\gamma^{(5;4)}\beta^{(4)}+(\alpha^{(3)}\gamma^{(3;2)}\beta^{(2)})'$ \\
                            & $\alpha^{(6)}\gamma^{(6;4)}\beta^{(4)}+(\alpha^{(2)}\gamma^{(2;1)}\beta^{(1)})'$ & $\alpha^{(5)}\gamma^{(5;4)}\beta^{(4)}+(\alpha^{(2)}\gamma^{(2;1)}\beta^{(1)})'$  \\
                            & $\alpha^{(6)}\gamma^{(6;4)}\beta^{(4)}+(\alpha^{(1)}\gamma^{(1;1)}\beta^{(1)})'$ & $\alpha^{(5)}\gamma^{(5;4)}\beta^{(4)}+(\alpha^{(1)}\gamma^{(1;1)}\beta^{(1)})'$ \\
                          \end{tabular}}
\end{center}
\section{Mathematical Details of the STU Model}\label{mdstu}
The explicit matrix form of a generic element $\phi^s\,T_s$  of the solvable Lie algebra parametrized by the $D=4$ scalar fields $z_i=\epsilon_i-i\,e^{\varphi_i}$, in the symplectic representation ${\bf R}$, reads:
\begin{align}
\phi^s\,T_s&=\sum_{i=1}^3\epsilon_i E_{\boldsymbol{\alpha}_i}+\varphi_i \frac{H_{\boldsymbol{\alpha}_i}}{2}=\left(\begin{matrix}A & B \cr {\bf 0} & -A^T\end{matrix}\right)\,,\nonumber\\
A&=\left(
\begin{array}{llll}
 \frac{\varphi _1}{2}+\frac{\varphi _2}{2}+\frac{\varphi _3}{2} & -\epsilon _1 & -\epsilon _2 & -\epsilon _3 \\
 0 & -\frac{\varphi _1}{2}+\frac{\varphi _2}{2}+\frac{\varphi _3}{2} & 0 & 0 \\
 0 & 0 & \frac{\varphi _1}{2}-\frac{\varphi _2}{2}+\frac{\varphi _3}{2} & 0 \\
 0 & 0 & 0 & \frac{\varphi _1}{2}+\frac{\varphi _2}{2}-\frac{\varphi _3}{2}
\end{array}
\right)\,,\nonumber\\
B&= \left(
\begin{array}{llll}
 0 & 0 & 0 & 0 \\
 0 & 0 & -\epsilon _3 & -\epsilon _2 \\
 0 & -\epsilon _3 & 0 & -\epsilon _1 \\
 0 & -\epsilon _2 & -\epsilon _1 & 0
\end{array}
\right)\,.
\end{align}
 If the STU model originates from Kaluza--Klein reduction from $D=5$, the resulting symplectic frame corresponds to the following ordering of the
    roots $\gamma_M$, $M=1,\dots,8$:
  \begin{align}
  (\Gamma_M)&=(\mathbb{C}_{MN}\Gamma^N)=(q_\Lambda,\,-p^\Lambda)\leftrightarrow\,\,(\gamma_M)\,,\nonumber\\
  (\vec{\gamma}_a)_{a=1,\dots,4}&=\left[\left(\frac{1}{2},-\frac{1}{2},-\frac{1}{2},-\frac{1}{2}\right),
  \left(\frac{1}{2},\frac{1}{2},-\frac{1}{2},-\frac{1}{2}\right),\,\left(\frac{1}{2},-\frac{1}{2},\frac{1}{2},-\frac{1}{2}\right),\,\left(\frac{1}{2},-\frac{1}{2},-\frac{1}{2},\frac{1}{2}\right)\right]\,,\nonumber\\
  (\vec{\gamma}_{a+4})_{a=1,\dots,4}&=\left[\left(\frac{1}{2},\frac{1}{2},\frac{1}{2},\frac{1}{2}\right),
  \left(\frac{1}{2},-\frac{1}{2},\frac{1}{2},\frac{1}{2}\right),\,\left(\frac{1}{2},\frac{1}{2},-\frac{1}{2},\frac{1}{2}\right),\,\left(\frac{1}{2},\frac{1}{2},\frac{1}{2},-\frac{1}{2}\right)\right]\,,\label{gammaord}
  \end{align}
  where we have represented each root $\gamma_M$ by its component vector $\vec{\gamma}_M$ in a Cartan subalgebra of
  $\mathfrak{so}(4,4)$:\footnote{The components of $\gamma_M$ in an orthonormal basis are obtained by multiplying $\vec{\gamma}_M$ by $\sqrt{2}$.} The first component is the grading $\gamma_M(H_0)$ with respect to the ${\rm O}(1,1)$ generator $H_0$ in the
  Ehlers group ${\rm SL}(2,\mathbb{R})_E$, the other entries are the components $\gamma_M(H_{\boldsymbol{\alpha}_i})/2$, with respect to the Cartan
  generators $H_{\boldsymbol{\alpha}_i}$ of $G_4$.\par
  The explicit form of the matrix $\mathcal{M}_{(4)\,MN}$ in (\ref{M4MN}) is:
  \begin{equation}
    \mathcal{M}_{(4)\,MN}=-e^{-(\varphi_1+\varphi_2+\varphi_3)}\,\left(\begin{matrix}{\bf M}_{11} & {\bf M}_{12}\cr {\bf M}_{12}^T & {\bf M}_{22} \end{matrix}\right)\,,
  \end{equation}
  where the $4\times 4$ blocks are:
 \begin{align}
 {\bf M}_{11}&= \left(
\begin{array}{llll}
 -|z_1|^2 |z_2|^2 |z_3|^2 & \epsilon _1
   |z_2|^2 |z_3|^2 & |z_1|^2 \epsilon _2
   |z_3|^2 & |z_1|^2 |z_2|^2 \epsilon _3 \\
 \epsilon _1 |z_2|^2 |z_3|^2 & -|z_2|^2
   |z_3|^2 & -\epsilon _1 \epsilon _2 |z_3|^2 & -\epsilon _1 |z_2|^2 \epsilon _3 \\
 |z_1|^2 \epsilon _2 |z_3|^2 & -\epsilon _1 \epsilon _2 |z_3|^2 & -|z_1|^2 |z_3|^2 & -|z_1|^2 \epsilon _2 \epsilon _3 \\
 |z_1|^2 |z_2|^2 \epsilon _3 & -\epsilon _1 |z_2|^2 \epsilon _3 & -|z_1|^2 \epsilon _2 \epsilon _3 & -|z_1|^2
   |z_2|^2
\end{array}
\right)\,,\nonumber\\
 {\bf M}_{12}&= \left(
\begin{array}{llll}
 -\epsilon _1 \epsilon _2 \epsilon _3 & -|z_1|^2 \epsilon _2 \epsilon _3 & -\epsilon _1 |z_2|^2 \epsilon _3 & -\epsilon _1 \epsilon _2 |z_3|^2 \\
 \epsilon _2 \epsilon _3 & \epsilon _1 \epsilon _2 \epsilon _3 & |z_2|^2 \epsilon _3 & \epsilon _2 |z_3|^2 \\
 \epsilon _1 \epsilon _3 & |z_1|^2 \epsilon _3 & \epsilon _1 \epsilon _2 \epsilon _3 & \epsilon _1 |z_3|^2 \\
 \epsilon _1 \epsilon _2 & |z_1|^2 \epsilon _2 & \epsilon _1 |z_2|^2 & \epsilon _1
   \epsilon _2 \epsilon _3
\end{array}
\right)\,,\nonumber\\
 {\bf M}_{22}&=\left(
\begin{array}{llll}
 -1 & -\epsilon _1 & -\epsilon _2 & -\epsilon _3 \\
 -\epsilon _1 & -|z_1|^2 & -\epsilon _1 \epsilon _2 & -\epsilon _1 \epsilon _3 \\
 -\epsilon _2 & -\epsilon _1 \epsilon _2 & -|z_2|^2 & -\epsilon _2 \epsilon _3 \\
 -\epsilon _3 & -\epsilon _1 \epsilon _3 & -\epsilon _2 \epsilon _3 & -|z_3|^2
\end{array}
\right)\,.
  \end{align}
A black hole solution with electric and magnetic quantized charges $p^\Lambda,\,q_\Lambda$ is described at radial infinity by $H_4$-covariant central and matter charges $Z,\,Z_1,\,Z_2,\,Z_3$, functions of $p^\Lambda,\,q_\Lambda$ and of the asymptotic values of the scalar fields. In terms of
$V^M$ and of its covariant derivatives $D_i$ ($D_iV:=\partial_i V+\frac{\partial_i K}{2}\,V$) we write the central and matter charges of the black hole solution:

\begin{align}
Z&=-V^T\mathbb{C}\,\Gamma
=e^{\frac{K}{2}}\,(-q_0-q_1 z_1-q_2 z_2+p^3 z_1 z_2-q_3 z_3+p^2 z_1 z_3+p^1 z_2 z_3-p^0 z_1 z_2 z_3)\,,\nonumber\\
 Z_1&=-e_{ 1}{}^i\,D_iV^T\mathbb{C}\,\Gamma=-i\,e^{\frac{K}{2}}\,\left(q_0+q_2 z_2+q_3 z_3-p^1 z_2 z_3+q_1 \bar{z}_1-p^3 z_2 \bar{z}_1-p^2 z_3 \bar{z}_1+p^0 z_2 z_3 \bar{z}_1\right)\,,\nonumber\\
 Z_2&=-e_{ 2}{}^i\,D_iV^T\mathbb{C}\,\Gamma=-i\,e^{\frac{K}{2}}\,\left(q_0+q_1 z_1+q_3 z_3-p^2 z_1 z_3+q_2 \bar{z}_2-p^3 z_1 \bar{z}_2-p^1 z_3 \bar{z}_2+p^0 z_1 z_3 \bar{z}_2\right)\,,\nonumber\\
 Z_3&=-e_{ 3}{}^i\,D_iV^T\mathbb{C}\,\Gamma=-i\,e^{\frac{K}{2}}\,\left(q_0+q_1 z_1+q_2 z_2-p^3 z_1 z_2+q_3 \bar{z}_3-p^2 z_1 \bar{z}_3-p^1 z_2 \bar{z}_3+p^0 z_1 z_2 \bar{z}_3\right)\,.
\end{align}
Let us also give the explicit form of the quartic invariant for the STU model:
\begin{align}
I_4(p,q)&=-(p^0)^2 q_0^2-2 \left(-2 p^1 p^2 p^3+p^0 q_3 p^3+p^0 p^1 q_1+p^0 p^2 q_2\right) q_0-(p^1)^2 q_1^2-\left(p^2 q_2-p^3 q_3\right)^2+\nonumber\\
&+2 q_1
   \left(p^1 p^3 q_3+q_2 \left(p^1 p^2-2 p^0 q_3\right)\right)\,.\label{qinv}
\end{align}
\subsection{The Scalar Manifold Geometry in $D=3$}\label{mdstu2}
The $\mathfrak{g}=\mathfrak{so}(4,4)$ algebra is described in its fundamental representation in terms of $8\times 8$ matrices leaving the following metric invariant:
\begin{equation}
\eta_0=e_{1,8}+e_{2,7}+e_{3,6}+e_{4,5}+e_{5,4}+e_{6,3}+e_{7,2}+e_{8,1}\,,
\end{equation}
where $e_{i,j}$ denotes a matrix whose only non vanishing entry is a $1$ in the $(i,j)$-position. The matrix $\eta_0$ should not be mistaken for the $H^*$-invariant matrix $\eta$ defining the $\sigma$-involution of Sect. \ref{sec1}:
\begin{equation}
\eta=-e_{1,1}-e_{2,2}+e_{3,3}+e_{4,4}+e_{5,5}+e_{6,6}-e_{7,7}-e_{8,8}\,.
\end{equation}
The solvable generators read (for the sake of simplicity we write $T_M$ as $T_1,\dots T_8$):
\begin{align}
H_0&=\frac{1}{2}\,\left(e_{1,1}+e_{2,2}-e_{7,7}-e_{8,8}\right)\,,\nonumber\\
H_{\boldsymbol{\alpha}_1}&=H_{{\alpha}_1}=e_{1,1}-e_{2,2}+e_{7,7}-e_{8,8}\,\,,\,\,\,H_{\boldsymbol{\alpha}_2}=H_{{\alpha}_3}=e_{3,3}+e_{4,4}-e_{5,5}-e_{6,6}\,,\nonumber\\
H_{\boldsymbol{\alpha}_3}&=H_{{\alpha}_4}=e_{3,3}-e_{4,4}+e_{5,5}-e_{6,6}\,,
\nonumber\\
E_{\boldsymbol{\alpha}_1}&=E_{{\alpha}_1}=e_{1,2}-e_{7,8}\,\,,\,\,\,E_{\boldsymbol{\alpha}_2}=E_{{\alpha}_3}=e_{3,5}-e_{4,6}
\,\,,\,\,\,E_{\boldsymbol{\alpha}_3}=E_{{\alpha}_4}=e_{3,4}-e_{5,6}\,,
\nonumber\\
T_1&=e_{6,7}-e_{2,3}\,\,,\,\,\,T_2=e_{6,8}-e_{1,3}\,\,,\,\,\,T_3=e_{2,5}-e_{4,7}\,\,,\,\,\,T_4=e_{2,4}-e_{5,7}\,,\nonumber\\
T_5&=e_{3,8}-e_{1,6}\,\,,\,\,\,T_6=e_{2,6}-e_{3,7}\,\,,\,\,\,T_7=e_{1,4}-e_{5,8}\,\,,\,\,\,T_8=e_{1,5}-e_{4,8}\,,\nonumber\\
T_\bullet&=e_{1,7}-e_{2,8}\,.
\end{align}
Constructing the matrix $\mathcal{M}(\phi^I)=\mathbb{L}(\phi^I)\eta \mathbb{L}(\phi^I)^\dagger$, with $\mathbb{L}$ given in (\ref{cosetr3}), we have the following identification of the $D=3$ scalar fields with the entries $m_{i,j}$ of $\mathcal{M}$:
\begin{align}
e^{-4{U}}&= m_{7,7} m_{8,8}-m_{8,7}^2\quad,
\end{align}
\begin{align}
{\varphi_{1}}&= -\frac{1}{2} \log \left(\frac{m_{8,8}^2}{m_{7,7}m_{8,8}-m_{8,7}^2}\right)\quad,
\end{align}
\begin{equation}
\begin{split}
\varphi_{2}&=-\frac{1}{2} \log \bigg[\frac{1}{m_{8,7}^2-m_{7,7} m_{8,8}}\big[-m_{8,7}^2 m_{6,5}^2+m_{7,7} m_{8,8} m_{6,5}^2-2 m_{7,7} m_{8,5} m_{8,6} m_{6,5}+\\
&+2 m_{7,5} m_{8,6} m_{8,7} m_{6,5}-m_{7,5}^2 m_{8,6}^2+m_{5,5} m_{7,7} m_{8,6}^2+m_{7,6}^2 \big(m_{5,5} m_{8,8}-m_{8,5}^2\big)+\\
&+m_{6,6} \big(m_{7,7} m_{8,5}^2-2 m_{7,5} m_{8,7} m_{8,5}+m_{5,5} m_{8,7}^2+\left(m_{7,5}^2-m_{5,5} m_{7,7}\right) m_{8,8}\big)+\\
&+2 m_{7,6} \big(\big(m_{6,5}m_{8,5}-m_{5,5} m_{8,6}\big) m_{8,7}+m_{7,5} \big(m_{8,5} m_{8,6}-m_{6,5} m_{8,8}\big)\big)\big]\bigg]\,,
\end{split}
\end{equation}
\begin{equation}
\begin{split}
\varphi_{3}&= \frac{1}{2} \log \bigg[\frac{1}{\big(m_{8,8} m_{7,6}^2-2 m_{8,6} m_{8,7} m_{7,6}+m_{6,6} m_{8,7}^2+m_{7,7} \big(m_{8,6}^2-m_{6,6}m_{8,8}\big)\big)^2}\cdot\\
&\cdot\left(m_{8,7}^2-m_{7,7} m_{8,8}\right) \bigg(-m_{8,7}^2 m_{6,5}^2+m_{7,7} m_{8,8} m_{6,5}^2-2 m_{7,7} m_{8,5} m_{8,6} m_{6,5}+2 m_{7,5} m_{8,6} m_{8,7} m_{6,5}+\\
&-m_{7,5}^2 m_{8,6}^2+m_{5,5} m_{7,7} m_{8,6}^2+m_{7,6}^2 \left(m_{5,5} m_{8,8}-m_{8,5}^2\right)+m_{6,6} \big(m_{7,7} m_{8,5}^2-2 m_{7,5} m_{8,7}m_{8,5}+\\
&+m_{5,5} m_{8,7}^2+\left(m_{7,5}^2-m_{5,5} m_{7,7}\right) m_{8,8}\big)+2 m_{7,6} \big(\left(m_{6,5} m_{8,5}-m_{5,5} m_{8,6}\right) m_{8,7}+\\
&+m_{7,5} \left(m_{8,5}m_{8,6}-m_{6,5} m_{8,8}\right)\big)\bigg)\bigg]\quad,
\end{split}
\end{equation}
\begin{align}
\epsilon_{1}= -\frac{m_{8,7}}{m_{8,8}}\quad,
\end{align}
\begin{align}
\epsilon_{2}=  \frac{m_{8,7} \left(m_{7,6} m_{8,4}+m_{7,4} m_{8,6}-m_{6,4} m_{8,7}\right)-m_{7,4} m_{7,6} m_{8,8}+m_{7,7} \left(m_{6,4}
   m_{8,8}-m_{8,4} m_{8,6}\right)}{m_{8,8} m_{7,6}^2-2 m_{8,6} m_{8,7} m_{7,6}+m_{6,6} m_{8,7}^2+m_{7,7} \left(m_{8,6}^2-m_{6,6}
   m_{8,8}\right)}\quad,
\end{align}
\begin{align}
\epsilon_{3}= \frac{m_{8,7} \left(m_{7,6} m_{8,5}+m_{7,5} m_{8,6}-m_{6,5} m_{8,7}\right)-m_{7,5} m_{7,6} m_{8,8}+m_{7,7} \left(m_{6,5}
   m_{8,8}-m_{8,5} m_{8,6}\right)}{m_{8,8} m_{7,6}^2-2 m_{8,6} m_{8,7} m_{7,6}+m_{6,6} m_{8,7}^2+m_{7,7} \left(m_{8,6}^2-m_{6,6}
   m_{8,8}\right)} \quad,
\end{align}
\begin{align}
\mathcal{Z}^{0}=\frac{m_{8,6} m_{8,7}-m_{7,6} m_{8,8}}{\sqrt{2} \left(m_{8,7}^2-m_{7,7} m_{8,8}\right)} \quad,\quad \mathcal{Z}_{0}= \frac{m_{7,3} m_{8,7}-m_{7,7} m_{8,3}}{\sqrt{2} \left(m_{8,7}^2-m_{7,7} m_{8,8}\right)}\quad,
\end{align}
\begin{align}
\mathcal{Z}^{1}= \frac{m_{7,6} m_{8,7}-m_{7,7} m_{8,6}}{\sqrt{2} \left(m_{8,7}^2-m_{7,7} m_{8,8}\right)} \quad,\quad \mathcal{Z}_{1}= \frac{m_{7,3} m_{8,8}-m_{8,3} m_{8,7}}{\sqrt{2} \left(m_{8,7}^2-m_{7,7} m_{8,8}\right)} \quad,
\end{align}
\begin{align}
\mathcal{Z}^{2}= \frac{m_{7,4} m_{8,8}-m_{8,4} m_{8,7}}{\sqrt{2} \left(m_{8,7}^2-m_{7,7} m_{8,8}\right)} \quad,\quad \mathcal{Z}_{2}=\frac{m_{7,7} m_{8,5}-m_{7,5} m_{8,7}}{\sqrt{2} \left(m_{8,7}^2-m_{7,7} m_{8,8}\right)} \quad,
\end{align}
\begin{align}
\mathcal{Z}^{3}= \frac{m_{7,5} m_{8,8}-m_{8,5} m_{8,7}}{\sqrt{2} \left(m_{8,7}^2-m_{7,7} m_{8,8}\right)} \quad,\quad \mathcal{Z}_{3}= \frac{m_{7,7} m_{8,4}-m_{7,4} m_{8,7}}{\sqrt{2} \left(m_{8,7}^2-m_{7,7} m_{8,8}\right)} \quad,
\end{align}
\begin{equation}
\begin{split}
a =&\frac{1}{2 m_{8,8} \left(m_{8,7}^2-m_{7,7} m_{8,8}\right)}\cdot\bigg[2 m_{8,7} \big(m_{8,4} m_{8,5}+m_{8,3} m_{8,6}+m_{8,2} m_{8,7}\big)-\big(2 m_{7,7} m_{8,2}+\\
&+m_{7,6} m_{8,3}+m_{7,5} m_{8,4}+m_{7,4} m_{8,5}+m_{7,3} m_{8,6}\big) m_{8,8}\bigg] \quad.
\end{split}
\end{equation}
The nilpotent elements in $\mathfrak{K}^*$ are described as eigenmatrices with respect to a suitable non-compact Cartan subalgebra in the coset space of $H^*/H_c$. Its generators are:
{\scriptsize \begin{align}
H_{\beta_1}&=\frac{1}{2}\,(e_{1,4}+e_{1,5}+e_{2,3}+e_{2,6}+e_{3,2}-e_{3,7}+e_{4,1}-e_{4,8}+e_{5,1}-e_{5,8}+e_{6,2}-e_{6,7
   }-e_{7,3}-e_{7,6}-e_{8,4}-e_{8,5})\,,\nonumber\\
H_{\beta_2}&=-\frac{1}{2}\,(-e_{1,4}-e_{1,5}+e_{2,3}+e_{2,6}+e_{3,2}-e_{3,7}-e_{4,1}+e_{4,8}-e_{5,1}+e_{5,8}+e_{6,2}-e_{6,
   7}-e_{7,3}-e_{7,6}+e_{8,4}+e_{8,5})\,,\nonumber\\
H_{\beta_3}&=-\frac{1}{2}\,(e_{1,4}-e_{1,5}+e_{2,3}-e_{2,6}+e_{3,2}+e_{3,7}+e_{4,1}+e_{4,8}-e_{5,1}-e_{5,8}-e_{6,2}-e_{6,7
   }+e_{7,3}-e_{7,6}+e_{8,4}-e_{8,5})\,,\nonumber\\
H_{\beta_4}&=-\frac{1}{2}\,(-e_{1,4}+e_{1,5}+e_{2,3}-e_{2,6}+e_{3,2}+e_{3,7}-e_{4,1}-e_{4,8}+e_{5,1}+e_{5,8}-e_{6,2}-e_{6,
   7}+e_{7,3}-e_{7,6}-e_{8,4}+e_{8,5})\,.\nonumber\\
\end{align}}
The chosen basis for $\mathfrak{K}^*$ is:
{\scriptsize \begin{align}
(+,+,+,+)&=\frac{1}{2} (-e_{1,2}-e_ {1,3}+e_ {1,6}-e_ {1,7}-e_ {2,1}+e_ {2,4}+e_ {2,5}+e_ {2,8}+e_ {3,1}-e_ {3,4}-e_ {3,5}-e_ {3,8}-e_ {4,2}-e_ {4,3}+e_ {4,6}-e_ {4,7}-e_ {5,2}-\nonumber\\&-e_ {5,3}+e_ {5,6}-e_ {5,7}-e_ {6,1}+e_ {6,4}+e_ {6,5}+e_ {6,8}-e_ {7,1}+e_ {7,4}+e_ {7,5}+e_ {7,8}+e_ {8,2}+e_ {8,3}-e_ {8,6}+e_ {8,7})\,,\nonumber\\
(+,+,-,-) &= \frac{1}{2} (e_ {1,2}-e_ {1,3}+e_ {1,6}+e_ {1,7}+e_ {2,1}-e_ {2,4}-e_ {2,5}-e_ {2,8}+e_ {3,1}-e_ {3,4}-e_ {3,5}-e_ {3,8}+e_ {4,2}-e_ {4,3}+e_ {4,6}+e_ {4,7}+e_ {5,2}-\nonumber\\&-e_ {5,3}+e_ {5,6}+e_ {5,7}-e_ {6,1}+e_ {6,4}+e_ {6,5}+e_ {6,8}+e_ {7,1}-e_ {7,4}-e_ {7,5}-e_ {7,8}-e_ {8,2}+e_ {8,3}-e_ {8,6}-e_ {8,7})\,,\nonumber\\
(+,-,+,-) &=  \frac{1}{2} (e_ {1,2}-e_ {1,3}-e_ {1,6}-e_ {1,7}+e_ {2,1}+e_ {2,4}-e_ {2,5}+e_ {2,8}+e_ {3,1}+e_ {3,4}-e_ {3,5}+e_ {3,8}-e_ {4,2}+e_ {4,3}+e_ {4,6}+e_ {4,7}+e_ {5,2}-\nonumber\\&-e_ {5,3}-e_ {5,6}-e_ {5,7}+e_ {6,1}+e_ {6,4}-e_ {6,5}+e_ {6,8}-e_ {7,1}-e_ {7,4}+e_ {7,5}-e_ {7,8}+e_ {8,2}-e_ {8,3}-e_ {8,6}-e_ {8,7})\,,\nonumber\\
(+,-,-,+) &=   \frac{1}{2} (e_ {1,2}-e_ {1,3}-e_ {1,6}-e_ {1,7}+e_ {2,1}-e_ {2,4}+e_ {2,5}+e_ {2,8}+e_ {3,1}-e_ {3,4}+e_ {3,5}+e_ {3,8}+e_ {4,2}-e_ {4,3}-e_ {4,6}-e_ {4,7}-e_ {5,2}+\nonumber\\&+e_ {5,3}+e_ {5,6}+e_ {5,7}+e_ {6,1}-e_ {6,4}+e_ {6,5}+e_ {6,8}-e_ {7,1}+e_ {7,4}-e_ {7,5}-e_ {7,8}+e_ {8,2}-e_ {8,3}-e_ {8,6}-e_ {8,7})\,,\nonumber\\
(+,-,-,-) &=  -e_{2,2}+e_ {2,3}-e_ {3,2}+e_ {3,3}-e_ {6,6}-e_ {6,7}+e_ {7,6}+e_ {7,7}\,,\nonumber\\
(+,-,+,+)  &= e_ {2,2}-e_ {2,6}+e_ {3,3}+e_ {3,7}+e_ {6,2}-e_ {6,6}-e_ {7,3}-e_ {7,7}\,,\nonumber\\
(+,+,-,+)&=  -e_{1,1}+e_ {1,4}-e_ {4,1}+e_ {4,4}-e_ {5,5}-e_ {5,8}+e_ {8,5}+e_ {8,8}\,,\nonumber\\
(+,+,+,-) &=  -e_{1,1}+e_ {1,5}-e_ {4,4}-e_ {4,8}-e_ {5,1}+e_ {5,5}+e_ {8,4}+e_ {8,8}\,,\nonumber\\
(-,+,+,+)&= -(+,-,-,-)^T\,\,,\,\,\,(-,+,-,-)=-(+,-,+,+)^T\,\,,\,\,\,(-,-,+,-)=- (+,+,-,+)^T\,\,,\,\,\,(-,-,-,+)=- (+,+,+,-)^T  \,,\nonumber\\
(-,-,-,-)&=(+,+,+,+)^T\,,\,\,\,(-,-,+,+)=(+,+,-,-)^T\,,\,\,\,(-,+,-,+)=(+,-,+,-)^T\,,\,\,\,(-,+,+,-)=(+,-,-,+)^T\,,
\end{align}}
where the $\pm$ gradings in the QUbit-basis refer to the above $H_{\beta_\ell}$ generators.

\begin{thebibliography}{1}

\bibitem{Gaillard:1981rj}
  M.~K.~Gaillard and B.~Zumino,
``Duality Rotations for Interacting Fields,''
  Nucl.\ Phys.\ B {\bf 193} (1981) 221.
  \bibitem{Breitenlohner:1987dg}
 P.~Breitenlohner, D.~Maison and G.~W.~Gibbons,
  ``Four-Dimensional black holes from Kaluza-Klein Theories,''
  Commun.\ Math.\ Phys.\  {\bf 120} (1988) 295.
   \bibitem{reviews} For reviews on black holes in superstring and supergravity theories see for example:
 J. M. Maldacena, ``Black-Holes in String Theory'', hep-th/9607235;
A.~W.~Peet, ``TASI lectures on black holes in string theory,''
  hep-th/0008241;
  B.~Pioline,
  ``Lectures on on black holes, topological strings and quantum attractors,''
  Class.\ Quant.\ Grav.\  {\bf 23} (2006) S981
  [arXiv:hep-th/0607227];
A.~Dabholkar,
  ``black hole Entropy And Attractors,''
  Class.\ Quant.\ Grav.\  {\bf 23} (2006) S957.
 L.~Andrianopoli, R.~D'Auria, S.~Ferrara and M.~Trigiante,
``Extremal black holes in supergravity,''
  Lect.\ Notes Phys.\  {\bf 737} (2008) 661
  [arXiv:hep-th/0611345];


 \bibitem{Cvetic:1995kv}M.~Cvetic and D.~Youm,
``All the static spherically symmetric black holes of heterotic string on a six torus,''
  Nucl.\ Phys.\ B {\bf 472} (1996) 249;
  M.~Cvetic and D.~Youm,
``Entropy of nonextreme charged rotating black holes in string theory,''
  Phys.\ Rev.\ D {\bf 54} (1996) 2612
  [hep-th/9603147].
 \bibitem{pioline}
M.~Gunaydin, A.~Neitzke, B.~Pioline  and A.~Waldron, ``BPS
black holes,
  quantum attractor flows and automorphic forms'', Phys. Rev. {\bf D73} (2006)
  084019;
 M.~Gunaydin, A.~Neitzke, B.~Pioline and A.~Waldron,
``Quantum Attractor Flows,''
  JHEP {\bf 0709} (2007) 056
  doi:10.1088/1126-6708/2007/09/056
  [arXiv:0707.0267 [hep-th]].
\bibitem{Gaiotto:2007ag}  D.~Gaiotto, W.~W. Li  and M.~Padi, ``Non-Supersymmetric
Attractor Flow in
  Symmetric Spaces'', JHEP {\bf 12} (2007) 093,
\bibitem{Bergshoeff:2008be}
  E. Bergshoeff, W. Chemissany, A.~Ploegh, M.~Trigiante and T.~Van Riet,
 ``Generating Geodesic Flows and Supergravity Solutions,''
  Nucl.\ Phys.\ {\bf B 812} (2009) 343
  [arXiv:0806.2310].
  \bibitem{Chemissany:2009hq}
  W.~Chemissany, J.~Rosseel, M.~Trigiante and T.~Van Riet,
  ``The Full integration of black hole solutions to symmetric supergravity theories,''
  Nucl.\ Phys.\ B {\bf 830} (2010) 391
  doi:10.1016/j.nuclphysb.2009.11.013
  [arXiv:0903.2777 [hep-th]].
\bibitem{Bossard:2009at}
  G.~Bossard, H.~Nicolai, K.~S.~Stelle,
 ``Universal BPS structure of stationary supergravity solutions'',
  JHEP {\bf 0907 } (2009)  003.
  \bibitem{Bossard:2009we}
   G.~Bossard, Y.~Michel and B.~Pioline,
 ``Extremal black holes, nilpotent orbits and the true fake superpotential,''
  JHEP {\bf 1001} (2010) 038
  [arXiv:0908.1742 [hep-th]].
 \bibitem{Kim:2010bf}
  S.~-S.~Kim, J.~Lindman Hornlund, J.~Palmkvist and A.~Virmani,
  ``Extremal Solutions of the S3 Model and Nilpotent Orbits of G2(2),''
  JHEP {\bf 1008} (2010) 072.
\bibitem{Chemissany:2010zp}
  W.~Chemissany, P.~Fre, J.~Rosseel, A.~S.~Sorin, M.~Trigiante and T.~Van Riet,
``Black holes in supergravity and integrability,''
  JHEP {\bf 1009} (2010) 080
  doi:10.1007/JHEP09(2010)080
  [arXiv:1007.3209 [hep-th]].
  \bibitem{Fre:2011uy}
  P.~Fre, A.~S.~Sorin and M.~Trigiante,
 ``Integrability of Supergravity black holes and New Tensor Classifiers of Regular and Nilpotent Orbits,''
  JHEP {\bf 1204} (2012) 015
  [arXiv:1103.0848 [hep-th]].

\bibitem{Dietrich:2016ojx}
  H.~Dietrich, W.~A.~de Graaf, D.~Ruggeri and M.~Trigiante,
 ``Nilpotent orbits in real symmetric pairs,''
  arXiv:1606.02611 [math.RT].

  \bibitem{bossard2}  G.~Bossard and C.~Ruef,
  ``Interacting non-BPS black holes,''
  Gen.\ Rel.\ Grav.\  {\bf 44} (2012) 21;
  G.~Bossard,
  ``Octonionic black holes,''
  JHEP {\bf 1205} (2012) 113.
  \bibitem{Bossard:2013nwa}
  G.~Bossard and S.~Katmadas,
``non-BPS walls of marginal stability,''
  JHEP {\bf 1310} (2013) 179.
\bibitem{Chemissany:2012nb}
  W.~Chemissany, P.~Giaccone, D.~Ruggeri and M.~Trigiante,
``Black hole solutions to the $F_4$-model and their orbits (I),''
  Nucl.\ Phys.\ B {\bf 863} (2012) 260.
  \bibitem{Fre:2012im}
  P.~Fre and A.~S.~Sorin,
 ``Extremal Multicenter Black Holes: Nilpotent Orbits and Tits Satake Universality Classes,''
  JHEP {\bf 1301} (2013) 003.
   \bibitem{Cvetic:2013cja}
  M.~Cvetic, M.~Guica and Z.~H.~Saleem,
``General black holes, untwisted,''
  JHEP {\bf 1309} (2013) 017.
  \bibitem{BPSmc}
  B.~Bates and F.~Denef,
``Exact solutions for supersymmetric stationary black hole composites,''
  JHEP {\bf 1111} (2011) 127;
   F.~Denef,
``On the correspondence between D-branes and stationary supergravity solutions of type II Calabi-Yau compactifications,''
  hep-th/0010222;
   F.~Denef,
``Supergravity flows and D-brane stability,''
  JHEP {\bf 0008} (2000) 050;
   K.~Behrndt, D.~Lust and W.~A.~Sabra,
``Stationary solutions of N=2 supergravity,''
  Nucl.\ Phys.\ B {\bf 510} (1998) 264;
   G.~Lopes Cardoso, B.~de Wit, J.~Kappeli and T.~Mohaupt,
``Stationary BPS solutions in N=2 supergravity with R**2 interactions,''
  JHEP {\bf 0012} (2000) 019.
     \bibitem{Andrianopoli:2013kya} L.~Andrianopoli, R.~D'Auria, A.~Gallerati and M.~Trigiante,
 ``Extremal Limits of Rotating black holes,''
  JHEP {\bf 1305} (2013) 071.
  \bibitem{Andrianopoli:2013jra}
  L.~Andrianopoli, A.~Gallerati and M.~Trigiante,
``On Extremal Limits and Duality Orbits of Stationary black holes,''
  JHEP {\bf 1401} (2014) 053
  [arXiv:1310.7886 [hep-th]].
\bibitem{Chow:2013tia}
  D.~D.~K.~Chow and G.~Compère,
 ``Seed for general rotating non-extremal black holes of $\mathcal {N}= 8$ supergravity,''
  Class.\ Quant.\ Grav.\  {\bf 31} (2014) 022001
  [arXiv:1310.1925 [hep-th]].
\bibitem{Chow:2014cca}
  D.~D.~K.~Chow and G.~Compère,
``Black holes in N=8 supergravity from SO(4,4) hidden symmetries,''
  Phys.\ Rev.\ D {\bf 90} (2014) 2,  025029
  [arXiv:1404.2602 [hep-th]].
  \bibitem{Bertini:2011ga}  S.~Bertini, S.~L.~Cacciatori and D.~Klemm,
  ``Conformal structure of the Schwarzschild black hole,''
  Phys.\ Rev.\ D {\bf 85} (2012) 064018.
\bibitem{Cvetic:2011dn}
  M.~Cvetic and F.~Larsen,
  ``Conformal Symmetry for black holes in Four Dimensions,''
  JHEP {\bf 1209} (2012) 076.

\bibitem{Virmani:2012kw}
 A.~Virmani,
 ``Subtracted Geometry From Harrison Transformations,''
  JHEP {\bf 1207} (2012) 086;
   A.~Chakraborty and C.~Krishnan,
 ``Attraction, with Boundaries,''
  arXiv:1212.6919 [hep-th].
 \bibitem{Andrianopoli:2012ee}  L.~Andrianopoli, R.~D'Auria, P.~Giaccone and M.~Trigiante,
  ``Rotating black holes, global symmetry and first order formalism,''
  JHEP {\bf 1212} (2012) 078.
\bibitem{deBoer:2014iba}
  J.~de Boer, D.~R.~Mayerson and M.~Shigemori,
``Classifying Supersymmetric Solutions in 3D Maximal Supergravity,''
  Class.\ Quant.\ Grav.\  {\bf 31} (2014) no.23,  235004.
\bibitem{Deger:2015tra}
  N.~S.~Deger, G.~Moutsopoulos, H.~Samtleben and Ö.~Sarioglu,
  ``All timelike supersymmetric solutions of three-dimensional half-maximal supergravity,''
  JHEP {\bf 1506} (2015) 147.
\bibitem{Rasheed}
  D.~Rasheed,
 ``The Rotating dyonic black holes of Kaluza-Klein theory,''
  Nucl.\ Phys.\ B {\bf 454} (1995) 379;
 \bibitem{Larsen}    F.~Larsen,
 ``Rotating Kaluza-Klein black holes,''
  Nucl.\ Phys.\ B {\bf 575} (2000) 211.
  \bibitem{Astefanesei}  D.~Astefanesei, K.~Goldstein, R.~P.~Jena, A.~Sen and S.~P.~Trivedi,
 ``Rotating attractors,''
  JHEP {\bf 0610} (2006) 058.
  \bibitem{Fre:2011uy}
  P.~Fre, A.~S.~Sorin and M.~Trigiante,
``Integrability of Supergravity black holes and New Tensor Classifiers of Regular and Nilpotent Orbits,''
  JHEP {\bf 1204} (2012) 015
  [arXiv:1103.0848 [hep-th]].
  \bibitem{Fre:2011ns}
  P.~Fre, A.~S.~Sorin and M.~Trigiante,
``black hole Nilpotent Orbits and Tits Satake Universality Classes,''
  arXiv:1107.5986 [hep-th].
  \bibitem{Chemissany:2012nb}
  W.~Chemissany, P.~Giaccone, D.~Ruggeri and M.~Trigiante,
 ``Black hole solutions to the $F_4$-model and their orbits (I),''
  Nucl.\ Phys.\ B {\bf 863} (2012) 260
  [arXiv:1203.6338 [hep-th]].

\bibitem{collingwood}The main reference on this subject is D. H. Collingwood and W.M. McGovern,  ``Nilpotent
Orbits in Semisimple Lie Algebras'', Van Nostrand Reinhold 1993.

   \bibitem{brown} R. B. Brown, ``Groups of Type $E_{7}$'', J. Reine
Angew. Math. \textbf{236}, 79 (1969);
 S. Ferrara, R. Kallosh, A. Marrani, ``
Degeneration of Groups of Type $E_{7}$ and Minimal Coupling in
Supergravity'', JHEP \textbf{1206}, 074 (2012).

\bibitem{Levay:2010ua}
  P.~Levay,
``STU black holes as Four Qubit Systems,''
  Phys.\ Rev.\ D {\bf 82} (2010) 026003
  [arXiv:1004.3639 [hep-th]].
\bibitem{Borsten:2012fx}
  L.~Borsten, M.~J.~Duff and P.~Levay,
``The black-hole/qubit correspondence: an up-to-date review,''
  Class.\ Quant.\ Grav.\  {\bf 29} (2012) 224008
  [arXiv:1206.3166 [hep-th]].
\bibitem{Borsten:2010db}
  L.~Borsten, D.~Dahanayake, M.~J.~Duff, A.~Marrani and W.~Rubens,
``Four-qubit entanglement from string theory,''
  Phys.\ Rev.\ Lett.\  {\bf 105} (2010) 100507
  [arXiv:1005.4915 [hep-th]].
\bibitem{Borsten:2011ai}
  L.~Borsten, M.~J.~Duff, S.~Ferrara, A.~Marrani and W.~Rubens,
  ``Small Orbits,''
  Phys.\ Rev.\ D {\bf 85} (2012) 086002.
\bibitem{parity}\bibitem{Ferrara:2013zga}
  S.~Ferrara, A.~Marrani, E.~Orazi and M.~Trigiante,
``Dualities Near the Horizon,''
  JHEP {\bf 1311} (2013) 056
  [arXiv:1305.2057 [hep-th]];
P. Aschieri and M. Trigiante, work in progress.
\bibitem{Trigiante:2016mnt}
  M.~Trigiante,
 ``Gauged Supergravities,''
  arXiv:1609.09745 [hep-th].
\bibitem{Goldstein:2008fq}
  K.~Goldstein and S.~Katmadas,
``Almost BPS black holes,''
  JHEP {\bf 0905} (2009) 058.
\bibitem{Bena:2009ev}
  I.~Bena, G.~Dall'Agata, S.~Giusto, C.~Ruef and N.~P.~Warner,
``Non-BPS Black Rings and black holes in Taub-NUT,''
  JHEP {\bf 0906} (2009) 015
  [arXiv:0902.4526 [hep-th]].
\bibitem{Dall'Agata:2010dy}
  G.~Dall'Agata, S.~Giusto and C.~Ruef,
 ``U-duality and non-BPS solutions,''
  JHEP {\bf 1102} (2011) 074
  [arXiv:1012.4803 [hep-th]].
\bibitem{Galli:2010mg}
  P.~Galli, K.~Goldstein, S.~Katmadas and J.~Perz,
``First-order flows and stabilisation equations for non-BPS extremal black holes,''
  JHEP {\bf 1106} (2011) 070
  [arXiv:1012.4020 [hep-th]].
\bibitem{Clement:2015cxa}
  G.~Clément, D.~Gal'tsov and M.~Guenouche,
``Rehabilitating space-times with NUTs,''
  Phys.\ Lett.\ B {\bf 750} (2015) 591.
\bibitem{Bellucci:2006xz}
  S.~Bellucci, S.~Ferrara, M.~Gunaydin and A.~Marrani,
 ``Charge orbits of symmetric special geometries and attractors,''
  Int.\ J.\ Mod.\ Phys.\ A {\bf 21} (2006) 5043
  doi:10.1142/S0217751X06034355
  [hep-th/0606209].
  \bibitem{Djo}D. Z. Djokovic, N. Lemire, and J. Sekiguchi, Tohoku.
Math J. 53, 395 (2000).
\bibitem{attractor}  S.~Ferrara, R.~Kallosh and A.~Strominger,
 ``N=2 extremal black holes,''
  Phys.\ Rev.\ D {\bf 52} (1995) R5412;
    A.~Strominger,
 ``Macroscopic entropy of N=2 extremal black holes,''
  Phys.\ Lett.\ B {\bf 383} (1996) 39;
    S.~Ferrara and R.~Kallosh,
``Supersymmetry and attractors,''
  Phys.\ Rev.\ D {\bf 54} (1996) 1514;
    S.~Ferrara and R.~Kallosh,
``Universality of supersymmetric attractors,''
  Phys.\ Rev.\ D {\bf 54} (1996) 1525;
    S.~Ferrara, G.~W.~Gibbons and R.~Kallosh,
``Black holes and critical points in moduli space,''
  Nucl.\ Phys.\ B {\bf 500} (1997) 75.

\bibitem{dfg2}
Heiko Dietrich, Paolo Faccin, and Willem~A. de~Graaf.
\newblock Regular subalgebras and nilpotent orbits of real graded {L}ie
  algebras.
\newblock {\em J. Algebra}, 423:1044--1079, 2015.
\bibitem{vinberg2}
{\`E}.~B. Vinberg.
\newblock Classification of homogeneous nilpotent elements of a semisimple
  graded {L}ie algebra.
\newblock {\em Trudy Sem. Vektor. Tenzor. Anal.}, (19):155--177, 1979.
\newblock English translation: Selecta Math. Sov. 6, 15-35 (1987).
\bibitem{hvl}
H.\ V{\^a}n L{\^e}.
\newblock Orbits in real {$\Z_m$}-graded semisimple {L}ie algebras.
\newblock {\em J.\ Lie Theory}, 21(2):285--305, 2011.

\bibitem{Crichigno:2016lac}
 P.~M.~Crichigno, F.~Porri and S.~Vandoren,
 ``Bound states of spinning black holes in five dimensions,''
  arXiv:1603.09729 [hep-th].


\end{thebibliography}

\def\cprime{$'$} \def\cprime{$'$} \def\Dbar{\leavevmode\lower.6ex\hbox to
  0pt{\hskip-.23ex \accent"16\hss}D} \def\cprime{$'$} \def\cprime{$'$}
  \def\cprime{$'$} \def\cprime{$'$} \def\cprime{$'$} \def\cprime{$'$}

\end{document}